%% file: article.tex
\RequirePackage{fix-cm}
\documentclass[natbib,smallextended]{svjour3mod}      

\smartqed  

\usepackage{graphicx}

\usepackage{amsmath}
\usepackage{amssymb}
\usepackage{amsfonts}
\usepackage{enumerate}
\usepackage[normalem]{ulem}
\usepackage[citecolor=blue,urlcolor=blue,breaklinks]{hyperref}
\hypersetup{hypertexnames=false}     

\numberwithin{equation}{section}
\numberwithin{figure}{section}

\newcommand{\eps}{\varepsilon}
\newcommand{\ph}{\varphi}
\newcommand{\tht}{\vartheta}
\newcommand{\realn}{{\mathbb{R}}}

\newcommand{\rmlab}[1]{{\scriptscriptstyle\mathrm{#1}}}
\newcommand{\NS}{\rmlab{NS}}
\newcommand{\PT}{\rmlab{PT}}
\newcommand{\KYbr}{\rmlab{KY}}

\newcommand{\defterm}[1]{\emph{#1}}              

\DeclareMathOperator{\sign}{sgn}
\DeclareMathOperator{\Det}{Det}
\DeclareMathOperator{\Tr}{Tr}

\renewcommand{\tens}[1]{{\boldsymbol{#1}}}       
\newcommand{\ts}[1]{{\boldsymbol{#1}}}         
\newcommand{\tp}[1]{{#1}}                  
\newcommand{\tpu}[1]{{#1}}                 
\newcommand{\tpd}[1]{{#1}}                 

\newcommand{\grad}{{\tens{d}}}                 
\newcommand{\coder}{{\tens{\delta}}}           
\newcommand{\twist}{{\mathbf{T}}}              
\newcommand{\twistc}{{\mathrm{T}}}             
\newcommand{\lied}{\pounds}                    
\newcommand{\covd}{{\tens{\nabla}}}            
\newcommand{\cv}[1]{{\tens{\partial}}_{#1}}    
\newcommand{\Ric}{{\mathbf{Ric}}}              
\newcommand{\pa}{\partial}                     

\newcommand{\A}[1]{A^{\!(#1)}}                 
\newcommand{\Aa}[1]{\mathcal{A}^{\!(#1)}}      
\newcommand{\Ja}{\mathcal{J}}                  
\newcommand{\Ua}{\mathcal{U}}                  

\newcommand{\xroot}[2]{{}^{\scriptscriptstyle{#1}\!}x_{#2}}     

\newcommand{\dg}{{{n}}}                           

\newcommand{\env}[1]{{\tens{e}_{#1}}}        
\newcommand{\enf}[1]{{\tens{e}^{#1}}}        
\newcommand{\ehv}[1]{{\hat{\tens{e}}_{#1}}}  
\newcommand{\ehf}[1]{{\hat{\tens{e}}^{#1}}}  
\newcommand{\ezv}{{\hat{\tens{e}}_{0}}}      
\newcommand{\ezf}{{\hat{\tens{e}}^{0}}}      

\newcommand{\px}[1]{p_{x_#1}}
\newcommand{\ppsi}[1]{p_{\psi_#1}}

\newcommand{\CCKY}[1]{\tens{h}^{\!(#1)}}              
\newcommand{\KY}[1]{\tens{f}^{(#1)}}                  
\newcommand{\KT}[1]{\tens{k}_{(#1)}}                  
\newcommand{\KV}[1]{\tens{l}_{(#1)}}                  
\newcommand{\CKT}[1]{\tens{Q}_{(#1)}}                 

\newcommand{\CCKYc}[1]{h^{\!(#1)}}              
\newcommand{\KYc}[1]{f^{(#1)}}                  
\newcommand{\KTc}[1]{k_{(#1)}}                  
\newcommand{\KVc}[1]{l_{(#1)}}                  
\newcommand{\CKTc}[1]{Q_{(#1)}}                 

\newcommand{\Qo}[1]{K_{#1}}                         
\newcommand{\Lo}[1]{L_{#1}}                         
\newcommand{\Qop}[1]{\mathcal{K}_{#1}}              
\newcommand{\Lop}[1]{\mathcal{L}_{#1}}              
\newcommand{\Qtop}[1]{\tilde{\mathcal{K}}_{#1}}     

\newcommand{\Qofc}{{\tilde{K}}}                     
\newcommand{\Lofc}{{\tilde{L}}}                     
\newcommand{\Mofc}{{\tilde{\ts{M}}}}                
\newcommand{\Xfc}{{\mathcal{X}}}                    

\newcommand{\sop}[1]{\mathcal{#1}}                  
\newcommand{\dirconj}{\widetilde}
\newcommand{\chrgconj}{{\scriptscriptstyle{c}}}

\newcommand{\psb}{\bar{\psi}}                                                                  

\newcommand{\Nb}{{\bar{\dg}}}

\newcommand{\kb}{{\bar{k}}}
\newcommand{\lb}{{\bar{l}}}
\newcommand{\mub}{{\bar\mu}}
\newcommand{\nub}{{\bar\nu}}

\newcommand{\Ab}[1]{\bar{A}{}^{\!(#1)}}            
\newcommand{\Aab}[1]{\bar{\mathcal{A}}{}^{\!(#1)}} 

\newcommand{\Ub}{\bar{U}}                              
\newcommand{\Uab}{\bar{\mathcal{U}}}                   

\newcommand{\Jb}{\bar{J}}             
\newcommand{\Jab}{\bar{\mathcal{J}}}  

\newcommand{\hamv}[1]{\ts{X}_{\!#1}}              
\newcommand{\hamvPT}[1]{\ts{\Xi}_{#1}}            
\newcommand{\levset}[1]{\mathcal{L}_{#1}}         

\newcommand{\be}{\begin{equation}}             
\newcommand{\ee}{\end{equation}}               

\newcommand{\details}[1]{
\medskip
\noindent\textsc{Remark:}
\begin{minipage}[t]{0.85\textwidth}\small
#1
\end{minipage}
\medskip
}

\journalname{Living Reviews in Relativity}
\journalref{Living Rev. Relativ. (2017) 20:6}
\arxivnumber{1705.5482 v3}
\begin{document}

\title{Black holes, hidden symmetries, and complete integrability}

\author{Valeri~P.~Frolov \and Pavel~Krtou\v{s} \and David~Kubiz\v{n}\'{a}k\hspace*{-5em}}

\institute{V.~P. Frolov%
\at
Theoretical Physics Institute,
Department of Physics, University of Alberta,\\
Edmonton, Alberta, T6G 2G7, Canada\\
\email{vfrolov@ualberta.ca}
\and
P. Krtou\v{s}%
\at
Institute of Theoretical Physics,
Faculty of Mathematics and Physics, Charles University,\\
V~Hole\v{s}ovi\v{c}k\'ach 2, Prague, Czech Republic\\
\email{Pavel.Krtous@utf.mff.cuni.cz}
\and
D. Kubiz\v{n}\'{a}k%
\at
Perimeter Institute,\\
31 Caroline St. N. Waterloo Ontario, N2L 2Y5, Canada\\
\email{dkubiznak@perimeterinstitute.ca}
}

\date{May 15, 2017 [last updated November 8, 2017]} 

\maketitle

\begin{abstract}
The study of higher-dimensional black holes is a subject which has recently attracted vast interest. Perhaps one of the most surprising discoveries is a realization that the properties of higher-dimensional black holes with the spherical horizon topology and described by the Kerr--NUT--(A)dS metrics are very similar to the properties of the well known four-dimensional Kerr metric. This remarkable result stems from the existence of a single object called the principal tensor. In our review we discuss explicit and hidden symmetries of higher-dimensional Kerr--NUT--(A)dS black hole spacetimes. We start with discussion of the Killing and Killing--Yano objects representing explicit and hidden symmetries. We demonstrate that the principal tensor can be used as a  ``seed object'' which generates all these symmetries.  It determines the form of the geometry, as well as guarantees its remarkable properties, such as special algebraic type of the spacetime, complete integrability of geodesic motion, and separability of the Hamilton--Jacobi, Klein--Gordon, and Dirac equations. The review also contains a discussion of different applications of the developed formalism and its possible generalizations.
\end{abstract}


\setcounter{secnumdepth}{2}
\setcounter{tocdepth}{2}
\tableofcontents


\newpage

\include{ch1-intro}    


\include{ch2-hskill}     


\include{ch3-kerr}     


\include{ch4-hdbh}     


\include{ch5-hshdbh}   


\include{ch6-intsep}   


\include{ch7-aplgen}   


\phantomsection
\begin{acknowledgements}
\label{sc:acknowledgements}
\addcontentsline{toc}{section}{Acknowledgements}
The authors thank Andrei Zelnikov for his help with the preparation of figures.
V.F.\ thanks the Natural Sciences and Engineering Research Council of Canada (NSERC) and the Killam Trust for their financial support, and thanks Charles University for hospitality during a stay and work on the review.
P.K.\ was supported by successive Czech Science Foundation Grants P203-12-0118 and 17-01625S, and thanks University of Alberta and Perimeter Institute for hospitality in various stages of work on the review.
D.K.\ acknowledges the Perimeter Institute for Theoretical Physics and the NSERC for their support. Research at Perimeter Institute is supported by the Government of Canada through the Department of Innovation, Science and Economic Development Canada and by the Province of Ontario through the Ministry of Research, Innovation and Science.
\end{acknowledgements}

\addtocontents{toc}{\protect\newpage}

\clearpage
\appendix\normalsize

\include{apx-notation}  

\include{apx-pscint}    

\include{apx-cky}       

\include{apx-mtrfcs}    


\include{apx-MP}        

\include{apx-spinor}    

\phantomsection
\addcontentsline{toc}{section}{References}


\bibliographystyle{spbasic}      
\bibliography{refs}   

\end{document}

%% file: ch1-intro.tex

\section{Introduction}
\label{sc:intro}

\subsubsection*{Black holes in four and higher dimensions}

The study of four-dimensional black holes has begun long time ago. Their detailed characteristics were obtained in the 1960s and 1970s, also known as the ``golden age'' of the general relativity.  A summary of the obtained results can be found, for example, in the books by \cite{MTW,Wald:book1984,HawkingEllis:book,Chandrasekhar:1983,FrolovNovikov:book,FrolovZelnikov:book}. According to the proven theorems, assuming the weak energy condition for the matter,  the black hole horizon has to have spherical topology. The most general stationary vacuum black hole solution of the Einstein equations is axially symmetric and can be described by the {\em Kerr metric}.

The interest in four-dimensional black holes is connected with the important role these objects play in modern astrophysics. Namely, there exist strong evidences that the stellar mass black holes manifest themselves in several \mbox{$X$-ray} binaries. Supermassive black holes were discovered in the centers of many galaxies, including our own Milky Way. Great discovery made by LIGO on September 14, 2015 gives first direct confirmation that strong gravitational waves have been emitted in the process of the coalescence of two black holes with masses around 30 solar mass \citep{Abbott:2016blz}. Three month later LIGO registered gravitational waves from another merging black hole binary. These events marked the beginning of the gravitational waves astronomy. In all previous observations the information concerning astrophysical black holes was obtained by registering the  electromagnetic waves emitted by the matter in the black hole vicinity. Such matter usually forms an accretion disc whose temperature and size are determined by the mass and angular momentum of the black hole. Before reaching a distant observer, the emitted radiation propagates in a strong gravitational field of the black hole; to extract the information contained in astrophysical observations one needs to solve the equations for particle and wave propagation in the Kerr spacetime. Fortunately, the remarkable properties of this geometry, namely the complete integrability of geodesics and the separability of wave equations, greatly simplify the required calculations. Based on these results there were developed powerful tools for studying physical effects in the black hole vicinity and their observational manifestation. Similar tools were also used for the study of quantum evaporation of mini-black holes.

In this review we mainly concentrate on black holes in dimensions greater than four, with a particular focus on their recently discovered remarkable geometric properties. Black holes in higher dimensions, see e.g., \cite{Emparan:2008eg, Horowitz:2012nnc} for extended reviews, have attracted  much attention for several reasons. A first reason is connected with the development of string theory and the demand for the corresponding black hole solutions. In order to make this theory consistent one needs to assume that besides usual four dimensions there exist (at least six) additional spatial dimensions.

A second reason stems from the (in past 20 years very popular) brane-world models \citep{Maartens:2010ar, Pavsic:book, Sridhar:book}. In these models the usual matter and the non-gravitational fields are confined to a four-dimensional brane, representing our world. This brane is embedded in higher-dimensional bulk spacetime where only gravity can propagate. Higher-dimensional black holes play a very special role in the brane-world models. Being just a clot of gravity, they can `live' both on and outside the brane. Mini-black-holes whose size is smaller than the size of extra dimensions thus play a role of probes of extra dimensions. One of the intriguing features of the brane-world models that is intensively discussed in the literature is a possibility of mini-black-hole formation in the collision of high energy particles in modern TeV colliders (see e.g., \cite{Landsberg:2014bya,Aad:2014gka} and references therein). Numerous discussions of this effect generated a great interest in the study of properties of higher-dimensional black holes.

A third main reason to study higher-dimensional black holes comes from the desire to better understand the nature of gravitational theory and in particular to identify which properties of gravitational fields are specific to four dimensions and which of them are valid more generally irrespective of the spacetime dimension \citep{Emparan:2008eg}.

\subsubsection*{Remarkable properties of the Kerr black hole}

The Kerr metric has the following remarkable properties: the equations of motion for a free particle in this geometry are completely integrable and the physically interesting field equations allow for the separation of variables. What stands behind these properties?

In a simpler case, when a black hole does not rotate, the answer is well known. The corresponding Schwarzschild solution is static and spherically symmetric. As a result of this symmetry, the energy of the particle  and the three components of its angular momentum are conserved. One can thus construct four integrals of geodesic motion that are functionally independent and mutually Poisson commute, choosing, for example, the (trivial) normalization of the four-velocity, the particle's energy, the square of its total angular momentum, and a projection of the angular momentum to an arbitrary `axis. According to the Liouville's theorem, the existence of such quantities makes the geodesic motion in the spherically symmetric black hole case completely integrable.

For rotating black holes the situation is more complicated since the total angular momentum is no longer conserved. Surprisingly, even in this case there exists another integral of motion, nowadays known as the {\em Carter's constant}. Obtained in 1968 by Carter by a method of separation of variables in the Hamilton--Jacobi equation \citep{Carter:1968rr,Carter:1968cmp}, this additional integral of motion is quadratic in momentum, and, as shown later by \cite{Walker:1970un}, it is in one-to-one correspondence with the rank 2 Killing tensor of the Kerr geometry. A rank 2 Killing tensor $k^{ab}$ is a symmetric tensor whose symmetrized covariant derivative vanishes, $\nabla^{(c}k^{ab)}=0$. It was demonstrated by Carter in the same papers \citep{Carter:1968rr,Carter:1968cmp} that not only the Hamilton--Jacobi equation but also the Klein--Gordon equation allows for a complete separation of variables in the Kerr spacetime.

\details{This fact may not be as surprising as it looks at first sight. In fact, the following 3 problems: complete integrability of geodesic equations, separability of the Hamilton--Jacobi equation, and separability of the Klein--Gordon equation are closely related. Namely, looking for a quasi-classical solution $\Phi\sim \exp(i S)$ of the Klein--Gordon equation $(g^{ab}\nabla_a\nabla_b -m^2)\Phi=0$, one obtains the Hamilton--Jacobi equation $g^{ab}\nabla_a S\nabla_b S +m^2=0$. By identifying $\nabla_a S$ with the momentum $p_a$, one reduces the problem of finding the action function $S$ to the problem of integrating the Hamilton equations of motion for a relativistic particle.}\medskip

Following Carter's success a boom of discoveries regarding the remarkable properties of the Kerr geometry has taken place. \cite{Teukolsky:1972,Teukolsky:1973} decoupled the equations for the electromagnetic and gravitational perturbations and separated variables in the obtained master equations. Equations for massless neutrinos were separated \cite{Unruh:1973} and \cite{Teukolsky:1973}, and the equations for the massive Dirac field were separated by \cite{Chandrasekhar:1976} and \cite{Page:1976}.

\cite{Penrose:1973} and \cite{Floyd:1973} demonstrated that in the Kerr geometry there exists a new fundamental object, the so called Killing--Yano tensor $f_{ab}$, which behaves as a `square root' of the Killing tensor. This object is a 2-form that obeys the following equation: $\nabla_{\!(c}f_{a)b}=0$.
If $f_{ab}$ is non-degenerate, the integrability conditions for this equation imply that the spacetime is algebraically special, of Petrov type D \citep{Collinson:1974}. \cite{HughstonSommers:1973} showed that the existence of such Killing--Yano tensor necessarily implies that the corresponding spacetime admits also two commuting Killing vectors, generating time translation and rotation.

It is interesting to note that some of the above described properties extend beyond the case of the vacuum Kerr geometry. Namely, in 1968  Carter obtained a 6-parametric solution of the Einstein--Maxwell equations with a cosmological constant $\Lambda$ that shares with the Kerr geometry  many of the remarkable properties \citep{Carter:1968cmp, Carter:1968pl}. Besides the cosmological constant $\Lambda$, the mass $M$ and the angular momentum $J$, this solution contains also an electric charge $Q$, a magnetic monopole $P$, and the NUT parameter $N$. The whole class of Carter's metrics admits the Killing--Yano tensor \citep{DemianskiFrancaviglia:1980, carter1987separability}.

Carter's solution is now called the charged {\em Kerr--NUT--(A)dS metric}. In the absence of the NUT parameter it is the most general regular solution describing a stationary isolated black hole in the four-dimensional asymptotically flat ($\Lambda=0$) or (anti) de Sitter ($\Lambda\ne 0$) space. The hidden symmetries of the four-dimensional Kerr--NUT--(A)dS metric and its generalization by  \cite{PlebanskiDemianski:1976} will be discussed in detail in chapter~\ref{sc:Kerr}.

\subsubsection*{Higher-dimensional black objects}

With the advent of interest in higher-dimensional black holes at the beginning of this century the following questions arose:
(i) How far can the results on the four-dimensional black holes be generalized to higher dimensions? (ii) What is the most general solution describing a stationary black hole in asymptotically flat and/or asymptotically (anti) de Sitter space? iii) What can one say about particle motion and field propagation in the gravitational field of such black holes? By now partial answers to some of these questions have been obtained.

The `zoo' of higher-dimensional black holes is vast: there exist extended objects such as as black strings and branes, and the topology of the horizon of an isolated stationary higher-dimensional black hole needs not to be spherical, see e.g. \cite{Emparan:2001wn, Elvang:2007rd, Kunduri:2014kja}. In particular, in 2002 Emparan and Reall obtained an exact solution of 5-dimensional vacuum Einstein equation which describes a stationary rotating black hole with toroidal horizon \citep{EmparanReall:2002a}. Later many new exact 5-dimensional vacuum stationary black hole solutions with a more complicated structure of the horizon were found. There are strong arguments that similar solutions do also exist in more than five dimensions, though the stability of all these objects is in question, e.g., \cite{Santos:2015iua}. Many useful references on this subject can be found in the remarkable review by \cite{Emparan:2008eg}, see also \cite{Emparan:2009vd, Kunz:2013jja,Kleihaus:2016kxj}.

The problem of uniqueness and stability of the higher dimensional black holes is far from its solution---see e.g. a review \cite{Hollands:2012xy} and references therein.

\subsubsection*{Higher-dimensional Kerr--NUT--(A)dS black holes}

Within this `zoo' of higher dimensional black objects there exists a large important family of black hole solutions which are natural generalizations of the four-dimensional Kerr--NUT--(A)dS solution. Called {\em higher-dimensional Kerr--NUT--(A)dS metrics}, these solutions will be in the main focus of this review. They have the spherical topology of the horizon, and in the absence of the NUT parameters, describe isolated rotating black holes in either asymptotically flat or asymptotically (A)dS spacetime.

\details{Let us emphasize, that even if the stationary black hole maintains the spherical horizon topology, its horizon may be `distorted'---the sizes of symmetric cycles on the horizon may vary non-monotonically with the polar angle. Such `bumpy' black holes were conjectured to exist in \cite{Emparan:2003sy} and later found numerically in \cite{Emparan:2014pra}. These black holes do not belong to the Kerr--NUT--(A)dS family and are not studied in this review. However, it might be an interesting problem for future studies to see whether some of the integrability results presented here for `smooth' Kerr--NUT--(A)dS black holes could be extended to bumpy black holes or other black holes as well.
}\medskip

Let us briefly recapitulate a history of study of the Kerr--NUT--(A)dS family of black hole solutions. Denote by $D=2n+\eps$ a total number of spacetime dimensions, with $\eps=0$ in even dimensions and $\eps=1$ in odd dimensions. A higher-dimensional generalization of the Schwarzschild black hole solution was readily obtained by \cite{tangherlini1963schwarzschild}. The Tangherlini solution is static and spherically symmetric, it admits $SO(D-1)$ group of rotational symmetries, and contains exactly one arbitrary parameter which can be identified with the gravitational radius and is related to the black hole mass. A task of finding  a higher-dimensional generalization of the Kerr geometry is much harder and was achieved by \cite{MyersPerry:1986}. The general solution contains, besides the mass $M$, up to $(n-1+\eps)$ independent rotation parameters.

In `our three-dimensional world' we are used to think about rotations as operations about a given axis and identify the angular momentum with a 3-vector $J^a$. In a general case, however, the angular momentum is described by a rank 2 antisymmetric tensor $J_{ab}$. In three dimensions one can write $J^a=\epsilon^{abc}J_{bc}$, where $\epsilon^{abc}$ is the totally antisymmetric tensor, and the usual  description is recovered. In higher dimensions such relation no longer exists. Nevertheless, one can always write $J_{ab}$ in a {\em canonical form} by finding a set of mutually orthogonal 2-planes such that the components of $J_{ab}$ vanish unless the two indices `belong' to the same 2-plane. Since the number of spatial dimensions is $D-1$, the largest possible number of mutually orthogonal 2-planes (and hence the number of independent components of the angular momentum tensor) is $(n-1+\eps)$. This is also the number of independent components of the angular momentum of the black hole which enters the general Myers--Perry solution.

It took another 20 years to find a generalization of the Myers--Perry metric which includes the cosmological constant. \cite{HawkingEtal:1999} found singly-spinning Kerr--(A)dS metrics in all dimensions. These metrics were then generalized by \cite{Gibbons:2004uw} and \cite{Gibbons:2004js} to the case of a general multiple spin. After several attempts to include NUT parameters \citep{Chong:2004hw,Chen:2006ea}, \cite{Chen:2006xh} finally found the most general higher-dimensional Kerr--NUT--(A)dS metric, generalizing the higher-dimensional Carter-like ansatz studied previously in \cite{Klemm:1998kd}. It is the purpose of this review to study the most general Kerr--NUT--(A)dS metric \citep{Chen:2006xh} and its remarkable symmetries.

\subsubsection*{Explicit and hidden symmetries}

Despite of being significantly more complicated, the most general Kerr--NUT--(A)dS metrics in all dimensions have very similar properties to their four-dimensional `cousin', the Kerr metric. A discussion of this similarity and its origin is the subject of the present review. Namely, we shall describe a fundamental geometric structure which is responsible for the remarkable properties of the Kerr--NUT--(A)dS metrics. These properties stem from the existence {a complete set (`tower')} of explicit and hidden symmetries that are `miraculously' present in these spacetimes. Moreover, the existence of such a Killing tower of symmetries is also a characteristic property of the Kerr--NUT--(A)dS spacetimes. It is possible that some of the hidden symmetries may also exist in other higher-dimensional black object spacetimes and their study is an open interesting problem. But we concentrate on the case when the metric possesses the complete tower of hidden symmetries.

What do we mean by a hidden symmetry? We say, that a spacetime possesses a symmetry if there exists a transformation which preserves its geometry. This means that the metric, as well as all other quantities constructed from it (for example curvature), remain unchanged by such a transformation. Continuous symmetry transformations are generated by Killing vector fields; we call the corresponding symmetries {\em explicit}. By famous Noether's theorem they generate conserved charges. Let us demonstrate this on an example of particle motion.

The motion of a free particle in a curved spacetime can be described using the Hamiltonian formalism. A state of the  particle is characterized by a point $(x^a,p_a)$ in the phase space. Its motion is defined by the Hamiltonian, which is quadratic in momenta. The explicit symmetry generated by the Killing vector $\xi^a$ implies that the quantity $p_a \xi^a$ remains constant along the particle's worldline, it is an integral of motion. Integrals of motion are phase space observables that Poisson commute with the Hamiltonian.

An important property of integrals of motion generated by spacetime symmetries is that they are linear in momentum. However, this does not exhaust all possibilities. There may exist integrals of motion that are higher-order polynomials in particle momenta. The existence of such integrals implies that the spacetime admits a special geometric structure, known as a Killing tensor. Killing tensors are in one-to-one correspondence with constants of geodesic motion that are homogeneous in particle momenta, namely, a rank $r$ Killing tensor gives rise to a homogeneous constant of motion of degree $r$ in momentum. Inhomogeneous polynomial integrals of geodesic motion can be decomposed into their homogeneous parts and are associated with Killing tensors of various ranks.

Perhaps the best known example of a Killing tensor is the spacetime metric itself. The corresponding conserved quantity is the Hamiltonian for the relativistic particle and its value is proportional to the square of particle's mass. Familiar Killing vectors, associated with the explicit spacetime symmetry, are Killing tensors of rank 1. To distinguish from this case, we call the geometric structure of the spacetime encoded in Killing tensors of rank 2 and higher a {\em hidden symmetry}.

\subsubsection*{Complete integrability of geodesic motion}

The existence of integrals of motion simplifies the study of dynamical systems. There exits a very special case, when the number of independent commuting integrals of motion of a dynamical system with $N$ degrees of freedom, described by a $2N$-dimensional phase space, is equal to $N$. Such a system is called {\em completely integrable} and its solution can be written in terms of integrals, a result known as the Liouville theorem \citep{liouville1855note}. Specifically, the equation of motion for a free relativistic particle in a $D$-dimensional spacetime can be explicitly solved if there exist $D$ independent Killing vectors and Killing tensors, including the metric, which are `in involution'.

\details{
For Killing tensor fields there exists an operation, a generalization of the Lie bracket, which allows one to construct from two Killing tensors a new one. This operation, called the Schouten--Nijenhuis commutator, will be defined in section~\ref{ssc:partcurvspc}. `In involution' then means that the Schouten--Nijenhuis commutator of the corresponding tensor fields mutually vanishes. On the level of the phase-space observables this is equivalent to the statement that the corresponding conserved quantities mutually Poisson-commute.
}\medskip

Consider, for example, a five-dimensional Myers--Perry metric with two independent rotation parameters. This metric has three Killing vectors:  one generates translations in time, and the other two correspond to rotations in the two independent 2-planes of rotation. Together with the normalization of the 5-velocity this gives 4 integrals of geodesic motion. For complete integrability, an additional integral corresponding to a Killing tensor is needed. This tensor can be found by Carter's method, that is by separating the variables in the Hamilton--Jacobi equation written in the standard Boyer--Lindquist coordinates \citep{FrolovStojkovic:2003b,FrolovStojkovic:2003a}, making the geodesic motion completely integrable in this case. Interestingly, in more than five dimensions and for general multiply-spinning black holes the Boyer--Lindquist type coordinates are not `nice' anymore and Carter's method no longer works. This mislead people to believe that the geodesic motion in these spacetimes is no longer integrable and black holes of smaller generality were studied.

\subsubsection*{Principal tensor and its Killing tower}

It turns out that the restriction on rotation parameters is not necessary and even for the most general multiply-spinning Kerr--NUT--(A)dS black holes one can find special coordinates in which the Hamilton--Jacobi equation separates, proving the geodesic motion completely integrable. A breakthrough in solving this problem occurred in 2007, when it was demonstrated that the Myers--Perry metric as well as the most general Kerr--NUT--(A)dS spacetime in any number of dimensions both admit a {\em non-degenerate closed conformal Killing--Yano \mbox{2-form}} \citep{Frolov:2007nt,Kubiznak:2006kt}. The claim is that the very existence of this single object implies complete integrability of geodesic motion in all dimensions.  Let us explain why this is the case.

Starting with the four-dimensional Kerr metric, we already know that the integrability is guaranteed by the existence of a Killing tensor $\tens{k}$, which in its turn is written as a square of the Killing--Yano 2-form $\tens{f}$. Its Hodge dual $\ts{h}=*\ts{f}$ is again a 2-form which obeys the following equation:
\be
\nabla_{\!c}h_{ab}=g_{ca}\xi_b -g_{cb}\xi_a\,,\qquad \xi_a=\frac{1}{D-1}\nabla_{\!b}h^b{}_a\, .
\ee
The object that satisfies such an equation is called a closed conformal Killing--Yano 2-form. Closed conformal Killing--Yano tensors of higher ranks obey a similar type of equation and they are Hodge dual to  Killing--Yano tensors. A remarkable property of closed conformal Killing--Yano tensors is that their wedge product is again a closed conformal Killing--Yano tensor \citep{Krtous:2006qy,Frolov:2007cb, Frolov:2008jr}. In particular, given a single closed conformal Killing--Yano 2-form in $(2n+\eps)$-dimensions, one can construct up to $n$ non-trivial closed conformal Killing--Yano tensors of increasing rank by taking its wedge powers.

In four dimensions this does not help much. Already at the first step of this procedure, one obtains a 4-form that is proportional to the totally antisymmetric tensor. In higher dimensions, however, the story is quite different: there is enough room to `accommodate' non-trivial higher-rank closed conformal Killing--Yano tensors. It is evident, that the smaller is the tensor-rank of the original form $\ts{h}$ the larger number of its non-trivial higher-rank ``successors" one can obtain. This makes the case of a  2-form $\tens{h}$ a special one. One can also assume that the matrix rank of this 2-form is the largest possible, that is, the 2-form is non-degenerate. In $(2n+\eps)$-dimensional spacetime the maximal matrix rank is $2n$. By `squaring' the Killing--Yano tensors obtained as Hodge duals of the so constructed `successors' of $\tens{h}$, one obtains the whole {\em Killing tower} of $n$ independent Killing tensors \citep{Krtous:2006qy}. Supplemented by the $(n+\eps)$ integrals of motion corresponding to explicit symmetries (as we shall see later such symmetries can also be generated from $\tens{h}$), one obtains a set of $D=2n+\eps$ (generically independent) mutually Poisson commuting constants of geodesic motion, making such a motion completely integrable. In the following chapters we discuss these results in very details, and give the corresponding references.

Since the expression ``non-degenerate closed conformal Killing--Yano 2-form'' is quite long, and since this object is a `main hero' of our review we shall simply call it a {\em principal tensor}. It happens that the existence of the principal tensor has consequences extending far beyond the above described property of complete integrability of geodesic motion. Being a maximal rank 2-form, the principal tensor can be written as
\be
\ts{h}=\sum_{\mu=1}^n x_{\mu}\ts{e}^{\mu}\wedge \hat{\ts{e}}^{\mu}\, ,
\ee
where 1-forms $\ts{e}^{\mu}$ and $\hat{\ts{e}}^{\mu}$ form an orthonormal basis. Let us include in the definition of the principal tensor one additional requirement. Namely, that all the eigenvalues $x_{\mu}$ are independent and different, and that this is valid not only at a point, but in some spacetime domain. In other words, $x_{\mu}$ are functionally independent scalar functions in this domain and they can be used as  coordinates. We shall demonstrate that the other $n+\eps$ coordinates $\psi_k$ can be chosen so that the Killing vectors, corresponding to explicit symmetries, take the form $\partial_{\psi_k}$. The coordinates $(x_\mu,\psi_k)$ are called the {\em canonical coordinates}.

Using canonical coordinates, internally connected with and determined by the principal tensor $\ts{h}$, greatly simplifies the study of properties of a spacetime which admits such an object. Namely, we demonstrate that the corresponding spacetime necessarily possesses the following remarkable properties \citep{Houri:2007xz, Krtous:2008tb, Houri:2008th}: {(i)} When the  Einstein equations are imposed one obtains the most general Kerr--NUT--(A)dS metric. {(ii)} The geodesic motion in such a space is completely integrable, and the Hamilton--Jacobi, Klein--Gordon, and Dirac equations allow for complete separation of variables. The separation of variables occurs in the canonical coordinates determined by the principal tensor.

\subsubsection*{`Hitchhikers guide' to the review}

The review is organized as follows. {In chapter~\ref{sc:CKY} we introduce the Killing vectors, Killing tensors, and the family of Killing--Yano objects and discuss their basic properties. In particular, the principal tensor is defined and its most important properties are overviewed. Chapter~\ref{sc:Kerr} contains a summary of the symmetry properties of the four-dimensional Kerr metric and its Kerr--NUT--(A)dS and Pleba\'nski--Demia\'nski generalizations. We demonstrate how the explicit and hidden symmetries of the Kerr spacetime arise from the principal tensor. Chapter~\ref{sc:hdbh} gives a comprehensive description of the higher-dimensional Kerr--NUT--(A)dS metrics. In chapter~\ref{sc:hshdbh}, starting from the principal tensor which exists in a general higher-dimensional Kerr--NUT--(A)dS spacetime, we construct a tower of Killing and Killing--Yano objects, responsible for the explicit and hidden symmetries of this metric. In chapter~\ref{sc:intsep} we discuss a free particle motion in the higher-dimensional Kerr--NUT--(A)dS spacetime and show how the existence of the principal tensor in these metrics leads to a complete integrability of geodesic equations. We also demonstrate the separability of the Hamilton--Jacobi, Klein--Gordon and Dirac equations in these spacetimes. Chapter~\ref{sc:aplgen} contains additional material and discusses further possible generalizations of the theory of hidden symmetries presented in this review.}

To help the reader with various concepts used in the review, we included some complementary material in appendices. Appendix \ref{apx:notation} summarizes our notation and conventions on exterior calculus; appendix \ref{apx:pscint} reviews the symplectic geometry, the concept of complete integrability, and the requirements for separability of the Hamilton--Jacobi equation.
Appendix \ref{apx:spinor} covers basic notions of a theory of spinors in a curved spacetime, discusses symmetry operators of the Dirac operator, as well as introduces Killing spinors and reviews their relationship to special Killing--Yano forms. Integrability conditions for the Killing--Yano objects are summarized in appendix~\ref{apx:cky}. Appendix \ref{apx:MyersPerry} discusses the Myers--Perry solutions in its original and Kerr--Schild forms and supplements thus material in section \ref{ssc:SubcasesLorentzian}. Finally, various identities and quantities related to the Kerr--NUT--(A)dS metric are displayed in appendix~\ref{apx:mtrfcs}.

Before we begin our exploration, let us mention several other review papers devoted to hidden symmetries and black holes that might be of interest to the reader \citep{Frolov:2008jr,Kubiznak:2008qp, Kubiznak:2009sm, Yasui:2011pr, Cariglia:2011uy, Frolov:2012ux, Cariglia:2014ysa, Chervonyi:2015ima}.

%% file: ch2-hskill.tex

\section{Hidden symmetries and Killing objects}
\label{sc:CKY}

In this chapter we discuss Killing vectors and Killig tensors, which are responsible for explicit and hidden symmetries of spacetime. We also introduce the Killing--Yano tensors which are generators  of hidden symmetries and discuss their basic properties. Named after K. Yano \citep{Yano:1952}, these new symmetries are  in some sense `more fundamental' than the  Killing tensors. A special attention is devoted to a subclass of closed conformal Killing--Yano tensors and in particular to the {\em principal tensor} which plays a central role for the theory of higher-dimensional black holes.

\subsection{Particle in a curved spacetime}
\label{ssc:partcurvspc}

Geometrical properties of a curved spacetime and its various symmetries can be studied through investigation of geodesic motion. For this reason we start with a short overview of the description of relativistic particle in a curved spacetime, formulated both from a spacetime perspective and in terms of the phase space language.

\subsubsection{Phase space description}

Let us consider a $D$-dimensional spacetime (\defterm{configuration space}) $M$ and a point-like particle moving in it. In the Hamilton approach the motion is described by a trajectory in the $2D$-dimensional \defterm{phase space}. A point in the phase space represents a position $x$ and a momentum $\ts{p}$ of the system. The momenta $\ts{p}$ are naturally represented as covectors (1-forms) on the configuration space, the phase space ${\Gamma}$ thus corresponds to a cotangent bundle over the configuration space.

The cotangent bundle has a natural symplectic structure ${\ts{\Omega}}$. Namely, let $x^a$ be coordinates on the configuration space $M$, then the components $p_a$ of the momentum $\ts{p}=p_a \grad x^a$ with respect to the co-frame $\grad x^a$ serve as remaining coordinates on the phase space, $(x^a,p_a)$. The natural symplectic structure takes the form:\footnote{{%
The sum over configuration space indices $a=1,\dots,D$ is assumed.}}
\begin{equation}\label{natsymplstr}
    \ts{\Omega} = \grad x^a \wedge \grad p_a\,,
\end{equation}
so $(x^a,p_a)$ are in fact \defterm{canonical coordinates} on the phase space. Although we used a particular choice of the spacetime coordinates, the symplectic structure $\ts{\Omega}$ is independent of such a choice.

Using the symplectic structure we can introduce the standard machinery of the symplectic geometry: we can define symplectic potential $\ts{\theta}$, the Poisson brackets $\{\ ,\ \}$, or the Hamiltonian vector field $\hamv{F}$ associated with an observable $F$. The overview of the symplectic geometry and the convention used in this review can be found in section~\ref{ssc:symplgeom} of the appendix, cf.\ also standard books \cite{Arnold:book,Goldstein:book}.

\subsubsection{Nijenhuis--Schouten bracket}

An observable $A$ is a function on the phase space. In what follows let us concentrate on observables that are {\em monomial} in momenta, also called the {\em tensorial powers} of momenta, that is observables of the form
\begin{equation}\label{monobs}
    A = a^{a_1\dots a_r}(x)\, p_{a_1}\dots p_{a_s}\,,
\end{equation}
where $a^{a_1\dots a_s} = a^{(a_1\dots a_s)}$ are components of a symmetric tensor field of rank $s$ on the configuration space.

It is straightforward to check that given two such observables $A$ and $B$, of orders $r$ and $s$, respectively, their Poisson bracket $C=\{A,B\}$ is again a tensorial power of order $r+s-1$ with the tensorial coefficient $\ts{c}$. The Poisson brackets of monomial observables thus define an operation $\ts{c}=[\ts{a},\ts{b}]_\NS$ on symmetric tensor fields, called the \defterm{Nijenhuis--Schouten bracket},
\begin{equation}\label{NSbrdef}
    C = \{A,B\} \quad\Leftrightarrow\quad \ts{c} = [\ts{a},\ts{b}]_\NS\,.
\end{equation}
It is explicitly given by
\be\label{NSbrcov}
c^{{a}_1\ldots{a}_{r-1}c{b}_1\ldots{b}_{s-1}}=
r\,a^{e({a}_1\ldots{a}_{r-1}}\,
\nabla_{\!e} b^{c{b}_1\ldots{b}_{s-1})}
-s\,b^{e({b}_1\ldots{b}_{s-1}}\,
\nabla_{\!e} a^{c{a}_1\ldots{a}_{r-1})}\;.
\ee

If one of the tensors, say ${\ts{a}}$, is of rank one, i.e., a vector field, the Nijenhuis--Schouten bracket reduces to the Lie derivative along ${\ts{a}}$,
\begin{equation}\label{NS=lied}
    [\ts{a},\ts{k}]_\NS = \lied_{\ts{a}}\ts{k}\,.
\end{equation}
In particular, for two vectors it reduces to the Lie bracket,
\begin{equation}\label{NS=liebr}
    [\ts{a},\ts{b}]_\NS = [\ts{a},\ts{b}]\,.
\end{equation}

\subsubsection{Time evolution and conserved quantities}

The time evolution in the phase space is determined by the Hamiltonian $H$. Namely, the time derivative of an observable $F$ is given by
\begin{equation}\label{obstimeder}
    \dot F = \{F,H\}\,.
\end{equation}
In particular, for canonical coordinates $(x^a,p_a)$ one gets the Hamilton canonical equations
\begin{equation}\label{HEM}
    \dot{x}^a 
      = \frac{\partial H}{\partial p_a} \,,\quad
    \dot{p}_a 
      = - \frac{\partial H}{\partial x^a} \,,
\end{equation}
which fully determine dynamical trajectories in phase space.

An observable $K$, which remains constant along the dynamical trajectories, is called a \defterm{conserved quantity} or an \defterm{integral/constant of motion}. Thanks to \eqref{obstimeder}, it must commute with the Hamiltonian~$H$,
\be
\{K,H\}=0\,.
\ee

\subsubsection{Relativistic particle and propagation of light}

The motion of a free relativistic particle in a curved spacetime is given by the following simple Hamiltonian:
\begin{equation}\label{RPHam}
    H = \frac{1}{2} g^{ab} p_a p_b\,.
\end{equation}
The Hamilton canonical equations read
\begin{equation}\label{RPHEM}
    \dot{x}^a =  g^{ab} p_b \,,\quad
    \dot{p}_a = - \frac12\, g^{bc}{}_{,a}\, p_b p_c \,,
\end{equation}
and lead to the \defterm{geodesic equation}
\begin{equation}\label{RPgeod}
    p^b\nabla_{b}p^a = 0\,,
\end{equation}
with the covariant derivative determined by the metric.

The value of the Hamiltonian \eqref{RPHam} remains constant, $H = - \frac12m^2$, and determines the mass $m$ of the particle. It gives the normalization of the momenta as
\begin{equation}\label{psqenorm}
    g^{ab} p_a p_b = - m^2\,.
\end{equation}
With this normalization, the affine time parameter $\sigma$ entering these equation is related to the proper time $\tau$ of the particle as
\begin{equation}\label{afparprtime}
    \tau = m \sigma\,.
\end{equation}

With minor modifications, the  above formalism can also describe the propagation of light, understood as a motion of massless particles. The only difference is that one has to consider solutions for which the value of the Hamiltonian \eqref{RPHam} vanishes.
Denoting by ${\ts{l}}$ the momentum in the massless case, we thus have
\be\label{masslessconstr}
g^{ab}l_al_b=0\,.
\ee
The corresponding Hamilton equations lead to the null geodesic equation
\be\label{nullGEO}
l^a\nabla_a l^b=0\,.
\ee

\details{
The normalization \eqref{psqenorm} fixes the norm of the momentum. The momentum thus has ${D-1}$ independent components. For a massive particle one can identify these with the spatial velocities, while the energy (the time component of the momentum) is computable from the normalization. In the massless case, one cannot chose an arbitrary magnitude of velocity, only a direction of the ray. At the same time, there exists an ambiguity in the choice of the affine parameter along the ray. Its rescaling results in the transformation $l^a \to \tilde{l}^a=\alpha l^a$, where $\alpha$ is constant. Although two such null particles differ just by a scale of their momenta and they follow geometrically the same path in the spacetime, they correspond to two physically different photons: they differ by their energy, or, intuitively, by their `color'. Instead of a freedom of choosing an arbitrary magnitude of velocity for massive particles, in the case of null particles we have thus a freedom choosing an arbitrary energy, i.e., an arbitrary `color'.
}

In the description of the relativistic particle above the configuration space is the whole ${D}$-dimensional spacetime, suggesting thus $D$ degrees of freedom. However, the correct counting of the physical degrees of freedom is ${D-1}$. The difference is related to the existence of the constraint $H=\text{const}$ and the remaining time-reparametrization freedom $\sigma\to\sigma+\text{const}$. For more details on the time-reparametrization symmetry and related constraints see, e.g., \cite{FrolovZelnikov:book,Sundermeyer:1982book,Thirring:1992book,Rohrlich:book}.

A charged relativistic particle under the influence of electromagnetic force can be described in a similar way, starting from the Hamiltonian
\begin{equation}\label{ChRPHamred}
    H = \frac{1}{2}\, g^{ab}\, (p_a- q A_a) (p_b - q A_b)\,.
\end{equation}
Combining the Hamilton canonical equations yields the equation of motion:
\begin{equation}\label{ChRPgeod}
    \dot{x}^b\nabla_{\!b}\dot{x}^a = q F^{a}{}_b\dot{x}^b\,,
\end{equation}
where
${F_{ab}=A_{b,a}-A_{a,b}}$ is the Maxwell's tensor. 

\subsection{Explicit and hidden symmetries}
\label{ssc:explhidsym}

If the spacetime has some symmetries they can be always `lifted up' to the phase space symmetries. The corresponding integrals of motion are observables in the phase space which are linear in momenta. However, the contrary is not true: not every phase space symmetry can be easily reduced to the configuration space. Symmetries which have the direct counterpart on the configuration space will be called the \defterm{explicit symmetries}, those which cannot be reduced to the configuration space transformation are called the \defterm{hidden symmetries}.

\subsubsection{Killing vectors}

We start with the description of explicit continuous symmetries of the spacetime geometry.  These are described by Killing vectors.
A curved spacetime with metric $\ts{g}$ admits a \defterm{continuous symmetry} (\defterm{isometry}) if there exists its continuous transformation into itself preserving the metric. Simply speaking, any measurement of the local spacetime properties (such as curvature) gives the same result before and after the symmetry transformation. Such a transformation is generated by the corresponding \defterm{Killing vector} $\ts{\xi}$  and the isometry condition can be written in the following form:
\be\label{Kil}
\lied_{\ts{\xi}}\ts{g}=0\,,
\ee
which is equivalent to the so called {\em Killing vector equation}
\be\label{Kill}
\nabla_{\!(a}\,\xi_{b)}=0\,.
\ee

Two isometries can be composed together, giving again an isometry; the symmetries of the metric form a Lie group called the \emph{isometry group}. Generators of the symmetries, the Killing vectors, form the corresponding Lie algebra, i.e., a linear space with antisymmetric operation given by the Lie bracket. Indeed, any linear  (with constant coefficients)  combination of Killing vectors is again a Killing vector, and for two Killing vectors $\ts{\xi}$ and $\ts{\zeta}$ their commutator $[\ts{\xi},\ts{\zeta}]$ is also a Killing vector.

The dynamics of a free relativistic particle is completely determined by the spacetime geometry, cf.\ Hamiltonian \eqref{RPHam}. According to Noether's theorem the continuous symmetry implies the existence of an integral of motion, which can be written in terms of the Killing vector $\ts{\xi}$ as ${I=\ts{\xi}\cdot\ts{p}=\xi^{a}p_a}$.
The corresponding Hamiltonian vector field reads
\be\label{XII}
\hamv{I}=\xi^a\,\cv{x^a}-\xi^b{}_{,a}p_b\,\cv{p_a}\,.
\ee
Upon a canonical projection to the spacetime manifold it reduces back to the Killing vector~${\ts{\xi}}$:
\begin{equation}\label{XIIb}
    \pi^* \hamv{I} = \xi^a\,\cv{x^a} = \ts{\xi}\;.
\end{equation}
When the canonical projection of a phase space symmetry to the spacetime reduces to a well defined spacetime transformation, which is a symmetry of the spacetime geometry, we say that the symmetry is \defterm{explicit}. Killing vectors thus generate explicit symmetries.

The well-definiteness of the projection requires that it is a quantity solely dependent on the spacetime variables, i.e., independent of the momentum. Clearly it means that the ${\cv{x^a}}$-term in the Hamiltonian vector field  ${\hamv{I}}$ must not depend on the momentum, which requires that the observable ${I}$ is linear in momentum. The integrals of particle motion in curved space that correspond to \emph{explicit} symmetries are thus \emph{linear} in particle's momentum.

\details{
The applicability of Killing vectors extends also to the infinite-dimensional dynamical systems, for example, those describing various fields. Namely, given a Killing vector $\ts{\xi}$ and a conserved energy momentum tensor $T^{ab}$, we have the following {\em conserved current}:
\begin{equation}\label{ja}
  J^a=T^{ab}\xi_b\,,
\end{equation}
which in its turn implies the existence of the corresponding conserved charge. Indeed, upon using the Killing equation \eqref{Kill} and the fact that $T^{ab}$ is symmetric, we have
\be\nonumber
\nabla_aJ^a=\nabla_a(T^{ab}\xi_b)=\xi_b\nabla_aT^{ab}+T^{ab}\nabla_a\xi_b=T^{ab}\nabla_{(a}\xi_{b)}=0\,.
\ee
}

\subsubsection{Killing tensors}

Besides the conserved quantities which are linear in momentum, there might also exist more complicated conserved quantities that indicate the existence of deeper and less evident symmetries. For the motion of relativistic particles these hidden symmetries are encoded in Killing tensors. Namely, the Killing tensors are in one-to-one correspondence with the integrals of geodesic motion that are monomial in momenta.

Let us assume that the system has an integral of motion $K$ of the monomial form \eqref{monobs}, $K=k^{{a}_1\ldots{a}_s}(x)\,p_{{a}_1}\ldots p_{{a}_s}$, where the tensor $\ts{k}$ is completely symmetric, ${k^{{a}_1\ldots{a}_s}=k^{({a}_1\ldots{a}_s)}}$.
Calculating the Poisson bracket
\be\label{Kdot}
\left\{K,H\right\}=
\frac{\partial{K}}{\partial x^{{c}}}
\frac{\partial{H}}{\partial p_{{c}}}-
\frac{\partial{H}}{\partial x^{{c}}}
\frac{\partial{K}}{\partial p_{{c}}}\,,
\ee
with the Hamiltonian \eqref{RPHam}, we obtain
\be\label{KHkg}
\left\{K,H\right\}=k^{{a}_1\ldots{a}_s}{}_{,{c}}\,g^{ca}p_{{a}}p_{{a}_1}\ldots p_{{a}_s}
-s\,\frac12 g^{{k}{l}}{}_{,{a_1}}\,
k^{{a}_1\ldots {a}_s}p_{{a}_2}\ldots
p_{{a}_{s}}p_{{k}}p_{{l}}\, .
\ee
Introducing the covariant derivative ${\covd}$ corresponding to the metric $\ts{g}$, the last expression can be rewritten in a covariant form
\be\label{KKKPPP}
\left\{K,H\right\}=(\nabla^{a_0}k^{{a}_1\ldots{a}_s})\,
  p_{{a}_0}p_{{a}_1}\ldots p_{{a}_s}\, .
\ee
Requiring that ${K}$ is the integral of motion, the condition $\left\{K,H\right\}=0$ must hold for an arbitrary choice of $p_{{a}}$, which gives that the tensor $\ts{k}$ has to obey
\be\label{KillTenscond}
\nabla^{(a_0}k^{{a}_1\ldots{a}_s)}=0\, .
\ee
This relation is called the {\em Killing tensor equation} and the symmetric tensor $\ts{k}$ that solves it is a \defterm{Killing tensor} of rank $s$ \citep{Stackel:1895}. A (trivial) example of a Killing tensor, which is present in every spacetime, is the metric itself. The Killing tensor of rank ${s=1}$ reduces to the Killing vector discussed above.

The condition $\left\{K,H\right\}=0$ can be also written in terms of the Nijenhuis--Schouten bracket
\begin{equation}\label{KillEqNS}
  [\ts{k},\ts{g}]_\NS = 0\;,
\end{equation}
which can be regarded as an alternative form of the Killing tensor equation.

The conserved quantity $K$ corresponds to a symmetry of the phase space which is generated by the Hamiltonian vector field:
\be
\hamv{K}=s\, k^{ac_2\dots c_s}p_{c_2}\dots p_{c_s}\,\cv{x^a}-k^{c_1\dots c_s}{}_{,a}\,p_{c_1}\dots p_{c_s}\,\cv{p_a}\,.
\ee
Its point-by-point projection into spacetime gives
\begin{equation}\label{KTproj}
\pi^*\hamv{K}=s\,k^{ac_2\dots c_s}p_{c_2}\dots p_{c_s}\,\cv{x^a}\,,
\end{equation}
which for ${s\ge2}$ explicitly depends on particle's momenta and cannot thus be regarded as a pure spacetime quantity. This means that the phase space symmetry generated by ${K}$ does not have a simple description in the spacetime. We call such symmetries the \defterm{hidden symmetries}.

In other words, Killing tensors of order ${s\ge2}$ represent symmetries that do not generate a spacetime diffeomorphism and in that sense they are not `encoded' in the spacetime manifold. Their presence, however, can be `discovered' by studying the particle dynamics in the spacetime. This is to be compared to the action of Killing vectors, $s=1$, for which the projection defines a spacetime isometry and the symmetry is explicit, cf. \eqref{XIIb}.

Given two constants of geodesic motion ${K_{(1)}}$ and ${K_{(2)}}$ of the type \eqref{monobs}, their Poisson bracket ${\{K_{(1)},K_{(2)}\}}$ is also an integral of motion of the same type. This immediately implies that provided $\ts{k}_{(1)}$ and $\ts{k}_{(2)}$ are two Killing tensors, so is their Nijenhuis--Schouten bracket $[\ts{k}_{(1)},\ts{k}_{(2)}]_\NS$. Slightly more generally, an integral of motion that is polynomial in the momentum corresponds to an inhomogeneous Killing tensor, defined as a formal sum of Killing tensors of different ranks. Such objects together form the Lie algebra under the Nijenhuis--Schouten bracket.

Similarly, given two monomial integrals of geodesic motion ${K_{(1)}}$ and ${K_{(2)}}$  of order $s_1$ and $s_2$, respectively, their product $K=K_{(1)}K_{(2)}$ is also a monomial constant of geodesic motion of order $s=s_1+s_2$. This means that $K$ corresponds to a Killing tensor $\ts{k}$ given by $k^{a_1\dots a_{s}}=k_{(1)}^{(a_1\dots a_{s_1}}\,k_{(2)}^{a_{s_1+1}\dots a_{s})}$. In other words, a symmetrized product of two Killing tensors is again a Killing tensor.

This hints on the following definition. A Killing tensor is called {\em reducible}, if it can be decomposed in terms of the symmetrized products of other Killing tensors and Killing vectors. Otherwise it is {\em irreducible}.

\details{An interesting generalization of Killing tensors has been recently proposed in \cite{Aoki:2016ift}. It follows from considering  `inconstructible rational first integrals' of the type $C=A/B$, where $A$ and $B$ are monomials of arbitrary orders. By requiring that the resultant ratio $C$ is an integral of geodesic motion, the corresponding tensor $\ts{a}$ (and similarly $\ts{b}$) has to obey the following {\em generalized Killing tensor equation}:
\be
\nabla^{\!(a}a^{a_1\dots a_s)}=\alpha^{(a}a^{a_1\dots a_s)}\,,
\ee
for some vector $\ts{\alpha}$. We refer the interested reader to \cite{Aoki:2016ift} for more details on this development.
}

\subsubsection{Conformal Killing vectors and Killing tensors}

So far we have discussed monomial integrals of relativistic particle motion (of order ${s}$ in momentum) and have shown that they correspond to Killing vectors ($s=1)$ and Killing tensors $(s\geq 2)$. Let us now briefly mention conformal generalizations of these objects that provide integrals for \emph{propagation of light}. These quantities are conserved only along {\em null} geodesics.
A \defterm{conformal Killing vector} $\ts{\xi}$ is a vector obeying the \emph{conformal Killing vector equation}
\be
\nabla^{(a}\xi^{b)}=\alpha g^{ab}\,,
\ee
for some function $\alpha$. Obviously, for $\alpha=0$ we recover a Killing vector. Given a conformal Killing vector $\ts{\xi}$, we can construct the observable ${I}$ conserved along null geodesics:
$I=\ts{\xi}\cdot \ts{l}=\xi^a l_a$.
Indeed, using the null geodesic equation \eqref{nullGEO} and the constraint \eqref{masslessconstr}, we have
\be
\dot{I}=l^a\nabla_aI=l^a\nabla_a(\xi^b l_b)=l_al_b\nabla^a\xi^b
  =l_al_b\nabla^{(a}\xi^{b)}=\alpha \,l_al_b g^{ab}=0\,.
\ee

\details{
Similarly to Killing vectors, conformal Killing vectors provide conserved quantities for any matter and fields whose energy momentum tensors $T^{ab}$ is (i) conserved: $\nabla_aT^{ab}=0$ and (ii) traceless: $T^{ab}g_{ab}=0$. Namely, the current
$J^a=T^{ab}\xi_b$
obeys the conservation law, $\nabla_a J^a=0$, as a result of the conformal Killing equation. Indeed, we have
\be\nonumber
\nabla_aJ^a=\nabla_a(T^{ab}\xi_b)=\xi_b\nabla_aT^{ab}+T^{ab}\nabla_a\xi_b=T^{ab}\nabla_{(a}\xi_{b)}=\alpha T^{ab}g_{ab}=0\,.
\ee
}

Considering next a monomial observable $K$ of rank $s$, \eqref{monobs}, we find that it is an integral of null geodesic motion if the symmetric tensor $\ts{k}$ satisfies the \emph{conformal Killing tensor equation} \citep{Walker:1970un, hughston1972quadratic}:
\be\label{KillTenscond2}
\nabla^{(a_0}k^{{a}_1\dots{a}_s)}=g^{(a_0a_1}\alpha^{a_2\dots a_s)}\,,
\ee
with ${\ts{\alpha}}$ being some (symmetric) tensor of rank $s-1$. For $\ts{\alpha}=0$ we recover Killing tensors.

It is obvious that symmetries generated by conformal Killing tensors (for $s\geq 2$) are again {\em hidden}. Moreover, by the same arguments as in the case of Killing tensors, it can be shown that the Nijenhuis--Schouten bracket of two conformal Killing tensors is again a conformal Killing tensor. Similarly, a symmetrized product of two conformal Killing tensors is again a conformal Killing tensor.

\subsection{Separability structures}
\label{ssc:separabilitystructures}

The geodesic motion in any number of spacetime dimensions can be also studied using the Hamilton--Jacobi equation. This approach is reviewed in section~\ref{ssc:HamJacEq} of the appendix. In this approach, the family of trajectories can be integrated as orbits of the momentum field, which is determined as the gradient of the \defterm{Hamilton's characteristic function} $S$. This function is the solution of the (time-independent) \defterm{Hamilton--Jacobi equation}
\begin{equation}
   H\Bigl(\tpu{q},\frac{\pa{S}}{\pa\tpu{q}}(\tpu{q})\Bigr) = E\,.
\end{equation}

The important case of interest is when this equation can be solved by a separation of variables. As explained in appendix \ref{apx:pscint}, this is closely related to the integrability of the given dynamical system. For a motion of free relativistic particle there exists a beautiful {\em intrinsic geometric characterization} for separability of the corresponding Hamilton--Jacobi equation. It is described by the theory of
\defterm{separability structures} \citep{benenti1979remarks, BenentiFrancaviglia:1980, DemianskiFrancaviglia:1980, KalninsMiller:1981}.

Separability structures are classes of separable charts for which the Hamil\-ton--Jacobi equation allows an additive separation of variables.
For each separability structure there exists such a family of separable coordinates
which admits a maximal number of, let us say $r$, ignorable coordinates.
Each system in this family is called a {\em normal separable system} of coordinates.
We call the corresponding structure the ${r}$-separability structure.
Its existence is governed by the following theorem:\\[0.5ex]
\noindent{\bf Theorem}:
{\it A manifold $M$ with a metric ${\ts{g}}$ admits an $r$-separability structure if and only if it admits $r$ Killing vectors $\tens{l}_{(i)}$ $(i=0,\dots,r-1)$ and $D-r$ rank 2 Killing tensors $\tens{k}_{(\alpha)}$ $(\alpha=1,\dots,D-r)$, all of them independent, which:\\
(i) all mutually (Nijenhuis--Schouten) commute:
\begin{equation}\label{PoissonK}
\bigl[\ts{k}_{(\alpha)},\ts{k}_{(\beta)}\bigr]_\NS=0\,,\quad
\bigl[\ts{l}_{(i)},\ts{k}_{(\beta)}\bigr]_\NS=0\,, \quad
\bigl[\ts{l}_{(i)}, \ts{l}_{(j)}\bigr]_\NS=0\,,
\end{equation}
(ii)\ Killing tensors $\tens{k}_{(\alpha)}$ have in common $D-r$ eigenvectors $\tens{m}_{(\alpha)}$,
such that
\begin{equation}\label{sepdeltacond}
[\tens{m}_{(\alpha)},\tens{m}_{(\beta)}]=0\,,\quad
[\tens{m}_{(\alpha)},\tens{l}_{(i)}]=0\,,\quad
\tens{g}(\tens{m}_{(\alpha)},\tens{l}_{(i)})=0\,.
\end{equation}
}

It is evident, that  the existence of a separability structure implies the complete integrability of geodesic motion. Indeed, the requirement of independence means that $r$ linear in momenta constants of motion $L_{(i)}$ associated with Killing vectors $\tens{l}_{(i)}$ and $(D-r)$ quadratic in momenta constants of motion $K_{(\alpha)}$ corresponding to Killing tensors $\tens{k}_{(\alpha)}$ are functionally independent. Moreover, equations \eqref{PoissonK} and the discussion in the appendix~\ref{apx:pscint} imply that all such constants are in involution, that is obey conditions~\eqref{ComplInt}. Hence the geodesic motion is completely integrable.

Let us mention yet another theorem which relates the (additive) separability of the
Hamilton--Jacobi equation with the (multiplicative) separability of the Klein--Gordon equation
\be
\Box\phi=m^2\phi\,,
\ee
with the wave operator $\Box = g^{ab}\nabla_{\!a}\nabla_{\!b}$. Following \cite{benenti1979remarks} we have the following:\\[0.5ex]
\noindent{\bf Theorem}:
{\it The Klein--Gordon equation allows a multiplicative separation of variables if and only if the manifold
possesses a separability structure in which the vectors $\tens{m}_{(\alpha)}$ are eigenvectors of the Ricci tensor.
In particular, if the manifold is an Einstein space, the Hamilton--Jacobi equation is separable if and only if the same holds for the wave equation.}\\[0.5ex]

The existence of a separable structure has strong consequences for the geometry: it restricts significantly a form of the metric in the normal separable coordinates. Namely, let $y^a=(\psi_i, x_\alpha)$ be separable coordinates, where we denoted by 
$\psi_i$ the ignorable coordinates associated with the Killing vectors $\tens{l}_{(i)}=\ts{\pa}_{\psi_i}$. The inverse metric $(\beta=1)$ and other Killing tensors $(\beta=2,\dots, D-r)$ then read
\be\label{KTsepStruc}
\tens{k}_{(\beta)}=\sum_{\alpha=1}^{D-r}\Bigl[(M^{-1})^\alpha{}_\beta(\pa_{x_\alpha})^2+ \sum_{i,j}N^{ij}_\alpha(x_\alpha) (M^{-1})^\alpha{}_\beta\pa_{\psi_i}\pa_{\psi_j}\Bigr]\,.
\ee
Here, $\mathrm{M}$ is a $(D-r)\times (D-r)$ {\em St\"ackel matrix}, that is a non-degenerate matrix whose each $\beta$-th column depends on a variable $x_\beta$ only, $M^\alpha{}_\beta=M^\alpha{}_\beta(x_\beta)$, and $\mathrm{N}_\alpha=\mathrm{N}_\alpha(x_\alpha)$ are $(D-r)$ of $r\times r$ matrices of one variable.

We will see a particular realization of this structure in chapter~\ref{sc:hshdbh} where we write down the metric consistent with a special example of separable structure, namely the off-shell Kerr--NUT--(A)dS metric. 
The separable structure of the Kerr--NUT--(A)dS spacetimes justifies the complete integrability of geodesic motion, as well as the fact that the Hamilton--Jacobi equation and the wave equations allow for a separation of variables, see chapter~\ref{sc:intsep}.

\subsection{Defining the Killing--Yano family}
\label{ssc:KYfamily}

\subsubsection{Motivation: parallel transport}

In the previous section we have discussed observables ${F(x,\ts{p})}$, depending on a position ${x}$ and momenta ${\ts{p}}$, which are conserved along geodesics. Namely, we have seen that the monomials in momenta, \eqref{monobs}, are in one-to-one correspondence with Killing tensors \eqref{KillTenscond}. Interestingly, this construction can be generalized to tensorial quantities. Let us consider a rank-${s}$ tensorial quantity
\begin{equation}\label{Bpppp}
    w_{a_1\dots a_s} = B_{c_1\ldots c_r a_1\ldots a_s} p^{c_1}\dots p^{c_r}\,,
\end{equation}
depending on the particle momenta ${\ts{p}}$ and the position ${x}$ through the tensor ${\ts{B}}$. Using the particle's equations of motion \eqref{RPgeod}, we can show that the quantity \eqref{Bpppp} is parallel-transported along geodesics if and only if the tensor ${\ts{B}}$ satisfies the {\em generalized Killing tensor equation}
\begin{equation}\label{GKTeq}
    \nabla_{\!(c_0} B_{c_1\ldots c_r)a_1\ldots a_s}=0\;,
\end{equation}
as discussed by \cite{collinson2000generalized}.

A special case occurs when ${r=1}$ and the tensor ${\ts{B}}$ is completely antisymmetric.
In such a case it is called a \defterm{Killing--Yano form} \citep{Yano:1952} and we denote it by  $\ts{f}$:
\be\label{KY1}
f_{{a}_0a_1\ldots {a}_s}=f_{[a_0{a}_1\ldots {a}_s]} \,,\qquad
\nabla_{\!(b}f_{c)a_1\ldots a_s}=0\, .
\ee
The corresponding conserved tensorial quantity ${\ts{w}}$
\begin{equation}\label{wfdef}
    {w_{a_1\ldots a_s}} = f_{ca_1\ldots a_s} p^c
\end{equation}
has now the special property that, apart from being parallel-transported, it is also `perpendicular' to particle's momentum at every index,
\be
w_{{a}_1\ldots{a}_j\ldots{a}_{s}}p^{{a}_j}=0\,.
\ee
This property has been used for an explicit construction of the parallel-trans\-ported frame in the Kerr geometry \citep{marck1983solution, marck1983parallel} and its higher-dimensional generalizations \citep{Connell:2008vn, Kubiznak:2008zs}, see chapter~\ref{sc:aplgen} for more details.
Conversely, any skew-symmetric quantity ${\ts{w}}$ that is liner in momenta and parallel-transported along and orthogonal to any geodesic, defines the Killing--Yano tensor $\tens{f}$.

\subsubsection{Decomposition of the covariant derivative}

One can arrive at the  definition of the Killing--Yano tensor, \eqref{KY1}, also by studying a general decomposition of the covariant derivative of an antisymmetric form into its irreducible parts, e.g., \cite{semmelmann2003conformal}. Such a covariant derivative belongs to the space ${\mathbf{T}^*\!\otimes\mathbf{\Lambda}\,M}$ of tensors with all but the first indices antisymmetric. This space naturally splits into three subspaces  given by the projectors ${\mathcal{A}}$, ${\mathcal{C}}$, and ${\mathcal{T}}$, defined as
\begin{align}\label{DECOMP}
    (\mathcal{A}\sigma)_{a{a}_1\ldots {a}_p} &=\sigma_{[a{a}_1\ldots {a}_p]}\,, \\
    (\mathcal{C}\sigma)_{a{a}_1\ldots {a}_p} &=\frac{p}{D{-}p{+}1}\, g_{a[a_1} \sigma^b{}_{|b|{a}_2\ldots {a}_p]}\, ,\\
    (\mathcal{T}\sigma)_{a{a}_1\ldots {a}_p} &= \sigma_{aa_1\ldots a_p}
       -\sigma_{[a{a}_1\ldots {a}_p]}
       -\frac{p}{D{-}p{+}1}\, g_{a[a_1} \sigma^b{}_{|b|{a}_2\ldots {a}_p]}\, ,
\end{align}
with ${\ts{\sigma}\in\mathbf{T}^*\!\otimes\mathbf{\Lambda}\,M}$, i.e., with ${\ts{\sigma}}$ satisfying ${\sigma_{aa_1,\ldots a_p} = \sigma_{a[a_1\ldots a_p]}}$. These projectors are orthogonal with respect to the natural scalar product given by the metric and close to the identity ${\mathrm{Id}=\mathcal{A}+\mathcal{C}+\mathcal{T}}$.

Using these projectors, the covariant derivative of an antisymmetric form $\ts{\omega}$ decomposes as
\be
\covd \ts{\omega}=
   \mathcal{A}\covd\ts{\omega}
  +\mathcal{C}\covd\ts{\omega}
  +\mathcal{T}\covd\ts{\omega}\, .
\ee
The first term is called an {\em antisymmetric part} and depends only on the exterior derivative ${\grad\ts{\omega}}$, the second term is called a {\em divergence} part and depends only on the divergence (co-derivative) ${\covd\cdot\ts{\omega}\equiv-\coder\ts{\omega}}$. The third term is given by the action of the so called \defterm{twistor operator}
\citep{semmelmann2003conformal, moroianu2003twistor, leitner2004normal}:
\begin{equation}\label{twistop}
\begin{split}
    \twistc_{\!a} \omega_{a_1\ldots a_p}
    &= (\mathcal{T}\nabla\omega)_{aa_1\ldots a_p}\\
    &= \nabla_{\!a}\omega_{a_1\ldots a_p}
      - \nabla_{\![a}\omega_{a_1\ldots a_p]}
      - \frac{p}{D{-}p{+}1}\, g_{a[a_1} \nabla^{b}\omega_{|b|{a}_2\ldots {a}_p]}\;.
\end{split}
\end{equation}

\details{
Note that we have defined here the twistor operator as an operator acting on the space of antisymmetric forms. Perhaps better known is the twistor operator defined on Dirac spinors which naturally complements the Dirac operator. Both twistor operators are closely related, but not identical. In particular, any $p$-form constructed from a twistor spinor (by sandwiching gamma matrices) belongs to the kernel of the above $p$-form twistor operator, see appendix~\ref{apx:spinor}.}

{Differential forms} with vanishing exterior derivative are called \defterm{closed forms}, forms with vanishing divergence are called \defterm{divergence-free} or \defterm{co-closed}. Such  forms play important role for example in the Hodge decomposition or in the de~Rham cohomology. Here we are mostly interested in forms for which the twistor operator vanishes. These forms are called \defterm{conformal Killing--Yano forms} \citep{kashiwada1968conformal, tachibana1969conformal}, see also \cite{Benn:1996su, Benn:1996ia, kress1997generalised, jezierski1997conformal, Cariglia:2003kf}, or \defterm{twistor forms} e.g. \cite{semmelmann2003conformal, moroianu2003twistor, leitner2004normal}. They satisfy the condition
\begin{equation}\label{CKYeqidx}
    \nabla_{\!a}\omega_{a_1\ldots a_p} =
      \nabla_{\![a}\omega_{a_1\ldots a_p]}
      + \frac{p}{D{-}p{+}1}\, g_{a[a_1} \nabla^{b}\omega_{|b|{a}_2\ldots {a}_p]}\;.
\end{equation}
The space of conformal Killing--Yano forms has two important subspaces: Killing--Yano and closed conformal Killing--Yano forms.

The form ${\ts{f}}$ is called a \defterm{Killing--Yano} form \citep{Yano:1952,yano1953curvature} if its covariant derivative is just given by the antisymmetric part. It obeys the condition:
\begin{equation}\label{KYeqidx}
    \nabla_{\!a} f_{a_1\ldots a_p} = \nabla_{\![a} f_{a_1\ldots a_p]}\;,
\end{equation}
and is clearly divergence-free.

The form ${\ts{h}}$ is called a \defterm{closed conformal Killing--Yano} form \citep{Krtous:2006qy,carter1987separability,semmelmann2003conformal,moroianu2003twistor,leitner2004normal} if its covariant derivative is given just by the divergence part. It obeys the following equation:
\begin{equation}\label{CCKYeqidx}
    \nabla_{\!a} h_{a_1\ldots a_p} =
    \frac{p}{D{-}p{+}1}\, g_{a[a_1} \nabla^{b}h_{|b|{a}_2\ldots {a}_p]}\;,
\end{equation}
and is obviously closed.

Finally, we could identify forms for which the covariant derivative is given by the twistor operator. Since for such objects both the exterior derivative and coderivative vanish, they are called \defterm{harmonic} forms. {A special subcase of all types of forms introduced above are \defterm{covariantly constant} forms.} All these definitions are summarized in the following table:

{\small%
\renewcommand{\arraystretch}{1.44}
\begin{equation}\label{A1}\nonumber
\begin{array}{|l|l|l|}
\hline
\multicolumn{3}{|c|}{\mbox{Decomposition of the covariant derivative of a form}\ \ts{\omega}} \\
\hline
\;\mbox{General form} & \multicolumn{2}{|l|}{
  \;\covd \ts{\omega}=\mathcal{A}\covd\ts{\omega}+\mathcal{C}\covd\ts{\omega}+\mathcal{T}\covd\ts{\omega}
  }\\
\hline
\;\mbox{Closed form} &
  \;\covd \ts{\omega}=\mathcal{C}\covd\ts{\omega}+\mathcal{T}\covd\ts{\omega} \;&
  \;\grad\ts{\omega} = 0
  \\
\hline
\;\mbox{Divergence-free co-closed form} &
  \;\covd \ts{\omega}=\mathcal{A}\covd\ts{\omega}+\mathcal{T}\covd\ts{\omega} \;&
  \;\coder\ts{\omega} = 0
  \\
\hline
\;\mbox{Conformal Killing--Yano form} &
  \;\covd \ts{\omega}=\mathcal{A}\covd \ts{\omega}+\mathcal{C}\covd \ts{\omega} \;&
  \;\twist\ts{\omega}=0
  \\
\hline
\;\mbox{Killing--Yano form} &
  \;\covd \ts{\omega}=\mathcal{A}\covd \ts{\omega} &
  \;\coder\ts{\omega} = 0\,,\;
  \twist\ts{\omega}=0
  \\
\hline
\;\mbox{Closed conformal Killing--Yano form} \; &
  \;\covd \ts{\omega}=\mathcal{C}\covd \ts{\omega} &
  \;\grad\ts{\omega} = 0\,,\;
  \twist\ts{\omega}=0
  \\
\hline
\;\mbox{Harmonic form}  &
  \;\covd \ts{\omega}=\mathcal{T}\covd \ts{\omega} &
  \;\grad\ts{\omega} = 0\,,\;
  \coder\ts{\omega} = 0
  \\
\hline
\;\mbox{Covariantly constant form} &
  \;\covd \ts{\omega}=0 &
  \;\grad\ts{\omega} = 0\,,\;
  \coder\ts{\omega} = 0\,,\;
  \twist\ts{\omega}=0\;
  \\
\hline
\end{array}
\end{equation}
\renewcommand{\arraystretch}{1}}

\subsubsection{Alternative definitions}

The definition of conformal Killing--Yano forms \eqref{CKYeqidx} can be reformulated in a slightly modified form:\\[0.5ex]
{\bf Theorem}:
{\it The antisymmetric form ${\ts{\omega}}$ is a conformal Killing--Yano form if and only if there exists antisymmetric forms ${\ts{\kappa}}$ and ${\ts{\xi}}$ such that the covariant derivative ${\covd\ts{\omega}}$ can be written as
\begin{equation}\label{CKYeqkapxiidx}
  \nabla_{\!a}\omega_{a_1\ldots a_p} =
    \kappa_{aa_1\ldots a_p} + p\, g_{a[a_1}\xi_{a_2\ldots a_p]}\;.
\end{equation}
The forms ${\ts{\kappa}}$ and ${\ts{\xi}}$ are then uniquely given by the following expressions:
\begin{equation}\label{kappaxidefidx}
    \kappa_{a_0a_1\ldots a_p} = \nabla_{\![a_0}\omega_{a_1\ldots a_p]}\,,\qquad
    \xi_{a_2\ldots a_p} = \frac{1}{D-p+1}\,\nabla^a\omega_{aa_2\ldots a_p}\;.
\end{equation}%
}\\[0.5ex]
Indeed, by antisymmetrizing \eqref{CKYeqkapxiidx} one obtains the first relation \eqref{kappaxidefidx}. Similarly, contracting the first two indices in \eqref{CKYeqkapxiidx} leads to the second relation \eqref{kappaxidefidx}. Substituting these two relations back to \eqref{CKYeqkapxiidx} one recovers the definition \eqref{CKYeqidx}.

Similarly, ${\ts{f}}$ is a Killing--Yano form if there exist a form ${\ts{\kappa}}$ such that
\begin{equation}\label{KYeqkapidx}
  \nabla_{\!a}f_{a_1\ldots a_p} = \kappa_{aa_1\ldots a_p}\;.
\end{equation}
A $p$-form ${\ts{h}}$ is a closed conformal Killing--Yano form if there exist a form ${\ts{\xi}}$ such that
\begin{equation}\label{CCKYeqxiidx}
  \nabla_{\!a}h_{a_1\ldots a_p} = p\, g_{a[a_1}\xi_{a_2\ldots a_p]}\;,
\end{equation}
with ${\ts{\kappa}}$ and ${\ts{\xi}}$ given by expressions analogous to \eqref{kappaxidefidx}.

Alternatively, the symmetrization of \eqref{CKYeqkapxiidx} in first two indices leads to
\begin{equation}\label{CKYeqidx2}
  \nabla_{\!(a_0}\omega_{a_1)a_2\ldots a_{p}} =
    g_{a_0a_1} \xi_{a_2\ldots a_{p}}
    -(p-1)\, g_{[a_2|(a_0} \xi_{a_1)|a_3\ldots a_{p}]}\;,
\end{equation}
which was originally postulated as a definition of conformal Killing--Yano forms \citep{kashiwada1968conformal, tachibana1969conformal}.
The equivalence of \eqref{CKYeqkapxiidx} and \eqref{CKYeqidx2} follows from the fact that one can reconstruct ${\nabla_{\!a_0}\omega_{a_1\ldots a_p}}$ from $\nabla_{\![a_0}\omega_{a_1\ldots a_p]}$ and $\nabla_{\!(a_0}\omega_{a_1)a_2\ldots a_p}$.
Killing--Yano forms are those for which $\ts{\xi}=0$, which gives
\be\label{KYeqidx2}
\nabla_{(a_0} f_{a_1)a_2\dots a_p}=0\,,
\ee
recovering the definition~\eqref{KY1}.

\subsubsection{Killing--Yano objects in a differential form notation}

Contracting \eqref{CKYeqkapxiidx} with a vector ${\ts{X}}$ we see that the ${p}$-form ${\ts{\omega}}$ is a conformal Killing--Yano form if and only if its covariant derivative can be written as (see appendix \ref{apx:notation} for notations on differential forms)
\be\label{CKYeq2}
  \covd_{\!\ts{X}} \ts{\omega} = \ts{X}\cdot \ts{\kappa} + \ts{X}\wedge \ts{\xi}\,,
\ee
for `some' $(p+1)$-form $\ts{\kappa}$ and `some' $(p-1)$-form $\ts{\xi}$. These forms are then given by
\be\label{kappaxidef}
  \ts{\kappa}=\frac{1}{p+1}\,\covd\wedge\ts{\omega}\,,\qquad
  \ts{\xi}=\frac{1}{D-p+1}\,\covd\cdot\ts{\omega}\,,
\ee
cf.\ \eqref{kappaxidefidx}, giving the following explicit definition, see \eqref{CKYeqidx}:
\be\label{CKYeq}
  \nabla_{\!\ts{X}} \ts{\omega}=
    \frac{1}{p{+}1}\,\ts{X} \cdot( \covd\wedge \ts{\omega})
    +\frac{1}{D{-}p{+}1}\,\ts{X}\wedge (\covd\cdot\ts{\omega})\,.
\ee
The Killing--Yano forms are then defined as objects obeying
\be\label{bezKY}
\covd_{\!\ts{X}} \ts{f} = \ts{X}\cdot \ts{\kappa} \,,
\ee
whereas closed conformal Killing--Yano tensors are those satisfying
\be\label{bezCCKY}
  \covd_{\!\ts{X}} \ts{h} = \ts{X}\wedge \ts{\xi}\,.
\ee

Definitions \eqref{CKYeq2}, \eqref{bezKY}, and \eqref{bezCCKY} remain equally valid for inhomogeneous (closed conformal) Killing--Yano forms, provided ${\ts{\kappa}}$ and ${\ts{\xi}}$ satisfy
\be\label{kappaxidefnonhom}
  \pi\,\ts{\kappa}=\covd\wedge\ts{f}\,,\qquad
  (D-\pi)\,\ts{\xi}=\covd\cdot\ts{h}\,,
\ee
using the rank operator $\pi$ introduced in \eqref{pieta}.

\subsection{Basic properties of conformal Killing--Yano forms}
\label{ssc:CKYproperties}

\subsubsection{Conformal Killing--Yano forms}

The conformal Killing--Yano tensors have a nice behavior under the Hodge duality. Namely, using the relations \eqref{Xhook}, \eqref{ddd}, it is easy to show that equation~\eqref{CKYeq} implies
\be\label{CKYHodge}
  \covd_{\!\ts{X}}(* \ts{\omega})=
    \frac{1}{p_*+1} \ts{X}\cdot (\covd\wedge*\ts{\omega})
    +\frac{1}{D-p_*+1}\ts{X}\wedge (\covd\cdot*\ts{\omega})\,,\quad\text{with ${p_*=D-p}$}\, .
\ee
This relation means that
\begin{itemize}
\item The Hodge dual of a conformal Killing--Yano tensor is again a conformal Killing--Yano tensor.
\item The Hodge dual of a closed conformal Killing--Yano tensor is a Killing--Yano tensor and vice versa.
\end{itemize}

The name ``conformal'' Killing--Yano tensor is connected with the behavior of these objects under a conformal rescaling. Namely,  if $\ts{\omega}$ is a conformal Killing--Yano $p$-form on a manifold with metric tensor $\ts{g}$, then 
\begin{equation}\label{conformalResc}
    \tilde{\ts{\omega}}=\Omega^{p+1}\ts{\omega}
\end{equation}
is a conformal Killing--Yano $p$-form with the conformally scaled metric ${\tilde{\ts{g}}=\Omega^2\ts{g}}$ \citep{Benn:1996ia}.

The existence of conformal Killing--Yano tensors implies the existence of a conformal Killing tensor and hence also a conserved quantity for null geodesics. Namely, having two conformal Killing--Yano $p$-forms $\ts{\omega}_1$ and $\ts{\omega}_2$, obeying \eqref{CKYeq2},
the following object:
\be\label{Kab1}
k{}^{ab}=\omega_{1}{}^{\!(a}{}_{c_2\ldots c_{p}}\;\omega_{2}{}^{b)c_2\ldots c_{p}}\,,
\ee
is a rank-2 conformal Killing tensor which gives rise to a quantity
\begin{equation}\label{Kconfcons}
    K = k^{ab} l_a l_b
\end{equation}
that is conserved along any null geodesic with momentum ${\ts{l}}$.
To prove this statement, let us calculate the symmetrized covariant derivative of ${\ts{k}}$,
\begin{equation}\label{kCKYproof1}
  \nabla{}^{(a}k{}^{bc)}
    = \omega_2{}^{(a}{}_{e_2\ldots e_p}\,\nabla{}^b \omega_1{}^{c)e_2\ldots e_p}\,
    + \omega_1{}^{(a}{}_{e_2\ldots e_p}\,\nabla{}^b \omega_2{}^{c)e_2\ldots e_p}\;.
\end{equation}
Substituting relation \eqref{CKYeqidx2} for the symmetrized covariant derivative of ${\ts{\omega}_1}$ and ${\ts{\omega}_2}$, only the first term in \eqref{CKYeqidx2} survives the contraction with the second form, and we obtain
\begin{equation}\label{kCKYproof2}
  \nabla{}^{(a}k{}^{bc)} = g^{(ab}\,
    \bigl(\omega_2{}^{c)}{}_{e_2\ldots e_p}\,\xi_{1}{}^{e_2\ldots e_p}
    + \omega_1{}^{c)}{}_{e_2\ldots e_p}\,\xi_2{}^{e_2\ldots e_p}\bigr)\;,
\end{equation}
which proves that ${\ts{k}}$ satisfies the conformal tensor equation \eqref{KillTenscond2}.

 The conservation of \eqref{Kconfcons} along null geodesics is related to the conservation of another tensorial quantity
\be
\ts{F} = \ts{l} \wedge ( \ts{l} \cdot \ts{\omega}),
\ee
which is parallel-transported along any null geodesic with momentum $\ts{l}$,
$\dot{\ts{F}}=\covd_{\!\ts{l}}\ts{F}=0$. Indeed, using the geodesic equation \eqref{nullGEO}, conformal Killing--Yano condition \eqref{CKYeq2}, Leibniz rule \eqref{insertionLeibniz}, and ${\ts{l}^2=0}$, we have
\be
\dot{\ts{F}}=\ts{l}\wedge (\tens{l}\cdot \covd_{\!\ts{l}}\ts{\omega})=
\ts{l}\wedge [\ts{l}\cdot (\ts{l}\cdot\ts{\kappa}+\ts{l}\wedge\ts{\xi})]=
\ts{l}\wedge [\ts{l}^2\ts{\xi}-\ts{l}\wedge (\ts{l}\cdot\tens{\xi})]=0\,.
\ee
Defining ${\ts{F}_1}$ and ${\ts{F}_2}$ for $\ts{\omega}_1$ and ${\ts{\omega}_2}$, any product of $\ts{F}$'s, $\ts{l}$'s, and the metric $\ts{g}$ is also parallel-propagated along null geodesics. In particular, this is true for
\be
  {F}_{1}{}_{a c_2\ldots c_{p}}\, {F}_{2}{}_{b}{}^{c_2\dots c_{p}}
  = l_a l_b\, K\,,
\ee
with $K$ given by \eqref{Kconfcons}. This means that $\dot{K}=0$, and we again obtained that $k_{ab}$ must be a conformal Killing tensor.

Contrary to conformal Killing tensors, conformal Killing--Yano tensors do not form in general a graded Lie algebra, though they do in constant curvature spacetimes \citep{Kastor:2007tg}. See \cite{Cariglia:2011yt, Ertem:2016fso, Ertem:2016fng} for  attempts to generalize this property using the suitably modified Schouten--Nijenhuis brackets.

\details{%
It is well known that skew-symmetric tensors form a (graded) Lie algebra with respect to the skew-symmetric Schouten--Nijenhuis (SSN) bracket \citep{Schouten:1940, Schouten:1954, Nijenhuis:1955}, defined as
\be\label{SSN_definition}
\begin{split}
&[\alpha,\beta]^{\rmlab{SSN}}_{a_1\dots a_{p+q-1}} =\\
  &\;=p\,\alpha{}_{b[a_1\dots a_{p-1}}\nabla^b \beta{}_{a_p\dots a_{p+q-1}]}
   +(-1)^{pq} q\,\beta{}_{b[a_1\dots a_{q-1}}\nabla^b \alpha{}_{a_q\dots a_{p+q-1}]}\,,
\end{split}\raisetag{6ex}\ee
for a $p$-form $\ts{\alpha}$ and a $q$-form $\ts{\beta}$. This fact led \cite{Kastor:2007tg} to investigate whether, similar to Killing vectors, Killing--Yano tensors form a subalgebra of this algebra. Unfortunately, such statement is not true in general, the authors were able to give counter examples disproving the conjecture. On the other hand, the statement is true in maximally symmetric spaces.
We also have the following property: let $\ts{\xi}$ be a conformal Killing vector satisfying $\lied_{\ts{\xi}}\ts{g}=2\lambda\ts{g}$, and $\ts{\omega}$ be a conformal Killing--Yano ${p}$-form. Then
\be\label{CKYsymop}
\tilde{\ts{\omega}}
  = [\ts{\xi},\ts{\omega}]^{\rmlab{SSN}}
  =\lied_{\ts{\xi}}\ts{\omega}-(p+1)\lambda\ts{\omega}\,
\ee
is a new conformal Killing--Yano $p$-form \citep{Benn:1996ia, Cariglia:2011yt}.
}

Let us finally mention that conformal Killing--Yano tensors are closely related to twistor spinors, and play a crucial role for finding symmetries of the massless Dirac operator. At the same time the subfamilies of Killing--Yano and closed conformal Killing--Yano tensors are responsible for symmetries of the massive Dirac equation \citep{CarterMcLenaghan:1979, Benn:1996ia, Cariglia:2003kf, Cariglia:2011yt}, see the discussion in section~\ref{ssc:DiracEquation} and appendix~\ref{apx:spinor}.

\subsubsection{Killing--Yano forms}

An important property of Killing--Yano tensors is that they `square' to Killing tensors. Namely, having two Killing--Yano $p$-forms $\ts{f}_1$ and $\ts{f}_2$, their symmetrized product
\be\label{Kab2}
k^{ab}=f_1{}^{(a}{}_{c_2\ldots c_{p}} \, f_2{}^{b)c_2\ldots c_{p}}
\ee
is a rank-2 Killing tensor. This property again follows by taking a symmetrized covariant derivative and employing the Killing--Yano condition \eqref{KYeqidx2}. It can also be obtained by contracting the associated forms  ${\ts{w}_1}$ and ${\ts{w}_2}$ defined by \eqref{wfdef}. Since they are both parallel-transported, the contracted quantity
\be
  K= w_{1c_1\ldots c_{p}}\,w_2{}^{c_2\ldots c_{p}} = k^{ab}\, p^a p^b
\ee
is also conserved and hence $\ts{k}$ is a Killing tensor.

It is obvious that this property can be immediately generalized to other cases. For example, let $\ts{f}_1$, $\ts{f}_2$, and $\ts{f}_3$ be three Killing--Yano 3-forms. Then
\be
k^{abc}=f_1{}^{\!(a|d|}{}_{e}\, f_2{}^{b|e|}{}_{f}\,f_3{}^{c)f}{}_{d}
\ee
is a rank-3 Killing tensor and gives rise to a constant of geodesic motion, given by the contracted product of associated forms $\ts{w}_i$ \eqref{wfdef},
\be
K=\Tr(\ts{w}_1\cdot \ts{w}_2\cdot \ts{w}_3)=k^{abc}p_ap_bp_c\,.
\ee
Similar is true for other products of the associated parallel transported $\ts{w}$'s.

\details{%
Similar to Killing vectors, Killing--Yano tensors provide conserved charges for the fields. For simplicity, let us consider a Killing--Yano 2-form $\ts{f}$. A naive generalization of \eqref{ja} would read
\be
j^{ab}=T^{ac}f_{c}{}^{b}\,.
\ee
It is easy to verify that by using the Killing--Yano equation \eqref{KYeqidx2}, the `current' $\ts{j}$ is again divergence free, $\ts{\nabla \cdot j}=0$. However, it is no longer completely antisymmetric and hence one cannot use the Stokes theorem to construct the corresponding conserved quantities.
For this reason, \cite{Kastor:2004jk} considered an `upgraded' current, given by
\be
j^{ab}=f^{cd}R_{cd}{}^{ab}-2f^{ac}R_c{}^b+2f^{bc}R_c{}^a+f^{ab}R\,,
\ee
which is both manifestly antisymmetric and divergence-free. This property can be immediately generalized for higher-rank Killing--Yano tensors and leads to a definition of `intensive' Yano--ADM charges, see \cite{Kastor:2004jk, Kastor:2004tq} for more details.
}

\subsubsection{Closed conformal Killing--Yano forms}
Closed conformal Killing--Yano tensors are conformal Killing--Yano tensors that are in addition closed with respect to the exterior derivative. This additional property implies the following two important results.

Consider a (non-null) geodesic with a momentum ${\ts{p}}$ and denote by
\be
P^a_{b}=\delta^a_{b}-\frac{p^a p_b}{p^2}\,
\ee
a projector to the space orthogonal to its tangent vector (which is proportional to the momentum). It satisfies
\be\label{COM}
\ts{P}\cdot \ts{p}=0\,,\quad \covd_{\!\ts{p}} \ts{P} = 0\, .
\ee
Let $\ts{h}$ be a rank-$s$ closed conformal Killing--Yano tensor. It allows us to define a new $s$-form
\be\label{Fpform}
F_{{a}_1\ldots {a}_s}=P_{{a}_1}^{b_1}\ldots P_{{a}_s}^{b_s}\, h_{{b}_1\ldots {b}_s}\, ,
\ee
which is parallel-transported along the geodesic. Indeed, using the properties \eqref{COM} and employing the defining property \eqref{bezCCKY} one has
\be
  p^a \nabla_{\!a} F_{a_1\ldots a_s}
  = P^{b_1}_{a_1}\ldots P^{b_s}_{a_s}\,p^a \nabla_{\!a}h_{b_1\ldots b_s}
  = s P^{b_1}_{a_1}\ldots P^{b_p}_{a_s}\, p_{[b_1}\xi_{b_2\ldots b_s]}=0\,.
\ee
In fact the converse is also true. When a form $\ts{F}$ defined by \eqref{Fpform} is parallel-transported along any geodesic, it implies that $\tens{h}$ is a closed conformal Killing--Yano form.

The second property of closed conformal Killing--Yano forms which plays a key role in the construction of hidden symmetries in higher-dimensional black hole spacetimes is the following statement \citep{Krtous:2006qy, Frolov:2007cb}:\\[0.5ex]
\noindent{\bf Theorem}:
{\it Let $\ts{h}_{1}$ and $\ts{h}_{2}$ be two closed conformal Killing--Yano $p$-form and $q$-form, respectively. Then their exterior product
\be\label{CCKYprod}
\ts{h}=\ts{h}_1\wedge \ts{h}_2
\ee
is a closed conformal Killing--Yano $(p+q)$-form.}\\[0.5ex]
This property can be considered as an `antisymmetric analogue' of the statement that a symmetrized product of (conformal) Killing tensors is again a (conformal) Killing tensor.
In order to prove this theorem, we take the covariant derivative of ${\ts{h}}$ along an arbitrary direction~${\ts{X}}$, use the Leibniz rule, and closed conformal Killing--Yano condition \eqref{bezCCKY} for both ${\ts{h}_1}$ and ${\ts{h}_2}$, to obtain
\be
  \covd_{\!\ts{X}} \ts{h}
  = (\covd_{\!\ts{X}} \ts{h}_1) \wedge \ts{h}_2
  + \ts{h}_1 \wedge (\covd_{\!\ts{X}} \ts{h}_2)
  = (\ts{X}\wedge\ts{\xi}_1)\wedge \ts{h}_2
  + \ts{h}_1 \wedge (\ts{X}\wedge\ts{\xi}_2)
  =\ts{X}\wedge \ts{\xi}\,,
\ee
with
\be
\ts{\xi}=\ts{\xi}_1\wedge \ts{h}_2 +(-1)^{p} \ts{h}_1\wedge \ts{\xi}_2\,,
\ee
which proves that ${\ts{h}}$ also satisfies the condition \eqref{bezCCKY} for closed conformal Killing--Yano forms.


\subsection{Integrability conditions and method of prolongation}
\label{ssc:Prolongation}

The conformal Killing--Yano equation \eqref{CKYeq} represents an {\em over-determined system} of partial differential equations \citep{dunajski2008overdetermined} and significantly restricts a class of geometries for which nontrivial solutions may exist. For this reason it is very useful to formulate and study integrability conditions for these objects. For example, it was shown in \cite{Mason:2010zzc} that the integrability condition for a non-degenerate conformal Killing--Yano 2-form implies  that the spacetime is necessary of type D of higher-dimensional algebraic classification \citep{Coley:2004jv, Milson:2004jx}. We refer to appendix~\ref{apx:cky} for detailed derivation of integrability conditions for (closed conformal) Killing--Yano tensors.

Taking into account that (closed conformal) Killing--Yano conditions impose severe restrictions on the spacetime geometry, it is natural to ask the following questions: (i) What is the maximum possible number of independent Killing--Yano symmetries that may in principle exist? (ii) Given a spacetime, is there an algorithmic procedure to determine how many Killing--Yano symmetries are present? Fortunately, the answers to both of these questions are known. Given a geometry, there is an effective method, called the \defterm{method of prolongation}, that provides an algorithmic tool for determining how many at most solutions of an over-determined system of Killing--Yano equations may exist \citep{Houri:2014hma}. See also \cite{Houri:2017tlk} for a recent study of the prolongation of the Killing tensor equation.

\subsubsection{Prolongation of the Killing vector equation}

Let us first consider the case of Killing vectors. For a Killing vector $\ts{\xi}$, the Ricci identities give the following integrability condition:
\begin{equation}\label{KVintegr}
\nabla_{\!a}\nabla_{\!b}\,\xi_c = -R_{bca}{}^e \xi_e\,,
\end{equation}
see \eqref{KY2ndder} for a more general formula and its proof. We can thus rewrite the Killing equation \eqref{Kill} and its integrability condition as a system of first-order partial differential equations for the 1-form $\xi_a$ and a 2-form $L_{ab}$:
\be\label{prolongation}
\nabla_a\xi_b=L_{ab}\,,\quad \nabla_a L_{bc}=-R_{bca}{}^d \xi_d\,.
\ee
These relations imply that all higher derivatives of ${\xi_a}$ and ${L_{ab}}$ at a given point ${x}$ are uniquely determined by the values of ${\xi_a}$ and ${L_{ab}}$ at this point. Hence, the Killing vector $\ts{\xi}$ in the neighborhood of ${x}$ is determined by the initial values of $\xi_a$ and $L_{ab}$ at $x$. The maximum possible number of Killing vector fields is thus given by the maximum number of these initial values. Since $L_{ab}$ is antisymmetric, the maximum number reads
\be\label{MaxDimKV}
N_{\rmlab{KV}}(D) 
=D+\frac{1}{2}D(D-1)=\frac{1}{2}D(D+1)\,.
\ee
As can be expected, the maximum number of Killing vectors exists in maximally symmetric spacetimes, see below.

Let us now explain the algorithm for finding the number of independent Killing vectors in a given spacetime. In the first step we take a derivative of the second equation \eqref{prolongation}, employ the Ricci identity and the first equation \eqref{prolongation},  to obtain
\be
\xi_e \nabla_a R_{bcd}{}^e-\xi_e\nabla_d R_{bca}{}^e-R_{adb}{}^eL_{ce}+R_{adc}{}^e L_{be}-
R_{bca}{}^e L_{de}+R_{bcd}{}^e L_{ae}=0\,.
\ee
Given the spacetime, the Riemmann tensor and its derivatives are known and this condition represents a system of $D^2(D^2-1)/12$ linear algebraic equations for $\xi_a$ and $L_{ab}$ at any point. Although some of these equations may be trivially satisfied, some will reduce the number of possible  independent Killing vector solutions. In the second step we differentiate this equation further, and employing the Ricci identity and equations~\eqref{prolongation} again, obtain another set of algebraic equations, and so on.
After a finite number of steps the algorithm terminates. This procedure determines the actual number of Killing vectors in our spacetime.

\subsubsection{Maximum number of (closed conformal) Killing--Yano forms}

The method of prolongation has been readily extended to (closed conformal) Killing--Yano tensors,  in which case one can use an elegant description in terms of the so called Killing connection. We refer to the work by \cite{Houri:2014hma} for details and state only the formulae for the maximum number of (closed conformal) Killing--Yano tensors, e.g. \cite{Kastor:2004jk}.

Using the Killing--Yano equation \eqref{KYeqkapidx} for a Killing--Yano $p$-form $\ts{f}$ and its integrability conditions \eqref{KY2ndder}, we find
\begin{equation}\label{kyprolong}
\begin{aligned}
  \nabla_{\!a} f_{a_1,\ldots a_p} &= \kappa_{aa_1\dots a_p}\,,\\
  \nabla_{\!a} \kappa_{a_0a_1,\ldots a_p}&=
    \frac{p+1}{2} R_{ca[a_0 a_1} f^{c}{}_{a_2,\ldots a_p]}\, .
\end{aligned}
\end{equation}
Since both ${f_{a_1\dots a_p}}$ and $\kappa_{a_0a_1\dots a_{p}}$ are completely antisymmetric, we have at most
\begin{equation}\label{MaxDimKY}
   N_{\rmlab{KY}}(D,p)= \binom{D}{p}+\binom{D}{p+1}=\binom{D+1}{p+1}=\frac{(D+1)!}{(D-p)!\,(p+1)!}
\end{equation}
Killing--Yano $p$-forms.

Similarly, using the equation \eqref{CCKYeqxiidx} for a closed conformal Killing--Yano $p$-form $\ts{h}$ and its integrability conditions \eqref{covderxi}, we find
\begin{equation}\label{cckyprolong}
\begin{aligned}
    \nabla_{\!a} h_{a_1,\ldots a_p} &= p\, g_{a[a_1}\xi_{a_2\dots a_p]}\,\\
    \nabla_a\xi_{a_2\dots a_p} &= \frac1{D-p}\Bigl(
    -R_{ba}\,h^{b}{}_{a_2\dots a_p} + \frac{p-1}{2} R_{bca[a_2}\, h^{bc}{}_{a_3\dots a_p]}
    \Bigr)\;.
\end{aligned}
\end{equation}
Again, since both ${h_{a_1\dots a_p}}$ and $\xi_{a_2\dots a_{p}}$ are completely antisymmetric, we have at most
\begin{equation}\label{MaxDimCCKY}
   N_{\rmlab{CCKY}}(D,p)= \binom{D}{p}+\binom{D}{p-1}=\binom{D+1}{p}=\frac{(D+1)!}{(D-p+1)!\,p!}
\end{equation}
closed conformal Killing--Yano forms. The same result can be obtained realizing that any closed conformal Killing--Yano tensor $\ts{h}$ of rank $p$  is given by a Hodge dual of a Killing--Yano $(D-p)$-form. We can thus substitute $p\to D-p$ in \eqref{MaxDimKY} obtaining again \eqref{MaxDimCCKY}.

We refer the reader to a recent paper by \cite{Batista:2014fpa}, where the integrability conditions are studied for a general conformal Killing--Yano tensor and to appendix~\ref{apx:cky} for the overview and derivations of the integrability conditions for Killing--Yano and closed conformal Killing--Yano forms.

\subsection{Killing--Yano tensors in maximally symmetric spaces}
\label{ssc:KYmaxsymm}

The maximally symmetric spaces possess the maximum number of Killing--Yano and closed conformal Killing--Yano tensors. Their special properties have been studied in \cite{Batista:2014fpa}. In what follows let us write explicitly a basis for these tensors in the simple case of a ${D}$-dimensional flat space, using the Cartesian coordinates.

Consider a set ${\mathcal{A}}$ of $p$ ordered indices,
\begin{equation}\label{ordset}
\mathcal{A}=\{a_1,\ldots a_p\} \quad\text{such that}\quad 1\le a_1 <a_2 <\dots <a_p\le D\,.
\end{equation}
Then the following $\binom{D}{p}$ objects:
\be\label{kflat}
  \ts{f}^{\{a_1,\ldots a_p\} }=
  \grad x^{a_1}\wedge \grad x^{a_2} \wedge \dots \wedge \grad x^{a_p}\,,
\ee
labeled by such a set, are (covariantly constant) {\em translational Killing--Yano} $p$-forms \cite{Kastor:2004jk}.

Furthermore, the following $\binom{D}{p+1}$ objects:
\be\label{hkflat}
  \ts{\hat{f}}^{\{a_0,\ldots a_{p}\} }=
  x^{[a_0}\grad x^{a_1}\wedge\grad x^{a_2} \wedge \dots \wedge \grad x^{a_{p}]}\,,
\ee
labeled by a set of ${p+1}$ indices, are the \defterm{rotational Killing--Yano} forms \cite{Kastor:2004jk}. Indeed, taking the covariant derivative, we have
\be
  \covd\ts{\hat{f}}^{\{a_0,\ldots a_{p}\} } =
  \frac{1}{p+1}\,\grad x^{[a_0}\wedge \grad x^{a_1}\wedge \grad x^{a_2} \wedge \dots \wedge \grad x^{a_{p}]}\,,
\ee
which proves the statement. The total number of Killing--Yano tensors \eqref{kflat} and \eqref{hkflat} is $\binom{D}{p}+\binom{D}{p+1}=N_{\rmlab{KY}}(D,p)$, giving thus a complete set of linearly independent Killing--Yano \mbox{$p$-forms} in flat space.

A basis in the space of closed conformal Killing--Yano $p$-forms can be constructed as the Hodge dual of the basis in the space of Killing--Yano $(D{-}p)$-forms. Let ${\bar{\mathcal{A}}}$ be a {\em complimentary set} to set ${\mathcal{A}}$, \eqref{ordset}, consisting of all integers ${1,\dots,D}$ which are different from those in ${\mathcal{A}}$. Clearly, the Hodge dual of the translation Killing--Yano form ${\ts{f}^{\mathcal{A}}}$ is, up to a sign, given again by the same type of the form, ${*\ts{f}^{\mathcal{A}}=\pm\ts{f}^{\bar{\mathcal{A}}}}$, only indexed by the complimentary set ${\bar{\mathcal{A}}}$. Ignoring the unimportant sign, we can thus define the (covariantly constant) {\em translational closed conformal Killing--Yano forms} by the same formula as above,
\be\label{hflat}
  \ts{h}^{\{a_1,\ldots a_p\} }=
  \grad x^{a_1}\wedge \grad x^{a_2} \wedge \dots \wedge \grad x^{a_p}\,,
\ee
labeled again by a set of ${p}$ indices. Since any closed form $\ts{h}$ can locally be written as $\ts{h}=\ts{d b}$, for our translation forms we may, for example, write
\be\label{bti}
\ts{h}^{\mathcal{A}}=\grad \ts{\hat{f}}{}^{\mathcal{A}}\,.
\ee
Hence, in flat space, the rotational Killing--Yano forms are potentials for the translational closed conformal Killing--Yano forms.

Similarly, we can define the {\em rotational closed conformal Killing--Yano} $p$-forms as Hodge duals of the rotational Killing--Yano ${(D{-}p)}$-forms. Let us consider ${\ts{\hat{f}}{}^{\mathcal{A}}}$ labeled by a set ${\mathcal{A}}$ of ${(D{-}p{+}1)}$ indices. Expanding the antisymmetrization in \eqref{hkflat} with respect to the first index and taking the Hodge dual gives
\begin{equation}\label{Hdualfhat}
    *\ts{\hat{f}}{}^{\mathcal{A}} =
    \pm\frac1{D-p+1}\Bigl(\sum_{a\in\mathcal{A}} x^a\grad x^a \Bigr)\wedge * \ts{f}{}^{\mathcal{A}}\;.
\end{equation}
Ignoring unimportant prefactors and renaming the labeling set $\bar{\mathcal{A}}\to\mathcal{A}$, we can define the following basis of rotational closed conformal Killing--Yano ${p}$-forms:
\begin{equation}\label{hathflat}
    \ts{\hat{h}}{}^{\mathcal{A}} =
    \Bigl(\sum_{a\in\bar{\mathcal{A}}} x^a\grad x^a\Bigr)
    \wedge\ts{h}{}^{\mathcal{A}}\,,
\end{equation}
labeled by a set ${\mathcal{A}}$ of ${p-1}$ indices. The divergences of these forms are
\begin{equation}\label{xihatflat}
    \ts{\hat{\xi}}{}^{\mathcal{A}} =
    \frac1{D-p+1}\,\covd\cdot\ts{\hat{h}}{}^{\mathcal{A}}
    =\ts{h}{}^{\mathcal{A}}\;.
\end{equation}
Each of the closed conformal Killing--Yano forms \eqref{hathflat} can be obtained from the potential
\be\label{bri}
    \ts{\hat{b}}{}^{\mathcal{A}} =
    \frac12\Bigl(\sum_{a\in\bar{\mathcal{A}}} (x^a)^2\Bigr)\,
    \ts{h}{}^{\mathcal{A}}\,.
\ee

\subsection{Principal tensor}
\label{ssc:PrincipalTensor}

There exists a very deep geometrical reason why the properties of higher-dimensional rotating black holes are very similar to the properties of their four-dimensional `cousins'. In both cases, the spacetimes admit a special geometric object which we call the {\em principal tensor}.
As we shall see, this tensor generates a complete set of explicit and hidden symmetries and uniquely determines the geometry, given by the off-shell Kerr--NUT--(A)dS metric. The purpose of this section is to introduce the principal tensor, a `superhero' of higher-dimensional black hole physics, and discuss its basic properties.

\subsubsection{Definition}

We define the \defterm{principal tensor} ${\ts{h}}$ as a \emph{non-degenerate} closed conformal Killing--Yano 2-form. Being a closed conformal Killing--Yano 2-form it obeys the equation
\be\label{PKYT}
\nabla_c h_{ab}=g_{ca}\xi_b-g_{cb}\xi_a\,,\quad \xi_a=\frac{1}{D-1}\nabla^b h_{ba}\,,
\ee
or in the language of differential forms
\be\label{defPKY}
\covd_{\!\ts{X}}\ts{h} = \ts{X}\wedge \ts{\xi}\,,\quad \ts{\xi}=\frac{1}{D-1} \covd\cdot\ts{h}\,.
\ee
Note that since $\ts{h}$ is closed, there exists, at least locally, a potential 1-form  $\ts{b}$ such that
\be
  \ts{h}=\grad \ts{b}\, .
\ee
The condition of non-degeneracy means that the principal tensor has the maximal possible (matrix) rank and possesses the maximal number of functionally independent eigenvalues.\footnote{%
In the Lorentzian signature we additionally assume certain generality of eigenvectors and eigenvalues, see the discussion below.}

\subsubsection{Darboux and null frames}

In order to explain the imposed condition of non-degeneracy in more details and to exploit the algebraic structure of the principal tensor we shall now introduce the \defterm{Darboux frame}. Consider a $(D=2n+\eps)$-dimensional manifold with (Riemannian---see later) metric $\ts{g}$. For any 2-form $\ts{h}$ in this space there exists an orthonormal frame $(\enf{\mu}, \ehf{\mu},\ezf)$, called the \defterm{Darboux frame}, so that we can write
\begin{gather}
\ts{h} = \sum_\mu x_\mu \enf{\mu}\wedge\ehf{\mu}\,,\label{PTcanform}\\
\ts{g} = \sum_\mu \bigl(
      \enf\mu\,\enf\mu + \ehf\mu\,\ehf\mu
      \bigr) + \eps\, \ezf\,\ezf\,.\label{orthonormmtrc}
\end{gather}
Here, the 1-forms ${\enf{\mu}}$ and ${\ehf{\mu}}$, $\mu=1\dots,\dg$, accompanied in odd dimensions with $\ezf$, are orthogonal to each other and normalized with respect to the metric, and the quantities $x_\mu$ are related to the `eigenvalues' of the 2-form $\ts{h}$ (see below).

The condition that the principal tensor is non-degenerate requires that there are exactly ${\dg}$
nonvanishing eigenvalues ${x_\mu}$, which, in a suitable neighborhood, give ${\dg}$ functionally independent (non-constant and with linearly independent gradients) functions.

The Darboux frame is closely related to eigenvectors of the principal tensor. Let us denote by ${{{}^{\sharp\!}\ts{h}}}$ a variant of the principal tensor with the first index raised by the metric,\footnote{%
{We will use this notation just in this section since we want to be more explicit here. Elsewhere we raise indices implicitly, without the sharp symbol.}}
$({{{}^{\sharp\!}{h}}})^a{}_b=g^{ac}h_{cb}$. This is a real linear operator on the tangent space which is antisymmetric {with respect to the transposition} given by the metric. As such, it has complex eigenvectors coming in complex conjugate pairs ${(\ts{m}_\mu,\,\bar{\ts{m}}_\mu)}$ with imaginary eigenvalues ${\pm i x_\mu}$,
\begin{equation}\label{eigenvectors}
    {{}^{\sharp\!}\ts{h}}\cdot\ts{m}_\mu = -i x_\mu\, \ts{m}_\mu\,,\quad
    {{}^{\sharp\!}\ts{h}}\cdot\bar{\ts{m}}_\mu = i x_\mu\, \bar{\ts{m}}_\mu\,,
\end{equation}
and a subspace of real eigenvectors with the vanishing eigenvalue. The maximal possible rank guarantees that in even dimensions there is no eigenvector with the vanishing eigenvalue and in odd dimensions there is exactly one eigenvector ${\ezv}$ with the vanishing eigenvalue,
\begin{equation}\label{zeroeigenvector}
    {{}^{\sharp\!}\ts{h}}\cdot\ezv = 0\,.
\end{equation}
The eigenvectors are null and satisfy the null-orthonormality conditions\footnote{%
The eigenvectors can be chosen orthonormal with respect to the hermitian scalar product on the complexification of the tangent space, ${\langle\ts{a},\ts{b}\rangle=\ts{g}(\bar{\ts{a}},\ts{b})}$. Such orthonormality relations translate into null-orthonormality conditions \eqref{nullorthonorm} written in terms of the metric.}
\begin{equation}\label{nullorthonorm}
    \ts{g}(\ts{m}_\mu,\ts{m}_\nu) = 0\,,\quad
    \ts{g}(\bar{\ts{m}}_\mu,\bar{\ts{m}}_\nu) = 0\,,\quad
    \ts{g}(\ts{m}_\mu,\bar{\ts{m}}_\nu) = \delta_{\mu\nu}\;.
\end{equation}
These eigenvectors can be used to define the Darboux frame. Namely, the vectors
\begin{equation}\label{Darbvect=eigenvec}
    \env{\mu} = -\frac{i}{\sqrt{2}}\bigl(\ts{m}_\mu-\bar{\ts{m}}_\mu\bigr)\,,\quad
    \ehv{\mu} = \frac{1}{\sqrt{2}}\bigl(\ts{m}_\mu+\bar{\ts{m}}_\mu\bigr)\,,
\end{equation}
together with ${\ezv}$ in odd dimensions, form an orthonormal basis and satisfy
\begin{equation}\label{Darbeigenvectors}
    {{}^{\sharp\!}\ts{h}}\cdot\env{\mu} = - x_\mu\, \ehv{\mu}\,,\quad
    {{}^{\sharp\!}\ts{h}}\cdot\ehv{\mu} = x_\mu\, \env{\mu}\,,\quad
    {{}^{\sharp\!}\ts{h}}\cdot\ezv = 0\;.
\end{equation}
The dual frame of 1-forms ${(\enf{\mu},\,\ehf{\mu},\,\ezf)}$ is exactly the Darboux frame in which the principal tensor takes the form \eqref{PTcanform}. At the same time, the null-or\-tho\-nor\-mality conditions \eqref{nullorthonorm} for the basis eigenvectors ${(\ts{m}_\mu,\,\bar{\ts{m}}_\mu,\,\ezv)}$ imply that the inverse metric can be written as
\begin{equation}\label{nullorthonormmtrc}
    \ts{g}^{-1}= \sum_\mu \bigl(
      \ts{m}_\mu\,\bar{\ts{m}}_\mu + \bar{\ts{m}}_\mu\,\ts{m}_\mu
      \bigr) + \eps\, \ezv\,\ezv\,,
\end{equation}
recovering \eqref{orthonormmtrc} upon the use of $(\enf{\mu}, \ehf{\mu},\ezf)$.

The Darboux basis can also be understood in terms of the eigenvectors of tensor $\ts{Q}$, defined as the square of the principal tensor,
\begin{equation}\label{Qdefhh}
    Q_{ab} = h_{a}{}^{c}\, h_{bc}\;.
\end{equation}
Being a particular case of definition \eqref{Kab1}, ${\ts{Q}}$ is a conformal Killing tensor. Since it can clearly be written as
\begin{equation}\label{Qinframe}
    \ts{Q} = \sum_\mu x_\mu^2 \bigl( \enf{\mu}\enf{\mu} + \ehf{\mu}\ehf{\mu}\bigr)\;,
\end{equation}
one gets the following eigenvector equations:
\begin{equation}\label{Qeigenvectors}
    \ts{Q}\cdot\env\mu = x_\mu^2\,\env\mu\,,\quad
    \ts{Q}\cdot\ehv\mu = x_\mu^2\,\ehv\mu\,,\quad
    \ts{Q}\cdot\ezv = 0\;.
\end{equation}

To summarize, the 2-form algebraic structure of the principal tensor splits the tangent space into orthogonal 2-planes, each of which is spanned on the pair of vectors ${(\env{\mu},\,\ehv{\mu})}$, in odd dimensions supplemented by an additional one-dimensional subspace spanned on ${\ezv}$.

\subsubsection{Metric signature}

All the formulae so far were adjusted to the Euclidean signature. For other signatures of the metric, most of the formulae can be written in the same way, only the reality of various quantities is different. In particular, for the Lorentzian signature one of the 1-forms in the Darboux frame is imaginary and two of the null eigenvectors are real (and not complex conjugate anymore). One can also perform a suitable `Wick rotation' and define real and properly normalized canonical frames. This will be done, for example, in the next chapter when discussing the canonical Darboux basis for the Kerr spacetime in four dimensions. In higher dimension, on other hand, we will use mostly the formal Euclidean definitions even in the case of Lorentzian signature (chapter~\ref{sc:hshdbh}) and we will perform the Wick rotation only for the coordinate form of the metric, see section~\ref{ssc:SubcasesLorentzian}.

However, the non-Euclidean signatures allow also other possibilities. The Darboux frame can take an exceptional `null form' when some of the vectors in \eqref{PTcanform} are null, cf.~\cite{Milson:2004wr}. It can also happen that some of the eigenvalues $x_\mu$ have a null gradient $\grad x_\mu$ which complicates the choice of the special Darboux frame discussed below, see~\eqref{specDarbouxframecond}. We do not consider such exceptional cases in our review. We assume that the principal tensor allows the choice of the Darboux frame in the form \eqref{specDarbouxframecond} and that eigenvalues $x_\mu$ are not globally null (although they can become null on special surfaces as, for example, at the horizon). A study and the classification of the exceptional null cases is an interesting open problem.

\subsubsection{Special Darboux frame}
In order to write down the Darboux frame above, we just exploited the algebraic properties of the principal tensor: that it is a maximally non-degenerate 2-form in the space with metric. Such a frame is not fixed uniquely. We still have a freedom which allows us to independently rotate  each 2-plane spanned on ${\env{\mu}}$, ${\ehv{\mu}}$:
\begin{equation}\label{framefreedom}
\begin{aligned}
    \env\mu &\to \cos\alpha\,\env\mu - \sin\alpha\,\ehv\mu\,,\quad
    & \ts{m}_\mu &\to \exp(-i\alpha)\,\ts{m}_\mu\,,\\
    \ehv\mu &\to \sin\alpha\,\env\mu + \cos\alpha\,\ehv\mu\,,\quad
    & \bar{\ts{m}}_\mu &\to \;\;\,\exp(i\alpha)\,\bar{\ts{m}}_\mu\,.\\
\end{aligned}
\end{equation}
This freedom allows one to further simplify the key objects related to the principal tensor, for example, to obtain a nice expression \eqref{xiinspecDarbouxfr} below for the 1-form $\ts{\xi}$.

Namely, by using the property that the principal tensor is a closed conformal Killing--Yano form, one can require \cite{Krtous:2008tb} that
\begin{equation}\label{specDarbouxframecond}
    \ehv\mu\cdot\grad x_\nu = 0
\end{equation}
for any ${\mu}$ and ${\nu}$. Moreover, with this condition the dual frame 1-forms ${\enf\mu}$ satisfy
\begin{equation}\label{specDarbouxframe}
    \grad x_\mu = \sqrt{Q_\mu}\, \enf\mu \,,
\end{equation}
where ${Q_\mu}$ is metric component ${Q_\mu=g^{\mu\mu}}$. We call such a frame the \defterm{special Darboux basis}. We will see that the special Darboux frame is used when the metric is specified, see chapter~\ref{sc:Kerr} for the case of four dimensions and chapter~\ref{sc:hdbh} for a general higher-dimensional case.

To justify that conditions \eqref{specDarbouxframecond} and \eqref{specDarbouxframe} can be enforced, we take the covariant derivative of the eigenvector equation \eqref{eigenvectors} along the direction ${\ts{m}_\nu}$. After employing the closed conformal Killing--Yano condition \eqref{PKYT}, one obtains
\begin{equation}\label{dereigenveceq}
    ({{}^{\sharp\!}\ts{h}}+ix_\mu\ts{I})\cdot\covd_{\ts{m}_\nu}\ts{m}_\mu
      + (\ts{m}_\mu\cdot\ts{\xi})\ts{m}_\nu + i (\ts{m}_\nu\cdot\grad x_\mu)\ts{m}_\mu=0\;.
\end{equation}
Taking component in ${\ts{m}_\mu}$ direction and using the eigenvector condition \eqref{eigenvectors} again, one finds
\begin{equation}\label{mdx1}
    \ts{m}_\nu\cdot\grad x_\mu=0\qquad\text{for ${\mu\neq\nu}$}\,,
\end{equation}
and
\begin{equation}\label{mdx2}
    \ts{m}_\mu\cdot\grad x_\mu = i\ts{m}_\mu\cdot\ts{\xi}
\end{equation}
when ${\mu=\nu}$. In odd dimensions, by a similar argument, one gets also
\begin{equation}\label{mdx3}
    \ezv\cdot\grad x_\mu=0\,.
\end{equation}
With the help of \eqref{nullorthonormmtrc}, \eqref{mdx1}, and \eqref{mdx3}, the function ${Q_\mu\equiv g^{\mu\mu}}={\grad x_\mu\cdot\ts{g}^{-1}\cdot\grad x_\mu}$ can be written as ${Q_\mu = 2 \left|\ts{m}_\mu\cdot\grad x_\mu\right|^2}$. It means that
\begin{equation}\label{mdxQphase}
    \ts{m}_\mu\cdot\grad x_\mu = \frac{i}{\sqrt2}\sqrt{Q_\mu} \exp(i\alpha)
\end{equation}
for some phase ${\alpha}$. Now we can take an advantage of the freedom \eqref{framefreedom} and fix the phase so that
\begin{equation}\label{mdxfixed}
    \ts{m}_\mu\cdot\grad x_\mu = \frac{i}{\sqrt2}\sqrt{Q_\mu} \;.
\end{equation}
Relations \eqref{Darbvect=eigenvec} then immediately imply
\begin{equation}\label{specDarbouxframedx}
    \env\mu\cdot\grad x_\mu = \sqrt{Q_\mu}\,,\quad
    \env\nu\cdot\grad x_\mu = 0\quad\text{for ${\nu\neq\mu}$}\,,\quad
    \ehv\kappa\cdot\grad x_\mu = 0\,,\quad
    \ezv\cdot\grad x_\mu = 0\,,
\end{equation}
which proves all the assertions given above.

As a bonus, the equation \eqref{mdx2} now yields
\begin{equation}\label{xiinspecDarbouxfr}
    \ts{\xi} = \sum_\mu \sqrt{Q_\mu}\,\ehv\mu + \sqrt{Q_0}\,\ezv\,,
\end{equation}
with yet unspecified function ${Q_0}$. Upon contracting with the principal tensor \eqref{PTcanform} and using \eqref{specDarbouxframe}, we obtain
\begin{equation}\label{xidoth}
    \ts{\xi}\cdot\ts{h} = - \sum_\mu x_\mu\sqrt{Q_\mu}\,\enf\mu
    =-\grad\Bigl(\frac12\sum_\mu\, x_\mu^2\Bigr)\;.
\end{equation}
Employing further the Cartan identity and the closeness of ${\ts{h}}$, we finally obtain
\begin{equation}\label{Liexih}
    \lied_{\ts{\xi}}\ts{h}
    = \ts{\xi}\cdot\grad\ts{h}+\grad(\ts{\xi}\cdot\ts{h})
    =0\;.
\end{equation}
Note that although we used the special Darboux frame to prove this relation, it is of course valid universally: 
the principal tensor is conserved along the flow generated by ${\ts{\xi}}$.

Importantly, by further studying the integrability conditions for the principal tensor, it can be shown that
\cite{Krtous:2008tb, Houri:2008ng, Yasui:2011pr, krtous:directlink}
\be\label{KVxig}
\lied_{\ts{\xi}}\ts{g}=0\,.
\ee
$\ts{\xi}$ is thus a Killing vector which we call the \defterm{primary Killing vector}.

The two properties \eqref{Liexih} and \eqref{KVxig} play a crucial role in the construction of the canonical metric admitting the principal tensor, see the discussion in chapter~\ref{sc:hshdbh} and original papers \cite{Houri:2007xz, Krtous:2008tb, Houri:2008ng}.

\subsubsection{Killing tower}

The special Darboux frame is only the first consequence of the existence of the principal tensor. One of the keystone properties of the principal tensor is that it can be used to generate a rich symmetry structure which we call the \defterm{Killing tower}. It is a sequence of various symmetry objects which, in turn, guarantee many important properties of the physical systems in spacetimes with the principal tensor. Here we only shortly sketch how the Killing tower is build to get an impression of this symmetry structure. We return to the Killing tower in chapter~\ref{sc:hshdbh}, where we explore its definitions and properties in much more detail, and in chapter~\ref{sc:intsep}, where we review its main physical consequences.

Starting with the principal tensor ${\ts{h}}$, we can build the following objects \cite{Krtous:2006qy, Frolov:2007cb,Frolov:2008jr}:
\begin{enumerate}[(i)]
\item
Closed conformal Killing--Yano forms ${\CCKY{j}}$ of rank ${2j}$:
\begin{equation}\label{previewCCKYj}
    \CCKY{j} = \frac1{j!}\,\ts{h}^{\wedge j}\;.
\end{equation}
\item
Killing--Yano forms ${\KY{j}}$ of rank $({D-2j})$:
\begin{equation}\label{previewKYj}
    \KY{j} = * \CCKY{j}\;.
\end{equation}
\item
Rank-2 Killing tensors~${\KT{j}}$,
\begin{equation}\label{previewKTj}
    \KTc{j}^{ab} = \frac1{(D{-}2j{-}1)!}\,
      \KYc{j}{}^{a}{}_{c_1\dots c_{D{-}2j{-}1}}\,\KYc{j}{}^{bc_1\dots c_{D{-}2j{-}1}}\,.
\end{equation}
\item
Rank-2 conformal Killing tensors ${\CKT{j}}$:
\begin{equation}\label{previewCKTj}
    \CKTc{j}^{ab} = \frac1{(2j{-}1)!} \,
      \CCKYc{j}{}^{a}{}_{c_1\dots c_{2j{-}1}}\CCKYc{j}{}^{bc_1\dots c_{2j{-}1}}\;.
\end{equation}
\item
Killing vectors ${\KV{j}}$:
\begin{equation}\label{previewKVj}
    \KV{j} = \KT{j}\cdot\ts{\xi}\;.
\end{equation}
\end{enumerate}

For ${j=0}$, the Killing tensor reduces to the metric, ${\KT{0} = \ts{g}}$, and the Killing vector $\KV{0}$ coincides with the primary Killing vector, ${\KV{0}=\ts{\xi}}$. We call the other Killing vectors ${\KV{j}}$ the \defterm{secondary Killing vectors}. Note also that for ${j=1}$, ${\CCKY{1}=\ts{h}}$, and the conformal Killing tensor reduces to the previously defined object \eqref{Qdefhh}, ${\CKT{1}=\ts{Q}}$.

\details{
To show that $\KV{j}$ are indeed Killing vectors, we note that taking covariant derivative of \eqref{previewKVj} and employing the Killing tensor equation \eqref{KillTenscond} for $\KT{j}$ and $\ts{\xi}$ gives \cite{Houri:2007xz}
\be
\nabla^{(a} \KVc{j}^{b)}=\frac{1}{2}\lied_{\ts{\xi}}\KTc{j}^{ab} - \xi^c\nabla_{\!c}\KTc{j}^{ab}\,.
\ee
Since the Killing tensor $\KT{j}$ is build up only using $\ts{h}$ and $\ts{g}$, the Lie derivative in the first term vanishes due to conditions \eqref{Liexih} and \eqref{KVxig}. Similarly, the covariant derivative in the second term vanishes thanks to $\covd_{\!\ts{\xi}}\ts{h}=0$ which is a direct consequence of the principal tensor equation \eqref{defPKY}. Let us, however, note that this proof relies on the condition \eqref{KVxig}, which is difficult to prove; see discussion in chapter~\ref{sc:hshdbh}, especially section~\ref{sec:commutationrel}. The character of the other objects in the Killing tower follows from the general properties of conformal Killing--Yano forms discussed previously in this section. See chapter~\ref{sc:hshdbh} for further discussion.
}

The objects in the Killing tower encode symmetry properties of the geometry. Killing vectors characterize its explicit symmetries, while Killing tensors describe the hidden symmetries. Together they generate a sufficient set of conserved quantities for a free particle motion, yielding such a motion completely integrable. They also define symmetry operators for the wave operator. The objects in the Killing--Yano tower enable one to separate the Dirac equation. We will discuss all these consequences in chapter~\ref{sc:intsep}.

\subsubsection{Geometry admitting the principal tensor}

As can be expected, the existence of the principal tensor imposes very restrictive conditions on the geometry. In fact, it determines the geometry: {\em the most general geometry consistent with the existence of the principal tensor is the off-shell Kerr--NUT--(A)dS geometry.} This geometry is the main object of our study in the following sections. Since it contains, as a special subcase, the metric for a general multiply-spinning black hole, it represents a generalization of the Kerr solution to an arbitrary dimension. For this reason, we start in the next section with a review of the properties of the four-dimensional Kerr solution.

In chapter~\ref{sc:hdbh} we introduce the general higher dimensional off-shell Kerr--NUT--(A)dS geometry. We define canonical coordinates in which the metric acquires a manageable form. With this machinery we shall return back to the discussion of the principal tensor in chapter~\ref{sc:hshdbh}.

The Killing tower can be build directly from the principal tensor, without referring to a particular form of the metric. This construction, sketched above, is discussed in detail in chapter~\ref{sc:hshdbh}. However, it is also useful to present these objects in an explicit coordinate form. This is the reason why we are postponing the further discussion of the Killing tower till chapter~\ref{sc:hshdbh}, only after we introduce the metric itself. Since the metric is determined by the existence of the principal tensor, the utilization of the metric in the discussion of the principal tensor does not mean a loss of generality.

%% file: ch3-kerr.tex

\section{Kerr metric and its hidden symmetries}
\label{sc:Kerr}

The main goal of this review is to describe properties of Kerr--NUT--(A)dS family of higher-dimensional black holes related to hidden symmetries. As we shall see many of these properties are similar to those of the Kerr metric. A deep reason for this is the existence of the principal tensor. In order to prepare a reader for `a travel' to higher dimensions, where all the formulas and relations look more complicated and the calculations are more technically involved, we summarize the results concerning the properties of the Kerr metric and its four-dimensional generalization described by the Kerr--NUT--(A)dS spacetime in this chapter. We also briefly discuss a related family of Pleba\'nski--Demia\'nski spacetimes which share with the Kerr metric some of its hidden symmetries.

\subsection{Kerr metric}
\label{ssc:KerrMetric}

The {\em Kerr metric} describes a rotating black hole. Found by \cite{Kerr:1963}, it is the most general stationary vacuum solution of Einstein's equations in an asymptotically flat spacetime with a regular event horizon.
The general properties of the Kerr metric are well known and can be found in many textbooks, see, e.g., \cite{MTW,HawkingEllis:book,Wald:book1984,Chandrasekhar:1983,FrolovNovikov:book, FrolovZelnikov:book}. In this section, we discuss the Kerr solution from a perspective of its {\em hidden symmetries}. 
As we shall demonstrate later, many of the remarkable properties of the Kerr geometry, that stem from these symmetries, are naturally generalized to
black holes of higher-dimensional gravity.

In the {\em Boyer--Lindquist coordinates} the Kerr metric takes the following form:
\be\begin{split}\label{kerr1}
\ts{g}&=-\!\left(1{-}\frac{2Mr}{\Sigma}\right)\grad t^2 -
\frac{4Mra\sin^2\theta}{\Sigma}\,\grad t\,\grad \phi
+\frac{A\,\sin^2\theta}{\Sigma}\,\grad \phi^2+\,\frac{\Sigma}{\Delta_r}\,\grad r^2+\Sigma\,\grad \theta^2,
\end{split}\ee
\be\begin{split}\label{4.1.3}
\Sigma&=r^2+a^2\,\cos^2\theta\,,\quad
\Delta_r=r^2-2Mr+a^2\,,\quad A=(r^2+a^2)^2 -\Delta_r\,a^2\,\sin^2\theta\,.
\end{split}\ee
The metric does not depend on coordinates $t$ and $\phi$, $ \ts{\xi}_{(t)}=\ts{\partial}_t$ and $\ts{\xi}_{(\phi)}= \ts{\partial}_{\phi}$ are two (commuting)
Killing vectors. The Killing vector $ \ts{\xi}_{(t)}$ is uniquely characterized by the property that it is timelike at infinity; the metric is {\em stationary}. The characteristic property of $ \ts{\xi}_{(\phi)}$ is that its integral lines are closed. In the black hole exterior the fixed points of $ \ts{\xi}_{(\phi)}$, that is the points where $ \ts{\xi}_{(\phi)}=0$, form a regular two-dimensional geodesic submanifold, called the axis of
symmetry---the metric is {\em axisymmetric}. The induced metric on the axis is
\be
\ts{\gamma}=-F \grad t^2+F^{-1} \grad r^2\,,\qquad F=\frac{\Delta_r}{r^2+a^2}\, .
\ee

The Kerr metric is characterized by two parameters:
$M$ and $a\,$. At far distances, for $r\to \infty$, the metric simplifies to
\be\label{kerrm}
\ts{g}\approx -\left(1-\frac{2M}{r}\right)\,\grad t^2 -
\frac{4Ma\sin^2\theta}{r}\,\grad t\,\grad \phi
+\grad r^2 +r^2 (\grad \theta^2+\sin^2\theta\,\grad \phi^2)\, .
\ee
From this asymptotic form one concludes that $M$ is the {\it
mass}, and $J=aM$ is the {\it angular momentum} of the black hole.
The parameter  $a$ 
is called the rotation
parameter. Similar to the mass $M$, it has a dimensionality of length. The
ratio of $a$ and $M$ is a dimensionless parameter $\alpha=a/M$, called
the {\em rotation rapidity}.
Similar to the case of the Schwarzschild black
hole, one can use $M$ as a scale parameter and write the
Kerr metric \eqref{kerr1} in the form
\be
\ts{g}=M^2 \ts{\tilde g}\, ,
\ee
where $\ts{\tilde g}$ is a dimensionless metric that contains only one non-trivial
dimensionless parameter: the rotation rapidity $\alpha$.




\subsection{Carter's canonical metric}
\label{ssc:CarterForm}

The Boyer--Lindquist coordinates naturally
generalize the Schwarzschild coordinates to the case of a rotating
black hole. We now present yet another form of the Kerr metric in which
its hidden symmetry is more evident. Let us perform the following coordinate transformation:
\be\label{kcan}
y=a\cos\theta\,,\qquad \psi=\phi/a\,,\qquad \tau=t-a \phi\,.
\ee
Then the Kerr metric \eqref{kerr1} takes the form
\be\label{canNOreference}
\ts{g}=\frac{1}{\Sigma}\left[ -\Delta_r(\grad \tau+y^2 \grad \psi)^2+\Delta_y(\grad \tau-r^2
\grad \psi)^2\right]+\Sigma\left[ \frac{\grad r^2}{\Delta_r}
+\frac{\grad y^2}{\Delta_y}\right]\, ,
\ee
\be\label{DD}
\Sigma=r^2+y^2\, ,\quad\Delta_r=r^2-2Mr+a^2\,,\quad\Delta_y=a^2-y^2\,.
\ee
As we shall see, similar coordinates will be very useful in higher dimensions. To stress this, we call $(\tau,r,y,\psi)$ the \emph{canonical coordinates}.

\subsubsection{Off-shell canonical metric}

In the new form of the metric \eqref{canNOreference} the parameters of the solution, mass $M$ and rotation parameter $a$, enter only through functions $\Delta_r$ and $\Delta_y$, both being quadratic polynomials in $r$ and $y$, respectively. It is often convenient not to specify functions $\Delta_r(r)$ and $\Delta_y(y)$ from the very beginning, but consider a metric with arbitrary functions instead:
\begin{equation}\label{can}
\begin{aligned}
\ts{g}&=-\frac{\Delta_r}{\Sigma}(\grad \tau+y^2 \grad \psi)^2+\frac{\Delta_y}{\Sigma}(\grad \tau-r^2
\grad \psi)^2+\frac{\Sigma}{\Delta_r}\grad r^2+\frac{\Sigma}{\Delta_y}\grad y^2\, ,\\
\Sigma&=r^2+y^2\,,\quad \Delta_r=\Delta_r(r)\,,\quad \Delta_y=\Delta_y(y)\,.
\end{aligned}
\end{equation}
We call such an ansatz the \emph{off-shell canonical metric}. This name emphasizes the fact that in a general case this metric is not a solution of  Einstein's equations.

It turns out that many calculations and results become more transparent and simpler when performed without specifying a concrete form of functions $\Delta_r(r)$ and $\Delta_y(y)$, that is, for the off-shell metric. For example, an important property of the off-shell metric is that its determinant $g$ does not depend on functions $\Delta_r(r)$ and $\Delta_y(y)$: 
\be\label{detcan}
\sqrt{-g}=\Sigma=r^2+y^2\, .
\ee
The inverse metric to \eqref{can} reads
\be\label{caninv}
\ts{g}^{-1}=\frac{1}{\Sigma}\Bigl[-\Delta_r^{-1} (r^2\ts{\pa}_\tau+\ts{\pa}_{\psi})^2+\Delta_y^{-1}(y^2\ts{\pa}_\tau-\ts{\pa}_{\psi})^2+\Delta_r(\ts{\partial}_r)^2+ \Delta_y (\ts{\partial}_y)^2
\Bigr]\, .
\ee

\subsubsection{Going on-shell: Kerr--NUT--(A)dS metric}

If one requires that the off-shell metric satisfies the Einstein equations, the functions $\Delta_r(r)$ and $\Delta_y(y)$ take a special form. We call the metric \eqref{can} with such functions $\Delta_r(r)$ and $\Delta_y(y)$ an {\em on-shell metric}.

For example, the on-shell metric with functions $\Delta_r(r)$ and $\Delta_y(y)$ given by \eqref{DD} reproduces the Kerr solution. However, one can easily check that this is not the most general vacuum on-shell metric. For example, one can add a linear in $y$ term, $2N y$, to the function $\Delta_y$. Such a generalization of the Kerr metric is known as the {\em Kerr--NUT solution}, and the parameter $N$ is called the NUT (Newmann--Tamburino--Unti) parameter \citep{newman1963empty}.

\details{There are many publications which discuss the physical meaning and interpretation of the NUT parameter.
In the presence of NUT parameters the spacetime is not regular and possesses a bad causal behavior, see, e.g., \cite{GriffithsEtal:2006,GriffithsPodolsky:2006a,GriffithsPodolsky:2006b,GriffithsPodolsky:2007}, see also \cite{Clement:2015cxa} for more recent developments.
}
\medskip

As we shall now demonstrate, the form \eqref{can} of the metric is very convenient for generalizing the Kerr--NUT geometry to the case of a non-vanishing cosmological constant: the functions $\Delta_r(r)$ and $\Delta_y(y)$ simply become fourth-order polynomials of their arguments. To show this, let us impose the vacuum Einstein equations with the cosmological constant $\Lambda$,
$R_{ab}-\frac{1}{2}Rg_{ab}+\Lambda g_{ab}=0$, implying
\be\label{Rab}
R_{a b }=\Lambda g_{a b }\, .
\ee
We consider first the trace equation
\be
R=4\Lambda\, ,
\ee
which takes a very simple form
\be
\partial^2_r \Delta_r+\partial^2_y \Delta_y=-4\Lambda (r^2+y^2)\, ,
\ee
and allows a separation of variables
\be
\partial^2_r \Delta_r+4\Lambda r^2= C\,,\qquad
\partial^2_y \Delta_y+4\Lambda y^2= -C\, .
\ee
The solution to each of these two equations contains 2 independent
integration constants. Thus, together with the separation constant $C$ one has 5 integration
constants. However, the metric \eqref{can} remains invariant under
the following rescaling:
\be
r\to p r,\quad y\to p y,\quad \tau\to p^{-1} \tau,\quad  \psi\to
p^{-3} \psi,\quad
\Delta_r\to p^{4}\Delta_r, \quad \Delta_y\to p^{4}\Delta_y\, .
\ee
This means that one of the five integration constants can be excluded by
means of this transformations. One more constant is excluded by one of the equations of the system \eqref{Rab}. After this all other equations \eqref{Rab} are identically satisfied.
We write the answer in the following standard form:
\be\begin{split}\label{Delr}
\Delta_r&=(r^2+a^2)(1-\Lambda r^2/3)-2Mr\, ,\\
\Delta_y&=(a^2-y^2)(1+\Lambda y^2/3)+2Ny\, .
\end{split}\ee

The four parameters in these functions are $\Lambda$, $M$, $N$, and
$a$. For $\Lambda=0$ and $N=0$ this metric coincides with the Kerr
metric, $M$ and $a$ being the mass and the rotation parameter,
respectively. In addition to these two parameters, a general solution
\eqref{Delr} contains the cosmological constant $\Lambda$, and the NUT parameter $N$. Solutions with non-trivial $N$ contain
singularities on the axis of symmetry in the black hole exterior. The solution with parameters
$M$, $a$, and $\Lambda$ describes a rotating black hole in the
asymptotically de Sitter (for $\Lambda>0$), anti de Sitter (for
$\Lambda<0$), or flat (for $\Lambda=0$) spacetime. A similar
solution containing the NUT parameter $N$ is known as the {\em Kerr--NUT--(A)dS metric}.

\details{The general form of the Kerr--NUT--(A)dS metric in four dimensions was first obtained by \cite{Carter:1968cmp}, and independently re-discovered by \cite{Frolov:1974} by using the Boyer--Lindquist-type coordinates. The charged generalization of the Kerr--NUT--(A)dS metric, which still takes the canonical form \eqref{can}, was studied in \cite{Carter:1968pl,Plebanski:1975}.
In 1976 Plebanski and Demianski considered a metric that is conformal to the Kerr--NUT--(A)dS one and demonstrated that such a class of metrics includes also the accelerating solutions, known as the $C$-metrics \citep{PlebanskiDemianski:1976} (see section~\ref{ssc:PD}).
}

The metric for the Kerr--NUT--(A)dS spacetime can be written in a more symmetric form by writing $x=ir$, $b_x=iM$, and $b_y=N$. This gives
\be\begin{split}\label{Delz}
\Delta_x&=(a^2-x^2)(1+\Lambda x^2/3)+2b_x x\, ,\\
\Delta_y&=(a^2-y^2)(1+\Lambda y^2/3)+2b_y y\,,
\end{split}
\ee
and the Kerr--NUT--(A)dS metric takes the following form:
\be\label{cansym}
\ts{g}=\frac{\Delta_y}{y^2-x^2}(\grad \tau+x^2\grad \psi)^2
+\frac{\Delta_x}{x^2-y^2}(\grad \tau+y^2 \grad \psi)^2
+\frac{y^2-x^2}{\Delta_y} \grad y^2  +\frac{x^2-y^2}{\Delta_x}\grad x^2\, ,
\ee
which is symmetric with respect to the formal substitution $x\leftrightarrow y$. It is this form of the Kerr--NUT--(A)dS metric, which will be generalized to higher dimensions.

\subsubsection{Hidden symmetries}

The off-shell metric \eqref{can} possesses the following property:

\noindent {\bf Theorem:}
{\it The (off-shell) canonical metric \eqref{can} admits a principal tensor
\be
\ts{h}= y \grad y\wedge(\grad\tau-r^2 \grad\psi)
          -r \grad r\wedge (\grad\tau+y^2 \grad\psi)\,, \label{f_ccky}
\ee
which can be generated from a potential $\ts{b}$, $\ts{h}=\ts{db}$, given by
\be\label{f_cckypot}
\ts{b}=-\frac{1}{2}\bigl[(r^2-y^2)\, \grad \tau +r^2y^2\, \grad \psi\bigr]\, .
\ee
}
\medskip

The fact that $\ts{h}$ obeys the closed conformal Killing--Yano equation \eqref{defPKY}
can be verified by a straightforward (but rather long) calculation, or perhaps more efficiently, by using the computer programs for analytic manipulations.
The condition of non-degeneracy follows from the discussion of the Darboux frame below, proving that $\ts{h}$ is a principal tensor. We may therefore apply the results of section~\ref{ssc:PrincipalTensor} and in particular construct the Killing tower associated with $\ts{h}$.

The Hodge dual of $\ts{h}$ is a Killing--Yano tensor $\ts{f}=*\ts{h}$
\be\label{ff_ccky}
\ts{f}= r \grad y\wedge (\grad\tau-r^2 \grad\psi)
          +y \grad r\wedge (\grad\tau+y^2 \grad\psi)\, .
\ee
Using $\ts{h}$ and $\ts{f}$, we can construct the corresponding conformal Killing tensor $Q_{ab}= h_{ac}h_b{}^{c}$ and the Killing tensor $k_{ab}= f_{ac}f_b{}^{c}$. They have the following form:
\begin{align}
  \ts{Q}&= \;\;\,\frac{1}{\Sigma}
    \left[ r^2\Delta_r(\grad \tau+y^2 \grad \psi)^2
    +y^2\Delta_y(\grad \tau-r^2\grad \psi)^2\right]
    +\Sigma\left[\frac{y^2\grad y^2}{\Delta_y}-\frac{r^2 \grad r^2}{\Delta_r}\right]
    \,,\label{Q_ccky}\\
  \ts{k}&= \frac{1}{\Sigma}
    \left[ y^2\Delta_r(\grad \tau+y^2 \grad \psi)^2
    +r^2\Delta_y(\grad \tau-r^2\grad \psi)^2\right]
    +\Sigma\left[\frac{r^2 \grad y^2}{\Delta_y}-\frac{y^2\grad r^2}{\Delta_r}\right]
    \,,\label{k_ccky}
\end{align}
or, in coordinates $(\tau,r,y,\psi)$:
\be\label{HHH}
Q^{a }{}_{b }=\left( \begin{array}{cccc}
y^2-r^2 &~~0~~&~~0~~&-r^2y^2\\
0&~~-r^2~~&~~0~~&0\\
0&~~0~~&~~y^2~~&0\\
-1&~~0~~&~~0~~&0
\end{array}
\right)\,,\;\;
\ee
\be\label{KKK_D}
k^{a }{}_{b }=\left( \begin{array}{cccc}
~~0~~~&~~0~~~&~~0~~~&-r^2y^2\\
~~0~~~&~~-y^2~~&~~0~~~&0\\
~~0~~~&~~0~~~&~~r^2~~~&0\\
~-1~~~&~~0~~~&~~0~~~&r^2-y^2
\end{array}
\right)\,.
\ee

A remarkable property of the off-shell metric  \eqref{can} is that the potential $\ts{b}$, the principal tensor $\ts{h}$, the Killing--Yano tensor $\ts{f}$, and $\ts{Q}$ and $\ts{k}$ in the form \eqref{HHH} and \eqref{KKK_D}, do not depend on functions $\Delta_r(r)$ and $\Delta_y(y)$. In particular this means that they have the same form as in the flat spacetime. Certainly, this property is valid only for the special choice of coordinates. However, the very existence of such coordinates is a non-trivial fact. As we shall see later, this is a generic property which remains valid also for higher-dimensional black holes.

The principal tensor generates the following primary $\ts{\xi}_{(\tau)}$ and secondary $\ts{\xi}_{(\psi)}$ Killing vectors:
\be\label{xixi}
\xi_{(\tau)}^a=\frac{1}{3}\nabla_bh^{ba}=\partial_\tau^a\,,\quad \xi_{(\psi)}^a =-k^a_{\ b}\xi_{(\tau)}^b=\partial_\psi^a\,.
\ee
The primary Killing vector is timelike at infinity, reflecting the fact that the metric is stationary. Moreover, a linear combination $\ts{\xi}_{(\phi)}=a^{-1}\ts{\xi}_{(\psi)}-a \ts{\xi}_{(\tau)}=\ts{\pa}_{\phi}$ has fixed points which form the axis of symmetry---the integral lines of this vector are closed cycles---making the metric axisymmetric.


The constructed Killing vectors $\ts{\xi}_{(\tau)}$ and $\ts{\xi}_{(\psi)}$, together with the Killing tensor $\ts{k}$ and the metric $\ts{g}$, are all independent and mutually (Nijenhuis--Schouten) commute. This means that the corresponding four integrals of motion for the geodesics are all independent and in involution, making the geodesic motion completely integrable.

\subsubsection{Darboux basis and canonical coordinates}

As discussed in section~\ref{ssc:PrincipalTensor}, in the presence of the principle tensor there exists a natural convenient choice of the tetrad, known as the Darboux basis, \eqref{PTcanform}. To illustrate its construction for the Kerr metric, we consider the eigenvalue problem \eqref{Qeigenvectors} for the conformal Killing tensor $\ts{Q}$:
\be
Q^a{}_{b} z^b=\lambda z^a\,,
\ee
where for different eigenvalues $\lambda$ eigenvectors $z^a$ are mutually orthogonal.
Using expression \eqref{HHH}, the characteristic equation
\be\label{dethh}
\det (Q^a_{\ b}-\lambda \delta^a_{\ b})=0\,
\ee
takes the following explicit form:
\be
(\lambda+r^2)^2(\lambda-y^2)^2=0\, ,
\ee
giving the following eigenvalues of $\ts{Q}$: $-r^2$ and $y^2$, where $(r,y)$
are the {\it canonical coordinates}.\footnote{Note that in the Lorentzian signature, the corresponding first eigenvalue is negative. This results in a slightly modified Darboux form of the principal tensor and the metric, see equations \eqref{h4dD} and \eqref{g4dD} below. Let us also notice that although the coordinate $r$ is spacelike or timelike in a generic point, it becomes null at the horizon in the Kerr--NUT--(A)dS spacetime. In a more general case, one of the eigenvalues of the tensor $\ts{h}$ might become null not only on a surface but in some domain \citep{dietz1981space,Taxiarchis:1985}. In what follows we do not consider this case.}
The eigenvectors for each eigenvalue form a two-dimensional plane. Whereas the 2-plane corresponding to $y$ is spacelike, the 2-plane associated with $r$ is timelike. It is then easy to check, that there exists such an orthonormal basis $\{\ts{n},\hat{\ts{n}},\ts{e},\hat{\ts{e}}\}$  which obeys the relations
\be\label{hry}
    h^a_{\ b} \hat{n}^b = -r {n}^a\, , \
    h^a_{\ b} {n}^b = -r \hat{n}^a\, , \
    h^a_{\ b} \hat{e}^b = y e^a\, ,\
    h^a_{\ b} e^b = -y \hat{e}^a\,.
\ee
This basis is defined up to 2-dimensional rotations in each of the 2-planes. We fix this ambiguity by the following  choice of the normalized (in the black hole exterior) Darboux basis:
\begin{equation}\label{Darboux4Dvec}
\begin{aligned}
  \ts{n} &= \sqrt{\frac{\Delta_r}{\Sigma}}\,\cv{r} \,, &
  \hat{\ts{n}} &=\frac1\Sigma\sqrt{\frac{\Sigma}{\Delta_r}}\,
    \bigl(\cv{\psi}+r^2\cv{\tau}\bigr)\,,\\
  \ts{e} &= \sqrt{\frac{\Delta_y}{\Sigma}}\,\cv{y} \,, &
  \hat{\ts{e}} &=\frac1\Sigma\sqrt{\frac{\Sigma}{\Delta_y}}\,
    \bigl(-\cv{\psi}+y^2\cv{\tau}\bigr)\, .\\
\end{aligned}
\end{equation}
The corresponding dual basis of 1-forms is
\begin{equation}\label{Darboux4Dfrm}
\begin{aligned}
  \ts{\nu} &= \sqrt{\frac{\Sigma}{\Delta_r}}\,\grad{r} \,, &
  \hat{\ts{\nu}} &=\sqrt{\frac{\Delta_r}{\Sigma}}\,\bigl(\grad\tau+y^2\grad\psi\bigr)\,,\\
  \ts{\epsilon} &= \sqrt{\frac{\Sigma}{\Delta_y}}\,\grad{y} \,, &
  \hat{\ts{\epsilon}} &=\sqrt{\frac{\Delta_y}{\Sigma}}\,\bigl(\grad\tau-r^2\grad\psi\bigr)\,.\\
\end{aligned}
\end{equation}
In this basis we have
\begin{align}
\ts{h}&=-r \ts{\nu}\wedge \hat{\ts{\nu}}+y \ts{\epsilon}\wedge \hat{\ts{\epsilon}}\,,\label{h4dD}\\
\ts{g}&=- \hat{\ts{\nu}}\hat{\ts{\nu}}+\ts{\nu}\ts{\nu}+\ts{\epsilon}\ts{\epsilon}+\hat{\ts{\epsilon}}\hat{\ts{\epsilon}}\,.\label{g4dD}
\end{align}
Moreover, since the conditions \eqref{specDarbouxframecond} are satisfied,
\be
\hat{\ts{n}}\cdot \grad r=0=\hat{\ts{n}}\cdot \grad y\,,\quad
\hat{\ts{e}}\cdot \grad r=0=\hat{\ts{e}}\cdot \grad y\,,
\ee
we have a {\em special Darboux frame.} For completeness, let us also express $\ts{Q}$ and $\ts{k}$ in this frame, giving
\begin{align}
\ts{Q}&=-r^2(-\hat{\ts{\nu}}\hat{\ts{\nu}}+\ts{\nu}\ts{\nu})+y^2(\ts{\epsilon}\ts{\epsilon}+\hat{\ts{\epsilon}}\hat{\ts{\epsilon}})\,,\\
\ts{k}&=-y^2(-\hat{\ts{\nu}}\hat{\ts{\nu}}+\ts{\nu}\ts{\nu})+r^2(\ts{\epsilon}\ts{\epsilon}+\hat{\ts{\epsilon}}\hat{\ts{\epsilon}})\,.
\end{align}

The principal tensor also naturally determines the {\em canonical coordinates}. This goes as follows.
\begin{itemize}
 \item
 The eigenvalues of the principal tensor, $r$ and $y$, determined by relations \eqref{hry}, are used as two of the canonical coordinates.

\item Since the principal tensor obeys $\lied_{\ts{\xi}} \ts{h}=0$ for both, the primary and secondary Killing vectors $\ts{\xi}_{(\tau)}$ and
$\ts{\xi}_{(\psi)}$, its eigenvalues $(r,y)$ are invariant under the action of $\tau$ and $\psi$ translations.

\item
 {Since the Killing vectors $\ts{\xi}_{(\tau)}$ and $\ts{\xi}_{(\psi)}$ commute, they} spread two-di\-men\-sional invariant surfaces; the values of $r$ and $y$ are constant on each such surface. One can hence use the Killing parameters $\tau$ and $\psi$ as coordinates on the invariant surfaces. This completes the construction of the canonical coordinates $(\tau,r,y,\psi)$.
\end{itemize}

\subsubsection{Principal tensor: immediate consequences}
Let us now summarize the properties of the off-shell metric \eqref{can} that are immediately related to the existence of the  principal tensor ${\ts{h}}$.

\begin{enumerate}

\item The principal tensor $\ts{h}$ exists for any off-shell metric \eqref{can}. It generates the Killing `turret' of symmetries: Killing--Yano tensor $\ts{f}$, conformal Killing tensor $\ts{Q}$, Killing tensor $\ts{k}$, and both generators of the isometries $\ts{\xi}_{(\tau)}$ and $\ts{\xi}_{(\psi)}$.

\item The integrability condition for $\ts{h}$ implies, generalizing the result of \cite{Collinson:1974}, that the spacetime is necessary of the special algebraic type D.  See \cite{Mason:2010zzc} for a higher-dimensional version of this statement.

\item The set  $\{\ts{g},\ts{k},\ts{\xi}_{(\tau)},\ts{\xi}_{(\psi)}\}$ forms a complete set of independent mutually (Nijenhuis--Schouten) commuting symmetries that guarantee complete integrability of geodesic motion, see section~\ref{ssc:Geodesics4D}.

\item The principal tensor also determines the preferred Darboux frame and the canonical coordinates $(\tau,r,y,\psi)$. Such geometrically defined coordinates are convenient for separating the Hamilton--Jacobi and the wave equation, while the Darboux frame is the one where the Dirac equation separates, see section~\ref{ssc:Separability4D}.

\item The  canonical metric \eqref{can} is the most general spacetime admitting the principal tensor, see also section~\ref{ssc:Kerr4duniqueness}.

\end{enumerate}

\subsection{Uniqueness of the Kerr metric}
\label{ssc:Kerr4duniqueness}

The Kerr metric was originally obtained by \cite{Kerr:1963} as `one of many' special algebraic type solutions  (see \cite{Teukolsky:2014vca} for a historical account). A few years later the solution was rediscovered by \cite{Carter:1968cmp} by imposing a special metric ansatz (assuming two commuting Killing vectors) and by requiring that both the Hamilton--Jacobi and wave equations should be solvable by a method of separation of variables (see also \cite{Debever:1971}). This not only allowed Carter to rederive the Kerr metric but to generalize it and to include the cosmological constant and the NUT parameter.

\details{Carter's derivation actually fits into the context of the theory of separability structures discussed in section~\ref{ssc:separabilitystructures}. Considering the $r=2$ separability structure in coordinates $(\tau,\psi,r,y)$, the separability of the Hamilton--Jacobi equation and, in an Einstein space, also of the Klein--Gordon equation is guaranteed for any $2\times 2$ St\"ackel matrix $\mathrm{M}$ and any two matrices $\mathrm{N}_r=\mathrm{N}_r(r)$ and $\mathrm{N}_y=\mathrm{N}_y(y)$ through relation \eqref{KTsepStruc}. In particular, the following choice leads to the Carter's canonical metric \eqref{can}:
\be\nonumber
\mathrm{M}=\left( \begin{array}{cc}
\frac{r^2}{\Delta_r}~&~\frac{y^2}{\Delta_y}\\
-\frac{1}{\Delta_r}~&~\frac{1}{\Delta_y}
\end{array}
\right),\;
\mathrm{N}_r=-\frac{1}{\Delta_r^2}\left( \begin{array}{cc}
r^4~&~r^2\\
r^2~&~1
\end{array}
\right),\;
\mathrm{N}_y=\frac{1}{\Delta_y^2}\left( \begin{array}{cc}
y^4~&~-y^2\\
-y^2~&~1
\end{array}\right).
\ee
See also \cite{Kolar:2015hea} for a higher-dimensional version of Carter's original argument.
}

It is well known that any stationary and asymptotically flat black hole solution  of the Einstein--Maxwell equations (with non-degenerate horizon) is the Kerr--Newman metric. The extended discussion of this uniqueness theorem and references can be found, e.g., in \cite{Mazur:2000pn,Hollands:2012xy}.
It is interesting that another version of the uniqueness theorem can be formulated:
\\[0.5ex]
{\noindent{\bf Theorem}:}
{\it The most general vacuum with $\Lambda$ solution of the Einstein equations that admits a principal tensor is the Kerr--NUT--(A)dS geometry.}
\\[0.5ex]
It is a special case of the higher-dimensional uniqueness theorem \citep{Houri:2007xz,Krtous:2008tb}, which will be discussed in chapter~\ref{sc:hshdbh}. See \cite{dietz1981space,Taxiarchis:1985} for earlier studies of this issue in four dimensions, where also exceptional metrics corresponding to the null forms of the principal tensor are discussed.

The proof of this statement proceeds in two steps. First, it can be shown that the most general off-shell metric that admits the principal tensor has to admit two commuting Killing vectors and takes the form \eqref{can}. Second, by imposing the Einstein equations, the remaining metric functions are uniquely determined and depend on 5 independent constants related to the mass, angular momentum, NUT charge, and the cosmological constant, yielding the Kerr--NUT--(A)dS spacetime. Note that if in addition we require regularity outside the horizon and in particular the absence of cosmic strings (see section~\ref{ssc:RemarksAngles}), the NUT charge has to vanish, and the Kerr--(A)dS geometry is recovered.

We have yet another observation.
Employing solely the principal tensor one can construct the principal electromagnetic field, given by \cite{Frolov:2017bdq}
\be\label{F4d}
\ts{F}=e\bigl(\ts{d\xi}+\frac{2}{3}\Lambda \ts{h}\bigr)\,,\quad \ts{\xi}=\frac{1}{3}\covd\cdot \ts{h}\,,
\ee
which solves the test Maxwell equations for the metric \eqref{can} obeying \eqref{Rab}. In four dimensions, this field can be backreacted on the geometry, provided a suitable choice of the metric functions $\Delta_r$ and $\Delta_y$ is done in \eqref{can}, to yield an electrically charged Kerr--NUT--(A)dS geometry. Moreover, if instead of \eqref{F4d} one considers the sourceless `aligned with $\ts{h}$' electromagnetic field studied by  \cite{Krtous:2007}, we recover the Kerr--NUT--(A)dS solution which is both electrically and magnetically charged. In this sense, the four-dimensional charged Kerr--NUT--(A)dS solution is uniquely defined by the principal tensor.

\subsection{Geodesics}
\label{ssc:Geodesics4D}

\subsubsection{Integrals of motion}

Since the off-shell metric \eqref{can} besides two Killing vectors $\ts{\xi}_{({\tau)}}$ and $\ts{\xi}_{(\psi)}$ possesses also the irreducible rank 2 Killing tensor $\ts{k}$, there exist the following four integrals of geodesic motion:
\begin{align} \label{CONS}
g^{a b} p_a p_b&=-m^2\,,\quad
k^{a b}p_a p_b=K \,,\quad
p_{\tau}\equiv \xi_{(\tau)}^a p_a=-E\,,\\
p_{\psi}&\equiv \xi_{(\psi)}^a p_a=L_{\psi}=aL_{\phi}-a^2E\,.
\end{align}
Here  $p_a$ is the four-momentum of the particle of  mass $m$, and $E$ and $L_{\phi}$ are its energy and angular momentum, respectively. The last conserved quantity, $K$, is the analogue of the Carter constant for the off-shell metric.  The existence of 4 independent commuting integrals of motion makes the geodesic motion completely integrable.

The last two relations of the system \eqref{CONS} can be used to express $p_r$ and $p_y$ as functions of the integrals of motion
\be\label{pry}
p_r=\pm\frac{\sqrt{{\cal X}_r}}{\Delta_r}\,,\qquad
p_y=\pm\frac{\sqrt{{\cal X}_y}}{\Delta_y}\, ,
\ee
where
\be \label{RY}
{\cal X}_r=(E r^2-L_{\psi})^2-\Delta_r(K+m^2 r^2)\,,\quad
{\cal X}_y=-(E y^2+L_{\psi})^2+\Delta_y(K- m^2 y^2)\, .
\ee
The signs $\pm$ in \eqref{pry} are independent; the sign change occurs at turning points where ${\cal X}_r=0$ and ${\cal X}_y=0$, respectively.

\subsubsection{First-order form of geodesic equations}

As a consequence of complete integrability, the geodesic equations can be written in a {\em first-order form}, that is, as a set of the first-order differential equations. Let us denote by the ``dot''  a derivative with respect to the affine parameter $\sigma$ (see section~\ref{ssc:partcurvspc}). Then using the relation
 \be\label{pmu}
 p_a=g_{ab}\dot{x}^b\, ,
 \ee
 we rewrite equations~\eqref{CONS}  in the form
\be\label{definitionofK1}
\xi_{(\tau) a} \dot{x}^a=-E\,,\quad
\xi_{(\psi) a}\dot{x}^a=L_{\psi}\,,\quad
g_{a b}\dot{x}^a\dot{x}^b=-m^2\,,\quad
K_{a b}\dot{x}^a\dot{x}^b=K \, .
\ee
These four equations for $\dot{x}^a=(\dot{\tau},\dot{r},\dot{y},\dot{\psi})$ can be solved to obtain the following set of the first order ordinary differential equations:
\begin{align}
\Sigma\,\dot{r}&=\pm \sqrt{{\cal X}_r}\, ,\label{FODE}\\
\Sigma\,\dot{y}&=\pm \sqrt{{\cal X}_y}\, ,\label{eth}\\
\Sigma\,\dot{\tau}&=\frac{ r^2( E r^2-L_{\psi})}{\Delta_r}
   -\frac{y^2( E y^2+L_{\psi})}{\Delta_y}\, ,\label{tteq}\\
\Sigma\,\dot{\psi}&=\frac{ E r^2-L_{\psi}}{\Delta_r}
   + \frac{E y^2+ L_{\psi}}{\Delta_y}\, ,\label{tteq22}
\end{align}
with ${\cal X}_r={\cal X}_r(r)$ and ${\cal X}_y={\cal X}_y(y)$ given by \eqref{RY}, and $\Sigma=r^2+y^2$. As earlier, signs $\pm$ in the equations \eqref{FODE} and  \eqref{eth} are independent. The change of the signs in these equations occurs at turning points, where ${\cal X}_r=0$ and ${\cal X}_y=0$, respectively. The convenience of the usage of the parameter $\sigma$ is that the equations of motion \eqref{FODE}--\eqref{tteq} allow for a simple limit $m\to 0$ (in ${\cal X}_r$ and ${\cal X}_y$) and hence can be used for {\em massless particles} as well.\footnote{In fact, for massless particles one could instead of the Killing tensor $\ts{k}$ use the conformal Killing tensor $\ts{Q}$, defining $Q^{a b}l_a l_b=K$ in \eqref{CONS} and similarly in \eqref{definitionofK1}. As can be expected, in the massless limit $m\to 0$, the resultant equations \eqref{FODE}--\eqref{tteq} remain the same.}

Instead of  the affine parameter $\sigma$, one can use another parameter $\tilde{\sigma}$, so that
\be\label{sigmarepar}
\frac{d{\sigma}}{d\tilde{\sigma}}=\Sigma\, .
\ee
For such a parametrization, the left hand side of the system of equations  \eqref{FODE}--\eqref{tteq} contains a derivative $dx^a/d\tilde{\sigma}$. This effectively decouples the first two equations \eqref{FODE} and \eqref{eth}, which can now be solved by integration. The result is then plugged to the last two equations \eqref{tteq22} and \eqref{tteq} which yield integrals for $\psi$ and $\tau$.\footnote{As we shall see, a similar trick does not work in higher dimensions. However, proceeding differently, the velocity equations can still be decoupled and in principle solved by integration, see section~\ref{ssc:IntegrGeodMot}.}

To translate \eqref{FODE}--\eqref{tteq} to the Boyer--Lindquist coordinates $(t,r,\theta, \phi)$, one should use the following relations:
\be\label{chvar}
t=\tau+a^2 \psi\,,\quad \phi=a \psi\,,\quad y=a\cos\theta\,,\quad L_{\psi}=a L_{\phi} -a^2 E\, .
\ee
Taking into account these remarks, it is easy to check that the equations \eqref{FODE}--\eqref{tteq} re-written in the Boyer--Lindquist coordinates take the standard form, which can be found, e.g. in  \cite{Carter:1968rr,Bardeen:1973tla,MTW}. Detailed discussion of particle and light motion in the four-dimensional Kerr--NUT--(A)dS spacetime can be found in \cite{Hackmann:2011wp,Grenzebach:2014fha}.

\subsubsection{Action-angle variables}

Instead of studying the details of particle's orbits, one might be interested in such `global' characteristics as, for example, the motion frequencies. A useful tool for this is provided by an {\em action-angle formalism}. This formalism is also useful for studying the adiabatic invariants and for the development of the perturbation theory when a system slightly differs from a completely integrable one. For the comprehensive discussion of this subject, we refer the reader to the remarkable books by \cite{Goldstein:book} and \cite{Arnold:book}. Here we just briefly discuss a construction of the action-angle variables for a free particle moving in the metric \eqref{can}. See appendix~\ref{apx:pscint} for a general introduction to this subject.

For our dynamical system the coordinate $\phi$ is cyclic while the value of the coordinate $y$  is bounded and changes in the interval $(y^-,y^+)$. The system admits different types of trajectories, depending on the concrete value of the integrals of motion $\{m^2, K, E,L_\phi\}$ so that the range of the coordinate $r$ may be unbounded. Let us here focus on the case of {\em bounded trajectories} for which the radial coordinate changes in the interval $(r^-,r^+)$. In such a case the corresponding level set for $(r,y,\phi)$ sector is a compact three-dimensional Lagrangian submanifold which, according to the general theorem, is a three-dimensional torus. One can choose three independent cycles on this torus as follows. Let us fix $y$ and $\phi$ and consider a closed path, which propagates from the minimal radius $r^-$ to the maximal radius $r^+$, and after this returns back to $r^-$ with opposite sign of the momentum. Another path is defined similarly for the $y$-motion. The third pass $r=$const, $y=$const is for the $\phi$-motion.

This allows us to introduce the following {\it action variables}, $I_i=(I_r,I_y,I_{\phi})$ for `spatial directions'
\begin{equation}\label{IIr}
\begin{aligned}
&I_r=I_r(m^2,K,E,L_\phi)
  =\frac{1}{\pi} \int_{r^-}^{r^{+}} dr \frac{\sqrt{{\cal X}_r}}{\Delta_r}\, ,\\
&I_y=I_y(m^2,K,E,L_\phi)
  =\frac{1}{\pi} \int_{y^-}^{y^{+}} dy \frac{\sqrt{{\cal X}_y}}{\Delta_y}\,,\\
&I_\phi= L_{\phi}\,.
\end{aligned}
\end{equation}
Here $r^{\pm}$ and  $y^{\pm}$ are turning points of ${r}$ and ${y}$, respectively, and we used the fact that $\phi$ is a cyclic coordinate with period $2\pi$.

Since the Hamiltonian is a function of integrals of motion, c.f. \eqref{HofP}, it can also be written in terms of the action variables as
\be
H=H(I_i,E)\, .
\ee
The \emph{angle variables} $\Phi_i$ are introduced as conjugates to $I_i$.
The Hamilton equations of motion in these variables take the form
\be
\dot{I}_i=0\,,\qquad \dot{\Phi}_i=\omega_i=\frac{\partial H}{\partial I_i}(I_i,E)\, .
\ee
The (constant) quantities $\omega_i$ are characteristic frequencies. If their ratios are not rational, the trajectories of the particle are not periodic.

We will return to the discussion of the action-angle variables in more details later, when discussing geodesics in the higher-dimensional Kerr--NUT--(A)dS spacetimes.

\subsubsection{Parallel transport}

There are many problems with interesting astrophysical applications that require solving the parallel transport equations in the Kerr metric. One of them is a study of a star disruption during its close encounter with a massive black hole, see, e.g., \cite{Frolov:1994jr} and references therein.

Let us consider a timelike geodesic in the Kerr geometry and denote by $\ts{u}$ its tangent vector. We have seen in section~\ref{ssc:KYfamily} that $\ts{w}=\ts{u}\cdot \ts{f}$, where $\ts{f}$ is the Killing--Yano tensor \eqref{ff_ccky}, is parallel-transported along the geodesic, $\covd_{\!\ts{u}}\ts{w}=0$, c.f. \eqref{wfdef}. This means that a bi-vector $\ts{*F}\equiv \ts{u}\wedge \ts{w}=\ts{u}\wedge (\ts{u}\cdot \ts{f})$ is also parallel-propagated,
$\covd_{\!\ts{u}}\ts{*F}=0\, .$
Since the Hodge duality operator $*$ commutes with the covariant derivative one also has
\be
\covd_{\!\ts{u}}\ts{F}=0\, .
\ee
That is, a 2-dimensional plane $\ts{F}=\ts{u}\cdot (\ts{u}\wedge \ts{h})$ is orthogonal to $\ts{*F}$ and parallel-transported along the geodesic. Let $\ts{e}_1$ and $\ts{e}_2$ be two orthonormal vectors which spread this 2-plane, and $\ts{m}$ be a complex null vector $\ts{m}=\frac{1}{\sqrt2}(\ts{e}_1+i \ts{e}_2)$. It is easy to show that one can find such a real function $\varphi$ so that $\ts{m}\exp{(i\varphi)}$ is parallel-transported along the geodesic. Thus one obtained a parallel-transported  basis $(\ts{u},\ts{w},\ts{m},\bar{\ts{m}}) $ \citep{marck1983solution}.
Similar procedure also works for constructing a parallel-transported basis along null geodesics, see \cite{marck1983parallel}. Interestingly, both these constructions
can be generalized to higher dimensions. We shall discuss this subject in section~\ref{ssc:PralTransp}.

The principal tensor also allows one to solve  an equation for a propagation of polarization of electromagnetic  waves in the spacetime with the metric (\ref{can}).
In the leading order of the geometric optics approximation  the Maxwell equations reduce to the equations for null geodesics. A vector of a linear polarization $\ts{q}$ is orthogonal to null geodesics and parallel-propagated along them.

Let us consider first an arbitrary geodesic  and denote by $\ts{u}$ its tangent vector. Let $\ts{q}$ be a parallel-propagated vector along this  geodesic, $\covd_{\!\ts{u}}\ts{q}=0$.
Then the quantity $ \ts{q}\cdot\ts{f}\cdot \ts{u}= -\ts{q}\cdot\ts{w}$ is obviously a constant along any timelike or null geodesic. For null geodesics there exists an additional conserved quantity defined by the principal tensor ${\ts{h}}$. Let $\ts{l}$ be a tangent vector to a null geodesic in an affine parametrization, $\covd_{\!\ts{l}}\ts{l}=0$, and let $\ts{q}$ be a parallel-propagated vector along it obeying $\ts{q}\cdot \ts{l}=0$. Then the following quantity: $\ts{q}\cdot \ts{h}\cdot\ts{l}$ is also conserved. Indeed,
\be\label{qqll}
  \covd_{\!\ts{l}}(\ts{q}\cdot \ts{h}\cdot \ts{l})
  =\ts{q}\cdot (\covd_{\!\ts{l}}\ts{h})\cdot \ts{l}
  =\ts{q}\cdot(\ts{l}\wedge\ts{\xi})\cdot\ts{l}=0\,,
\ee
where we used the closed conformal Killing--Yano condition \eqref{bezCCKY}, $\ts{l}^2=0$, and $\ts{q}\cdot \ts{l}=0$.

 Denoting by $\ts{z}=\ts{h}+i * \ts{h}$, we just showed
that the following complex number:
\be\label{shift}
\ts{q}\cdot \ts{z}\cdot \ts{l}\,
\ee
is constant along the null ray \citep{Walker:1970un}. This result allows one to easily find a polarization of a photon after its scattering by a rotating black hole and determine the angle of the corresponding Faraday rotation \citep{Connors:1977,ConnorsEtal:1980,Ishihara:1987dv}.

\subsection{Separation of variables in the canonical metric}
\label{ssc:Separability4D}

In this section we show that the fundamental physical equations do separate in the (off-shell) canonical spacetime \eqref{can}. We also discuss the intrinsic characterization of such separability, linked to the existence of the principal tensor. In particular, we concentrate on the Hamilton--Jacobi, Klein--Gordon, and Dirac equations, and do not discuss the electromagnetic and gravitational perturbations. Whereas for the Maxwell equations the link between separability and the principal tensor still can be found, e.g. \cite{Benn:1996su, Araneda:2016iwr}, this is not obvious for the gravitational perturbations.

\subsubsection{Hamilton--Jacobi equation}

Equations \eqref{FODE}--\eqref{tteq} allow one to find trajectories of massive particles in the
Kerr spacetime. This problem can be alternatively studied by using the Hamilton--Jacobi equation, following Carter's original paper \citep{Carter:1968rr}.

Using the inner time variable $\sigma$, related to the particle proper time $\tau=m\sigma$, see section~\ref{ssc:partcurvspc}, the Hamiltonian of a free particle with mass $m$ reads
\be
H=\frac{1}{2} g^{ab}p_a p_b\, .
\ee
Since this is an autonomous system ($H$ does not explicitly depend on $\sigma$), the time-dependent Hamilton--Jacobi equation
\be
\frac{\pa {\bar S}}{\pa {\sigma}}+H(x^a,p_a)\big|_{p_a={\bar S}_{,a}}=0\,
\ee
can be solved by the ansatz
\be
{\bar S}(x^a,\sigma)=\frac{1}{2} m^2 \sigma +S(x^a)\,  .
\ee
This results in the following time-independent Hamilton--Jacobi equation:
\be\label{HJm}
g^{a b} S_{,a}  S_{,b} +m^2=0\,,
\ee
for the Hamilton's principal function $S(x^a)$, see section~\ref{ssc:HamJacEq} for more details.
A solution of this equation, which contains 4 independent constants, is a complete integral.

Let us now study the Hamilton--Jacobi equation \eqref{HJm} in the canonical spacetime \eqref{can}. Since the coordinates $\tau$ and $\psi$ are cyclic, the Hamilton's function $S$ can be written in the form
\be\label{SSS0}
S(x^a)=-E \tau+L_{\psi}\psi +\hat{S}(r,y)\, .
\ee
It is a remarkable property of canonical coordinates $(\tau, r, y, \psi)$ that a further additive separation of variables is possible.  Namely,  by substituting
\be\label{SSS}
S(x^a)=-E \tau+L_{\psi}\psi +S_r(r)+S_y(y)\,
\ee
into \eqref{HJm} one finds a consistent equation provided that functions $S_r$ and $S_y$ satisfy the following ordinary differential equations:
\be \label{SRY}
\Delta_r(\partial_r S_r)^2-\frac{{\cal X}_r}{\Delta_r}=0\, ,\quad
\Delta_y (\partial_y{ S_y})^2-\frac{{\cal X}_y}{\Delta_y}=0\, .
\ee
Here, ${\cal X}_r$ and ${\cal X}_y$ are given by \eqref{RY} and the quantity $K$ in these functions plays a role of the separation constant.
The solution to \eqref{SRY} can be calculated in terms of the  elliptic integrals,
\be\label{integralsSrSy}
S_r(r)=\pm\int^r_{r_0} dr \frac{\sqrt{{\cal X}_r}}{\Delta_r}\,,\qquad
S_y(y)=\pm\int^y_{y_0} dy \frac{\sqrt{{\cal X}_y}}{\Delta_y}\, .
\ee
The choice of the initial coordinates $r_0$ and $y_0$ is not important, since their change just adds a constant to $S$; in the case when the motion has turning points it is convenient to choose $r_0$ and $y_0$ to coincide with them. Since the solution $S$ given by \eqref{SSS} depends on coordinates $x^a$ and four independent constants $P_a=(m^2,K,E,L_{\psi})$, it is a {\em complete integral}. As discussed in section~\ref{ssc:HamJacEq}, its existence implies complete integrability of geodesic motion in canonical spacetimes.

\details{The separability of the Hamilton--Jacobi equation \eqref{HJm} is intrinsically characterized by the existence of the {\em separability structure}, see section~\ref{ssc:separabilitystructures}. Namely, the Killing tensors $\ts{g}$ and $\ts{k}$, together with the Killing vectors $\ts{\xi}_{(\tau)}$ and $\ts{\xi}_{(\psi)}$ satisfy \eqref{PoissonK}. Moreover, the Killing tensors have in common the following eigenvectors: $\ts{\partial}_r$ and $\ts{\partial}_y$ that together with $\ts{\xi}_{(\tau)}$ and $\ts{\xi}_{(\psi)}$ obey \eqref{sepdeltacond}. Hence all the requirements of the theorem in section~\ref{ssc:separabilitystructures} are satisfied and the separability of the Hamilton--Jacobi equation is justified.
}

It turns out that in four dimensions the Hamilton--Jacobi equation separates also in the standard Boyer--Lindquist coordinates, giving a complete integral in the form
\be\label{SSSt}
S(x^a)=-E t +L_{\phi} \phi +S_r(r)+S_\theta(\theta)\,,
\ee
where $S_r$ is formally given by the same integral \eqref{integralsSrSy}, with $L_{\psi}=a (L_{\phi}-a E)$.

The parameters $P_a$ can be identified with new momenta in the phase space. We denote the canonically conjugate coordinates by $Q^a$. The Hamilton's  function $S(x^a,P_a)$ is a generating function of the canonical transformation $(x^a,p_a)\to (Q^a,P_a)$,
\be
p_a=\frac{\partial S}{\partial x^a}\,,\qquad Q^a=\frac{\partial S}{\partial P_a}\, .
\ee
The new coordinates are of the form
\begin{equation}\label{Q1234}
\begin{aligned}
Q^1&=\frac{\partial S}{\partial m^2}
    =\frac{\partial S_r}{\partial m^2}+\frac{\partial S_y}{\partial m^2}\,,\quad&
Q^3&=\frac{\partial S}{\partial E}
    =-\tau +\frac{\partial S_r}{\partial E}
     +\frac{\partial S_y}{\partial E}\,,\\
Q^2&=\frac{\partial S}{\partial K}
    =\frac{\partial S_r}{\partial K}
    +\frac{\partial S_y}{\partial K}\,,\quad&
Q^4&=\frac{\partial S}{\partial L_{\psi}}
    ={\psi +\frac{\partial S_r}{\partial L_{\psi}}
    +\frac{\partial S_y}{\partial L_{\psi}}}\,  .
\end{aligned}
\end{equation}
The first two equations allow one to write the `old' coordinates $r$ and $y$ in terms of $P_a$ and the `new' coordinates $Q^1$ and $Q^2$. After this the last two equation define $\tau$ and $\psi$ as functions of $Q^a$ and $P_a$.

The parameters {$(m^2,K,E,L_{\psi})$} denote values of the integrals of motion $P_a$ on the phase space. A four-dimensional subspace of the phase space, determined by the fixed values of these parameters, is a Lagrangian submanifold (see section~\ref{ssc:complintgrb}). The coordinates $Q^a$ conjugate to $P_a$ have simple evolution
\be
\frac{dQ^a}{d\sigma}=\frac{\partial H}{\partial P_a}\, .
\ee
Thus the equation of motion in the new coordinates are
\be
{Q^1=\frac{1}{2}\sigma+\mbox{const}} \,,\quad
Q^2=\mbox{const}\,,\quad
Q^3=\mbox{const}\,,\quad
Q^4=\mbox{const}\,.
\ee
Let us notice, that these equations can also be written in the form
$\partial {\bar S}/\partial P_a=\text{const}$.

\subsubsection{Separability of the Klein--Gordon equation}

Let us next concentrate on the massive Klein--Gordon equation  in the spacetime \eqref{can}. Denote by $\Box$ the scalar wave operator
\be
\Box=g^{ab}\nabla_a\nabla_b\, .
\ee
Then the massive Klein--Gordon equation (which is essentially an eigenfunction equation for the wave operator) reads
\be\label{box}
(\Box -m^2)\Phi=\frac{1}{\sqrt{-g}}\partial_a\bigl(\sqrt{-g}g^{ab}\partial_b\Phi\bigr)-m^2\Phi=0\,,
\ee
where the latter expression for $\Box$ is a well known identity.
Using the expression \eqref{detcan} for the determinant of the canonical metric, and the
formula \eqref{caninv} for the inverse metric, we write this equation in the following form:
\begin{equation}\label{gbox}
\begin{split}
  &\sqrt{-g}(\Box -m^2)\Phi=\pa_r(\Delta_r\pa_r\Phi)+\pa_y(\Delta_y\pa_y\Phi)\\
    &-\frac{1}{\Delta_r}(r^2\pa_\tau+\pa_{\psi})^2\Phi
     + \frac{1}{\Delta_y}(y^2\pa_\tau-\pa_{\psi})^2\Phi -m^2(r^2+y^2)\Phi=0\,.\quad
\end{split}
\end{equation}
This equation allows the multiplicative separation of variables
\be\label{sepf}
\Phi=e^{-i E \tau} e^{i L_{\psi} \psi} R(r)Y(y)\,,
\ee
giving the following ordinary differential equations for functions $R(r)$ and $Y(y)$:
\be\label{KGSEP}
\pa_r(\Delta_r\pa_r R)+\frac{{\cal X}_r}{\Delta_r} R=0\, ,\quad
\pa_y(\Delta_y\pa_y Y)+\frac{{\cal X}_y}{\Delta_y}Y=0\, .
\ee
Here, functions ${\cal X}_r$ and ${\cal X}_y$ are the same as in equations \eqref{RY}. They contain all the parameters $(m^2,K,E,L)$, with parameter $K$ playing the role of a separation constant.

Both equations \eqref{KGSEP} have a similar form---they can be written as the second-order ordinary differential equations with polynomial coefficients. However, there is an essential difference between them. The coordinate $y$ is restricted to the interval $y\in [-a,a]$ and the endpoints of this interval, $y=\pm a$, are singular points of the $y$-equation. Regularity of $Y$ at these points cannot be satisfied for an arbitrary value of the parameter $K$, therefore, for regular solutions $K$ has a discrete spectrum. In other words, one needs to solve the Sturm--Liouville boundary value problem. The solutions of this problem for the scalar field are called \emph{spheroidal wave functions}. They  were studied in detail by \cite{Flammer:book}. Similar spherical harmonics for fields of higher spin are called \emph{spin-weighted spheroidal harmonics}, see, e.g. \cite{Fackerell:1977}.

The separability of the Klein--Gordon equation \eqref{box} can be intrinsically characterized by the existence of the following complete set of mutually commuting operators:
$\{\Box, \mathcal{K}, \mathcal{L}_{\tau}, \mathcal{L}_\psi\}$, where
\begin{equation}
  \Box=\nabla_{\!a} g^{ab}\nabla_{\!b}\,,\quad
  \mathcal{K}=\nabla_{\!a}k^{ab}\nabla_{\!b}\,,\quad
  \mathcal{L}_\tau={i}\xi^a_{(\tau)}\nabla_{\!a}\,,\quad
  \mathcal{L}_\psi={i}\xi^a_{(\psi)}\nabla_{\!a}\,.
\end{equation}
The separated solution \eqref{sepf} is simply the `common eigenfunction' of these operators.
Let us note that whereas the operators constructed from Killing vectors always commute with the box operator, those constructed from a Killing tensor result in a general case in `anomalies'  obstructing this commutation. General conditions under which the anomalies vanish were studied by \cite{Carter:1977pq} (see also \cite{Kolar:2015cha} for a recent study in a general dimension). In particular, it turns out that when the Killing tensor is constructed as a square of a Killing--Yano tensor (as in our case) the anomalies vanish and the commutation is guaranteed. We finally mention that since the canonical metric \eqref{can} admits a separability structure with common eigenevectors of the Killing and Ricci tensors, the theorem discussed in section~\ref{ssc:separabilitystructures} applies and the separability of the Klein--Gordon equation is guaranteed.

\subsubsection{Separability of the Dirac equation}

As we already mentioned  the equations for massless fields with non-zero spin in the Kerr metric allow complete separation of variables. This was discovered  by \cite{Teukolsky:1972,Teukolsky:1973}. Namely, he demonstrated that these equations can be decoupled and reduced to one scalar (master) equation, which in its turn allows a complete separation of variables. Later Wald showed that the solution of the master equation allows one to re-construct a solution of the original many-component equation \citep{Wald:1978vm}.

To separate variables in the {\em massive Dirac equation} in the Kerr metric, \cite{Chandrasekhar:1976,Chandrasekhar:1983} used another approach. Namely, he used a special ansatz for the spinor solution, and demonstrated that this allows one to obtain the separated equations for the functions which enter this ansatz. It turns out that the separability of the massive Dirac equation in the Kerr spacetime is also connected with its hidden symmetry, and, as a result, it also takes place in the canonical metric \eqref{can} for an arbitrary choice of the metric functions $\Delta_r(r)$ and $\Delta_y(y)$. Let us now demonstrate this result.

The Dirac equation in curved spacetime writes as
\be\label{DiracSen}
\bigl(\gamma^a\nabla_a+m\bigr)\psi=0\, .
\ee
Here $\gamma^a$ are gamma matrices, $\gamma^a=(\gamma^0,\gamma^1,\gamma^2,\gamma^3)$, obeying {$\{\gamma^a,\gamma^b\}=2g^{ab}$}, and
$\nabla_a$ stands for the spinorial covariant derivative, defined as
\be
\nabla_a=\pa_a+\frac{1}{4}{\omega}_{abc}\gamma^b\gamma^c\, .
\ee
We denoted by $\pa_{a}=\ts{e}_a\cdot\pa$ a derivative in the direction of $\ts{e}_a$ and $\omega_{abc}$ are the standard spin coeficients with respect to frame $\ts{e}^a$. The 1-forms of the curvature, $\ts{\omega}^b{}_{c}=\ts{e}^a\omega_{a}{}^{b}{}_{c}$, obey the Cartan equations
$\tens{de}^a+\tens{\omega}^a_{\ b}\wedge \tens{e}^b=0$.

To study the Dirac equation \eqref{DiracSen} in the canonical spacetime \eqref{can}, let us chose the basis of 1-forms as $\ts{e}^a=(\hat{\ts{\nu}},  \ts{\nu}, \hat{\ts{\epsilon}},  \ts{\epsilon})$, \eqref{Darboux4Dfrm}, and the dual basis of vectors as
$\ts{e}_a=(\hat{\ts{n}},  \ts{n}, \hat{\ts{e}},  \ts{e})$, \eqref{Darboux4Dvec}.
The spin connection is then obtained from
the Cartan's equation 
and is given as follows:
\begin{equation}\label{spcon}
\begin{aligned}
\tens{\omega}_{\hat \nu \nu}&=-A\hat{\ts{\nu}}-B\hat{\ts{\epsilon}}\,,\quad
\tens{\omega}_{\hat \nu \hat \epsilon}=-B\ts{\nu}+C\ts{\epsilon}\,,\quad
\tens{\omega}_{\hat \nu \epsilon}=-D\hat{\ts{\nu}}-C\hat{\ts{\epsilon}}\,,\\
\tens{\omega}_{\nu \hat \epsilon}&=B\hat{\ts{\nu}}
-E \hat{\ts{\epsilon}}\,,\quad
\tens{\omega}_{\nu {\epsilon}}=D\ts{\nu}-E\ts{\epsilon}\,,\quad
\tens{\omega}_{\hat \epsilon {\epsilon}}=-C\hat{\ts{\nu}}-F\hat{\ts{\epsilon}}\,,
\end{aligned}
\end{equation}
where
\begin{equation}
\begin{aligned}
A&=\frac{d}{dr}\Bigl(\sqrt{\frac{\Delta_r}{\Sigma}}\Bigr)\,,\quad
B=\frac{r}{\Sigma}\sqrt{\frac{\Delta_y}{\Sigma}}\,, \quad
C=-\frac{y}{\Sigma}\sqrt{\frac{\Delta_r}{\Sigma}}\,,\\
D&=\frac{y}{\Sigma}\sqrt{\frac{\Delta_y}{\Sigma}}\,,\quad
E=\frac{r}{\Sigma}\sqrt{\frac{\Delta_r}{\Sigma}}\,,\quad
F=-\frac{d}{dy}\Bigl(\sqrt{\frac{\Delta_y}{\Sigma}}\Bigr)\,.
\end{aligned}
\end{equation}
Using the connection \eqref{spcon} and the inverse basis \eqref{Darboux4Dvec}, we thus find the following explicit form of the Dirac equation:
\begin{equation}\label{Diracexplicit}
\begin{split}
\Big[&\frac{\gamma^0}{\Sigma}\sqrt{\frac{\Sigma}{\Delta_r}}\Bigl(\partial_\psi+r^2\partial_\tau\Bigr)+
\gamma^1\Bigl(\frac{A}{2}+E+\sqrt{\frac{\Delta_r}{\Sigma}}\partial_r\Bigr)+
\frac{\gamma^2}{\Sigma}\sqrt{\frac{\Sigma}{\Delta_y}}\,
    \Bigl(-\partial_\psi+y^2\partial_\tau\Bigr)\\
    &\qquad\qquad\quad+   \gamma^3\Bigl(D-\frac{F}{2}+\sqrt{\frac{\Delta_y}{\Sigma}}\partial_y\Bigr)+
    \frac{B}{2}\gamma^{0}\gamma^1\gamma^2+\frac{C}{2}\gamma^0\gamma^2\gamma^3+m\Big]\psi=0\,.
\end{split}
\end{equation}

To proceed further, we use the following representation of gamma matrices:
\be
\gamma^0=\left(
\begin{array}{cc}
0 & -I\\
I& 0
\end{array}
\right)\, ,\quad
\gamma^1=\left(
\begin{array}{cc}
\ 0 & \ I\\
\ I& \ 0
\end{array}
\right)\,, \quad
\gamma^2=\left(
\begin{array}{cc}
\sigma^2 & 0\\
0& -\sigma^2
\end{array}
\right)\, ,\quad
\gamma^3=\left(
\begin{array}{cc}
\sigma^1 & 0\\
0& -\sigma^1
\end{array}
\right)\, ,
\ee
where $\sigma^i$ are the Pauli matrices.
In this representation, the separation of the Dirac equation can be achieved with the ansatz
\be
\psi=\left(
\begin{array}{c}
(r-iy)^{-1/2}R_+Y_+\\
(r+iy)^{-1/2}R_+Y_-\\
(r+iy)^{-1/2}R_-Y_+\\
(r-iy)^{-1/2}R_-Y_-
\end{array}
\right)\,e^{i(L_\psi\psi-E\tau)}\,,
\ee
where functions $R_{\pm}=R_{\pm}(r)$ and $Y_{\pm}=Y_{\pm}(y)$.
Inserting this ansatz in \eqref{Diracexplicit}, we obtain eight equations with four separation constants.
The consistency of these equations requires that only one of the separation constants is independent, we denote it by $K$. Hence
we recovered the following four coupled first order ordinary differential equations for $R_{\pm}$ and $Y_{\pm}$:
\begin{equation}
\begin{aligned}
\frac{d R_\pm}{dr}+R_\pm\frac{\Delta_r'\pm V_r}{4\Delta_r}+R_{\mp}\frac{mr\mp K}{\sqrt{\Delta_r}}&=0\,,\\
\frac{d Y_\pm}{dy}+Y_\pm\frac{\Delta_y'\pm V_y}{4\Delta_y}-Y_\mp\frac{K\pm imy}{\sqrt{\Delta_y}}&=0\,,
\end{aligned}
\end{equation}
where
\be
V_r=4i(L_\psi-Er^2)\,,\quad V_y=4(L_\psi+Ey^2)\,.
\ee
As we shall see in section~\ref{ssc:DiracEquation}, this approach can be generalized to the case of higher-dimensional Kerr--NUT--(A)dS spacetimes.

\details{Similar to the Klein--Gordon case, the separability of the Dirac equation can be intrinsically characterized by the existence of the corresponding set of mutually commuting operators whose common eigenfunction is the separated solution.
The set consists of
$\{\sop{D},\, \sop{K},\,\sop{L}_\tau,\, \sop{L}_\psi\}$\,, where $\sop{D}=\gamma^a\nabla_a$
is the Dirac operator,
\be
\sop{K}=\gamma^{abc}h_{bc}\nabla_{a}+\frac{2}{3}\gamma^a(\nabla\cdot h)_a\,
\ee
is the symmetry operator corresponding to the principal tensor, and
\be
\sop{L}_\tau=\xi_{(\tau)}^a\nabla_a+\frac{1}{8}\gamma^{ab}(d\xi_{(\tau)})_{ab}\,,\quad
\sop{L}_\psi=\xi_{(\psi)}^a\nabla_a+\frac{1}{8}\gamma^{ab}(d\xi_{(\psi)})_{ab}\,
\ee
are the symmetry operators associated with the explicit symmetries. Here, $\gamma^{a_1\dots a_p}$ is the antisymmetrized product of $p$ gamma matrices, $\gamma^{a_1\dots a_p}=\gamma^{[a_1}\dots \gamma^{a_p]}$.  We refer to section~\ref{ssc:DiracEquation} and references \cite{CarterMcLenaghan:1979, Cariglia:2011yt, Cariglia:2011qb} for more details.
}

\subsection{Special limits of the Kerr metric}
\label{ssc:KerrSpecialCases}

\subsubsection{Flat spacetime limit: ${M=0}$}

Let us discuss now special limiting cases of the Kerr geometry.
In the absence of mass, that is when $M=0$, the curvature vanishes and the spacetime is flat. The  Kerr metric \eqref{kerr1} then takes the following form:
\be\label{flat}
\ts{g}=-\grad t^2+\left(r^2+a^2\,\cos^2\theta\right)\left[\frac{\grad r^2}
{r^2+a^2}+\grad \theta^2\right]+\left(r^2+a^2\right)\,\sin^2\theta\,
\grad \phi^2\,.
\ee
By changing the coordinates according to
\be\label{TXYZ}
T=t\,,\;\;
Z=r\cos\theta\,,\;\;
X=\sqrt{r^2+a^2}\,\sin\theta\,\cos\phi\,,\;\;
Y=\sqrt{r^2+a^2}\,\sin\theta\,\sin\phi\,,
\ee
the metric is transformed into the Minkowski metric
\be
\ts{g}=-\grad T^2+\grad X^2+\grad Y^2+\grad Z^2\,. \label{4.1.6}
\ee
A surface $r=\text{const}$ is an \emph{oblate ellipsoid of rotation}
\be\label{oblel}
\frac{X^2+Y^2}{r^2+a^2}+\frac{Z^2}{r^2}=1\, .
\ee

The $M\to 0$ limit of the Kerr metric in {canonical coordinates} is also quite straightforward.
The metric maintains the same form \eqref{can}, with $\Delta_r=r^2+a^2$.
Since the expressions \eqref{f_ccky}--\eqref{ff_ccky} for $\ts{h}, \ts{b}, \ts{f}$, and the expressions \eqref{HHH}, \eqref{KKK_D} for $Q^a_{\ b}$, $k^a_{\ b}$  do not contain the mass parameter at all, they remain unchanged.

Let us find an expression for the potential $\ts{b}$, \eqref{f_cckypot} in  Cartesian coordinates .
For this purpose we first use the transformation \eqref{kcan} from canonical coordinates $(\tau,r,y,\psi)$ to the Boyer--Lindquist coordinates $(t,r,\theta,\phi)$, to recover
\be
\ts{b}=-\frac{1}{2}\left[ (r^2-a^2 \cos^2\theta) \grad t-a(r^2\sin^2\theta -a^2\cos^2\theta) \grad \phi\right]\, .
\ee
After this we make the coordinate transformation \eqref{TXYZ} and omitting trivial constant terms, we find
\be\label{bbbb}
\ts{b}=-\frac{1}{2}\left[R^2 \grad T-a (X \grad Y-Y \grad X)\right]\, ,
\ee
where $R^2=X^2+Y^2+Z^2$.
It is easy to check that the potential \eqref{bbbb} is a special linear combination of the potentials \eqref{bti} and \eqref{bri}. One then finds
\be\label{tshf}
\ts{h}=d\ts{b}=\grad T\wedge (X\grad X+Y\grad Y+Z\grad Z)+a \grad X\wedge \grad Y\, .
\ee
Using the terminology of section~\ref{ssc:KYmaxsymm} one can say that $\ts{h}$ consists of two parts, the translational part, $\grad X\wedge \grad Y$, and a rotational 2-form, $\grad T\wedge (X\grad X+Y\grad Y+Z\grad Z)$. For $a=0$ the potential $\ts{b}$ is static and spherically symmetric, that is, it has the property $\lied_{\xi} \ts{b}=0$ valid for the Killing vectors $\ts{\xi}$ generating the time-translation and three-dimensional rotations. The term proportional to $a$ spoils the spherical symmetry. It singles out a two-plane $(X,Y)$ and preserves the invariance of $\ts{b}$ only with respect to rotations in this two plane. In other words, $\ts{b}$ is axisymmetric.

Using the following notations for flat space Killing vectors, generators of the Poincare group:
\be
\begin{gathered}
\ts{L}_X=Y\ts{\pa}_Z-Z\ts{\pa}_Y\,,\quad  {\ts{L}_Y=Z\ts{\pa}_X-X\ts{\pa}_Z\,,\quad \ts{L}_Z=X\ts{\pa}_Y-Y\ts{\pa}_X}\,,\\
  \ts{P}_T=\ts{\pa}_T\,,\quad  \ts{P}_Z=\ts{\pa}_Z\, ,
\end{gathered}
\ee
one finds
\be\label{KFLAT}
k^{ab}=f^{ac}f^b_{\  c}=L^{a b}+a (P_T^a L_Z^b+L_Z^aP_T^b)+a^2 (P_T^a P_T^b -P_Z^a P_Z^b)\, ,
\ee
where
\be
L^{a b}=L_X^a L_X^b+L_Y^a L_Y^b+L_Z^a L_Z^b\, .
\ee
The relation \eqref{KFLAT} implies that the Killing tensor $k^{ab}$ in the flat spacetime is reducible, and the corresponding
conserved quantity is
\be
k^{a b}p_{a}p_{b}=L^2+2a p_T L_Z+a^2 (p_T^2-p_Z^2)\, ,
\ee
where $L^2$ is the square of the total angular momentum.

Let us finally note that the primary Killing vector is $\ts{\xi}_{(\tau)}=\frac{1}{3}\ts{\nabla}\cdot \ts{h}=\ts{P}_T$, while the secondary Killing vector reads $\ts{\xi}_{(\psi)}=-\ts{k}\cdot \ts{\xi}_{(\tau)}=a^2\ts{P}_T+a\ts{L}_Z$.

\subsubsection{Extremal black hole: $M=a$}

In the limit $a=M$, the event and inner horizons have the same radius $r_+=r_-=M$. Such a rotating black hole is called {\em extremal}. The spatial distance to the horizon in the limit $a\to M$ infinitely grows. It is interesting that some of the hidden symmetries in the vicinity of the horizon of extremal black holes become explicit. Two connected effects take place: the eigenvalues of the principal tensor become functionally dependent, and, besides $\ts{\pa}_t$ and $\ts{\pa}_{\phi}$, two new additional Killing vectors arise. Let us discuss the case of the extremal black hole in more detail.

We start by noticing that in the extremal limit the function $\Delta_r$, \eqref{DD}, which enters the Kerr metric \eqref{kerr1}, takes the form $\Delta_r=(r-M)^2$. As a result $r$ becomes a `bad coordinate' in the vicinity of the horizon. To obtain a regular metric near the extremal horizon we first make the following coordinate transformation:
\be
r=M(1+\epsilon \rho)\,,\quad \tau=MT/\epsilon\,,\quad y=Mz\,,\quad \psi=(\varphi+T/\epsilon)/M \, .
\ee
After writing the Kerr metric \eqref{kerr1} in new coordinates $(T,\rho,z,\varphi)$, taking the limit $\epsilon\to 0$, and rescaling by a constant factor, $\ts{g}\to M^{-2}\ts{g}$ (just to simplify expressions), one obtains the following metric \citep{Bardeen:1999px}:
\be
\ts{g}=(1+z^2)\left(-\rho^2 \grad T^2+\frac{\grad \rho^2}{\rho^2}+\frac{\grad z^2}{1-z^2}\right)
   + \frac{1-z^2}{1+z^2} (2\rho \grad T+\grad \ph)^2\, .\label{exts}
\ee
It is again a solution of the vacuum Einstein equations.
The limiting metric $\ts{g}$ has two obvious Killing vectors
\be
\ts{\xi}=\ts{\pa}_{\ph}  \,,\qquad \ts{\eta}=\ts{\pa}_T\, ,
\ee
which can be obtained by taking the limit of the following Killing vectors of the original Kerr metric: $-M\ts{\pa}_{\tau}$ and $\epsilon^{-1}(M\ts{\pa}_{\tau}+M^{-1}\ts{\pa}_{\psi})$.

In the same limit,  the potential $\ts{b}$, \eqref{f_cckypot}, after omitting an infinite constant, ignoring the overall sign, and making rescaling $\ts{b}\to M^{-3}\ts{b}$, takes the form
\be
{\ts{b}}=\rho(1+z^2) \grad T +\frac{1}{2}z^2\, \grad \ph\, .
\ee
This yields the following (closed conformal) Killing--Yano quantities for the metric $\ts{g}$ \eqref{exts}:
\begin{align}
&{\ts{h}}=\grad {\ts{b}}=(1+z^2) \grad \rho\wedge \grad T +2\rho z \grad z\wedge \grad T +z \grad z\wedge \grad \ph\, ,\\
&{\ts{f}}=-z(1+z^2) \grad \rho\wedge \grad T +2\rho \grad z\wedge \grad T + \grad z\wedge \grad \ph\, .
\end{align}
The primary Killing vector is
\be
\ts{\xi}= \frac{1}{3} \ts{\nabla}\cdot \ts{h}=\ts{\partial}_\varphi\,.
\ee
However, the action of the Killing tensor {${k}^{a }{}_{b }={f}^a_{\ c}{f}^{\ c}_{b}$} on $\ts{\xi}$ does not produce a new Killing vector, as one has
\be
{k}^a_{\ b} \xi^b = -\xi^a\, .
\ee

It is easy to check that besides Killing vectors $\ts{\xi}$ and $\ts{\eta}$ the metric \eqref{exts} allows two additional  Killing vectors
\be
\ts{\zeta}_1=T\ts{\pa}_T-\rho\ts{\pa}_{\rho}\,,\quad
\ts{\zeta}_2=(T^2+\rho^{-2})\ts{\pa}_T-2T\rho\ts{\pa}_{\rho}-4\rho^{-1}\ts{\pa}_{\ph}\, .
\ee
Thus the original group  of symmetries of the Kerr spacetime is enhanced in the extremal near-horizon geometry and becomes $U(1)\times SL(2,1)$ \citep{Bardeen:1999px}.
This is the origin of the Kerr/CFT correspondence \citep{Guica:2008mu}. Moreover, the Killing tensor is reducible \citep{Galajinsky:2010zy,Rasmussen:2010rw,AlZahrani:2010qb} and can be presented in the form
\be
{k}^{ab}=\eta^{(a} \zeta_2^{b)}-\zeta_1^{a}\zeta_1^{b}+4\xi^a\xi^b+ g^{ab}\, .
\ee

\subsubsection{Non-rotating black hole: $a=0$}

The last limiting case of the Kerr metric which we are going to consider here is that of a non-rotating black hole.
The limit $a\to 0$ can be easily taken in the Kerr metric in the Boyer--Lindquist coordinates \eqref{kerr1}. It gives the Schwarzschild metric
\be
ds^2=-F dt^2+\frac{dr^2}{F}+r^2 (d\theta^2+\sin^2\theta d\phi^2)\,,\qquad F=1-\frac{2M}{r}\, .
\ee
The same limit in the canonical coordinates is slightly more involved. The reason is that the range of coordinate $y$ is chosen such that the function $\Delta_y=a^2-y^2$ is non-negative. In the limit $a\to0$ it implies that this range would become degenerate.  In order to escape this problem one should rescale $y$, for example, by setting $y=a \cos\theta$. Range of $\theta$ remains regular under the limit.

The best way to study the fate of  hidden symmetries in the limit $a\to 0$ is to return from canonical to the Boyer--Lindquist coordinates first, using \eqref{kcan}, and then take the limit $a\to 0$. The  leading in $a$ terms give
\be
\ts{b}=-\frac{1}{2} r^2 \grad t\,
\ee
for the potential $\ts{b}$.  One also has
\be
\ts{h}=r \grad t \wedge \grad r \,,\qquad \ts{f}=r^3 \sin\theta \grad \phi\wedge \grad \theta\, .
\ee
The resultant closed conformal Killing--Yano tensor is degenerate,
\be
\ts{h}\wedge \ts{h}=0\,,
\ee
and so are tensors  $\ts{Q}$ and $\ts{k}$.  Moreover, the Killing tensor $\ts{k}$ is reducible. Denoting by $(\ts{L}_X,\ts{L}_Y,\ts{L}_Z)$ the three Killing vectors that generate the spherical symmetry of the Schwarzschild metric:
\be
\ts{L}_X=-\cos\phi \ts{\pa}_{\theta}+\cot\theta \sin\phi \ts{\pa}_{\phi}\,,\quad
\ts{L}_Y=\sin\phi \ts{\pa}_{\theta}+\cot\theta \cos\phi \ts{\pa}_{\phi}\,,\quad
\ts{L}_Z=\ts{\pa}_{\phi}\,,
\ee
one has
\be
k^{ab}=L_X^a L_X^b+L_Y^a L_Y^b+L_Z^a L_Z^b\, .
\ee
The primary Killing vector is $\ts{\xi}=\ts{\pa}_t$, while the secondary Killing vector vanishes, $k^a{}_b\xi^b=0$.


\subsection{Kerr--Schild form of the Kerr metric}
\label{ssc:KerrSchild4D}

It is a remarkable property of the Kerr metric that it can be written in the Kerr--Schild form, that is, as a linear in $M$ deformation of flat spacetime \citep{kerr1965some, Debney:1969zz}. This property is intrinsically related to the special algebraic type of the Weyl tensor and the existence of hidden symmetries.

Starting from the canonical form of the metric \eqref{canNOreference} we may write
\begin{equation}\label{KerrtoKSch}
\begin{aligned}
\ts{g}&=-\frac{\Delta_r}{\Sigma}\Bigl((\grad \tau+y^2\grad \psi)^2-\frac{\Sigma^2}{\Delta_r^2}\grad r^2\Bigr)
+\frac{\Delta_y}{\Sigma}(\grad \tau-r^2
\grad \psi)^2+\frac{\Sigma}{\Delta_y}\grad y^2\\
&=
-\frac{\Delta_r}{\Sigma}\,\ts{l}\,\ts{l}+\grad r\,\ts{l}+\ts{l}\,\ts{dr}+
\frac{\Delta_y}{\Sigma}(\grad \tau-r^2
\grad \psi)^2+\frac{\Sigma}{\Delta_y}\grad y^2\,,
\end{aligned}
\end{equation}
where we introduced a null vector
\be
\ts{l}\equiv\grad \tau+y^2\grad \psi+\frac{\Sigma}{\Delta_r}\grad r
   =\sqrt{\frac{\Sigma}{\Delta_r}}(\ts{\nu}+\ts{\hat\nu})\,.
\ee
Defining new coordinates
\be
\grad \hat \tau=\grad \tau+\frac{r^2}{\Delta_r}\grad r-\frac{y^2}{\Delta_y}\grad y\,,\quad
\grad \hat \psi=\grad \psi+\frac{\grad r}{\Delta_r}+\frac{\grad y}{\Delta_y}\,,
\ee
we find that
\be
\ts{l}=\grad \hat \tau+y^2\grad \hat \psi\,,
\ee
and the term $(\grad\tau-r^2\grad\psi)$ in the metric \eqref{KerrtoKSch} reads $(\grad\hat\tau-r^2\grad\hat\psi+\frac\Sigma{\Delta_y}\grad y)$. Upon recalling the form \eqref{4.1.3} of the metric function $\Delta_r$, the Kerr metric then rewrites in the Kerr--Schild form
\be\label{KSchild4dmetric}
\ts{g}=\mathring{\ts{g}}+\frac{2Mr}{\Sigma}\,\ts{l}\,\ts{l}\,,
\ee
where
\begin{equation}\label{flatKerrSch}
\begin{gathered}
   \mathring{\ts{g}}=-\frac{\mathring{\Delta}_r}{\Sigma}\ts{l}^2
   +\grad r\,\ts{l}+\ts{l}\,\ts{dr}+
   \frac{\mathring{\Delta}_y}{\Sigma}\Bigl(\grad \hat\tau-r^2\grad \hat\psi+\frac{\Sigma}{\mathring{\Delta}_y}\grad y\Bigr)^2
   +\frac{\Sigma}{\mathring{\Delta}_y}\grad y^2\,,\\
   \mathring{\Delta}_r=r^2+a^2\,,\quad \mathring{\Delta}_y=\Delta_y=a^2-y^2\,,\quad \Sigma=r^2+y^2\,
\end{gathered}
\end{equation}
is the flat metric. Indeed, introducing `flat' canonical coordinates $(\mathring{\tau},r,y,\mathring{\psi})$ as
\be
\grad \hat\tau=\grad \mathring{\tau}+\frac{r^2}{\mathring{\Delta}_r}\grad r-\frac{y^2}{\mathring{\Delta}_y}\grad y\,,\quad
\grad \hat\psi=\grad \mathring{\psi}+\frac{\grad r}{\mathring{\Delta}_r}+\frac{\grad y}{\mathring{\Delta}_y}\,,
\ee
brings the metric $\mathring{\ts{g}}$ into the `canonical form' of the Kerr metric
\be
\mathring{\ts{g}}=\frac{1}{\Sigma}
\Bigl[ -\mathring{\Delta}_r(\grad \mathring{\tau}+y^2 \grad \mathring{\psi})^2+\mathring{\Delta}_y(\grad \mathring{\tau}-r^2
\grad \mathring{\psi})^2\Bigr]
+\Sigma\Bigl[ \frac{\grad r^2}{\mathring{\Delta}_r}+\frac{\grad y^2}{\mathring{\Delta}_y}\Bigr]\, ,
\ee
with $M=0$. We can also check that $\ts{l}=\grad \mathring{\tau}+y^2\grad \mathring{\psi}+\frac{\Sigma}{\mathring{\Delta}_r}\grad r$ is a null vector with respect to the flat metric $\mathring{\ts{g}}$.

The principal tensor can be written as
\be
\ts{h}
=-r\grad r\wedge \ts{l}+y \ts{\epsilon}\wedge \hat{\ts{\epsilon}}\,.
\ee
The vector $\ts{l}$ is an eigenvector of the principal tensor. At the same time it is a principal null direction of the metric, and a vector that plays a special role for the Kerr--Schild structure \eqref{KSchild4dmetric}. This nicely illustrates how all such properties: hidden symmetries, special algebraic type of the Weyl tensor, and the Kerr--Schild form, are interconnected. As we shall see, this remains true also
for higher-dimensional Kerr--NUT--(A)dS spacetimes.

\subsection{Remarks on the choice of angle variable}
\label{ssc:RemarksAngles}

In the next chapter, we shall discuss higher-dimensional metrics that generalize the four-dimensional Kerr metric \eqref{kerr1}. We shall see that there exists a natural canonical form for such metrics, where the coordinates are determined by the principal tensor. Part of these coordinates are Killing parameters associated with the corresponding primary and secondary Killing vectors. These angle coordinates are similar to the angle $\psi$, used in \eqref{can}. A natural question is how these angles are related to the other set of angle variables, similar to $\phi$ in \eqref{kerr1}. In order to clarify this point, let us make here a few remarks, which will be useful later.

\subsubsection{Axis of rotational symmetry}

In a general case, one says that a $D$-dimensional manifold is \emph{cyclicly symmetric} (or just \emph{cyclic}) if it is invariant under an action of the one-parametric cyclic group $SO(2)$. It requires that the Killing vector generating this symmetry has closed orbits.

Fixed points of a Killing vector field are points where the Killing vector vanishes. These points intuitively correspond to an axis of symmetry. In general, the manifold does not have to be smooth at these points or the metric does not have to be regular (a well known example is a conical singularity). In such cases we speak about a generalized axis of symmetry.

\begin{figure}[h]
\bigskip
\hfill\includegraphics[width=5cm]{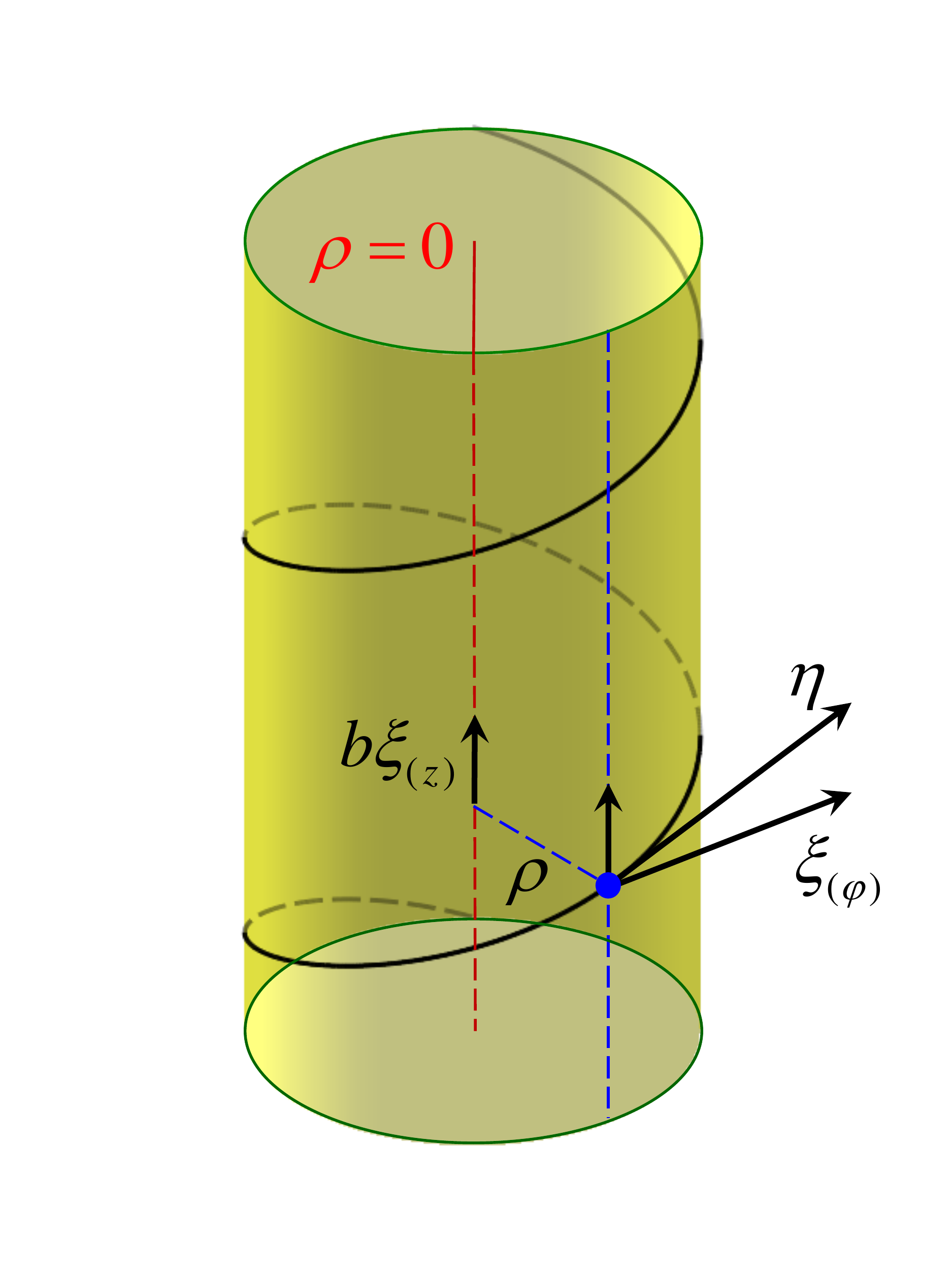}\quad\includegraphics[width=5cm]{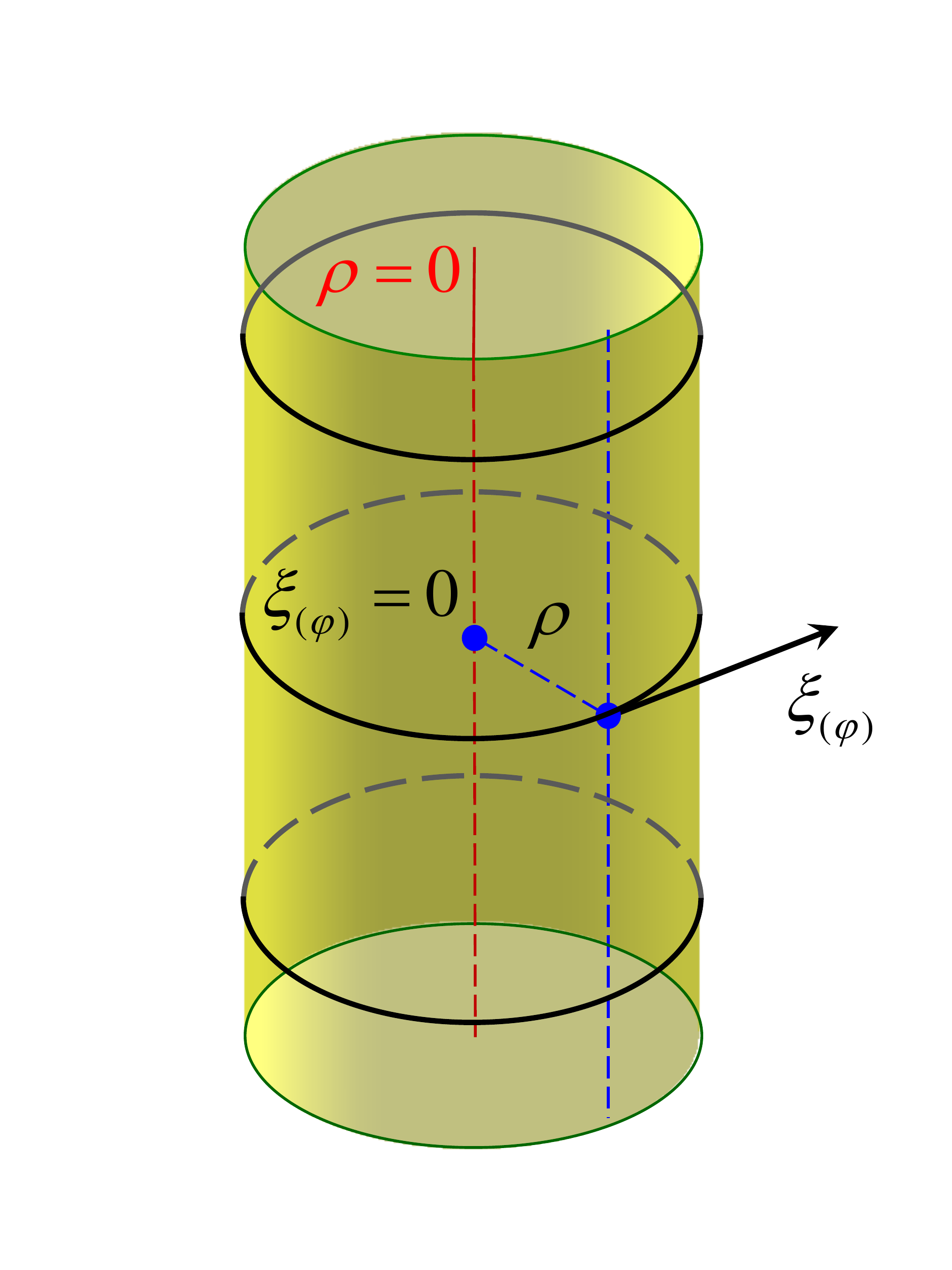}
\hfill\hfill
\caption{\label{fig:cgract}%
{\bf Killing vectors with closed and open orbits.}
Left figure shows the action of the symmetry  with non-closed orbits. The corresponding Killing vector does not have fixed points. The right figure illustrates the action of the cyclic group with closed orbits. The corresponding Killing vector has fixed points which form the axis of symmetry $\rho=0$.}
\end{figure}

As an example, consider a flat four-dimensional spacetime equipped with cylindrical coordinates $(T,Z,\rho,\phi)$. In order to have a regular metric  at $\rho=0$, the coordinate $\ph$ must  be periodic, with period $2\pi$. The point $\rho=0$  is a fixed point of the Killing vector $\ts{\xi}_{(\ph)}$ in the plane $T,Z=\text{const}$. In a general case, however, a Killing vector field may not have fixed points.  For example, consider a Killing vector  $\ts{\eta}=\ts{\xi}_{(\ph)}+ \alpha\ts{\xi}_{(Z)}$. One finds $\ts{\eta}^2=\rho^2 +\alpha^2>0$ for a non-vanishing value of $\alpha$. Thus, the Killing vector $\ts{\eta}$ neither has fixed points nor it is cyclic. However, fixed points exist for $\ts{\xi}_{(\ph)}$. These two cases are illustrated in figures~\ref{fig:cgract}. The left figure shows a symmetry in a three-dimensional flat space generated by the Killing vector $\ts{\eta}$. The orbits are not closed and this vector field does not vanish anywhere. The case when the symmetry is cyclic group, orbits  are closed, and there is an axis of symmetry, is shown in the right figure. The symmetry is generated by the Killing vector $\ts{\xi}_{(\ph)}$, which vanishes at the axis of symmetry, $\rho=0$.

\subsubsection{{Twisting construction}}

In principle, one could modify the flat spacetime by making the orbits of the Killing vector $\ts{\eta}$ cyclic. This can be achieved by cutting the spacetime along the half-plane $\ph=0$ and re-gluing it shifted by $\alpha$ in the $Z$-direction. In other words, we identify points with coordinates $(T,\rho,\ph,Z)$ given by values $(T,\rho,0,Z)$ and $(T,\rho,2\pi,Z+2\pi\alpha)$. This can be reformulated in coordinates adapted to the Killing vector $\ts{\eta}$. If we define
\begin{equation}\label{zetapsidef}
    \zeta = Z-\alpha\ph\,,\quad \psi=\ph\,,
\end{equation}
we have $\ts{\eta}=\cv{\psi}$, and the identifications of the spacetime can be formulated as a periodicity of the coordinate $\psi$, i.e., the identification of $\psi=0$ and $\psi=2\pi$ with the same values of $T,\rho,\zeta$. In such a way we obtain what we call a \emph{twisted} flat spacetime.

In this spacetime the Killing vector $\ts{\eta}$ is cyclic but (as in the previous case) it does not have fixed points. On other hand, the Killing vector $\ts{\xi}_{(\ph)}$ has still fixed points but it is not cyclic anymore. Its orbits are not closed and the corresponding symmetry group is not $SO(2)$ but $\realn$. The twisted spacetime thus has only a generalized axis of the symmetry at $\rho=0$. This axis does not form a regular submanifold of the full twisted spacetime.

\subsubsection{{Rotating string and conical singularity}}

Similar `twisting' construction can be done with time-like Killing vector $\ts{\xi}_{(T)}$ instead of  $\ts{\xi}_{(Z)}$. The Killing vector $\ts{\eta}=\ts{\xi}_{(\ph)}+ \alpha\ts{\xi}_{(T)}$ generates a cyclic symmetry in a spacetime which is obtained by cutting the flat spacetime along half-plane $\ph=0$ and re-gluing it so that the coordinates
\begin{equation}\label{taupsidef}
    \tau = T-\alpha\ph\,,\quad \psi=\ph\,,
\end{equation}
are identified as $(\tau,\rho,\psi=0,Z)\leftrightarrow(\tau,\rho,\psi=2\pi,Z)$. This spacetime corresponds to a thin straight spinning cosmic string, cf., e.g., section~3.4.1 in \cite{GriffithsPodolsky:book}. In this spacetime $\ts{\eta}$ is cyclic but does not have fixed points and $\ts{\xi}_{(\ph)}$ has fixed points but it is not cyclic. The cosmic string is located on the irregular generalized axis of symmetry at $\rho=0$.

Because of the time-like nature of the Killing vector $\ts{\xi}_{(T)}$, a new phenomenon occurs in this case. The Killing vector $\ts{\eta}$ is spacelike far from the axis, for $\rho>|\alpha|$, and timelike near the axis, for $\rho<|\alpha|$. The surface $\rho=|\alpha|$, where $\ts{\eta}^2=0$, is an \emph{ergosurface} of the Killing vector $\ts{\eta}$. Let us emphasize that, although $\ts{\eta}^2=0$, these are not fixed points of the Killing vector $\ts{\eta}$ since $\ts{\eta}$ is not vanishing here. The ergosurface contains orbits of the symmetry which are null closed curves, they correspond to light rays orbiting the axis in closed trajectories. Inside the ergosurface, where the Killing vector is timelike, the orbits of the symmetry are closed time-like curves. Clearly, such a behavior is not very physical. However, it seems that in a generic case it may not be escaped.

There is yet another aspect related to the identification of the axis of the symmetry and its regularity. Let us consider an axisymmetric spacetime with coordinate $\ph\in(0,2\pi)$ which parameterizes orbits of the cyclic symmetry. In general, the metric may not be regular on the axis---it can contain a conical singularity. Such a singularity can  be eliminated choosing a different range of periodicity for coordinate $\ph$. It can be achieved by introducing a rescaled coordinate $\phi=\beta\ph$ which is required to be periodic on the interval $(0,2\pi)$. Physically, the conical singularity corresponds to a static thin string on the axis \citep{vilenkin2000cosmic,GriffithsPodolsky:book}.

\subsubsection{{Kerr geometry}}

For the canonical metric \eqref{can} of the rotating black hole spacetime, the coordinates $\tau$ and $\psi$ are directly connected with the principal tensor of this spacetime. Namely, they are the proper Killing coordinates for the primary $\ts{\xi}_{(\tau)}$ and secondary $\ts{\xi}_{(\psi)}$ Killing vectors. If one makes the coordinate $\psi$ to be cyclic by identifying $\psi=0$ and $\psi=2\pi$, the corresponding spacetime is not axisymmetric, since the Killing vector $\ts{\xi}_{(\psi)}$ does not have fixed points.

We can then ask if one can find the correct axisymmetric coordinate. For that we need to find a Killing vector which has fixed points. Let us consider a vector $\ts{\eta}=\ts{\xi}_{(\psi)}+ \alpha \ts{\xi}_{(\tau)}$ and require that it vanishes at some points.  It can be shown that this happens only if the following two conditions are met: $\Delta_y=0$ and $\alpha=-a^2$. The vector $\ts{\eta}=\ts{\xi}_{(\psi)}-a^2\ts{\xi}_{(\tau)}$ thus has fixed points at roots of $\Delta_y$. The coordinates adapted to this Killing vector are $t=\tau+a^2\psi$, ${\ph=\psi}$. However, if one makes $\ph$ periodic on the interval $(0,2\pi)$, there would be a conical singularity on the axis. One has to make an additional rescaling ${\phi=a\ph}$ leading to the Boyer--Lindquist coordinates $(t,r,\theta,\phi)$ given by \eqref{kcan}. If the coordinate $\phi$ is made periodic on interval $(0,2\pi)$, the axis is regular: the spacetime contains a cyclic Killing vector $\ts{\xi}_{(\phi)}=\frac1a\ts{\xi}_{(\psi)}-a\ts{\xi}_{(\tau)}$ with fixed points identifying the axis and there is no conical singularity on this axis.

\subsubsection{{Effect of NUT charges}}

Let us briefly comment on a more complicated metric with non-trivial NUT parameters for which the metric functions ${\Delta_r}$ and ${\Delta_y}$ are given by \eqref{Delz} with ${\Lambda=0}$. The polynomial ${\Delta_y}$ has now two nontrivially different roots ${{}^\pm y}$ and the coordinate ${y}$ runs between these roots, ${y\in({}^-y,{}^+y)}$. In this case one can find two candidates for the Killing vector with  fixed points: ${\ts{\eta}_+}$ and ${\ts{\eta}_-}$, with fixed points at ${y={}^+y}$ and ${y={}^-y}$, respectively. One can choose one of the properly rescaled corresponding coordinates, say ${\phi_+}$, to be periodic with period ${2\pi}$. With such a choice, the submanifold ${y={}^+y}$ becomes the regular axis. Physically, it corresponds only to a semi-axis of the spacetime. The other semi-axis ${y={}^-y}$ is not regular, the cyclic  Killing vector ${\ts{\eta}_+}$ does not have fixed points here. Of course, one can assume periodicity of the  other coordinate ${\phi_-}$, making thus the semi-axis ${y={}^-y}$ regular. However, the semi-axis ${y={}^+y}$ becomes now non-regular. So, it is not a priori guaranteed that one can chose a unique Killing vector which makes the spacetime globally axisymmetric.

\subsection{Hidden symmetries of the Pleba\'nski--Demia\'nski metric}
\label{ssc:PD}

The Pleba\'nski--Demia\'nski metric \citep{PlebanskiDemianski:1976} is the most general four-dimensional electrovacuum solution of Einstein's equations that is stationary, axisymmetric, and whose Weyl tensor is of the special algebraic type D. It describes a wide family of spacetimes that generalize the Kerr--NUT--(A)dS family described in previous sections. Besides the cosmological constant, mass, rotation, and NUT parameter it also admits electric and magnetic charges and the acceleration parameter. As we shall discuss now, the Pleba\'nski--Demia\'nski metric admits a `weaker' (conformal) form of hidden symmetries of the Kerr geometry.

\subsubsection{Solution}

Generalizing the canonical form of the Kerr--NUT--(A)dS spacetime \eqref{can}, the
Pleba\'nski--Demia\'nski solution reads
\begin{equation}\label{PD4d}
\begin{gathered}
\ts{g}=\Omega^{2}\Bigl[-\frac{\Delta_r}{\Sigma}(\ts{d}\tau+y^2 \ts{d}\psi)^2+\frac{\Delta_y}{\Sigma}(\ts{d}\tau-r^2
\ts{d}\psi)^2+\frac{\Sigma}{\Delta_r}\ts{d}r^2+\frac{\Sigma}{\Delta_y}\ts{d}y^2\Bigr]\, ,\\
\ts{F}=\ts{dA}\,,\quad \ts{A}=-\frac{er}{\Sigma}\bigl(\grad \tau + y^2\,\grad\psi\bigr)-\frac{gy}{\Sigma}\bigl(\grad \tau - r^2\,\grad\psi\bigr)\;,
\end{gathered}
\end{equation}
where $\Sigma=r^2+y^2$.  It obeys the Einstein--Maxwell equations with the electric and magnetic charges $e$ and $g$ and the cosmological
constant $\Lambda$ provided the functions $\Delta_y=\Delta_y(y)$ and $\Delta_r=\Delta_r(r)$ take the following form:
\begin{equation}\label{functionsQP_A}
\begin{split}
\Delta_r&=k+e^2+g^2-2mr+\epsilon r^2-2nr^3-(k+\Lambda/3)r^4\;,\\
\Delta_y&=k+2ny-\epsilon y^2+2my^3-(k+e^2+g^2+\Lambda/3)y^4\,,
\end{split}
\end{equation}
while the conformal factor $\Omega$ reads
\begin{equation}\label{Omega_A}
\Omega^{\!-1}=1-yr\;.
\end{equation}
Constants $k, m, \epsilon, n$ are free parameters that are related to mass, rotation, NUT parameter, and acceleration. The Kerr--NUT--(A)dS geometry belongs to this class, but it can be identified only after a proper redefinition of coordinates and parameters. We refer to \cite{GriffithsPodolsky:2006b} for details and for a discussion and the interpretation of special cases of the Pleba\'nski--Demia\'nski metric.
For a recent progress on understanding the thermodynamics of accelerating black holes see \cite{Appels:2017xoe, Astorino:2016ybm}.

\subsubsection{Hidden symmetries}

The Pleba\'nski--Demia\'nski metric admits a hidden symmetry of a non-de\-gen\-er\-ate rank-2 conformal Killing--Yano 2-form:\footnote{This 2-form is no longer closed. In consequence the structure of hidden symmetries is weaker than that of the Kerr--NUT--(A)dS spacetime. For example, only null but not timelike geodesics are integrable in the Pleba\'nski--Demia\'nski background.}
\begin{equation}\label{bconf_A}
\tens{h}=\Omega^{3}\Bigl[y \grad y\wedge(\grad\tau-r^2 \grad\psi)
          -r \grad r\wedge (\grad\tau+y^2 \grad\psi)\Bigr]\;,
\end{equation}
obeying
\begin{equation}\label{CKY_4rank2_A}
\nabla_{a}h_{bc}= \nabla_{[a}h_{bc]}+ {2}\, g_{a[b} \xi_{c]}\;,\quad
\xi_{a}=\frac{1}{3}\,\nabla_{c}h^{c}{}_{a}\;.
\end{equation}
This property remains true also for the off-shell metric \eqref{PD4d}, characterized by arbitrary functions $\Delta_r(r), \Delta_y(y)$ and an arbitrary conformal factor $\Omega(r,y)$ \citep{Kubiznak:2007kh}.

The corresponding Hodge dual, $\ts{f}=\ts{*h}$, is yet another non-degenerate conformal Killing--Yano 2-form, given by
\be
\ts{f}= \Omega^3\Bigl[r \grad y\wedge (\grad\tau-r^2 \grad\psi)
          +y \grad r\wedge (\grad\tau+y^2 \grad\psi)\Bigr]\, .
\ee
These 2-forms generate both isometries of the metric according to
\be
\tens{\xi}\equiv \frac{1}{3}\ts{\nabla}\cdot \ts{h}=\tens{\partial}_{\tau}\;,\qquad
\tens{\eta}\equiv \frac{1}{3}\ts{\nabla}\cdot \ts{f}=\tens{\partial}_{\psi}\;,
\ee
as well as give rise to the corresponding conformal Killing tensors. Namely, $Q^{(h)}_{ab}=h_{ac}h_b{}^c$ reads
\be\label{Qh_A}
\tens{Q}^{(h)}=\Omega^4\Bigl[\frac{r^2\Delta_r}{\Sigma}(\grad\tau+y^2\grad\psi)^2+\frac{y^2\Delta_y}{\Sigma}(\grad\tau-r^2\grad\psi)^2
+\frac{\Sigma}{\Delta_y}y^2\grad y^2-\frac{\Sigma}{\Delta_r} r^2\grad r^2\Bigr]\,,
\ee
while for $Q^{(f)}_{ab}=f_{ac}f_b{}^c$ we have
\begin{equation}\label{QhQk}
  \tens{Q}^{(f)} = \tens{Q}^{(h)} + \Omega^2(r^2-y^2)\, \tens{g}\;.
\end{equation}

The existence of either of these conformal Killing tensors guarantees complete integrability of null geodesic motion. Namely, we have the following constants of null geodesic equations: 
\be\label{PDnullInt}
{\xi_{a} \dot{x}^a=-E\,,\quad
\eta_{a}\dot{x}^a=L\,,}\quad
Q^{(h)}_{a b}\dot{x}^a\dot{x}^b=K \,,\quad
g_{a b}\dot{x}^a\dot{x}^b=0\,.
\ee
These four equations can be solved for $\dot{x}^a=(\dot{\tau},\dot{r},\dot{y},\dot{\psi})$, giving:
\begin{align}
\Omega^2\Sigma\,\dot{r}&=\pm \sqrt{{\cal X}_r}\, ,\\
\Omega^2\Sigma\,\dot{y}&=\pm \sqrt{{\cal X}_y}\, ,\\
\Omega^2\Sigma\,\dot{\psi}&=\frac{ E r^2-L}{\Delta_r}+ \frac{E y^2+ L}{\Delta_y}\, ,\\
\Omega^2\Sigma\,\dot{\tau}&=\frac{ r^2( E r^2-L)}{\Delta_r}-
    \frac{y^2( E y^2+L)}{\Delta_y}\, ,
\end{align}
where
\be \label{RYPD}
{\cal X}_r=(E r^2-L)^2-K\Delta_r\,,\quad
{\cal X}_y=-(E y^2+L)^2+K\Delta_y\, .
\ee
cf. the expressions for null geodesics ($m^2=0)$ in Kerr--NUT--(A)dS spacetimes, \eqref{FODE}--\eqref{tteq}. Similar to the discussion therein,
the equations for $\dot r$ and $\dot  y$ can be decoupled by introducing the convenient geodesic parameter.

For a discussion of the integrability of a charged particle motion in the Pleba\'nski--Demia\'nski metric see \cite{DuvalValent:2005}. As a consequence of the existence of the conformal Killing--Yano 2-form $\ts{h}$, also the massless Hamilton--Jacobi, Klein--Gordon, and Dirac equations separate in the Pleba\'nski--Demia\'nski backgrounds. We do not review here the corresponding calculations. The first two are easy to perform and we refer to the original papers \citep{kamran1983separation, kamran1984separation} for the separability of massless Dirac equation; see also  \cite{torres1988separability, silva1995killing} for a discussion of electromagnetic and Rarita--Swinger perturbations.

\subsubsection{Higher-dimensional generalizations}

As we shall see in the next chapter, the four-dimensional Kerr--NUT--(A)dS metrics can be generalized to higher dimensions. However, similar attempts for the Pleba\'nski--Demia\'nski metric have failed so far. In particular, people have tried to obtain a higher-dimensional generalization of an accelerated black hole described by the so called C-metric, which is a special case of the Pleba\'nski--Demia\'nski class.

\details{The C-metric typically describes a pair of black holes moving in the opposite direction with constant acceleration caused either by a cosmic string of negative energy density between them or by two positive-energy strings pulling the black holes from infinity. As the string is present, the corresponding solution does not represent, strictly speaking, a regular isolated black hole.}

A straightforward method of multiplying the higher-dimensional Kerr--NUT--(A)dS spacetime \eqref{KerrNUTAdSmetric} with a properly chosen conformal factor $\Omega$, accompanied by a proper adjustment of metric functions $X_\mu$, turned out to be very naive and does not work, e.g., \cite{Kubiznak:2007kh}. However, a partial success has been achieved in five dimensions, where two different factors, rescaling various parts of the Kerr--NUT--(A)dS spacetime, have been used to construct a new metric whose limits lead to the black holes of spherical horizon topology on one side and to the black rings with toroidal horizon topology on the other side \citep{Lu:2008js, Lu:2008ze, Lu:2014sza}.


%% file: ch4-hdbh.tex

\section{Higher-dimensional Kerr--NUT--(A)dS metrics}
\label{sc:hdbh}

Higher-dimensional Kerr--NUT--(A)dS metrics \citep{Chen:2006xh} describe a large family of geometries of various types and signatures that solve the vacuum Einstein equations with and without the cosmological constant. Parameterized by a set of free parameters that can be related to mass, rotations, and NUT parameters, they directly generalize the four-dimensional Carter's canonical metric \eqref{can} studied in the previous chapter. The general rotating black holes of Myers and Perry \eqref{MPM} \citep{MyersPerry:1986}, their cosmological constant generalizations due to \cite{Gibbons:2004js,Gibbons:2004uw}, the higher-dimensional Taub-NUT spaces \citep{Mann:2003zh, Mann:2005ra, Clarkson:2006zk, Chen:2006ea}, or the recently constructed deformed and twisted black holes \citep{Krtous:2015zco}, all emerge as certain limits or subcases of the Kerr--NUT--(A)dS spacetimes. All such geometries inherit hidden symmetries of the Kerr--NUT--(A)dS metrics.

In this chapter, we perform a basic analysis of the Kerr--NUT--(A)dS metrics, discussing their signature, coordinate ranges, scaling properties, and meaning of free metric parameters. We also identify their several special subcases, namely, the sphere, the Euclidean instanton, and various black hole solutions. The discussion of hidden symmetries is postponed to the next chapter.

\subsection{Canonical form of the metric}
\label{ssc:KerrNUTAdS}

\subsubsection{Metric}

The canonical metric describing the \defterm{Kerr--NUT--(A)dS geometry} in ${D=2\dg+\eps}$ number of dimensions (with $\eps=0$ in even and $\eps=1$ in odd dimensions) reads
\begin{equation}\label{KerrNUTAdSmetric}
\tens{g}
  =\sum_{\mu=1}^\dg\;\biggl[\; \frac{U_\mu}{X_\mu}\,{\grad x_{\mu}^{2}}
  +\, \frac{X_\mu}{U_\mu}\,\Bigl(\,\sum_{j=0}^{\dg-1} \A{j}_{\mu}\grad\psi_j \Bigr)^{\!2}
  \;\biggr]
  +\eps\frac{c}{\A{\dg}}\Bigl(\sum_{k=0}^\dg \A{k}\grad\psi_k\!\Bigr)^{\!2}\;.
\end{equation}
The employed coordinates naturally split into two sets: \defterm{Killing coordinates} ${\psi_k}$ (${k}={0,\,\dots,\,\dg{-}1{+}\eps}$) associated with the explicit symmetries, and \defterm{radial and longitudinal coordinates} ${x_\mu}$  ($\mu=1,\,\dots,\,\dg$) labeling the orbits of Killing symmetries.

\details{
As we shall see in the next chapter, both types of canonical coordinates are uniquely determined by the principal tensor $\ts{h}$. Namely, $x_{\mu}$'s are the eigenvalues of the principal tensor and $\psi_j$'s are the Killing coordinates associated with the primary (${j=0}$) and secondary (${j>0}$) Killing vectors generated by this tensor. Such a choice of coordinates, internally connected with the principal tensor, makes the canonical form of the metric \eqref{KerrNUTAdSmetric} quite simple. It is also directly `linked to' the separability properties of the geometry.
}\medskip

The functions ${\A{k}}$, ${\A{j}_\mu}$, and ${U_\mu}$ are `symmetric polynomials' of coordinates ${x_\mu}$:
\begin{equation}\label{AUdefs}
  \A{k}=\!\!\!\!\!\sum_{\substack{\nu_1,\dots,\nu_k=1\\\nu_1<\dots<\nu_k}}^\dg\!\!\!\!\!x^2_{\nu_1}\dots x^2_{\nu_k}\;,
\qquad
  \A{j}_{\mu}=\!\!\!\!\!\sum_{\substack{\nu_1,\dots,\nu_j=1\\\nu_1<\dots<\nu_j\\\nu_i\ne\mu}}^\dg\!\!\!\!\!x^2_{\nu_1}\dots x^2_{\nu_j}\;,\qquad
  U_{\mu}=\prod_{\substack{\nu=1\\\nu\ne\mu}}^\dg(x_{\nu}^2-x_{\mu}^2)\;,
\end{equation}
and each metric function ${X_\mu}$ is a function of a single coordinate ${x_\mu}$:
\begin{equation}\label{Xfcdependence}
    X_\mu=X_\mu(x_\mu)\;.
\end{equation}
If these functions are unspecified, we speak about the \emph{off-shell metric}.
The vacuum Einstein equations with a cosmological constant restrict these functions into a polynomial form (see \eqref{Xsol} below). With this choice we call \eqref{KerrNUTAdSmetric} the {\em on-shell metric}. We see that the metric components of the on-shell Kerr--NUT--(A)dS metric are rational functions of the coordinates $x_{\mu}$. Constant $c$ that appears in odd dimensions is a free parameter.

The metric \eqref{KerrNUTAdSmetric} is written in the most symmetric form adjusted to the Euclidean signature and is very convenient for the analysis of explicit and hidden symmetries. This most symmetric form is naturally broken when one describes the black hole case: in order to guarantee the Lorentzian signature, one needs to assume that some of the coordinates and parameters take imaginary values. In what follows we shall call this procedure a \emph{`Wick rotation'}. We should also mention that coordinates $\psi_j$ are different from the `standard azimuthal' angles $\phi_\mu$, used in the Boyer--Lindquist form of the Myers--Perry metric (see next section).

The inverse metric takes the following form:
\begin{equation}\label{KerrNUTAdSinvmetric}
  \tens{g}^{-1}
  =\sum_{\mu=1}^\dg\;\biggl[\; \frac{X_\mu}{U_\mu}\,{\cv{x_{\mu}}^2}
  + \frac{U_\mu}{X_\mu}\,\Bigl(\,\sum_{k=0}^{\dg-1+\eps}
    {\frac{(-x_{\mu}^2)^{\dg-1-k}}{U_{\mu}}}\,\cv{\psi_k}\Bigr)^{\!2}\;\biggr]
  +\eps\,\frac1{c\A{\dg}}\,\cv{\psi_n}^2\;.
\end{equation}
The determinant of the metric reads
\begin{equation}\label{detmetric}
    \det[g_{ab}] = \bigl(c\A{\dg}\bigr)^\eps\, V^2\;,\quad
    V\equiv\prod_{\substack{\mu,\nu=1\\\mu<\nu}}^\dg(x^2_\mu-x^2_\nu)=\det[\A{j}_\mu]\; .
\end{equation}
As in four dimensions, it is independent of the choice of arbitrary functions $X_{\mu}(x_{\mu})$. Correspondingly, the Levi-Civita tensor is given by
\begin{equation}\label{LeviCivita}
  \tens{\eps}= \bigl(c\A{\dg}\bigr)^{\frac\eps2}\, V\;
    \grad x_1\wedge\dots\wedge\grad x_\dg\wedge
    \grad\psi_0\wedge\dots\wedge\grad\psi_{\dg-1+\eps}\;.
\end{equation}

\subsubsection{Special Darboux frame}

The metric and its inverse can be obtained by employing the natural orthonormal frame of 1-forms ${\enf\mu,\, \ehf\mu}$ (${\mu=1,\dots,\dg}$), and ${\ezf}$ (in odd dimensions): \begin{equation}\label{Darbouxformfr}
\enf\mu = {\Bigl(\frac{U_\mu}{X_\mu}\Bigr)^{\!\frac12}}\grad x_{\mu}\;,\;
\ehf\mu = {\Bigl(\frac{X_\mu}{U_\mu}\Bigr)^{\!\frac12}}
  \sum_{j=0}^{\dg-1}\A{j}_{\mu}\grad\psi_j\;,\;
\ezf = {\Bigl(\frac{c}{\A{\dg}}\Bigr)^{\frac12}}\,\sum_{k=0}^{\dg}\A{k}\grad\psi_k\,,
\end{equation}
and the dual frame  of vectors ${\env\mu,\,\ehv\mu,\,\ezv}$,
\begin{equation}\label{Darbouxvecfr}
\env\mu = {\Bigl(\frac{X_\mu}{U_\mu}\Bigr)^{\!\frac12}}{\cv{x_\mu}}\;,\;
\ehv\mu = {\Bigl(\frac{U_\mu}{X_\mu}\Bigr)^{\!\frac12}}\sum_{k=0}^{\dg-1+\eps}
  {\frac{(-x_{\mu}^2)^{\dg-1-k}}{U_{\mu}}}\,{\cv{\psi_{k}}}\;,\;
\ezv= \bigl(c \A{\dg}\bigr)^{\!-\frac12}{\cv{\psi_{\dg}}}\,.
\end{equation}
The duality follows from important properties of the metric functions ${\A{j}_\mu}$ and ${U_\mu}$ listed in appendix~\ref{ssc:mtrfcs} (see \eqref{Aid1}--\eqref{Aid2}).
In this frame the metric and its inverse take the trivial diagonal forms:
\begin{equation}
\ts{g} =  \sum_{\mu=1}^\dg\,\bigl(\,\enf\mu \enf\mu \,+\, \ehf\mu \ehf\mu\,\bigr)\;
  +\eps\,\ezf\ezf\;,\qquad
\ts{g}^{-1} =  \sum_{\mu=1}^\dg\,\bigl(\,\env\mu \env\mu \,+\, \ehv\mu \ehv\mu\,\bigr)\;
  +\eps\,\ezv\ezv\;.
\end{equation}
It is explicitly seen here that we use a Euclidean normalization of the frame and we do so even in the Lorentzian case, in which case some of the frame vectors become imaginary. We shall provide a detailed discussion of the signature and suitable choices of coordinates and signs of the metric functions in the next section.

In this frame the principal tensor takes the following simple form:
\be
\ts{h} = \sum_{\mu=1}^n x_\mu\, \enf\mu\wedge\ehf\mu\;,
\ee
which is exactly the form \eqref{PTcanform} discussed in section~\ref{ssc:PrincipalTensor}. Moreover, one can easily check that the additional condition \eqref{specDarbouxframecond} is satisfied. For this reason, the frame $\{\enf\mu, \ehf\mu, \ezf\}$ is nothing but the {\em special Darboux frame} introduced in section~\ref{ssc:PrincipalTensor}.

\subsubsection{Curvature}

The curvature of the metric \eqref{KerrNUTAdSmetric} has been calculated in \cite{Houri:2007xz}.
The important property of the Ricci tensor is that it is diagonal in the frame \eqref{Darbouxformfr}, a property that complements a rich symmetry structure of the geometry. It reads
\begin{equation}\label{Ricci}
  \Ric = -\sum_{\mu=1}^\dg \; r_\mu\; \bigl(\enf\mu\enf\mu+\ehf\mu\ehf\mu\bigl)
        \;-\; r_0\; \ezf\ezf\; .
\end{equation}
In even dimensions the components $r_\mu$ are
\begin{equation}
  r_\mu
    = \frac12\frac{X_\mu''}{U_\mu}
      +\sum_{\substack{\nu=1\\\nu\neq\mu}}^\dg\frac{x_\nu X_\nu' {-} x_\mu X_\mu'}{U_\nu(x_\nu^2{-}x_\mu^2)}
      -\sum_{\substack{\nu=1\\\nu\neq\mu}}^\dg\frac{X_\nu - X_\mu}{U_\nu(x_\nu^2{-}x_\mu^2)}
    = \frac\partial{\partial x_\mu^2}\Biggl[\,\sum_{\nu=1}^{\dg}\frac{x_\nu^2\bigl(x_\nu^{-1}X_\nu\bigr)_{\!,\nu}}{U_\nu}\Biggr]\,,
\end{equation}
while in odd dimensions we have
\begin{equation}
  r_\mu
    = \frac12\frac{\bar{X}_\mu''}{U_\mu}
      +\frac1{2x_\mu}\frac{\bar{X}_\mu'}{U_\mu}
      +\sum_{\substack{\nu=1\\\nu\neq\mu}}^\dg\frac{x_\nu \bar{X}_\nu' {-} x_\mu \bar{X}_\mu'}{U_\nu(x_\nu^2{-}x_\mu^2)}
    = \frac\partial{\partial x_\mu^2}\Biggl[\,\sum_{\nu=1}^{\dg}\frac{x_\nu\bar{X}_\nu'}{U_\nu}\Biggr]\,,\;
  r_0 = \sum_{\nu=1}^\dg\frac{\bar{X}_\nu'}{x_\nu U_\nu}\,.
\end{equation}
In the latter relations we used the shifted metric functions
\begin{equation}\label{shiftedX}
  \bar{X}_\mu = X_\mu + \frac{\eps c}{x_\mu^2}\;.
\end{equation}
The scalar curvature simplifies to
\begin{equation}\label{sccur}
  R = -\sum_{\nu=1}^\dg \; \frac{\bar{X}_\nu''}{U_\nu}
       - 2\,\eps\sum_{\nu=1}^\dg \; \frac1{x_\nu}\frac{\bar{X}_\nu'}{U_\nu}\;.
\end{equation}
In the above expressions, the prime denotes a differentiation with respect to the (single) argument
of the metric function, e.g., ${X_\mu'=X_{\mu,\mu}}$.

\subsubsection{On-shell metric}

Imposing the vacuum Einstein equations, $R_{ab}-\frac12R g_{ab} + \Lambda g_{ab}=0$,  results in the following form of the metric functions \citep{Chen:2006xh,Houri:2007xz}:
\begin{equation}\label{Xsol}
  X_\mu =
  \begin{cases}
    {\displaystyle -2b_\mu\, x_\mu + \sum_{k=0}^{\dg}\, c_{k}\, x_\mu^{2k}}
      \qquad &\text{for $D$ even}\;,\\
    {\displaystyle -\frac{c}{x_\mu^2} - 2b_\mu + \sum_{k=1}^{\dg}\, c_{k}\, x_\mu^{2k}}
      \qquad &\text{for $D$ odd}\;.
  \end{cases}
\end{equation}
The parameter $c_\dg$ is related to the cosmological constant as
\be\label{cnpar}
\Ric=(-1)^\dg(D-1)c_\dg \ts{g}\,\quad \Leftrightarrow\quad \Lambda=\frac{1}{2}(-1)^\dg(D-1)(D-2)c_\dg\,.
\ee

\details{It is interesting to note that, similar to four dimensions, a single equation corresponding to the trace of the Einstein equations, ${R=\frac{2D}{D-2}\Lambda}$, almost fully determines relations \eqref{Xsol}. Once this equation is valid, all other Einstein's equations require just equality of the absolute terms in all polynomials ${X_\mu}$ and otherwise they are identically satisfied \citep{Houri:2007xz}.}

\subsection{Parameters and alternative form of the metric}

Before we proceed to discussing various special cases of the on-shell Kerr--NUT--(A)dS spacetimes, let us comment on a different, more convenient for its interpretation,  form of the metric, and the parameters of the solution.
For simplicity, in the rest of this section we restrict our discussion to even dimensions ${D=2\dg}$, that is ${\eps=0}$, analysis in odd dimensions would proceed analogously.

\subsubsection{Parametrization of metric functions}

In even dimensions, the metric \eqref{KerrNUTAdSmetric} simplifies to\footnote{%
The Greek indices always take values ${\mu,\nu,{\dots}=1,\dots,\dg}$ and, in even dimensions, the Latin indices from the middle of alphabet take values ${j,k,l,{\dots}=0,\dots,\dg-1}$. We do not use the Einstein summation convention for them but also we do not indicate limits in sums and products explicitly, i.e., ${\sum_\mu\equiv\sum_{\mu=1}^\dg}$, ${\sum_k\equiv\sum_{k=0}^{\dg-1}}$.}
\begin{equation}\label{KerrNUTAdSpsi}
    \ts{g} = \sum_\mu\biggl[\,\frac{U_\mu}{X_\mu}\,\grad x_\mu^2+\frac{X_\mu}{U_\mu}\biggl(\sum_k \A{k}_\mu\grad\psi_k\biggr)^{\!\!2}\,\biggr]\;.
\end{equation}

Inspecting the on-shell metric functions $X_{\mu}$, \eqref{Xsol}, we see that they are given by a common even polynomial ${\Ja}$  modified by ${\mu}$-dependent linear terms:
\begin{equation}\label{Xsol2n}
    X_\mu = \lambda \Ja(x_\mu^2)-2b_\mu x_\mu\; .
\end{equation}
The parameter $\lambda$ is trivially related to $c_\dg$ in \eqref{Xsol} according to $\lambda=(-1)^\dg c_\dg$. Instead of other coefficients ${c_k}$, it will be useful to characterize the common polynomial ${\Ja}$ using its roots. Assuming they are all real the polynomial can be written as
\begin{equation}\label{Jadef}
    \Ja(x^2) = \prod_\nu (a_\nu^2-x^2) = \sum_{k=0}^\dg \Aa{k} (-x^2)^{\dg-k}\;,
\end{equation}
where the constants ${\Aa{k}}$ can be expressed in term of new parameters ${a_\mu^2}$ in a similar way as the functions ${\A{k}}$ in terms of ${x_\mu^2}$ in \eqref{AUdefs}, cf.~\eqref{Aa} in appendix~\ref{ssc:mtrfcs}.
We shall give the interpretation of all the parameters below.
However before that, let us start with a remark on two types of angular variables.

\subsubsection{Two types of angular variables}

As we already mentioned, the canonical `angles' $\psi_k$ in the metric \eqref{KerrNUTAdSpsi} are the Killing parameters for the primary and secondary Killing vectors constructed from the principal tensor. In a general case, such Killing vectors do not have fixed points and the angles do not correspond to azimuthal angles in independent rotation 2-planes. However, there may exist other angular variables such that the corresponding Killing vectors have fixed points and, hence, they define axes of symmetry and planes of rotation.

\details{
The same thing happens with the Kerr metric written in the canonical form \eqref{can}. As explained in section~\ref{ssc:RemarksAngles}, the axisymmetry of the Kerr metric implies that, aside the Killing coordinate $\psi$, there exists another angular variable~$\phi$, such that the Killing vector $\cv{\phi}$ has fixed points and corresponds to the azimuthal angle in the 2-plane of rotation.
}

We can indeed introduce new higher-dimensional angular variables $\phi_{\alpha}$, that have a desired property (at least for the special case, when ${b_\mu=0}$ for ${\mu<\dg}$, see below). These new angular coordinates $\phi_{\alpha}$ are linear combinations of $\psi_k$:
\begin{equation}\label{phipsirel}
    \phi_\alpha = \lambda a_\alpha \sum_k \Aa{k}_\alpha\psi_k\;
    \quad\Leftrightarrow\quad
    \psi_k = \sum_\alpha\frac{(-a_\alpha^2)^{\dg{-}1{-}k}}{\Ua_\alpha}\frac{\phi_\alpha}{\lambda a_\alpha}\;.
\end{equation}
Since they are just linear combinations of $\psi$'s with constant coefficients, they are also Killing coordinates.
Using these angles, the metric can be written in the form\footnote{%
Note that in this form of the metric, one cannot straightforwardly set $\lambda$, related to the `radius of the deformed sphere', to zero. The `vacuum limit' $\lambda\to 0$ is discussed in the Lorentzian signature in section~\ref{ssc:SubcasesLorentzian}. Lorentzian version of \eqref{phipsirel} is given by \eqref{psiphirelLor} below.}
\begin{equation}\label{KerrNUTAdSphi}
    \ts{g} = \sum_\mu\biggl[\,\frac{U_\mu}{X_\mu}\,\grad x_\mu^2 + \frac{X_\mu}{U_\mu}
    \biggl(\sum_\alpha \frac{J_\mu(a_\alpha^2)}{\Ua_\alpha}
    \frac{1}{\lambda a_\alpha}\grad\phi_\alpha\biggr)^{\!\!2}\,\biggr]\;,
\end{equation}
where ${J_\mu}$, ${\A{k}_\mu}$,  ${\Ja_\mu}$, ${\Aa{k}_\mu}$, ${U_\mu}$, and ${\Ua_\mu}$ are defined and related as
\begin{equation}\label{JJadef}
\begin{aligned}
    J_\mu(a^2)&=\prod_{\substack{\nu\\\nu\neq\mu}}(x_\nu^2-a^2)
              =\sum_k \A{k}_\mu (-a^2)^{\dg{-}1{-}k}\;,\\
    \Ja_\mu(x^2)&=\prod_{\substack{\nu\\\nu\neq\mu}}(a_\nu^2-x^2)
              =\sum_k \Aa{k}_\mu (-x^2)^{\dg{-}1{-}k}\;,
\end{aligned}
\end{equation}
and
\begin{equation}\label{UUadef}
   U_\mu = J_\mu(x_\mu^2)\;,\qquad
   \Ua_{\mu} = \Ja_\mu(a_\mu^2)
     \;,
\end{equation}
cf.\ appendix~\ref{ssc:mtrfcs}.

\subsubsection{Parameters of the solution}
The on-shell geometries \eqref{KerrNUTAdSpsi} and \eqref{KerrNUTAdSphi} are labeled by parameters ${a_\mu}$, ${b_\mu}$, and ${\lambda}$. As we have already said, the clearest interpretation has the parameter ${\lambda}$. After plugging the metric into the Einstein equations, ${R_{ab}-\frac12R g_{ab}+\Lambda g_{ab}=0}$, one finds that $\lambda$ is related to the cosmological constant ${\Lambda}$ according to
\begin{equation}\label{Lambdalambda}
  \Lambda=(2\dg-1)(\dg-1)\lambda\;,
\end{equation}
cf.\ \eqref{cnpar}. A general wisdom tells us that ${a}$'s should be related to rotations (at least in the weak field limit), and ${b}$'s to the mass and NUT charges. However, the exact interpretation depends on various other choices that have to be made before interpreting the meaning of the parameters.

First we realize that the parameters ${a_\mu}$ and ${b_\mu}$ are not independent.  There exists a one-parametric  freedom in rescaling coordinates, metric functions, and parameters which leaves the metric in the same form:
\begin{equation}\label{rescaling}
\begin{gathered}
   x_\mu \to s x_\mu\;,\quad
     \phi_\alpha\to\phi_\alpha\;,\quad
     \psi_k\to s^{-(2k{+}1)}\psi_k\;,\\
   a_\mu \to s a_\mu\;,\quad
     b_\mu\to s^{2\dg-1} b_\mu\;,\quad
     \lambda\to\lambda\;,\\
   X_\mu\to s^{2\dg} X_\mu\;,\quad
     U_\mu\to s^{2(\dg{-}1)}U_\mu\;,\quad
     \A{k}_\mu\to s^{2k}\A{k}\;.
\end{gathered}
\end{equation}
This transformation simply rescales dimensional coordinates ${x_\mu}$ and parameters ${a_\mu}$, properly rescales NUT parameters ${b_\mu}$, and leaves untouched dimensionless angles ${\phi_\alpha}$. Using this transformation, one of the parameters $a_\mu$ can be set to a suitable value. Later we shall fix this freedom by imposing the condition \eqref{aNnorm}.

Taking into account this freedom, we find that for a fixed cosmological constant the on-shell Kerr--NUT--(A)dS  metric in ${D=2\dg}$ dimensions contains $2\dg-1$ independent parameters. In the black hole case they are connected with mass, $(\dg-1)$ rotations parameters, and $(\dg-1)$ NUT charges.

Similar counting would proceed in odd dimensions, where the analogous scaling freedom reduces the number of independent free parameters in $D=2n+1$ dimensions to $2\dg-1$, giving mass, $\dg$ rotations parameters, and $(\dg-2)$ NUT parameters for the black hole case, see \cite{Chen:2006xh}.

\subsection{Euclidean signature: instantons}
\label{ssc:SubcasesEuclidean}
The Kerr--NUT--(A)dS metric can describe various geometries. Depending on a choice of coordinate ranges and values of parameters it can have both Euclidean and Lorentzian signatures. We will see in the next chapter that common feature of the solution independent of a particular interpretation of the geometry is the presence of a rich symmetry structure. If one is interested mainly in the symmetries of the Kerr--NUT--(A)dS geometry and its integrability and separability properties, the general form of the metric presented above is sufficient to proceed directly to chapters~\ref{sc:hshdbh} and \ref{sc:intsep}.

In the rest of this chapter we make a short overview of several important special cases of the Kerr--NUT--(A)dS metric. In this section we explain appropriate coordinate ranges for Euclidean version of the geometry, in the next section we discuss the Wick rotations of coordinates appropriate for the Lorentzian signature.

\subsubsection{Sphere}

Let us begin with a `trivial' example of a $D$ dimensional sphere. This homogeneous and isotropic metric is a very special case of Kerr--NUT--(A)dS geometry. The corresponding metric is obtained by setting the NUT and mass parameters equal to zero, $b_\mu=0$, while keeping the parameters $a_\mu$ arbitrary, and $\lambda>0$. The on-shell metric functions $X_\mu$ then simplify and take the form of a common polynomial $\lambda\Ja(x^2)$ in the corresponding variable:
\begin{equation}\label{Xmaxsym}
  X_\mu=\lambda\Ja(x_\mu^2)\;.
\end{equation}
The roots of this polynomial are exactly the parameters~$a_\mu^2$ whose interpretation is discussed below. With this choice we can employ the orthogonality relations \eqref{Jortrel3}
in the angular part of the metric \eqref{KerrNUTAdSphi} and transform it to the following form:
\begin{equation}\label{KerrNUTAdSphiMS}
    \ts{g} = \sum_\mu\Bigg[\,\frac{U_\mu}{\lambda \Ja(x_\mu^2)}\,\grad x_\mu^2
      -\frac{J(a_\mu^2)}{\Ua_\mu}\frac{1}{\lambda a_\mu^2}\grad\phi_\mu^2\,\Biggr]\;.
\end{equation}
Here ${J(a^2)=\prod_{\nu}(x_\nu^2-a^2)}$ is given by definition \eqref{metricpolysx} analogous to \eqref{Jadef} above.

Let us introduce $\dg+1$ new coordinates $\rho_\mu$, $\mu=0,\,1,\dots,\,\dg$, instead of $\dg$ coordinates $x_\mu$, and apply the {\em Jacobi transformation}
\begin{equation}\label{rhodef}
    \lambda\rho_\mu^2=\frac{J(a_\mu^2)}{-a_\mu^2\,\Ua_\mu}
      = \frac{\prod_{\nu}(x_\nu^2-a_\mu^2)}{-a_\mu^2\prod_{\nu\neq\mu}(a_\nu^2-a_\mu^2)}\;,\quad
    \lambda\rho_0^2 = \frac{\A{\dg}}{\Aa{\dg}}
      = \frac{\prod_\nu x_\nu^2}{\prod_\nu a_\nu^2}\; .
\end{equation}
Then one can show that the new coordinates $\rho_{\mu}$ are restricted by the constraint
\begin{equation}\label{rhoconstraint}
    \sum_{\mu=0}^\dg\rho_\mu^2 = \frac{1}{\lambda}\,,
\end{equation}
and the ${x}$-part of the metric can be written as
\begin{equation}\label{dx2=drho2}
    \sum_\mu\frac{U_\mu}{\lambda \Ja(x_\mu^2)}\,\grad x_\mu^2
       = \grad\rho_0^2 + \sum_\mu\grad\rho_\mu^2\;.
\end{equation}
Using these relations we obtain the following simple form of the  metric $  \ts{g}$
\begin{equation}\label{multicyl}
    \ts{g} = \grad\rho_0^2 + \sum_{\mu} \Bigl[ \grad\rho_\mu^2 +  \rho_\mu^2\,\grad\phi_\mu^2\Bigr]\;,
\end{equation}
with coordinates $\rho_\mu$ constrained by \eqref{rhoconstraint}.
Clearly, $(\rho_0$, $\rho_\mu$, $\phi_\mu)$ are multi-cylindrical coordinates on  a $2\dg$-dimensional sphere embedded in a $(2\dg{+}1)$-dimensional flat space. The sphere is given by the constrain equation \eqref{rhoconstraint}.

It is interesting to observe that this metric describes the maximally symmetric geometry of the sphere of the same radius $1/\sqrt{\lambda}$ for any choice of parameters ${a_\mu}$. Going in the opposite direction, from the spherical geometry \eqref{multicyl}, expressed in coordinates $({\rho_0,\,\rho_\mu,\,\phi_\alpha})$, to the Kerr--NUT--(A)dS metric \eqref{KerrNUTAdSphiMS}, and then to \eqref{KerrNUTAdSphi}, expressed in the coordinates $({x_\mu,\,\phi_\alpha})$, it turns out that the parameters ${a_\mu}$ characterize a freedom in implicit definitions \eqref{rhodef} of variables ${x_\mu}$ obeying the constrain \eqref{rhoconstraint}. Jacobi coordinates ${x_\mu}$ are sort-of elliptic coordinates (the surfaces of given ${x_\mu}$ being elliptical or hyperbolic surfaces) with an exact shape governed by parameters ${a_\mu}$.

To specify the ranges of coordinates in details, let us start with ${\lambda>0}$, ${\rho_0\in\realn}$, ${\rho_\mu\in\realn^+}$ and ${\phi_\alpha\in(-\pi,\pi)}$ for which the metric \eqref{multicyl} is the homogeneous geometry on the sphere. Assuming further 
\begin{equation}\label{aord}
    0<a_1<\dots<a_\dg\;,
\end{equation}
the ranges of the coordinates ${x_\mu}$ should be chosen as
\begin{equation}\label{xrangesMS}
    -a_1<x_1<a_1\;,\qquad a_{\mu-1}<x_\mu<a_\mu\;,\quad\mu=2,\dots,\dg\;,
\end{equation}
which guarantees that ${U_\mu}$ are nonsingular and ${X_\mu/U_\mu>0}$. The boundaries of ${x_\mu}$-ranges coincide with the roots of the metric functions ${X_\mu}$ and correspond to symmetry axes. Inspecting \eqref{multicyl}, we see that the axes are given by ${\rho_\nu=0}$. In terms of coordinates $x_\mu$, Jacobi transformation \eqref{rhodef} gives that ${x_\mu=a_\mu}$ identifies with ${\rho_\mu=0}$, and, for ${\mu>1}$, ${x_\mu=a_{\mu-1}}$ corresponds to ${\rho_{\mu-1}=0}$. Each of the axes ${\rho_\mu=0}$ (for ${\mu<\dg}$) thus splits into two regions described by ${x_\mu=a_\mu}$ and ${x_{\mu+1}=a_\mu}$, respectively. Finally, a sign of ${x_1}$ is the same as the sign of ${\rho_0}$.

For non-vanishing parameters ${b_\mu}$ one cannot use {the orthogonality relation} \eqref{Jortrel3} and transform the Kerr--NUT--(A)dS metric \eqref{KerrNUTAdSphi} to the form \eqref{KerrNUTAdSphiMS}. However, we have at least learned that coordinates ${x_\mu}$ take values between the roots of metric functions ${X_\mu}$, and these roots represents the axes of the Killing symmetry. This property survives in the generic case.

Let us finally note that the metric \eqref{multicyl} or the corresponding Kerr--NUT--(A)dS form \eqref{KerrNUTAdSphi} can also describe a pseudo-sphere of various signatures, obtainable by a suitable Wick rotation of coordinates. We will discuss this below after we introduce the black hole solutions.

\subsubsection{Euclidean instantons}
Let us now describe the choice of coordinate ranges and parameters for which the Kerr--NUT--(A)dS metric describes a non-trivial geometry of the Euclidean signature.

\details{For briefness we call such metrics \defterm{Euclidean instantons} or simply \defterm{instantons}. In fact, in order to be a `proper instanton', the space must be regular and the corresponding gravitational action finite. These properties can impose additional restrictions on the parameters of the solution, which we do not study here and refer the interested reader to a vast literature on the subject of gravitational instantons, e.g. \cite{hawking1977gravitational, Page:1979zv, Page:1979aj, gibbons1979classification, eguchi1980gravitation, hunter1998action, Mann:1999pc, Chamblin:1998pz, Mann:2003zh, Mann:2005ra, Clarkson:2006zk, Chen:2006ea, Yasui:2011pr}.
}

Let us assume that ${\lambda>0}$ and all coordinates $x_\mu$, $\psi_k$ and parameters ${a_\mu}$, ${b_\mu}$ are real. We further order parameters ${a_\mu}$ as in~\eqref{aord} and ${x_\mu}$ so that
\begin{equation}\label{xord}
    x_1<x_2<\dots<x_\dg\;.
\end{equation}
This guarantees that ${U_\mu}$ are nonsingular and their signs are ${\sign U_\mu = -(-1)^\mu}$. As we have seen above, when all $b_{\mu}$ vanish, the ranges of $x_{\mu}$ coordinates are given by \eqref{xrangesMS}. If some parameters $b_{\mu}$ do not vanish, the ranges of $x_{\mu}$ must be modified. Since the signature of the metric \eqref{KerrNUTAdSpsi} is determined by the signs of metric functions $X_\mu/U_\mu$, to obtain a Euclidean metric we thus need ${\sign X_\mu = -(-1)^\mu}$. Therefore, the ranges of coordinates ${x_\mu}$,
\begin{equation}\label{xrange}
    \xroot-\mu<x_\mu<\xroot+\mu\;,
\end{equation}
should be chosen between the roots ${\xroot\pm\mu}$ of metric functions ${X_\mu}$ such that the suitable sign of ${X_\mu}$ is guaranteed.

\begin{figure}
\centerline{\includegraphics[scale=0.9]{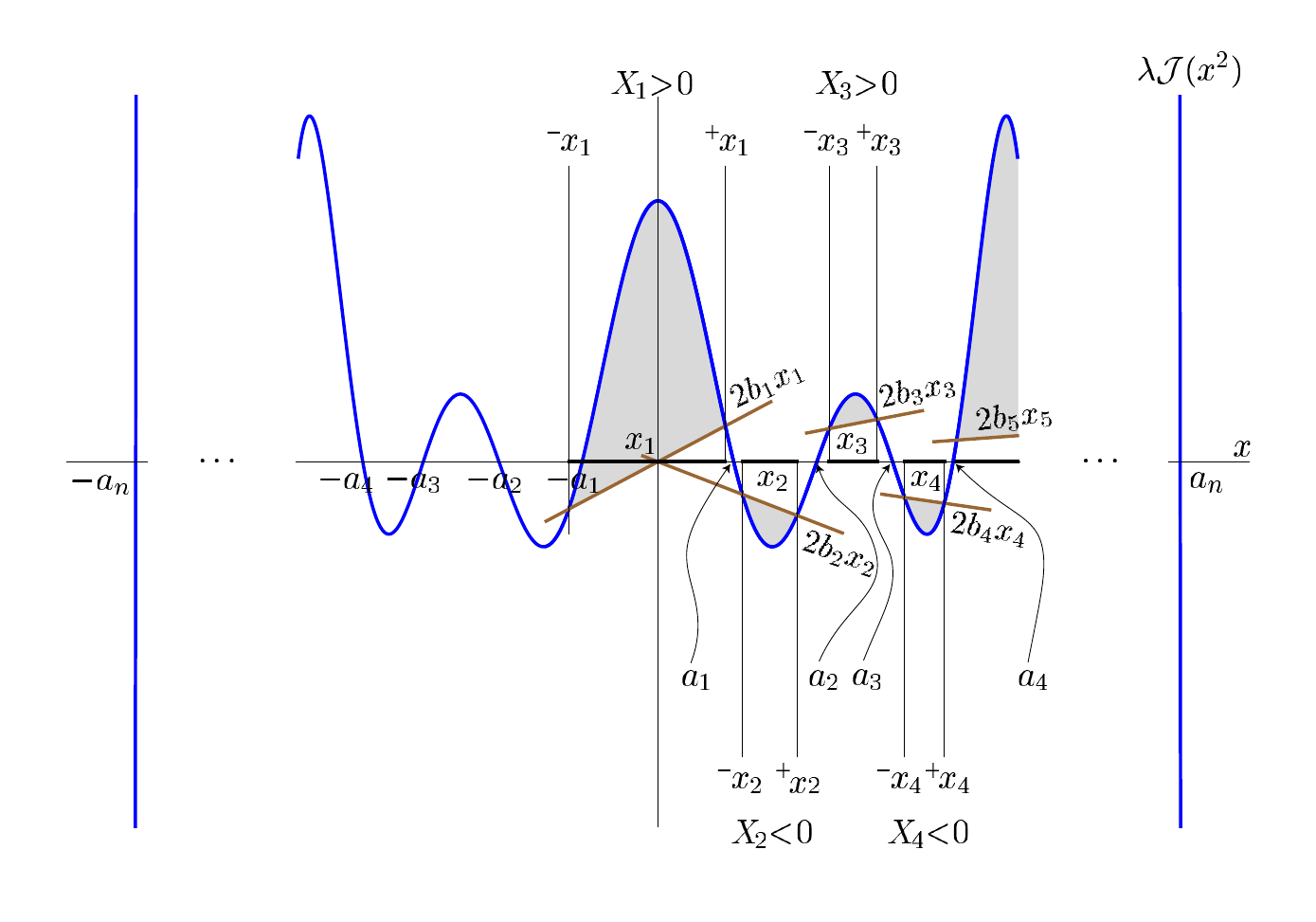}}\vspace*{-5mm}
\caption{\label{fig:XrangesEI}\small%
{\bf Euclidean instanton.}
The graph of the common polynomial ${\lambda\Ja(x^2)}$ combined with various linear contributions ${2b_\mu x_\mu}$. Intersections of the polynomial with these linear lines correspond to roots of the metric functions ${X_\mu}$, cf.~\eqref{Xsol2n}. The shaded areas indicate regions where ${\sign X_\mu = -(-1)^\mu}$. These regions can be chosen as ranges of coordinates ${x_\mu}$, cf.~\eqref{xrange}.}
\end{figure}

For small values of the NUT parameters ${b_\mu}$ these roots will be `close' to the roots ${a_\mu}$ of the common polynomial ${\lambda\Ja(x^2)}$. As one can see in figure~\ref{fig:XrangesEI}, if
\begin{equation}\label{bsigns}
    \sign b_\mu = -(-1)^\mu\;,
\end{equation}
the relevant roots ${\xroot\pm\mu}$ of ${X_\mu}$  are
\begin{equation}\label{XrootsEI}
    X_\mu(\xroot\pm\mu) = 0\;,\qquad a_{\mu-1}<\xroot-\mu<\xroot+\mu<a_\mu\,,
\end{equation}
with the only exception of ${\xroot-1}$ which is the largest root of ${X_1}$ \emph{smaller} than ${-a_1}$.

For such a choice the metric \eqref{KerrNUTAdSpsi} represents the {\em Euclidean instanton} of signature ${(++\dots+)}$. Parameters $b_\mu$ encode deformations of the geometry, namely how it deviates from the geometry of the sphere. For non-vanishing $b_\mu$, parameters $a_\mu$ become essential. That is they do not just label  a choice of coordinates, as in the maximally symmetric case, but their change results in the  change of the geometry (e.g. its curvature).

The global definition and regularity of the geometry described by metrics \eqref{KerrNUTAdSpsi} or \eqref{KerrNUTAdSphi} has to be established by specifying which Killing angles should be cyclic and what are the periods of these cyclic angles. In the maximally symmetric case, which we discussed above, there was a natural choice of cyclic coordinates ${\phi_\alpha\in(-\pi,\pi)}$ with their natural identification at ${\phi_\alpha=\pm\pi}$. However, in general, any linear combination of Killing coordinates (with constant coefficients) forms again a Killing coordinate and it is not a~priory clear which of the Killing coordinates should be periodic. Learning a lesson from the maximally symmetric case, the angles $\psi_k$ are typically \emph{not} those which should be periodic. Since Killing coordinates are non-trivially coupled in the metric (the metric is not diagonal in these directions), a particular choice of the periodicity of Killing coordinates can introduce a non-trivial twisting of the geometry, as well as possible irregularities on the axes. We will not discuss these characteristics in more detail as this is still an open problem awaiting its complete solution. For our purposes it is sufficient to simply remember that the Euclidean instanton describes a deformed and twisted spherical-like geometry. Other examples of compact Riemannian manifolds that can be obtained as special limits of the Kerr--NUT--(A)dS  metrics include the most general explicitly known Einstein--K\"ahler and Einstein--Sasaki metrics, see e.g. \cite{Yasui:2011pr} and references therein.

\subsection{Lorentzian signature: black holes}
\label{ssc:SubcasesLorentzian}

Let us now discuss the Kerr--NUT--(A)dS metrics with the Lorentzian signature.  For vanishing NUT parameters such metrics describe an isolated rotating higher-dimensional black hole in either asymptotically flat or asymptotically (anti-)de~Sitter spacetime. We start our discussion with the case of non-vanishing NUT parameters and proceed to the Kerr-(A)dS and Myers--Perry black holes in the the next step.

\subsubsection{General multiply-spinning black holes with NUTs}
The Lorentzian signature can be achieved by a suitable Wick rotation of coordinates and parameters. Different choices can lead to physically different spacetimes. We concentrate on the case where the coordinate $x_\dg$ is Wick-rotated to a radial coordinate $r$ and the angular coordinate $\phi_\dg$ to a time coordinate~$t$:
\begin{equation}\label{Wick}
    x_\dg = i r\;,\quad
    \phi_\dg = \lambda a_\dg t\;,
\end{equation}
with $r$ and $t$ real, while the remaining ${x}$'s and ${\phi}$'s retain their original character.
We also define the (real) mass parameter ${M}$ by
\begin{equation}\label{mass}
    b_\dg = i M\;.
\end{equation}
To obtain the correct signature we also need to correlate the sign of the cosmological constant with the sign of ${a_\dg^2}$. By employing the  scaling transformations \eqref{rescaling} we can use this freedom to impose the following condition:
\begin{equation}\label{aNnorm}
    a_\dg^2 = -\frac1\lambda\;.
\end{equation}
Thank to this choice, the temporal coordinate ${\phi_\dg}$ is Wick-rotated by an imaginary factor only for ${\lambda>0}$. Namely, introducing the cosmological scale ${\ell}$, we get
\begin{equation}\label{Wickbyslambda}
\begin{aligned}
    \lambda &=\frac1{\ell^2}>0\;:\quad  & a_\dg &= i\ell\;,\quad & \phi_\dg &= i\frac t\ell\;,\\
    \lambda &=-\frac1{\ell^2}<0\;:\quad & a_\dg &= \ell\;,\quad  & \phi_\dg &= -\frac t\ell\;.
\end{aligned}
\end{equation}

Let us now introduce a notation which will allow us to separate the angular sector from the temporal and radial ones. For the angular sector we employ the barred indices. Using ${\Nb=\dg-1}$ we can thus write the ranges for barred Greek indices: ${\mub, \nub = 1,\dots,\Nb}$ and barred Latin indices ${\kb,\lb=0,\dots,\Nb-1}$. We also use quantities ${\Ab{\kb}_\mub}$, ${\Ub_\mub}$, ${\Jab(x^2)}$, ${\Aab{\kb}}$, etc.\ to denote the same expressions as ${\A{k}_\mu}$, ${U_\mu}$, ${\Ja(x^2)}$, ${\Aa{k}}$, only with appropriately modified ranges of coordinates.

Using this notation and after the Wick rotation the Kerr--NUT--(A)dS metric \eqref{KerrNUTAdSphi} takes the following form:
\begin{equation}\label{KerrNUTAdSphiWick}
\begin{split}
    \ts{g} &=
    -\frac{\Delta_r}{\Sigma}\biggl(
      \prod_\nub\frac{1+\lambda x_\nub^2}{1+\lambda a_\nub^2} \;\grad t
      - \sum_\nub \frac{\Jb(a_\nub^2)}{a_\nub(1+\lambda a_\nub^2)\Uab_\nub}\grad\phi_\nub\biggr)^{\!\!2}\\
    &\quad+\frac{\Sigma}{\Delta_r}\,\grad r^2
    +\sum_\mub \frac{(r^2{+}x_\mub^2)}{\Delta_\mub/\Ub_\mub}\,\grad x_\mub^2\\
    &\quad+\sum_\mub\frac{\Delta_\mub/\Ub_\mub}{(r^2{+}x_\mub^2)}\biggl(
      \frac{1{-}\lambda r^2}{1{+}\lambda x_\mub^2}
      \prod_\nub\frac{1{+}\lambda x_\nub^2}{1{+}\lambda a_\nub^2} \,\grad t
      + \sum_\nub \frac{(r^2{+}a_\nub^2)\Jb_\mub(a_\nub^2)}
      {a_\nub(1{+}\lambda a_\nub^2)\,\Uab_\nub} \grad\phi_\nub \biggr)^{\!\!2}\;,
\end{split}\raisetag{14ex}
\end{equation}
where the metric functions read
\begin{align}
    \Delta_r= -X_\dg &= \bigl(1{-}\lambda r^2\bigr)\prod_\nub\bigl(r^2{+}a_\nub^2\bigr) - 2Mr\;,
       &\;
    U_\dg &=\Sigma=\prod_\nub(r^2+x_\nub^2)\;,
       \notag\\
    \Delta_\mub=-X_\mub &= \bigl(1{+}\lambda x_\mub^2\bigr) \Jab(x_\mub^2) + 2 b_\mub x_\mub \;,
       &\;
    \Ub_\mub &=\prod_{\substack{\nub\\\nub\neq\mub}}(x_\nub^2-x_\mub^2)\;.\label{physmtrcfc}
\end{align}
We call the coordinates $(t,r, x_\mub, \phi_\mub)$ the {\em generalized Boyer--Lindquist coordinates} and the form \eqref{KerrNUTAdSphiWick} with \eqref{physmtrcfc} the generalized Boyer--Lindquist form of the Kerr--NUT--(A)dS black hole geometry.\footnote{%
As opposed to the Myers--Perry coordinates $(t,r,\mu_k, \phi_k)$, the coordinates $(t,r, x_\mub, \phi_\mub)$ are already all independent, c.f. constraint \eqref{constr}. For this reason the metric \eqref{KerrNUTAdSphiWick} is `closer' to the Boyer--Lindquist form of the Kerr geometry in four dimensions than the Myers--Perry form \eqref{MPM}.}

Alternatively, it is useful to write the Lorentzian metric in the Carter-like form. To do this, we split the set of Killing coordinates ${\psi_k}$ into temporal coordinate ${\tau\equiv\psi_0}$ and angular coordinates
${\psb_{\kb}\equiv\psi_{\kb+1}}$. After Wick rotation, relations \eqref{phipsirel} become
\begin{equation}\label{psiphirelLor}
\begin{aligned}
    \tau \equiv \psi_0 &= \frac{1}{\prod_\mub (1+\lambda a_\mub^2)} t
        - \sum_\mub \frac{(-a_\mub^2)^\Nb}{(1+\lambda a_\mub^2)\Uab_\mub}
          \frac{\phi_\mub}{a_\mub}\;,
    \\
    \psb_\kb \equiv \psi_{\kb+1} &= \frac{\lambda^{\kb+1}}{\prod_\mub (1+\lambda a_\mub^2)} t
        - \sum_\mub \frac{(-a_\mub^2)^{\Nb-1-\kb}}{(1+\lambda a_\mub^2)\Uab_\mub}
          \frac{\phi_\mub}{a_\mub}\;,
\end{aligned}
\end{equation}
giving
\begin{equation}\label{phipsirelLor}
    t = \tau + \sum_\kb\Aab{\kb+1}\psb_\kb\;,
    \qquad
    \frac{\phi_\mub}{a_\mub} =
      \lambda \tau - \sum_\kb\bigl(\Aab{\kb}_\mub-\lambda\Aab{\kb+1}_\mub\bigr)\psb_\kb\;
\end{equation}
for the inverse expressions.
With these definitions, the metric \eqref{KerrNUTAdSpsi} takes the following {\em Carter-like form}:
\begin{equation}\label{KerrNUTAdSpsiWick}
\begin{split}
    \ts{g} &=
    -\frac{\Delta_r}{\Sigma}
      \Bigl( \grad \tau + \sum_\kb \Ab{\kb+1}\grad\psb_\kb \Bigr)^{\!2}
    +\frac{\Sigma}{\Delta_r}\,\grad r^2
    \\
    &\quad+\sum_\mub \frac{(r^2{+}x_\mub^2)}{\Delta_\mub/\Ub_\mub}\,\grad x_\mub^2
    +\sum_\mub\frac{\Delta_\mub/\Ub_\mub}{(r^2{+}x_\mub^2)}\Bigl(
      \grad \tau + \sum_\kb\bigl(\Ab{\kb+1}_\mub-r^2\Ab{\kb}_\mub\bigr)\grad\psb_\kb
      \Bigr)^{\!2}\;,
\end{split}\raisetag{12ex}
\end{equation}
generalizing \eqref{canNOreference} in four dimensions.

\begin{figure}
\centerline{\includegraphics[width=\textwidth]{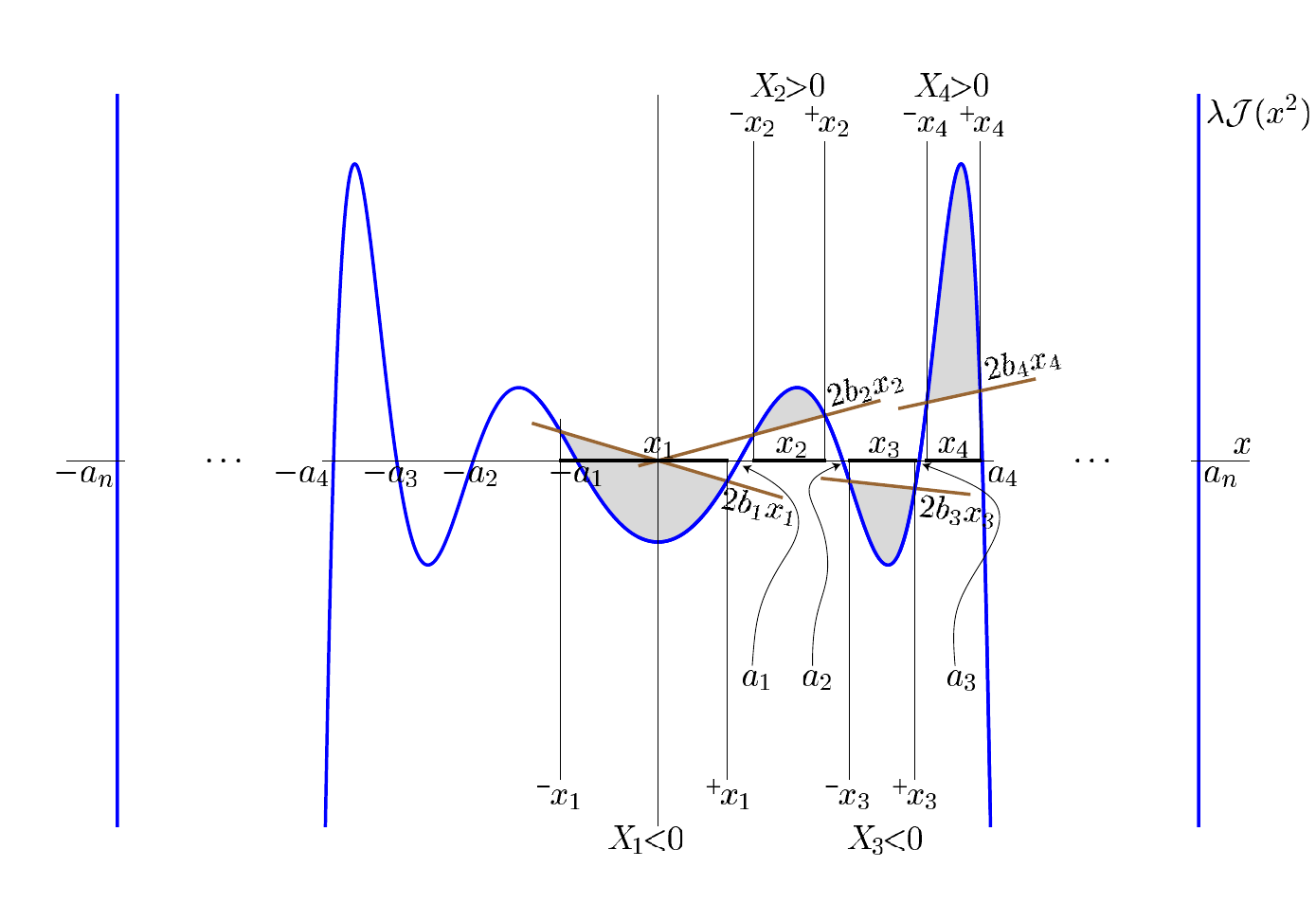}}\vspace*{-5mm}
\caption{\label{fig:XrangesBH}\small%
{\bf Black hole geometries.}
The graph of the common polynomial ${\lambda\Ja(x^2)=-(1+\lambda x^2)\Jab(x^2)}$ is combined with various linear contributions ${2b_\mub x_\mub}$. Intersections of the polynomial with these linear lines correspond to roots of the metric functions $\Delta_\mub=-X_\mub$, 
cf.~\eqref{physmtrcfc}. The shaded areas indicate regions where ${\sign \Delta_\mub = -(-1)^\mub}$. These regions can be chosen as ranges of coordinates ${x_\mub}$, cf.~\eqref{xrangeBH}.}\bigskip
\centerline{\includegraphics{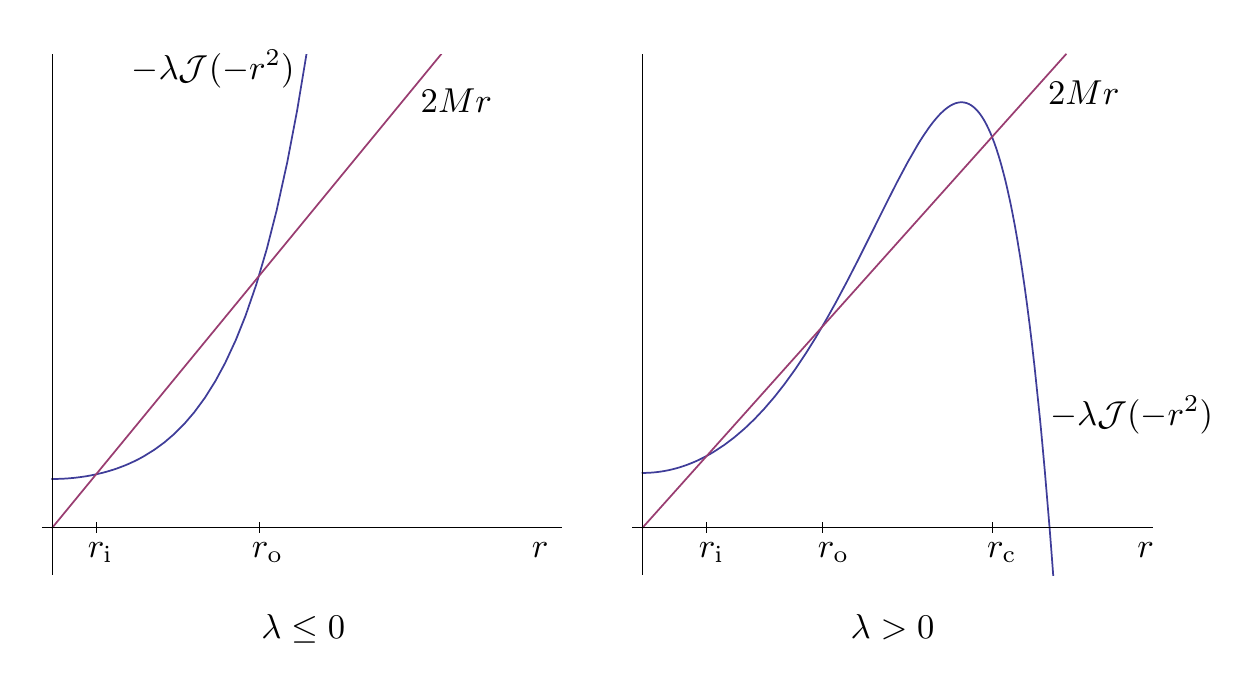}}\vspace*{-5mm}
\caption{\label{fig:horfc}\small%
{\bf Black hole horizons.} The roots of the metric function $\Delta_r = - \lambda \Ja(-r^2)- 2 M r$, cf.~\eqref{physmtrcfc}, determine horizons of the black hole. The diagrams show graphs of the even order polynomial ${- \lambda \Ja(-r^2)}$ and of the linear term $2 M r$. Their intersections define the horizons. For $\lambda\leq0$ there can be two intersections (outer and inner horizons), one touching intersection (extremal horizon) or no intersections (naked singularity). For $\lambda>0$ there is one additional intersection corresponding to the cosmological horizon.}
\end{figure}

Let us now discuss the suitable ranges  of coordinates. We assume ordering of the parameters ${a_\mub}$ as
\begin{equation}\label{aordBH}
    0<a_1<\dots<a_\Nb\; .
\end{equation}%
When all the NUT parameters vanish, each $x_{\mub}$ takes its values in the interval bounded by two neighbours of the corresponding $a_{\mub}$. One can also identify the proper ranges of coordinates when NUT parameters ${b_\mub}$ do not vanish, provided they satisfy additional requirements. Namely, they should have signs
\begin{equation}\label{bsignsBH}
    \sign b_\mub = (-1)^\mub\;,
\end{equation}
and the metric functions ${X_\mub}$ should have roots close to ${a}$'s,
\begin{equation}\label{XrootsBH}
    X_\mub(\xroot\pm\mub) = 0\;,\qquad a_{\mub-1}<\xroot-\mub<\xroot+\mub<a_\mub
\end{equation}
(with the exception of ${\xroot-1}$ which is the largest root of ${X_1}$ \emph{smaller} than ${-a_1}$). The coordinates ${x_\mub}$ then take the following values:
\begin{equation}\label{xrangeBH}
    \xroot-\mub<x_\mub<\xroot+\mub\;,
\end{equation}
and satisfy
\begin{equation}\label{xordBH}
    x_1<x_2<\dots<x_\Nb\;,
\end{equation}
see figure~\ref{fig:XrangesBH} describing this situation.  The ranges and periodicity of coordinates ${\phi_\mub}$ and ${\psb_\kb}$ have to be specified to meet some kind of regularity on the axes. These conditions highly depend on the values of parameters ${a_\mu}$ and NUT parameters ${b_\mub}$; a complete discussion of this problem has not yet been performed in the literature.
The temporal coordinates ${\tau}$ and ${t}$ are real, and so is the radial coordinate ${r}$. The metric function ${\Delta_r}$ determines the horizon structure. Depending on the sign of the cosmological constant it has typically two or three roots ${r_{\mathrm{i}}}$, ${r_{\mathrm{o}}}$, and ${r_{\mathrm{c}}}$ that identify the inner horizon, the outer horizon, and (for $\lambda>0$) the cosmological horizon. The form of the metric function is illustrated in figure~\ref{fig:horfc}.

\subsubsection{Vacuum rotating black holes with NUTs}

For the vanishing cosmological constant, ${\lambda=0}$, the black hole metric \eqref{KerrNUTAdSphiWick} significantly simplifies and reads\footnote{Note that the fact that the vacuum limit, $\lambda\to 0$, can be taken in \eqref{KerrNUTAdSphiWick} has its origin in the gauge choice \eqref{aNnorm}.}
\begin{equation}\label{KerrNUTAdSphiWicklambda0}
\begin{split}
    \ts{g} &=
    -\frac{\Delta_r}{\Sigma}\Bigl(
      \;\grad t
      - \sum_\nub \frac{\Jb(a_\nub^2)}{a_\nub\Uab_\nub}\grad\phi_\nub\Bigr)^{\!2}
    +\frac{\Sigma}{\Delta_r}\,\grad r^2
    \\
    &\quad+\sum_\mub \frac{(r^2{+}x_\mub^2)}{\Delta_\mub/\Ub_\mub}\,\grad x_\mub^2
    +\sum_\mub\frac{\Delta_\mub/\Ub_\mub}{(r^2{+}x_\mub^2)}\Bigl(
      \grad t
      + \sum_\nub (r^2{+}a_\nub^2) \frac{\Jb_\mub(a_\nub^2)}
      {a_\nub\Uab_\nub} \grad\phi_\nub \Bigr)^{\!2}\;,
\end{split}
\end{equation}
with the metric functions
\begin{equation}\label{physmtrcfclambda0}
\begin{aligned}
    \Delta_r\;= -X_\dg &= \prod_\nub\bigl(r^2{+}a_\nub^2\bigr) - 2 M r \;,
       \\
    \Delta_\mub=-X_\mub &= \Jab(x_\mub^2) + 2 b_\mub x_\mub \;,
\end{aligned}
\end{equation}
and other metric functions unchanged.

The metric \eqref{KerrNUTAdSpsiWick} does not change its form for the vanishing cosmological constant, apart from the simplification of the metric functions \eqref{physmtrcfclambda0}. Note also that the  relations \eqref{psiphirelLor} and \eqref{phipsirelLor} between temporal coordinates and angles partially decouple:
\begin{equation}\label{psiphirelLorlambda0}
    \tau = t - \sum_\mub \frac{(-a_\mub^2)^{\Nb}}{\Uab_\mub} \frac{\phi_\mub}{a_\mub}\;,
    \qquad
    \psb_\kb = - \sum_\mub \frac{(-a_\mub^2)^{\Nb-1-\kb}}{\Uab_\mub}
          \frac{\phi_\mub}{a_\mub}\;,
\end{equation}
\begin{equation}\label{phipsirelLor2}
    t = \tau + \sum_\kb\Aab{\kb+1}\psb_\kb\;,
    \qquad
    \frac{\phi_\mub}{a_\mub} =
       - \sum_\kb\Aab{\kb}_\mub\psb_\kb\;.
\end{equation}

\subsubsection{Kerr--(A)dS,  Myers--Perry, and Tangherlini metrics}

Let us now focus on the physically interesting case of black hole geometries for which all the NUT parameters $b_\mu$, apart from  the mass parameter ${b_\dg}$, vanish. In this case only the metric function ${\Delta_r=-X_\dg}$ differs from the simple form \eqref{Xmaxsym}. One can thus employ the orthogonality transformation \eqref{Jortrel3} of the angular part of the metric \eqref{KerrNUTAdSphi}, as we did in the case of a sphere, obtaining so the terms  \eqref{KerrNUTAdSphiMS} augmented with an extra piece proportional to mass ${M}$:
\begin{equation}\label{KerrAdS1}
\begin{split}
    \ts{g} &= \sum_\mub\,\frac{U_\mub}{\lambda \Ja(x_\mub^2)}\,\grad x_\mub^2
    +\frac{\Sigma}{\Delta_r}\grad r^2
    -\sum_\mub\frac{J(a_\mub^2)}{\Ua_\mub}\frac{1}{\lambda a_\mub^2}\grad\phi_\mub^2
    -\lambda\frac{J(a_\dg^2)}{\Ua_\dg}\grad t^2\\
    &\mspace{180mu}+\frac{2Mr}{\Sigma}\Biggl(
      \sum_\mub\frac{J_\dg(a_\mub^2)}{\Ua_\mub}\frac{1}{\lambda a_\mub}\grad\phi_\mub
      + \frac{J_\dg(a_\dg^2)}{\Ua_\dg}\grad t \Biggr)^{\!\!2}\;.
\end{split}
\end{equation}
Here, we have split the sums to angular terms ${\mub=1,\dots,\Nb}$ and temporal/radial terms ${\mu=\dg}$, employed the Wick rotation and the gauge fixing \eqref{Wick}--\eqref{Wickbyslambda}, and introduced metric functions \eqref{physmtrcfc}.

To write down the metric in Myers--Perry coordinates, we next employ the {\em Jacobi transformation}, to
transform ${\Nb}$ variables ${x_\mub}$ to ${\Nb+1}$ variables~${\mu_\nub}$:
\begin{equation}\label{mudef}
    \mu_\nub^2=\frac{\Jb(a_\nub^2)}{-a_\nub^2\,\Uab_\nub}
       = \frac{\prod_{\bar{\alpha}}(x_{\bar{\alpha}}^2-a_\nub^2)}
         {-a_\nub^2\prod_{\bar{\alpha}\neq\nub}(a_{\bar{\alpha}}^2-a_\nub^2)}\;,\quad
    \mu_0^2 = \frac{\Ab{\Nb}}{\Aab{\Nb}}
       = \frac{\prod_{\bar{\alpha}} x_{\bar{\alpha}}^2}{\prod_{\bar{\alpha}} a_{\bar{\alpha}}^2}\;,
\end{equation}
subject to a constrain
\begin{equation}\label{muconstraint}
    \sum_{\nub=0}^\Nb \mu_\nub^2 = 1\;.
\end{equation}

The new coordinates ${\mu_\nub}$ are related to the coordinates ${\rho_\nub}$ introduced in \eqref{rhodef} by
\begin{equation}\label{rhomurel}
    \lambda\rho_\nub^2 = \frac{a_\nub^2+r^2}{a_\nub^2-a_\dg^2}\mu_\nub^2\,,
\end{equation}
and
\begin{equation}\label{Rdef}
    1-\lambda R^2 \equiv \lambda \rho_\dg^2
      = (1-\lambda r^2) \Bigl(\mu_0^2+\sum_\nub\frac{\mu_\nub^2}{1+\lambda a_\nub^2}\Bigr)\;.
\end{equation}
Employing these relations and other non-trivial identities for the Jacobi transformation,
the metric \eqref{KerrAdS1} can be written in the following form:
\begin{equation}\label{KerrAdS}
\begin{split}
  \ts{g}&=-(1-\lambda R^2)\grad t^2
    + \frac{2M r}{\Sigma}\Bigl(\grad t+\sum_\nub\frac{a_\nub\mu_\nub^2}{1+\lambda a_\nub^2}
      \grad(\phi_\nub-\lambda a_\nub t)\Bigr)^{\!\!2}\\
    &\qquad\qquad
    + \frac{\Sigma}{\Delta_r}\grad r^2
    + r^2 \grad \mu_0^2 + \sum_\nub\frac{r^2+a_\nub^2}{1+\lambda a_\nub^2}
      \Bigl( \grad \mu_\nub^2+\mu_\nub^2\grad\phi_\nub^2\Bigr)\\
    &\qquad\qquad
    +\frac{\lambda}{1-\lambda R^2}\Bigl(r^2 \mu_0\grad\mu_0 +
      \sum_\nub\frac{r^2+a_\nub^2}{1+\lambda a_\nub^2}\mu_\nub\grad\mu_\nub\Bigr)^{\!\!2}\;,
\end{split}
\end{equation}
with the metric functions given by
\begin{equation}\label{KerrAdSmtrcfc}
\begin{aligned}
    \Delta_r &= (1-\lambda r^2)\prod_\nub(r^2+a_\nub^2) - 2 M r\;,\\
    \Sigma &= \Bigl(\mu_0^2+\sum_\nub\frac{r^2\mu_\nub^2}{r^2+a_\nub^2}\Bigr)
       \prod_\mub(r^2+a_\mub^2)\;.
\end{aligned}
\end{equation}
This is the {\em Kerr--(A)dS metric} derived by \cite{Gibbons:2004js,Gibbons:2004uw}.
We remind that in these expressions the coordinates ${\mu_\nub}$ are constrained by \eqref{muconstraint}. For vanishing ${b_\mub}$, the parameters ${a_\mub}$ are directly related to rotations of the black hole.

If also the parameters $a_\mub$ vanish, we obtain the {\em Schwarzschild--Tangherlini--(A)dS black hole} \citep{tangherlini1963schwarzschild}
\begin{equation}\label{TangherliniAdS}
\ts{g}=- f\grad t^2
    + \frac{\grad r^2}{f}
    + r^2 \ts{d\Omega}_\Nb^2\;,\quad
    f=1 - \lambda r^2 - 2M r^{3-2\dg}\,,
\end{equation}
where one can use, for example, the following parametrization of the homogeneous spherical metric in ${\Nb}$ dimensions:
\begin{equation}\label{dOmega}
     \ts{d\Omega}_\Nb^2 = \grad \mu_0^2
      + \sum_\nub \bigl( \grad \mu_\nub^2+\mu_\nub^2\grad\phi_\nub^2\bigr)\;,
\end{equation}
using the coordinates ${\mu_\nub}$ and ${\phi_\nub}$. Other parameterizations of $\ts{d\Omega}_\Nb^2$, suitable for a given problem, are of course possible.

If on the other hand the cosmological constant vanishes, ${\lambda=0}$, the Kerr--(A)dS metric \eqref{KerrAdS} yields the (even-dimensional) {\em Myers--Perry solution} \citep{MyersPerry:1986}
\begin{equation}\label{MyersPerry}
\begin{gathered}
\begin{aligned}
  &\ts{g}=-\grad t^2
    + \frac{2M r}{\Sigma}\Bigl(\grad t+\sum_\nub a_\nub \mu_\nub^2 \grad\phi_\nub \Bigr)^{\!\!2}
    + \frac{\Sigma}{\Delta_r}\grad r^2\\
    &\mspace{180mu}+ r^2 \grad \mu_0^2 + \sum_\nub (r^2+a_\nub^2)
      \Bigl( \grad \mu_\nub^2+\mu_\nub^2\grad\phi_\nub^2\Bigr)\;,
\end{aligned}\\
\Delta_r = \prod_\nub(r^2+a_\nub^2) - 2 M r\;,\quad
    \Sigma = \Bigl(\mu_0^2+\sum_\nub\frac{r^2\mu_\nub^2}{r^2+a_\nub^2}\Bigr)
       \prod_\mub(r^2+a_\mub^2)\;,
\end{gathered}
\end{equation}
discussed in more details in appendix~\ref{apx:MyersPerry}. Indeed, if we identify indices $\nub=1,\dots,\Nb$ with indices $i=1,\dots,m$ of appendix \ref{apx:MyersPerry} for coordinates $\mu_\nub,\phi_\nub$ and parameters $a_\nub$, and if we relate metric functions as $\Sigma=rU$, $\Delta_r=r(V-2M)$, we recover metric \eqref{MPM} with \eqref{MPmfc} in even dimensions ($\eps=0$).

\subsection{Multi-Kerr--Schild form}
\label{ssc:KerrSchild}

In section~\ref{ssc:KerrSchild}, we have seen that the Myers--Perry metric can be cast as a linear in mass deformation of the flat space, that is in the Kerr--Schild form \eqref{KS_MP}. The same remains true for the higher-dimensional Kerr-(A)dS solutions \eqref{KerrAdS} of \cite{Gibbons:2004js,Gibbons:2004uw}, replacing the flat space with the corresponding maximally symmetric geometry. Remarkably, in the presence of NUT charges, the on-shell metric \eqref{KerrNUTAdSmetric} can be written in the {\em multi-Kerr--Schild form} \citep{Chen:2007fs}, that is as a multi-linear deformation of the maximally symmetric space, with deformation terms proportional to generalized masses, see \eqref{multiKerrSchild} below. The modified construction goes as follows.

Introducing the following complex null 1-forms $\ts{\mu}^\nu$ and $\bar{\ts{\mu}}^\nu$:
\begin{align}
    \ts{\mu}^\nu =
       \sum_{j=0}^{\dg-1}\A{j}_\nu\grad\psi_j
       +i\frac{U_\nu}{X_\nu}\grad x_\nu
       \,,\label{KSmudef}\\
    \bar{\ts{\mu}}^\nu =
       \sum_{j=0}^{\dg-1}\A{j}_\nu\grad\psi_j
       -i\frac{U_\nu}{X_\nu}\grad x_\nu
       \,,\label{KSmubdef}
\end{align}
complemented with
\begin{equation}\label{KSep0def}
    \hat{\ts{\epsilon}}^0 = \sum_{j=0}^{\dg}\A{j}\grad \psi_j
\end{equation}
in odd dimensions, the canonical metric \eqref{KerrNUTAdSmetric} reads
\begin{equation}\label{metricmumub0}
    \ts{g} = \sum_\nu \frac12\frac{X_\nu}{U_\nu}
      \bigl(\ts{\mu}^\nu \bar{\ts{\mu}}^\nu + \bar{\ts{\mu}}^\nu \ts{\mu}^\nu \bigr)
      +\eps \frac{c}{\A{\dg}}\, \hat{\ts{\epsilon}}^0 \hat{\ts{\epsilon}}^0
      \,.
\end{equation}
When all coordinates $x_\nu$ and $\psi_j$ are real, the null 1-forms $\ts{\mu}^\nu$ and $\bar{\ts{\mu}}^\nu$ are complex conjugate. If some $x$'s are imaginary, say $x_\dg=ir$, the corresponding 1-forms are real and independent.

Now we break the symmetry between $\ts{\mu}^\nu$ and $\bar{\ts{\mu}}^\nu$ and eliminate $\bar{\ts{\mu}}^\nu$ using the following relation:
\begin{equation}\label{mub=mu+}
    \bar{\ts{\mu}}^\nu = \ts{\mu}^\nu - 2 i \frac{U_\nu}{X_\nu}\grad x_\nu\,.
\end{equation}
The metric \eqref{metricmumub0} can be rewritten as
\begin{equation}\label{metricmumub}
    \ts{g} = \sum_\nu \frac{X_\nu}{U_\nu} \ts{\mu}^\nu \ts{\mu}^\nu
      -i \sum_\nu \frac{X_\nu}{U_\nu}
        \bigl(\ts{\mu}^\nu \grad x_\nu + \grad x_\nu \ts{\mu}^\nu \bigr)
      +\eps \frac{c}{\A{\dg}}\, \hat{\ts{\epsilon}}^0 \hat{\ts{\epsilon}}^0
      \,.
\end{equation}
Expressing the on-shell metric functions $X_\nu$, \eqref{Xsol}, as a deformation of the background functions~$\mathring{X}_\nu$:
\begin{equation}\label{Xsoldeform}
    X_\nu = \mathring{X}_\nu - 2b_\nu x_\nu^{1-\eps}\,,\qquad
    \mathring{X}_\nu = X_\nu|_{b_\kappa=0}\,,
\end{equation}
allows one to re-write the Kerr--NUT--(A)dS metric as
\begin{equation}\label{multiKerrSchildcplx}
    \ts{g} = \mathring{\ts{g}}
    - 2\sum_\nu\frac{b_\nu x_\nu^{1-\eps}}{U_\nu}\,\ts{\mu}^\nu \ts{\mu}^\nu\,,
\end{equation}
where the `background' metric $\mathring{\ts{g}}$ is given by the same expression \eqref{metricmumub}, just with the metric functions $\mathring{X}_\nu$,
\begin{equation}\label{metricmumubbkg}
    \mathring{\ts{g}} = \sum_\nu \frac{X_\nu}{U_\nu} \ts{\mu}^\nu \ts{\mu}^\nu
      -i \sum_\nu \frac{\mathring{X}_\nu}{U_\nu}
        \bigl(\ts{\mu}^\nu \grad x_\nu + \grad x_\nu \ts{\mu}^\nu \bigr)
      +\eps \frac{c}{\A{\dg}}\, \hat{\ts{\epsilon}}^0 \hat{\ts{\epsilon}}^0
      \,.
\end{equation}
In order to be able to interpret the metric $\mathring{\ts{g}}$ as the (A)dS metric written in the Kerr--NUT--(A)dS form with metric functions $\mathring{X}_\nu$, one has to be able to write 1-forms $\ts{\mu}^\nu$ and $\hat{\ts{\epsilon}}^0$ in terms of the background coordinates in a way analogous to \eqref{KSmudef} and \eqref{KSep0def},
\begin{equation}\label{KSmuep0bkg}
    \ts{\mu}^\nu =
       \sum_{j=0}^{\dg-1}\A{j}_\nu\grad\mathring{\psi}_j
       +i\frac{U_\nu}{\mathring{X}_\nu}\grad x_\nu
       \,,\quad
    \hat{\ts{\epsilon}}^0 = \sum_{j=0}^{\dg}\A{j}\grad \mathring{\psi}_j
    \,.
\end{equation}
These conditions can be formally solved for $\mathring{\psi}_j$
\begin{equation}\label{KSpsibkgcplx}
    \grad\mathring{\psi}_j = \grad\psi_j
      + i \sum_\nu (-x_\nu^2)^{\dg{-}1{-}j}\,
      \frac{2b_\nu x_\nu^{1-\eps}}{X_\nu\mathring{X}_\nu}\,\grad x_\nu
      \,.
\end{equation}
One can even introduce the Kerr--Schild coordinates
\begin{equation}\label{KSpsi}
    \grad\hat{\psi}_j = \sum_\nu\frac{(-x_\nu^2)^{\dg{-}1{-}j}}{U_\nu}\ts{\mu}^\nu
      \,,
\end{equation}
in terms of which
\begin{equation}\label{KSpsipsipsi}
   \grad\hat{\psi}_j
   = \grad\psi_j + i \sum_\nu \frac{(-x_\nu^2)^{\dg{-}1{-}j}}{X_\nu}\grad x_\nu
   = \grad\mathring{\psi}_j + i\sum_\nu \frac{(-x_\nu^2)^{\dg{-}1{-}j}}{\mathring{X}_\nu}\grad x_\nu
   \,.
\end{equation}

Unfortunately, this construction is spoiled by complex character of various quantities. We have shown that the `background' metric $\mathring{\ts{g}}$ has the same form as the original metric with metric functions $\mathring{X}_\nu$. However, this metric is, in general, complex. Indeed, in \eqref{multiKerrSchildcplx} $\ts{g}$ is real, but null 1-forms $\ts{\mu}^\nu$ are complex and thus $\mathring{\ts{g}}$ is complex. It corresponds to the fact that coordinates $\mathring{\psi}_\nu$ are, in general complex, as can be seen from \eqref{KSpsibkgcplx}, e.g., with $x_\mu,\,\psi_j$, and $b_\nu$ real.

Interestingly, there exists an important subcase when this construction gives a real result. Let us assume that some of the coordinates $x_\mu$ are Wick-rotated into imaginary values. We have seen, that such a Wick rotation is needed for the Lorentzian signature, when $x_\dg=ir$. More generally, let us assume the Wick rotation for the last $D-\Nb$ coordinates $x_\nu$, for some $\Nb$,
\begin{equation}\label{multiWickrot}
    x_\nu = i r_\nu \qquad\text{for $\nu>\Nb$}\,.
\end{equation}
Let us also assume that the corresponding NUT charges are also Wick-rotated and the remaining NUT charges vanish,
\begin{equation}\label{multiWickb}
\begin{aligned}
    &b_\nu = 0 &\quad&\text{for $\nu=1,\dots,\Nb$}\,,\\
    &b_\nu = i M_\nu &\quad&\text{for $\nu=\Nb+1,\dots\dg$ and $D$ even}\,,\\
    &b_\nu = - M_\nu &\quad&\text{for $\nu=\Nb+1,\dots\dg$ and $D$ odd}\,.\\
\end{aligned}
\end{equation}
In this case the metric \eqref{multiKerrSchildcplx} takes the real multi-Kerr--Schild form
\begin{equation}\label{multiKerrSchild}
    \ts{g} = \mathring{\ts{g}}
    + 2\sum_{\nu=\Nb+1}^{\dg}\frac{M_\nu r_\nu^{1-\eps}}{U_\nu}\,\ts{\mu}^\nu \ts{\mu}^\nu\,,
\end{equation}
where the background metric $\mathring{\ts{g}}$ and 1-forms $\ts{\mu}^\nu$ and $\hat{\ts{\epsilon}}^0$ are given by \eqref{metricmumubbkg} and \eqref{KSmuep0bkg} in terms of real background coordinates
\begin{equation}\label{KSpsibkg}
    \grad\mathring{\psi}_j = \grad\psi_j
      + \sum_\nu (r_\nu^2)^{\dg{-}1{-}j}\,
      \frac{2M_\nu r_\nu^{1-\eps}}{X_\nu\mathring{X}_\nu}\,\grad r_\nu
      \,.
\end{equation}
Clearly, for $\nu>\Nb$, 1-forms $\ts{\mu}^\nu$ are real, and therefore also the background metric $\mathring{\ts{g}}$ is real.

The coordinates $\hat{\psi}_j$ need more attention. One has to modify their definition in such a way that in \eqref{KSpsipsipsi} the sum runs only over Wick rotated coordinates.

The case when only one coordinate, $x_\dg=ir$, is Wick rotated covers the Lorentzian signature. It demonstrates, that the black hole solution (with vanishing NUT charges) can be written in the standard Kerr--Schild form. The four-dimensional case discussed in section~\ref{ssc:KerrSchild4D} is an example of this case, as well as the Kerr--Schild form of the Myers--Perry solution discussed in appendix \ref{apx:MyersPerry} (after some additional effort of identifying Myers--Perry and canonical coordinates).

The opposite case of the multi-Kerr--Schild form of the metric, when all $x_\nu$ coordinates are Wick-rotated, can be related to an analogous  discussion in \cite{Chen:2007fs}, where all coordinates $\psi_j$ have been Wick-rotated and the multi-Kerr--Schild form has been obtained.

Finally, let us note that the principal tensor $\ts{h}$ is the same for the full metric $\ts{g}$ as for the background metric $\mathring{\ts{g}}$ and reads
\begin{equation}\label{hKerSchild}
    \ts{h} = \sum_\nu x_\nu\, \grad x_\nu \wedge \ts{\mu}^\nu\,.
\end{equation}
In other words, the vector variants of 1-forms $\ts{\mu}^\nu$ and $\hat{\ts{\epsilon}}^0$ are the eigenvectors of the principal tensor with eigenvalues $-ix_\nu$ and $0$, respectively. They differ from the eigenvectors $\ts{m}_\mu$ and $\ezv$ discussed in section~\ref{ssc:PrincipalTensor} just by normalization,
\begin{equation}\label{mmurel}
    \ts{m}_\nu = \frac1{\sqrt{2}}\sqrt{\frac{X_\nu}{U_\nu}}\,\ts{\mu}^\nu\,,\qquad
    \ezv = \sqrt{\frac{c}{\A{\dg}}}\,\hat{\ts{\epsilon}}^0\,.
\end{equation}
They correspond to principal null directions of the Weyl tensor (WANDs) \citep{Hamamoto:2006zf, Krtous:2008tb, Mason:2010zzc, Kubiznak:2008qp}. The expressions \eqref{multiKerrSchild} and \eqref{hKerSchild} nicely demonstrate the connection between the existence of the principal tensor and the form of the corresponding Kerr--Schild structure. See also \cite{Ortaggio:2008iq} for a more general discussion on higher-dimensional Kerr--Schild spacetimes, and \cite{Monteiro:2014cda, Luna:2015paa} for a recent new twist on applications of the Kerr--Schild form.

%% file: ch5-hshdbh.tex

\section{Hidden symmetries of Kerr--NUT--(A)dS spacetimes}
\label{sc:hshdbh}

In the previous chapter the Kerr--NUT--(A)dS metric and its interpretation were discussed. Let us consider now the  symmetries of this geometry in more detail. Namely, we shall show that, similar to its four-dimensional counterpart, this metric admits the {\em principal tensor}. The latter generates the whole tower of explicit and hidden symmetries. In fact, the geometry itself is uniquely determined by the principal tensor. Let us begin exploring this remarkable geometric construction.

\subsection{Principal tensor}
\label{ssc:PTKerrNUTAdS}

It was shown in \cite{Frolov:2007nt, Kubiznak:2006kt} that the Kerr--NUT--(A)dS geometry \eqref{KerrNUTAdSmetric} in any number of dimensions  admits the \defterm{principal tensor}. According to the definition given in chapter~\ref{sc:CKY}, this is a non-degenerate closed conformal Killing--Yano 2-form $\ts{h}$ obeying
\begin{equation}\label{PCCKYeq}
\nabla_\ts{X}\ts{h}=\ts{X}\wedge \ts{\xi}\,,\quad \quad \Leftrightarrow \quad
  \nabla_a h_{bc} = g_{ab}\,\xi_{c} - g_{ac}\,\xi_{b}\;,
\end{equation}
where $\ts{\xi}$ is given by
\begin{equation}\label{primKV}
  \ts{\xi} = \frac{1}{D-1}\covd\cdot \ts{h}
  \quad \Leftrightarrow \quad
  \xi^a=\frac{1}{D-1}\nabla_bh^{ba}\;.
\end{equation}
The non-degeneracy means that in ${D=2\dg+\eps}$ dimensions, $\ts{h}$ has a maximal possible matrix rank ${2\dg}$ with ${\dg}$ pairs of conjugate eigenvectors and associated imaginary eigenvalues ${\pm i x^\mu}$ that are all functionally independent (and hence also non-constant).

The principal tensor reads
\begin{equation}\label{PCCKY}
  \ts{h} = \sum_{\mu=1}^{\dg} x_\mu\, \grad x_\mu \wedge
    \Bigl(\sum_{k=0}^{\dg-1}\A{k}_\mu \grad\psi_k\Bigr)
         = \sum_\mu x_\mu\, \enf\mu\wedge\ehf\mu\;.
\end{equation}
The latter expression means that the frame $(\enf\mu,\,\ehf\mu)$ (and ${\ezf}$ in odd dimensions) introduced in the previous chapter, \eqref{Darbouxformfr}, is the \emph{special Darboux frame}; the eigenvalues ${x_\mu}$ supplemented with the Killing coordinates ${\psi_k}$, are  the \defterm{canonical coordinates}.

Since the principal tensor  ${\ts{h}}$ is closed, there exists a local potential ${\ts{b}}$,
\begin{equation}\label{PCCKYpot}
\ts{b} = \frac12 \sum_{k=0}^{\dg-1}\A{k+1}\grad\psi_k\;,
\end{equation}
such that ${\ts{h}=\grad\ts{b}}$.

It turns out that $\tens{\xi}$, given by \eqref{primKV}, is a Killing vector,\footnote{
Using the integrability relation \eqref{covderxi}, it is easy to show  that
\begin{equation}
  2(D-2)\nabla_{\!(a}\xi_{b)} = h_{ac}R^c{}_b - R_{ac}\,h^{c}{}_{b}\, .\nonumber
\end{equation}
Thus for the on-shell metric, when $ R_{ac}=\frac{2}{D-2}\Lambda g_{ac}$, the vector $\ts{\xi}$ obeys the Killing equation. The same conclusion remains also true for any off-shell Kerr--NUT--(A)dS metric. One way to demonstrate this is to use the explicit form of the off-shell metric.
However, it is also possible to prove this result without referring to the metric, using only the properties of the principal tensor.
This proof is involved and we present it later, at the end of this chapter.
}
\begin{equation}\label{xi=KV}
    \ts{\xi} = \cv{\psi_0} = \sum_\mu \Bigl(\frac{X_\mu}{U_\mu}\Bigr)^{\!\frac12}\ehv\mu
      +\eps\Bigl(\frac{c}{\A{\dg}}\Bigr)^{\!\frac12}\ezv\,.
\end{equation}
Since it will be used as a `seed' for the constructions of other Killing vectors in the Kerr--NUT--(A)dS spacetime, we call it a {\em primary Killing vector.} Thanks to \eqref{primKV}, the principal tensor ${\ts{h}}$ plays a role of a co-potential for the primary Killing vector ${\ts{\xi}}$.

\subsection{Killing tower}
\label{ssc:KillingTower}

As we have already revealed in chapter~\ref{sc:CKY}, from the principal tensor $\ts{h}$ one can generate the whole tower of explicit and hidden symmetries of the Kerr--NUT--(A)dS geometry. We call this {set} a \defterm{Killing tower}. In what follows we shall review two methods for generating such a tower: a \emph{direct method} of construction (based on theorems of chapter~\ref{sc:CKY}) and the method of a {\em generating function}.

\subsubsection{Direct method of construction}

The construction of the Killing tower goes as follows \citep{Krtous:2006qy, Frolov:2007cb,Frolov:2008jr}:
\begin{enumerate}[(i)]
\item
By employing the theorem \eqref{CCKYprod} of chapter~\ref{sc:CKY}, one can construct  a tower of closed conformal Killing--Yano tensors by taking various wedge products of the principal tensor $\ts{h}$ with itself \citep{Krtous:2006qy}. Since $\ts{h}$ is a non-degenerate 2-form, this gives the following ${\dg+1}$ closed conformal Killing--Yano forms ${\CCKY{j}}$ of increasing rank ${2j}$ $({j=0,\dots,\dg}$): 
\begin{equation}\label{CCKYj}
    \CCKY{j} = \frac1{j!}\,\ts{h}^{\wedge j}\;.
\end{equation}

Note that for ${j=0}$ we have a trivial 0-form ${\CCKY{0}=1}$. We also have ${\CCKY{\dg}=\sqrt{\A{\dg}}\,\ts{\eps}}$ and ${\CCKY{\dg} = \sqrt{\A{\dg}}\,\ts{\eps}\cdot\ezv}$ for even and odd dimensions, respectively. Here, as earlier, $\ts{\eps}$ is the Levi-Civita tensor.

\item
{As discussed after \eqref{CKYHodge}, the Hodge dual of a closed conformal Killing--Yano $2j$-form ${\CCKY{j}}$ is a Killing--Yano $({D{-}2j})$-form, which we call ${\KY{j}}$:}
\begin{equation}\label{KYj}
    \KY{j} = * \CCKY{j}\;.
\end{equation}

In particular, this gives the Levi-Civita tensor ${\KY{0}=\ts{\eps}}$ for $j=0$. In even dimensions one has
${\KY{\dg} = \sqrt{\A{\dg}}}$.
{In odd dimensions, ${\KY{\dg}=\sqrt{\A{\dg}}\,\ezf}$. The vector version of ${\KY{\dg}}$ has to be a Killing vector. Namely, we get
\be
\KY{\dg}=\frac{1}{\sqrt{c}}\cv{\psi_\dg}\,.
\ee}


\item
Partial contractions of squares of Killing--Yano forms ${\KY{j}}$ define the following rank-2 Killing tensors~${\KT{j}}$, cf.~\eqref{Kab2},
\begin{equation}\label{KTj}
    \KTc{j}^{ab} = \frac1{(D{-}2j{-}1)!}\,
      \KYc{j}{}^{a}{}_{c_1\dots c_{D{-}2j{-}1}}\,\KYc{j}{}^{bc_1\dots c_{D{-}2j{-}1}}\,.
\end{equation}

For ${j=0}$, the Killing tensor reduces to the metric
\begin{equation}\label{KT0=mtrc}
 k_{(0)}^{ab}= g^{ab}\;.
\end{equation}
For odd dimensions the top Killing tensor is reducible, ${\KT{\dg}=\A{\dg}\,\ezv\,\ezv}={c^{-1}\cv{\psi_\dg}\cv{\psi_\dg}}$, whereas in even dimensions we define ${\KT{\dg}=\ts{0}}$.

\item
Similarly, partial contractions of closed conformal Killing--Yano forms ${\CCKY{j}}$ give rank-2 conformal Killing tensors ${\CKT{j}}$:
\begin{equation}\label{CKTj}
    \CKTc{j}^{ab} = \frac1{(2j{-}1)!} \,
      \CCKYc{j}{}^{a}{}_{c_1\dots c_{2j{-}1}}\CCKYc{j}{}^{bc_1\dots c_{2j{-}1}}\;.
\end{equation}

We define ${\CKT{0}=\ts{0}}$, and introduce a simpler notation ${\ts{Q}}$ for the first conformal Killing tensor:
\begin{equation}\label{CKT}
    Q^{ab} \equiv \CKTc{1}^{ab} = h^{a}{}_{c}\,h^{bc}\;.
\end{equation}

The conformal Killing tensors ${\CKT{j}}$ contain essentially the same information as the Killing tensors ${\KT{j}}$. Namely, for all ${j=1,\dots,\dg}$ it holds
\begin{equation}\label{KTCKTrel}
    \KT{j}+\CKT{j} = \A{j}\,\ts{g}\; .
\end{equation}
where the scalar function ${\A{j}}$ can be expressed as
\begin{equation}\label{traceKTCKT}
    \A{j} = \CCKY{j}\bullet\CCKY{j} = \KY{j}\bullet\KY{j}
          = \frac1{2j}\,\CKTc{j}{}^{n}_{n} = \frac1{D{-}2j}\,\KTc{j}{}^{n}_{n}\; .
\end{equation}
Here we used the scalar product \eqref{bulletdef}. It turns out that functions ${\A{j}}$ are exactly the symmetric polynomials \eqref{AUdefs} introduced earlier.
The conformal Killing tensors and the Killing tensors are also related by
\begin{equation}\label{CKTKYhrel}
    \CKTc{j}^{ac} = h^{a}{}_{b}\, h^{c}{}_{d}\, \KTc{j{-}1}^{bd}
      = Q^{a}{}_{b}\,\KTc{j{-}1}^{bc}\;.
\end{equation}

\item
We conclude our construction by defining the following vectors:
\begin{equation}\label{KVj}
    \KV{j} = \KT{j}\cdot\ts{\xi}\;.
\end{equation}
They turn out to be Killing vectors related to coordinates ${\psi_j}$, namely $\KV{j}=\partial_{\psi_j}$. Since $\KV{j}$  are constructed using the primary Killing vector ${\ts{\xi}}$, we call them \defterm{secondary Killing vectors}.
For ${j=0}$ we have ${\KV{0}=\ts{\xi}}$. Since in even dimensions the top Killing tensor ${\KT{\dg}}$ vanishes by definition, we have ${\KV{\dg}=0}$. On the other hand in odd dimensions the top Killing vector is non-trivial and reads
\be
\KV{\dg} = \sqrt{c\A{\dg}}\,\ezv = \cv{\psi_\dg}\,.
\ee

These Killing vectors can be generated from the {\em Killing co-potentials} ${\ts{\omega}^{(j)}}$, e.g., \cite{Kastor:2009wy,Cvetic:2010jb},
\be
   \KV{j}=\covd\cdot\ts{\omega}^{(j)}\,,
\ee
where
\be\label{omegajjjjj}
\omega^{(j)}_{ab}=\frac{1}{D{-}2j{-}1}\,\KTc{j}{}_{a}{}^n \,h_{nb}\,
\ee
for $j=0,\dots, \dg-1$, and $\ts{\omega}^{(n)}=-\frac{1}{n!}\sqrt{c}\,\ts{*}(\ts{b}\wedge \ts{h}^{\wedge(n-1)})$ in odd dimensions. Note that, apart from $\ts{\omega}^{(0)}$, the Killing co-potentials are not closed, $\ts{d\omega}^{(j)}\neq 0$.

Let us also  mention the following useful relation:
\begin{equation}\label{hKV=dA}
    \ts{h}\cdot  \KV{j} = \frac12\grad\A{j+1}\;,
\end{equation}
which implies---through the Cartan identity and closeness of ${\ts{h}}$---that the principal tensor ${\ts{h}}$ is conserved along the vector fields ${\KV{j}}$,
\begin{equation}\label{LieKVh}
    \lied_{\KV{j}}\ts{h} = 0\;.
\end{equation}

\end{enumerate}

All the `Killing objects' in the tower are generated from a single object, the principal tensor. As a result they form an abundant structure, with many special algebraic and differential relations among them. In particular, the  Killing tensors ${\KT{j}}$ and the Killing vectors ${\KV{j}}$ commute in the sense of the Nijenhuis--Schouten brackets:
\be\label{KTKVNS}
[\KT{i},\KT{j}]_\NS = 0\;,\quad
    [\KT{i},\KV{j}]_\NS = 0\;,\quad
    [\KV{i},\KV{j}] = 0\;.
\ee
It means that the corresponding observables on the phase space
\begin{equation}\label{KillObs}
    \Qo{j} = \KTc{j}^{ab}\,p_a\,p_b\;,\quad
    \Lo{j} = \KVc{j}^{a}\,p_a\,,
\end{equation}
are in involution, which is the key observation behind the complete integrability of geodesic motion in Kerr--NUT--(A)dS spacetimes discussed in the next section. Note also that since ${\KT{0}=\ts{g}}$, the relations \eqref{KTKVNS} directly imply that ${\KT{j}}$ and ${\KV{j}}$ are Killing tensors and Killing vectors, respectively.

\details{For propagation of light, one can instead of the set $\{\KT{i},\KV{j}\}$ use a different set of Killing symmetries $\{\CKT{i},\KV{j}\}$, where each Killing tensor $\KT{i}$ is replaced by the corresponding conformal Killing tensors $\CKT{i}$. By relation \eqref{KTCKTrel}, $\CKT{i}$ differs from $\KT{i}$ only by a term proportional to the metric. Consequently, the two tensors give the same value of conserved quantities for null rays. Note also that \eqref{KTCKTrel} can be used together with \eqref{KTKVNS} to extract the commutation relations of the objects in the new set. Namely, we find
\be\label{KTKVNSnull}
[\ts{Q}_{(i)}, \ts{Q}_{(j)}]_\NS^{abc} ={\alpha}^{(a}_{(ij)} g^{bc)}\;,\quad
    [\ts{Q}_{(i)},\ts{l}_{(j)}]_\NS =0\;,\quad
    [\KV{i},\KV{j}] = 0\; .
\ee
Here
\be
\ts{\alpha}_{(ij)}=[A^{(i)}, \ts{g}A^{(j)}-\ts{k}_{(j)}]_\NS+[\ts{g}A^{(i)}-\ts{k}_{(i)}, A^{(j)}]_\NS\,,
\ee
and we used that $[A^{(i)},l_{(j)}]_\NS= -\lied_{\ts{l}_{(j)}}A^{(i)}=0\,.$ To obtain observables in the phase space one needs to multiply the objects, which enter \eqref{KTKVNSnull}, by null vectors tangent to the null ray. As a result the Poisson bracket algebra of the conserved quantities corresponding to \eqref{KTKVNSnull} becomes trivial. This justifies the complete integrability of null geodesic motion.
}

Some properties of the above constructed Killing tower are simpler to prove than others. Namely, by theorems of chapter~\ref{sc:CKY} we know that ${\CCKY{j}}$, ${\KY{j}}$, ${\KT{j}}$ and ${\CKT{j}}$ are closed conformal Killing--Yano forms, Killing--Yano forms, Killing tensors, and conformal Killing tensors, respectively. However, to show that ${\KV{j}}$ are Killing vectors (and in particular that ${\ts{\xi}}$ given by  \eqref{primKV} is indeed a primary Killing vector) and to demonstrate the commutation relations \eqref{KTKVNS} poses a  more difficult task. Of course, one way to show these is a `brute force' calculation, employing the explicit form of the Kerr--NUT--(A)dS metric and the induced covariant derivative. However, it turns out that it is possible to prove all these relations directly from the integrability conditions of the principal tensor ${\ts{h}}$, without referring to a particular form of the metric \citep{krtous:directlink}. We will sketch the corresponding line of reasoning in section~\ref{sec:commutationrel}.

\subsubsection{Method of generating functions}

There exists  another (more compact) way for constructing the Killing tower \citep{krtous:directlink}. Namely, it is possible to define a $\beta$-dependent Killing tensor ${\ts{k}(\beta)}$ and a $\beta$-dependent Killing vector ${\ts{l}(\beta)}$, both functions of a real parameter ${\beta}$, such that the Killing tensors $\KT{j}$ and the Killing vectors $\KV{j}$ in the Killing tower above emerge as coefficients of the \mbox{${\beta}$-expansion} of ${\ts{k}(\beta)}$ and ${\ts{l}(\beta)}$, respectively. This procedure is related to \cite{Krtous:2006qy, Houri:2007uq}, where generating function for conserved observables is studied.

Starting with the conformal Killing tensor ${\ts{Q}}$ introduced in \eqref{CKT}, we define a ${\beta}$-dependent conformal Killing tensor
\begin{equation}\label{qfcdef}
    \ts{q}(\beta) = \ts{g} + \beta^2\ts{Q}\,,
\end{equation}
and a scalar function
\begin{equation}\label{Afcdef}
   A(\beta) = \sqrt{\frac{\Det\ts{q}(\beta)}{\Det\ts{g}}}\;.
\end{equation}
Using these definitions we introduce two more objects\footnote{%
The inverse ${\ts{q}^{-1}}$ of a non-degenerate symmetric rank-2 tensor ${\ts{q}}$ is defined in a standard way: $\ts{q}\cdot \ts{q}^{-1}=\ts{1}$, or in components, ${q_{ac} (q^{-1}){}^{bc}=\delta^b_a}$\;.}
\begin{equation}\label{KTfcdef}
    \ts{k}(\beta) = A(\beta)\,\ts{q}^{-1}(\beta)\,,
\end{equation}
and
\begin{equation}\label{KVfcdef}
    \ts{l}(\beta) = \ts{k}(\beta)\cdot\ts{\xi}\;.
\end{equation}
One can show that these functions generate the objects from the Killing tower
\begin{gather}
    \ts{k}(\beta) = \sum_{j}\beta^{2j}\,\KT{j}\;,\label{KTfcexp}\\
    \ts{l}(\beta) = \sum_{j}\beta^{2j}\,\KV{j}\;.\label{KVfcexp}
\end{gather}
One also has
\be
 A(\beta)  = \sum_{j}\beta^{2j}\,\A{j}\; .\label{Afcexp}
 \ee
Since for a fixed $\beta$, $\ts{k}(\beta)$ is a linear combination of Killing tensors, it is itself a Killing tensor, and similarly $\ts{l}(\beta)$ is a Killing vector.
The commutativity relations \eqref{KTKVNS} can be reformulated as a requirement that the Killing tensors ${\ts{k}(\beta)}$ and the Killing vectors ${\ts{l}(\beta)}$ commute for different $\beta$:
\begin{equation}\label{NSgenfc}
    [\,\ts{k}(\beta_{1}),\,\ts{k}(\beta_{2})\,]_\NS = 0\;,\quad
    [\,\ts{k}(\beta_{1}),\,\ts{l}(\beta_{2})\,]_\NS = 0\;,\quad
    [\,\ts{l}(\beta_{1}),\,\ts{l}(\beta_{2})\,]_\NS = 0\;.
\end{equation}

Similar generating functions can also be constructed for the tower of closed conformal Killing--Yano and Killing--Yano forms, respectively. However, since such objects are of increasing rank, the corresponding generating functions are inhomogeneous forms, i.e. a mixture of forms of various ranks. Concretely, one can define ${\ts{h}(\beta)}$  as a wedge exponential of the principal tensor ${\ts{h}}$,
\begin{equation}\label{CCKYfc}
    \ts{h}(\beta) = \hat\exp(\beta\ts{h}) \equiv \sum_j \frac1{j!} \beta^j \ts{h}^{\wedge j}\;,
\end{equation}
and ${\ts{f}(\beta)}$ as its Hodge dual,
\begin{equation}\label{KYfc}
    \ts{f}(\beta) = * \ts{h}(\beta)\;.
\end{equation}

Since the definition \eqref{CCKYfc} contains just a sum of wedge-powers of the principal tensor, ${\ts{h}(\beta)}$ is a closed conformal Killing--Yano form, cf.\ \eqref{CCKYprod}. Its Hodge dual  ${\ts{f}(\beta)}$ then must be a Killing--Yano form. They satisfy the (closed conformal) Killing--Yano conditions \eqref{bezCCKY} and \eqref{bezKY} adapted to inhomogeneous forms, namely
\begin{equation}\label{CCKYKYnonhom}
  \covd_{\!\ts{X}}\ts{h}(\beta) = \ts{X}\wedge\ts{\xi}(\beta)\,,\qquad
    \covd_{\!\ts{X}}\ts{f}(\beta) = \ts{X}\cdot\ts{\kappa}(\beta)\,,
\end{equation}
where ${\ts{\xi}(\beta)}$ and  ${\ts{\kappa}(\beta)}$ are ${\beta}$-dependent inhomogeneous forms satisfying
\begin{equation}\label{XiPhiCond}
  (D-\pi)\, \ts{\xi}(\beta) = \covd\cdot\ts{h}(\beta)\,,\qquad
    \pi\, \ts{\kappa}(\beta)=\covd\wedge\ts{f}(\beta)\,,
\end{equation}
with ${\pi}$ being the rank operator \eqref{pieta}. Surprisingly, in this case they can be written as
\begin{equation}\label{XiPhirel}
   \ts{ \xi}(\beta) = \beta\,\ts{\xi}\wedge\ts{h}(\beta)\;,\qquad
   \ts{ \kappa}(\beta) = -\beta\,\ts{\xi}\cdot\ts{f}(\beta)\;.
\end{equation}

The right-hand sides of \eqref{CCKYKYnonhom} are thus algebraic expressions in ${\ts{\xi}}$, ${\ts{h}(\beta)}$, and ${\ts{f}(\beta)}$, which shows that all the non-trivial information about the covariant derivative of ${\ts{h}(\beta)}$ and ${\ts{f}(\beta)}$ is hidden in the primary Killing vector~${\ts{\xi}}$.
Inspecting the expansion of the wedge exponential in \eqref{CCKYfc} (which is finite due to fact that the rank of a form is bounded by the spacetime dimension) and taking into account the linearity of the Hodge dual in \eqref{KYfc}, we easily realize that ${\ts{h}(\beta)}$ and ${\ts{f}(\beta)}$ are generating functions
 for ${\CCKY{j}}$  and ${\KY{j}}$, respectively. It means that they satisfy  relations similar to \eqref{KTfcexp} and \eqref{KVfcexp},
\begin{gather}\label{CKYfcexp}
    \ts{h}(\beta) = \sum_{j}\beta^{j}\,\CCKY{j}\;,\qquad
    \ts{f}(\beta) = \sum_{j}\beta^{j}\,\KY{j}\;.
\end{gather}
One could also establish relations analogous to \eqref{KTj} and \eqref{Afcexp}. However, one would have to properly define partial and total contractions for inhomogeneous forms, which is possible, but will not be needed here.

\subsubsection{Killing tower in a Darboux frame}

The link between definitions \eqref{qfcdef}--\eqref{KVfcdef} and expansions \eqref{KTfcexp}--\eqref{Afcexp} can be established by writing down all the quantities in the Darboux frame determined by the principal tensor $\ts{h}$. {For that, it is sufficient to specify the Darboux frame just in terms of the principal tensor, without refereing to its explicit coordinate form \eqref{Darbouxformfr}. However, if one seeks the expressions in terms of canonical coordinates, one can easily substitute relations \eqref{Darbouxformfr} and \eqref{Darbouxvecfr}.}

As discussed in section~\ref{ssc:PrincipalTensor}, the canonical Darboux frame is determined by the following two equations:
\begin{align}
   \ts{g} &= \sum_\mu \bigl( \enf{\mu} \enf{\mu} + \ehf{\mu} \ehf{\mu} \bigr)
      + \eps\, \ezf \ezf\,, \label{mtrcfr}\\
   \ts{h} &= \sum_\mu x_\mu \, \enf{\mu} \wedge \ehf{\mu} \,. \label{hfr}
\end{align}
{We also know that in terms of the dual frame of vectors $(\env{\mu},\,\ehv{\mu}\,, {\ezv})$, the principal tensor has eigenvalues $\pm i x_\mu$, corresponding to the eigenvectors $\ts{m}_\mu=\ehv{\mu} + i \env{\mu}$, and $\bar{\ts{m}}_\mu=\ehv{\mu} - i \env{\mu}$, namely}\footnote{%
When speaking about eigenvectors of a rank-2 tensor $\ts{h}$, we always assume a proper adjustment of its indices (by using the metric) to form a linear operator $h^{a}{}_{b}$.}
\begin{equation}\label{heigenvec}
    \ts{h}\cdot (\ehv{\mu}\mp i \env{\mu}) = \pm i x_\mu\,  (\ehv{\mu}\mp i \env{\mu})\;,
\end{equation}
in odd dimensions accompanied by an additional trivial eigenvalue:
\begin{equation}\label{heigenvecodd}
    \ts{h}\cdot \ezv=0\;.
\end{equation}
The principal tensor thus splits the tangent space into $\dg$ 2-planes spanned on pairs of vectors $\env{\mu},\,\ehv{\mu}$ and, in odd dimensions, one degenerate direction $\ezv$.

Using the equations \eqref{mtrcfr} and \eqref{hfr} we can now establish the following results. First, the definition \eqref{CCKYj} yields the explicit form for the closed conformal Killing--Yano tower: 
\begin{equation}\label{CCKYfr}
    \CCKY{j} = \!\!\sum_{\substack{\nu_1,\dots,\nu_j\\\nu_1<\dots<\nu_j}}\!\!\!
       x_{\nu_1}\dots x_{\nu_j}\;
      \enf{\nu_1}\wedge\ehf{\nu_1}\wedge\dots\wedge\enf{\nu_j}\wedge\ehf{\nu_j}\;.
\end{equation}
In particular, this gives
\begin{equation}\label{CCKYnorm}
     \CCKY{j}\bullet\CCKY{j} = \!\!\sum_{\substack{\nu_1,\dots,\nu_j\\\nu_1<\dots<\nu_j}}\!\!\!
       x^2_{\nu_1}\dots x^2_{\nu_j} = \A{j}\;,
\end{equation}
establishing the first equality in \eqref{traceKTCKT}.
Next, calculating the partial traces \eqref{CKTj}, we obtain
\begin{equation}\label{CKTfr}
    \CKT{j} = \sum_\mu x^2_\mu\,\A{j{-}1}_\mu\,\bigl(\env{\mu}\env{\mu}+\ehv{\mu}\ehv{\mu}\bigr)\;,
\end{equation}
and in particular
\begin{equation}\label{CKTfr1}
    \ts{Q} = \sum_\mu x^2_\mu\,\bigl(\env{\mu}\env{\mu}+\ehv{\mu}\ehv{\mu}\bigr)\;.
\end{equation}
To write down the tower of Killing--Yano forms $\KY{j}$ we need to distinguish the cases of even and odd dimensions. In even dimensions the Levi--Civita tensor reads
\begin{equation}\label{LCTeven}
    \ts{\eps} = \enf{1}\wedge\ehf{1}\wedge \dots\wedge\enf{\dg}\wedge\ehf{\dg}\;,
\end{equation}
giving the following expression for the Hodge duals \eqref{KYj}:
\be\label{KYevenfr}
\KY{j} = \sum x_{\nu_1}\dots x_{\nu_j}\;
      \enf{\mu_{j{+}1}}\wedge\ehf{\mu_{j{+}1}}\wedge\dots\wedge\enf{\mu_{\dg}}\wedge\ehf{\mu_{\dg}}\;,
\ee
where the sum is over all splittings of indices $1,\dots,\dg$ into two disjoint ordered sets $\nu_1,\dots,\nu_j$ and $\mu_{j{+}1},\dots,\mu_{\dg}$. 
In odd dimensions the Levi--Civita tensor contains also the degenerate direction
\begin{equation}\label{LCTodd}
  \ts{\eps} = \ezf\wedge\enf{1}\wedge\ehf{1}\wedge \dots\wedge\enf{\dg}\wedge\ehf{\dg}\;,
\end{equation}
giving
\begin{equation}\label{KYoddfr}
    \KY{j} = \sum
       x^{\nu_1}\dots x^{\nu_j}\;
      \enf{\mu_{j{+}1}}\wedge\ehf{\mu_{j{+}1}}\wedge\dots\wedge\enf{\mu_{\dg}}\wedge\ehf{\mu_{\dg}}\wedge\ezf\;
\end{equation}
for the Killing--Yano tensors, where the sum has the same meaning as in even dimensions.
In both cases we immediately see that square-norms of $\KY{j}$ are
\begin{equation}\label{KYnorm}
    \KY{j}\bullet\KY{j} = \!\!\sum_{\substack{\nu_1,\dots,\nu_j\\\nu_1<\dots<\nu_j}}\!\!\!
       x^2_{\nu_1}\dots x^2_{\nu_j}  = \A{j}\;,
\end{equation}
which proves the second equality in \eqref{traceKTCKT}. The partial trace \eqref{KTj} gives the following expressions for the Killing tensors $\KT{j}$:
\begin{equation}\label{KTjfr}
    \KT{j} = \sum_\mu \A{j}_\mu\,\bigl(\env{\mu}\env{\mu}+\ehv{\mu}\ehv{\mu}\bigr)
      +\eps \A{j}\,\ezv\ezv\;.
\end{equation}
Using a simple identity $\A{j} = \A{j}_\mu +x^2_\mu\A{j{-}1}_\mu$, we obtain relation \eqref{KTCKTrel}.
Finally, taking into account the extra information
\eqref{xi=KV} and orthogonality \eqref{hfr}, we infer the following form of the Killing vectors
\eqref{KVj}:
\begin{equation}\label{KVjfr}
    \KV{j} = \sum_\mu \A{j}_\mu \Bigl(\frac{X_\mu}{U_\mu}\Bigr)^{\!\frac12}\ehv{\mu}
       +\eps \A{j}\Bigl(\frac{c}{\A{\dg}}\Bigr)^{\!\frac12}\ezv\;.
\end{equation}
The corresponding Killing co-potentials \eqref{omegajjjjj} take a very simple form \citep{Cariglia:2011yt}
\be
\ts{\omega}^{(j)}=\frac{1}{D-2j-1}\sum_{\mu}x_\mu A_\mu^{(j)}\enf{\mu} \wedge \ehf{\mu}\,.
\ee
Note that the relationship between $\ts{\omega}^{(j)}$ and the principal tensor $\ts{h}$, \eqref{hfr}, is `formally analogous' to  the relationship between the Killing tensor $\KT{j}$, \eqref{KTjfr}, and the metric $\ts{g}$, \eqref{mtrcfr}.

Using the explicit form of the Darboux frame of the Kerr--NUT--(A)dS spacetime \eqref{Darbouxvecfr}, we can write the expressions for the Killing objects in terms of canonical coordinates $({x_\mu, \psi_k})$. It turns out that the expressions for the (conformal) Killing tensors do not take a particularly nice form, they are of the same order of complexity as expression \eqref{KerrNUTAdSinvmetric} for the inverse metric. For example,
\begin{equation}\label{KTjcoor}
  \KT{j}
  =\sum_{\mu=1}^\dg\; \A{j}_\mu\biggl[\; \frac{X_\mu}{U_\mu}\,{\cv{x_{\mu}}^2} 
  + \frac{U_\mu}{X_\mu}\,\Bigl(\,\sum_{k=0}^{\dg-1+\eps}
    {\frac{(-x_{\mu}^2)^{\dg-1-k}}{U_{\mu}}}\,\cv{\psi_k}\Bigr)^{\!2}\;\biggr]
  +\eps\,\frac{\A{j}}{\A{\dg}}\cv{\psi_n}^2\,.
\end{equation}
However, the expressions for Killing vectors simplify significantly,
\begin{equation}\label{KVjcoor}
    \KV{j} = \cv{\psi_j}\;.
\end{equation}
Killing coordinates ${\psi_j}$ are thus associated directly with the Killing vectors ${\KV{j}}$ defined through the contraction \eqref{KVj} of the Killing tensors with the primary Killing vector.

Let us conclude this section by writing explicitly down the $\beta$-dependent quantities. Definitions \eqref{qfcdef} and \eqref{Afcdef} give
\begin{align}
 \ts{q}(\beta) &= \sum_\mu ( 1 + \beta^2 x^2_\mu ) \bigl(\enf{\mu}\enf{\mu}+\ehf{\mu}\ehf{\mu}\bigr)
        +\eps \,\ezf\ezf\;,\label{qfcfr}\\
 A(\beta) &= \prod_\nu ( 1 + \beta^2 x^2_\nu ) = \sum_{j=0}^\dg \A{j}\,\beta^{2j}\;,\label{Afcfr}
\end{align}
which justifies the expansion \eqref{Afcexp}.
Since the conformal Killing tensor ${\ts{q}(\beta)}$ is expressed in the diagonal form \eqref{qfcfr}, we can easily substitute its inversion into the definition \eqref{KTfcdef} of $\ts{k}(\beta)$ to obtain
\begin{equation}\label{KTfcfr1}
    \ts{k}(\beta)
      = \biggl(\prod_\nu ( 1 + \beta^2 x^2_\nu )\biggr)
        \biggl(\sum_\mu\frac1{1+\beta^2x^2_\mu}
        \bigl(\env{\mu}\env{\mu}+\ehv{\mu}\ehv{\mu}\bigr)
        +\eps\,\ezv\ezv\biggr)\;.
\end{equation}
Employing expansion \eqref{Afcfr}, and similar expression for ${\A{j}_\mu}$, we arrive at
\begin{equation}\label{KTfcfr}
    \ts{k}(\beta)
      =\sum_{j=0}^{\dg}\Bigl(\sum_\mu\A{j}_\mu\bigl(\env{\mu}\env{\mu}+\ehv{\mu}\ehv{\mu}\bigr)
        + \eps \A{j}\,\ezv\ezv\Bigr)\;\beta^{2j}\;,
\end{equation}
which is the expansion \eqref{KTfcexp}, cf.~\eqref{KTjfr}. With the help of \eqref{KVj}, we also immediately get the expansion \eqref{KVfcexp} for the Killing vectors, which thanks to \eqref{KVjcoor} reads
\begin{equation}\label{KVfccoor}
    \ts{l}(\beta) = \sum_j^\dg \beta^{2j} \cv{\psi_j} \;.
\end{equation}

\subsection{Uniqueness theorem}
\label{ssc:UniquenessTh}

It is obvious from the above construction of the Killing tower that the principal tensor $\ts{h}$ determines uniquely a set of canonical coordinates. Namely,  the set of $n$ functionally independent eigenvalues $x_\mu$ is supplemented by a set of $\dg+\eps$ Killing coordinates $\psi_j$ associated with the Killing vectors ${\KV{j}}$. It is then no such a wonder that the principal tensor uniquely defines the corresponding geometry.\footnote{%
As we mentioned in section \ref{ssc:PrincipalTensor}, we do not consider the exceptional null form of the principal tensor which is possible for the Lorentzian signature. Such a case would lead, in general, to its own special geometry.} This geometry has a local form of the off-shell Kerr--NUT--(A)dS metric and is determined up to $\dg$ arbitrary metric functions of a single variable, $X_\mu=X_\mu(x_\mu)$. It also possess a number of remarkable geometric properties. Namely the following central theorem has been formulated in \cite{Krtous:2008tb}, culminating the previous results from \cite{Houri:2007xz}:\\[0.5ex]

\noindent{\bf Uniqueness theorem}:
{\it The most general geometry which admits a principal tensor can be locally written in the off-shell Kerr--NUT--(A)dS form \eqref{KerrNUTAdSmetric}. When the Einstein equations
are imposed, the geometry is given by the on-shell Kerr--NUT--(A)dS metric described by the metric functions \eqref{Xsol}.}\\[0.5ex]

Moreover, this metric possesses the following properties:\\[0.5ex]

\noindent{\bf Theorem}:
{\it The off-shell Kerr--NUT--(A)dS metric is of the special type D of higher-dimensional algebraic classification. The geodesic motion in this spacetime is completely integrable, and the  Hamilton--Jacobi, Klein–-Gordon, and Dirac equations allow a separation of variables.}\\[0.5ex]

We refer to the literature \citep{Houri:2007uq, Houri:2007xz, Krtous:2008tb, Yasui:2008zz, Houri:2008ng,Yasui:2011pr, krtous:directlink} for various versions of the proof of the uniqueness theorem.
The fact that the metric is of the type D \citep{Hamamoto:2006zf} of higher-dimensional algebraic classification \citep{Coley:2004jv, Ortaggio:2012jd, Pravda:2007ty} follows directly from studying the integrability conditions of a non-degenerate conformal Killing--Yano 2-form \citep{Mason:2010zzc}.
The separability and integrability properties of the Kerr--NUT--(A)dS geometry will be demonstrated in the next chapter.

\details{Perhaps the `shortest route' to the uniqueness theorem and the Kerr--NUT--(A)dS metric is through the separability structure theory for the Hamilton--Jacobi and Klein--Gordon equations, see section~\ref{ssc:separabilitystructures}. Namely, the existence of the principal tensor implies the existence of a Killing tower of symmetries, which in its turn implies the separability of the Hamilton--Jacobi and Klein--Gordon equations. It then follows that one can use
the canonical metric constructed in \cite{benenti1979remarks} admitting such separability structure. In the spirit of Carter's derivation of the four-dimensional Kerr--NUT--(A)dS metric \citep{Carter:1968cmp}, this then directly leads to the higher-dimensional Kerr--NUT--(A)dS geometry \citep{Houri:2007uq, Houri:2007xz, Yasui:2011pr}, see also \cite{Kolar:2015hea}.
}\medskip

Since the existence of a principal tensor $\ts{h}$ uniquely determines the off-shell Kerr--NUT--(A)dS geometry, when discussing the Killing tower one does not need to strictly distinguish among the properties that follow from general considerations with Killing--Yano tensors, the properties that follow from the existence of a (general) Darboux basis, and the properties that use the explicit form of the Darboux basis of the Kerr--NUT--(A)dS geometry. However,  all the properties of the Killing tower can be derived directly from the properties of the principal tensor, without referring to the explicit form of the metric.

Let us finally note that when the non-degeneracy condition on the principal tensor is relaxed, one obtains a broader class of geometries that has been named the \defterm{generalized Kerr--NUT--(A)dS geometry} \citep{Houri:2008th, Houri:2008ng, Oota:2008uj}. This class will be briefly reviewed in section~\ref{ssc:GenKNA}.

\subsection{Proof of commutation relations}
\label{sec:commutationrel}

By now we have established most of the properties of objects in the Killing tower. However, we have not yet proved the commutation relations \eqref{KTKVNS} or \eqref{NSgenfc}. Since ${\KT{0}=\ts{g}}$, these relations in particular imply that ${\KT{j}}$ and ${\KV{j}}$ are Killing tensors and Killing vectors, respectively. The fact that ${\KT{j}}$ are Killing tensors follows directly from their construction. However, that ${\KV{j}}$ are Killing vectors we observed only using the identity \eqref{xi=KV} and its consequences \eqref{KVjcoor}. In other words, we have used the explicit form \eqref{KerrNUTAdSmetric} of the Kerr--NUT--(A)dS metric. However, as mentioned above,  it is possible to demonstrate the commutativity \eqref{NSgenfc} directly from the existence of the principal tensor and without any reference to canonical coordinates, proving in particular that ${\KV{j}}$ are Killing vectors. We give here a brief overview of such a procedure, for details see \cite{krtous:directlink}.

\subsubsection{Commutation relations}

Using the fundamental property \eqref{PCCKYeq} of the principal tensor and definitions \eqref{qfcdef}, \eqref{Afcdef}, and \eqref{KTfcdef}, one can express covariant derivatives of the conformal Killing tensor ${\ts{q}(\beta)}$, the function ${A(\beta)}$, and the Killing tensor ${\ts{k}(\beta)}$ as follows:
\begin{align}
  \nabla_{\!c}\, q_{ab} &= 2\beta^2\,\bigl(g_{c(a}\,h_{b)n}+h_{c(a}\,g_{b)n}\bigr)\,\xi^n\,,\\
  \nabla_{\!a} A\; &= 2\beta^2\, h_{am}\, k^{mn}\,\xi_n\,,\label{gradAb}\\
  \nabla^{c} k^{ab} & =
    \frac{2\beta^2}{A}\bigl(
    k^{ab}\,k^{cn}\,h_{n}{}^{m}+h^{m}{}_{n}\,k^{n(a}\,k^{b)c}
    +k^{m(a}\,k^{b)n}\,h_{n}{}^{c}\bigr)\,\xi_m\,.\label{covdKTb}
\end{align}
To shorten the expressions, here and in the rest of this section we omit the argument ${\beta}$; to distinguish two different values of ${\beta}$ we write ${\ts{k}_1=\ts{k}(\beta_1)}$, ${\ts{k}_2=\ts{k}(\beta_2)}$, and similarly for ${\ts{l}_1}$ and ${\ts{l}_2}$ (do not confuse with ${\KT{j}}$ and ${\KV{k}}$). In the following we shall also sometimes work with rank 2 tensors as with matrices: denoting by ${\ts{A}\cdot\ts{B}}$ the matrix multiplication and by the following square bracket the commutator:
\be\label{commutatorAB}
{[\ts{A},\ts{B}]\equiv \ts{A}\cdot\ts{B}-\ts{B}\cdot\ts{A}}\,.
\ee

The explicit form \eqref{NSbrcov} of the Nijenhuis--Schouten brackets \eqref{NSgenfc} in terms of the covariant derivative reads
\begin{align}
\bigl[k_1,k_2\bigr]_\NS^{abc}
  &= 3\bigl(k_1^{e(a}\nabla_{\!e}k_2^{bc)}-k_2^{e(a}\nabla_{\!e}k_1^{bc)}\bigr)
  \,,\\
\bigl[k_1,\,l_2\,\bigr]_\NS^{ab\;}
  &= 2\;k_1^{e(a}\nabla_{\!e}l_2^{b)}-l_2^{e}\nabla_{\!e}k_1^{ab}
  \,,\\
\bigl[\,l_1\,,\,l_2\,\bigr]_\NS^{a\;\;}
  &= l_1^{e}\nabla_{\!e}l_2^{a}-l_2^{e}\nabla_{\!e}l_1^{a}
  \,.
\end{align}
Upon substituting the definition \eqref{KVfcdef} and the expression \eqref{covdKTb} to these equations, the straightforward long calculation yields
\begin{align}
\bigl[k_1,k_2\bigr]_\NS^{abc} &= 0
  \,,\label{NSkkexpl}\\
\bigl[k_1,\,l_2\,\bigr]_\NS^{ab\;}
  & = k_1^{am}\,(\nabla_{\!m}\xi_n)\,k_2^{nb}
  + k_2^{am}\,(\nabla_{\!n}\xi{}_m)\,k_1^{nb}
  \,,\label{NSklexpl}\\
\bigl[\,l_1\,,\,l_2\,\bigr]_\NS^{a\;\;}
  & = \bigl(k_1^{am}\,(\nabla_{\!m}\xi_n)\,k_2^{nb}
  - k_2^{am}\,(\nabla_{\!m}\xi{}_n)\,k_1^{nb}\bigr)\,\xi_b
  \,.\label{NSllexpl}
\end{align}
The Killing tensors ${\ts{k}_1}$ and ${\ts{k}_2}$ are diagonal in the same basis, cf.\ \eqref{KTjfr}, so they commute as linear operators, ${[\ts{k}_1,\ts{k}_2]=\ts{k}_1\cdot \ts{k}_2 - \ts{k}_2\cdot \ts{k}_1=0}$. If they also commute with ${\covd\ts{\xi}}$
\begin{equation}\label{[kbetacovdxi]}
    [k,\nabla\xi]{}^{a}{}_{b}=
    k^{a}{}_e\,(\nabla^{e}\xi_b)-(\nabla^{a}\xi_{e})\,k^{e}{}_{b} = 0\,,
\end{equation}
the last Nijenhuis--Schouten bracket \eqref{NSllexpl} vanishes. The same is true for the second bracket \eqref{NSklexpl}, if, additionally, ${\covd\ts{\xi}}$ is antisymmetric,
\begin{equation}\label{covdxiasym}
    \nabla_{\!(a}\xi{}_{b)}=0\;,
\end{equation}
which is clearly the Killing vector condition for ${\ts{\xi}}$.

To summarize, the proof of the Nijenhuis--Schouten commutativity \eqref{KTKVNS} reduces to proving the properties \eqref{[kbetacovdxi]} and \eqref{covdxiasym} for the primary Killing vector ${\ts{\xi}}$. It turns out, that both these conditions follow in a complicated way from the integrability conditions for the principal tensor. We discus this in more details in the next section.

\subsubsection{Structure of the curvature}
To complete the proof of the commutativity \eqref{KTKVNS} we need first to  discuss the integrability conditions for the principal tensor and establish their implications for the structure of the curvature tensor.

Applying the integrability relation \eqref{covderxi} to the principal tensor ${\ts{h}}$, we obtain the following expression for the covariant derivative of ${\ts{\xi}}$:
\begin{equation}\label{covdxi}
  (D-2)\nabla_{\!a}\xi_{b} = -R_{ac}\,h^{c}{}_{b} + \frac12 h_{cd}R^{cd}{}_{ab}\;,
\end{equation}
which upon symmetrization gives
\begin{equation}\label{covdxisym}
  2(D-2)\nabla_{\!(a}\xi_{b)} = h_{ac}R^c{}_b - R_{ac}\,h^{c}{}_{b}\;.
\end{equation}
The aim is to show that the right hand side vanishes, that is the principal tensor always commutes with the Ricci tensor, and hence ${\ts{\xi}}$ is a primary Killing vector. This
is trivial if the spacetime satisfies the vacuum Einstein equations with a cosmological constant, since the Ricci tensor is then proportional to the metric. More generally we have the following construction.

Writing down the integrability condition \eqref{CCKYcurvaturecond} for the principal tensor ${\ts{h}}$, one obtains
\begin{equation}\label{hRintcond}
  (D{-}2)R^{ab}{}_{e[c}\, h^{e}{}_{d]}
    - h_{ef} R^{ef[a}{}_{[c}\,\delta^{b]}_{d]}
    - 2 R^{[a}{}_{e}\,\delta^{b]}_{[c}\, h^{e}{}_{d]}=0\,.
\end{equation}
This condition puts rather strong restrictions on the curvature. In order to express them in a compact way, let as introduce shortcuts for tensors obtained by various combinations of the principal tensor ${\ts{h}}$, the Riemann tensor ${\ts{R}}$ and the Ricci tensor ${\Ric}$.

First, let us denote by ${\ts{h}^p}$ the ${p}$-th matrix power of ${\ts{h}}$,
\begin{equation}\label{htop}
    {h^p{\,}^a{}_b} = h^a{}_{c_1} h^{c_1}{}_{c_2}\cdots h^{c_{p{-}1}}{}_b\;.
\end{equation}
Let us emphasize obvious, it is a different operation than the wedge power used in the definition of ${\CCKY{j}}$, \eqref{CCKYj}. Next we define the tensor ${\mathbf{Rh}^{(p)}}$ as the contraction of ${\ts{h}^p}$ with the Riemann tensor in the first two indices
\begin{equation}\label{Rhpdef}
    \mathrm{Rh}^{(p)}{}_{ab} = h^p{}_{cd}\, R^{cd}{}_{ab}\;.
\end{equation}
Similarly, we define the tensor ${\mathbf{Rich}^{(p)}}$ as the contraction of ${\ts{h}^p}$ with the Riemann tensor in other pair of indices
\begin{equation}\label{Ricpdef}
    \mathrm{Rich}^{(p)}{}_{ab} = h^p{}_{cd}\, R^{c}{}_{a}{}^{d}{}_{b}\;.
\end{equation}
The notation is motivated by the fact that for ${p=0}$ we get just the Ricci tensor, ${\mathbf{Rich}^{(0)}=\Ric}$.

Rather non-trivial calculations \citep{krtous:directlink} show that the integrability condition \eqref{hRintcond} implies that all contractions of the Riemann tensor with an arbitrary power of ${\ts{h}}$ commute, in the sense of \eqref{commutatorAB}, with ${\ts{h}}$ itself,
\begin{equation}\label{Rhcommut}
  \bigl[\mathbf{Rh}^{(p)},\ts{h}\bigr] = 0\;,\qquad
    \bigl[\mathbf{Rich}^{(p)},\ts{h}\bigr] = 0\;.
\end{equation}
Moreover, for odd ${p}$ the tensors ${\mathbf{Rich}^{(p)}}$ trivially vanish.

In particular, the commutativity \eqref{Rhcommut} tells us that  $[\,\Ric,\ts{h}\,] = 0$, which guarantees that the right hand side of \eqref{covdxisym} vanishes, proving thus that ${\ts{\xi}}$ is a Killing vector. Moreover, \eqref{Rhcommut} also implies that all the tensors ${\mathbf{Rh}^{(p)}}$ and ${\mathbf{Rich}^{(p)}}$ are diagonal in the Darboux frame. Indeed, the vectors of the Darboux frame are eigenvectors of $\ts{h}^p$, which guarantees that the Ricci tensor has to have the structure \eqref{Ricci}.

Taking a commutator of \eqref{covdxi} with ${\ts{h}}$, one obtains
\begin{equation}\label{covdxih}
  (D{-}2)\bigl[\covd\ts{\xi},\ts{h}\bigr]=
     \bigl[\ts{h},\Ric\bigr]\cdot\ts{h}+\frac12\bigl[\mathbf{Rh}^{(1)},\ts{h}\bigr]\;.
\end{equation}
Employing the commutativity \eqref{Rhcommut}, we find that the covariant derivative ${\covd\ts{\xi}}$ of the primary Killing vector commutes with ${\ts{h}}$,
\begin{equation}\label{[covdxih]}
    \bigr[\covd\ts{\xi},\ts{h}\bigl]=0\;.
\end{equation}
However, the Killing tensor ${\ts{k}(\beta)}$ is defined as a function of ${\ts{Q}}$, see definitions \eqref{qfcdef} and \eqref{KTfcdef}, which is just ${\ts{Q}=-\ts{h}\cdot\ts{h}}$, cf.\ \eqref{CKT}. Therefore, we also have
\begin{equation}\label{[covdxikbeta]}
    \bigl[\covd\ts{\xi},\ts{k}(\beta)\bigr]=0\;,
\end{equation}
which proves the condition \eqref{[kbetacovdxi]}.

Both conditions \eqref{[kbetacovdxi]} and \eqref{covdxiasym} thus follow from the integrability condition \eqref{hRintcond} for the principal tensor. That concludes the proof of the Nijenhuis--Schouten commutativity \eqref{NSgenfc}, respectively \eqref{KTKVNS}.

In section~\ref{ssc:AlignedEM} we will see that ${\ts{\xi}}$ can be used as a vector potential for a special electromagnetic field which leads to an integrable motion of charged particles. The result \eqref{[covdxih]} thus shows, that its Maxwell tensor ${\ts{F}=\grad\ts{\xi}=2\covd\ts{\xi}}$ commutes with the principal tensor ${\ts{h}}$ and can also be skew-diagonalized in the Darboux basis.

\subsection{Principal tensor as a symplectic structure}
\label{ssc:PTSymplStruct}

\subsubsection{Motivation}

In the construction of the Killing tower from the principal tensor we have defined the Killing vectors ${\KV{j}}$ by \eqref{KVj}, or in terms of a generating function by \eqref{KVfcdef}. We then claimed that we can associate Killing coordinates with these Killing vectors and it turns out that those are exactly coordinates ${\psi_j}$ in the canonical metric \eqref{KerrNUTAdSmetric}, namely ${\KV{j}=\cv{\psi_j}}$. However, we also mentioned that this last equality is not obvious and appears only after one reconstructs the full form of the metric, employing  the uniqueness theorem in section~\ref{ssc:UniquenessTh}.

Indeed, immediately after the definition \eqref{KVj}, it is not clear that one can introduce the common Killing coordinates ${(\psi_0,\dots,\psi_{\dg{-}1})}$, concentrating on even dimensions, where each ${\psi_j}$ would be constant along ${\KV{k}}$ for ${k\neq j}$ and ${\KV{j}\cdot\grad\psi_j=1}$. For that, it is sufficient to show that the  Killing vectors Lie commute,
\begin{equation}\label{KVLieComm}
    \bigl[\KV{i},\KV{j}\bigr]=0\,,
\end{equation}
and that the Killing vectors leave coordinates ${x_\mu}$ constant, ${\KV{j}\cdot\grad x_\mu=0}$.

The Lie commutativity \eqref{KVLieComm} has been shown (in the terms of Nijenhuis--Schouten brackets) in the previous section. However, this result can also be established by a slightly different argument which possesses a beauty on its own and to this argument we devote this section.

\subsubsection{Symplectic structure on the spacetime}

The principal tensor $\ts{h}$ is a closed non-degenerate 2-form on the configuration space
${M}$. As such it defines a symplectic structure on this space for $M$ even-dimensional, and a contact structure on $M$ in the case of odd dimensions.
To explore this idea, and since we have not introduced \emph{contact manifolds}, let us restrict to the case of even number of dimensions. (The discussion in odd dimensions would proceed analogously.)

In even dimensions the principal tensor ${\ts{h}}$ thus plays a role of the symplectic structure on the space ${M}$ in the sense of the theory described in section~\ref{ssc:symplgeom}. We can define its inverse ${\ts{h}^{-1}}$, which in the Darboux basis reads
\begin{equation}\label{PTinv}
    \ts{h}^{-1}=\sum_\mu x_\mu^{-1} (\env{\mu}\ehv{\mu}-\ehv{\mu}\env{\mu})\;.
\end{equation}

For any function ${f}$ on ${M}$, we can define the associated Hamiltonian vector field ${\hamvPT{f}}$ as, cf.\ \eqref{hamvect},
\begin{equation}\label{PThamvect}
    \hamvPT{f} = \ts{h}^{-1}\cdot\grad f\, .
\end{equation}
Note that we denoted this vector field in the $D$-dimensional spacetime $M$ by $\hamvPT{}$ in order to distinguish it from a similar Hamiltonian vector field $\ts{X}$ on a $2D$ phase space. We introduce the Poisson bracket of two functions ${f}$ and ${g}$ as follows
\begin{equation}\label{PTPB}
    \{f,g\}_\PT = \grad f \cdot \ts{h}^{-1}\cdot \grad g\;,
\end{equation}
where ``PT'' stands for the \emph{principal tensor} generated Poisson bracket.

\details{%
Let us stress that these operations are not related to analogous operations on the relativistic particle phase space, which is realized as the cotangent space  ${\mathbf{T}^*M}$, see section~\ref{ssc:PhaseSpcCotBundle}. The Hamiltonian vector ${\hamvPT{f}}$ is an ordinary vector field on ${M}$. The bracket \eqref{PTPB} expects as arguments ordinary functions depending just on the position in ${M}$. The dynamics of a relativistic particle is governed by the Hamiltonian on the phase space ${\mathbf{T}^*M}$ and cannot be translated in a straightforward way to the language of the symplectic geometry generated by the principal tensor ${\ts{h}}$.}\medskip

Let us state a couple of observations. First, from the discussion of the special Darboux frame \eqref{specDarbouxframecond} in section~\ref{ssc:PrincipalTensor} it follows that ${\ehv{\mu}\cdot\grad x_\nu=0}$, and therefore
\begin{equation}\label{xxPTcomm}
    \{x_\mu,x_\nu\}_\PT =0\;.
\end{equation}
As a consequence we see that the principal tensor Poisson bracket of any two functions, which depend just on ${x_\mu}$ coordinates, vanishes.

Next, the relations\footnote{%
Let us mention that relations \eqref{hKV=dA} follow from the relation \eqref{gradAb}, together with the definition \eqref{KVfcdef} and expansions \eqref{Afcexp} and \eqref{KVfcexp}. They can thus be deduced directly from the properties of the principal tensor, without referring to canonical coordinates. Also, here and below we use the relation ${\grad\A{j+1}=2\sum_\mu x_\mu\A{j}_\mu\grad x_\mu}$, which follows from \eqref{Aid0}, for example.}
\eqref{hKV=dA} actually mean that ${\KV{j} = - \hamvPT{\frac12\A{j{+}1}}}$. It motivates us to introduce functions ${\alpha^j}$
\begin{equation}\label{alphadef}
    \alpha^j = \frac12\A{j+1}\,,\qquad j=0,\,1,\,\dots,\,\dg-1\;,
\end{equation}
which can serve as coordinates instead of functions ${x_\mu}$. We thus have
\begin{equation}\label{alphaPTPB}
    \{\alpha^i,\alpha^j\}_\PT=0\;,
\end{equation}
and
\begin{equation}\label{KVishamv}
   \KV{j} = - \hamvPT{\alpha^j} \,.
\end{equation}
As an immediate consequence we obtain
\begin{equation}\label{KVdotdalpha}
\KV{i}\cdot\grad\alpha^j=\{\alpha^i,\alpha^j\}_\PT=0\,,
\end{equation}
cf.\ \eqref{FG}. Hence, the Killing vectors leave coordinates ${\alpha^j}$, as well as ${x_\mu}$, constant. Similarly, using \eqref{LB-PB} we get
\begin{equation}\label{KVLiePTcomm}
    \bigl[\KV{\alpha^i},\KV{\alpha^j}\bigr]=-\hamvPT{\{\alpha^i,\alpha^j\}_\PT}=0\;,
\end{equation}
which proves that the Killing vectors Lie commute.

One can thus expect that it is possible to introduce associated coordinates ${\psi_j}$. This is actually provided by the Liouville's procedure described in section~\ref{ssc:complintgrb}. The coordinates ${\alpha^j}$ commute with each other, \eqref{alphaPTPB}, and the Liouville's procedure teaches us that they can thus be complemented  into a canonical set of coordinates ${(\alpha^0,\dots,\alpha^{\dg{-}1},\psi_{0},\dots,\psi_{\dg{-}1})}$ in which the symplectic form ${\ts{h}}$ reads
\begin{equation}\label{PTalphapsi}
    \ts{h} = \sum_j \grad\alpha^j\wedge\grad\psi_j\,.
\end{equation}
With the help of \eqref{HVFC}, it implies
\begin{equation}\label{cvalphapsi}
    \KV{j}=-\hamvPT{\alpha^j}=\cv{\psi_j}\,,\qquad \hamvPT{\psi_j} = \cv{\alpha^j}\,.
\end{equation}

Killing vectors ${\KV{j}}$ thus indeed define Killing coordinates ${\psi_j}$. As we already discussed in section~\ref{ssc:UniquenessTh}, this observation is an important piece of the uniqueness theorem. We established that the principal tensor defines canonical coordinates. The uniqueness theorem additionally provides the explicit form of the metric in these coordinates.

Let us conclude this section with some related observations. The symplectic potential \eqref{PCCKYpot} for the principal tensor can be written as
\begin{equation}\label{PTsymplpot}
    \ts{b} = \sum_j \alpha^j \grad \psi_j\,.
\end{equation}
Using the coordinates ${x_\mu}$ instead of ${\alpha^j}$, see \eqref{alphadef},  one can rewrite the principal tensor in the following  form \eqref{PCCKY}:
\begin{equation}\label{PTxpsi}
    \ts{h} = \sum_\mu x_\mu\grad x_\mu \wedge \Bigl(\sum_j\A{j}_\mu \grad \psi_j\Bigr)\,.
\end{equation}
We see that ${x_\mu}$ and ${\psi_j}$ are not canonically conjugate in the sense of the principal tensor symplectic geometry. However, since ${\alpha}$'s are functions of only ${x}$'s, the coordinate vectors ${\cv{\psi_j}}$ introduced in \eqref{cvalphapsi} coincide with coordinate vectors ${\cv{\psi^j}}$ of the coordinate set ${(x_\mu,\psi_j)}$.

We could ask what are the coordinates canonically conjugate to ${x^\mu\equiv x_\mu}$. It is easy to check that in terms of coordinates
\begin{equation}\label{pidef}
    \pi_\mu = x^\mu\sum_j\A{j}_\mu\psi_j\,,
\end{equation}
a 1-form ${\tilde{\ts{b}}}$ defined as
\begin{equation}\label{PTsymplpotalt}
    \tilde{\ts{b}} = - \sum_\mu \pi_\mu\grad x^\mu = - \sum_j \psi_j \grad\alpha^j\,,
\end{equation}
is also the symplectic potential for the principal tensor, ${\ts{h}=\grad\tilde{\ts{b}}}$. It implies that
\begin{equation}\label{PTxpi}
    \ts{h} = \sum_\mu \grad x^\mu\wedge\grad\pi_\mu\;,
\end{equation}
and ${(x^1,\dots,x^\dg,\pi_1,\dots,\pi_\dg)}$ are canonical coordinates in the sense of the principal tensor symplectic geometry.

%% file: ch6-intsep.tex

\section{Particles and fields: Integrability and separability}
\label{sc:intsep}

In this chapter, we study particles and fields in the vicinity of higher-dimen\-sional rotating black holes. As can be expected their behavior reflects the rich structure of hidden symmetries discussed in the previous chapter: the motion of particles and light is completely integrable and the fundamental physical equations allow separation of variables. Let us start our discussion with a brief overview of the discovery of these unexpected properties.


The Kerr--NUT--(A)dS metric in four spacetime dimensions possesses a number of remarkable properties related to hidden symmetries. In particular, those discovered by \cite{Carter:1968rr, Carter:1968cmp, Carter:1968pl} include the complete integrability of geodesic equations and the separability of the Hamilton--Jacobi and Klein--Gordon equations. A natural question is whether and if so how far these results can be extended to higher dimensions.

A first successful attempt on such a generalization, employing non-trivial hidden symmetries, was made by \cite{FrolovStojkovic:2003a,FrolovStojkovic:2003b}. In these papers the authors generalized Carter's approach to five-dimensional Myers--Perry metrics with two rotation parameters, and demonstrated that the corresponding Hamilton--Jacobi equation in the Myers--Perry coordinates allows a complete separation of variables. This enabled to obtain an explicit expression for the second-rank irreducible Killing tensor present in these spacetimes.\footnote{Let us note here that Carter's method had been used to study higher-dimensional black hole spacetimes prior to the works \citep{FrolovStojkovic:2003a,FrolovStojkovic:2003b}. For example, in \cite{Gibbons:1999uv, Herdeiro:2000ap} the authors demonstrated the separability of the Hamilton--Jacobi and Klein--Gordon equations for the five-dimensional so called BMPV black hole \citep{Breckenridge:1996is}. In this case, however, the corresponding Killing tensor is reducible and the explicit symmetries of the spacetime are enough to guarantee the obtained results.}

These results were later generalized by \cite{Kunduri:2005fq} to the case of a five-dimensional Kerr--(A)dS metric. Fields and quasinormal modes in five-dimensional black holes are studied in \cite{FrolovStojkovic:2003a, Cho:2011pb, Cho:2011yp}. Page and collaborators \citep{Vasudevan:2004mr,Vasudevan:2004ca,Vasudevan:2005js} discovered that particle equations are completely integrable and the Hamilton--Jacobi and Klein--Gordon equation are separable in the higher-dimensional Kerr--(A)dS spacetime, provided it has a special property: its spin is restricted to two sets of equal rotation parameters.   A similar result was obtained slightly later for the higher-dimensional Kerr--NUT--(A)dS spacetimes  subject to the same restriction on rotation parameters \citep{Davis:2006hy,Chen:2006ui}. With this restriction the Kerr--NUT--(A)dS metric becomes of cohomogeneity-two and possesses an enhanced symmetry which ensures the corresponding integrability and separability properties.

Attempts to apply Carter's method for general rotating black holes in six and higher dimensions have met two obstacles. First, the explicit symmetries of the Kerr--NUT--(A)dS metrics are, roughly speaking,  sufficient to provide only half of the required integrals of motion. This means, that already in six dimensions one needs not one, but two independent Killing tensors, and the number of required independent Killing tensors grows with the increasing of number of spacetime dimensions. Second, more serious problem is that the separation of variables in the Hamilton--Jacobi equation may exist only in a very special coordinate system. However, how to choose the convenient coordinates was of course unknown. In particular, the widely used Myers--Perry coordinates have an unpleasant property of having a constraint \eqref{constr}, which makes them inconvenient for separation of variables in more than five dimensions.

The discovery of the principal tensor for the most general higher-dimen\-sional Kerr--NUT--(A)dS \citep{Frolov:2007nt,Kubiznak:2006kt} spacetimes made it possible to solve both these problems. Namely, the associated Killing tower contains a sufficient number of hidden symmetries complementing the isometries to make the geodesic motion integrable. Moreover, the eigenvalues of the principal tensor together with the additional Killing coordinates, give the geometrically preferred canonical coordinates in the Kerr--NUT--(A)dS spacetime. It turns out that exactly in these coordinates the Hamilton--Jacobi as well as the Klein--Gordon equations separate. The following sections are devoted to a detailed discussion of these results.

\subsection{Complete integrability of geodesic motion}
\label{ssc:IntegrGeodMot}

The geodesic motion describing the dynamics of particles and the propagation of light in the Kerr--NUT--(A)dS spacetimes is completely integrable. In this section we prove this result, discuss how to obtain particles' trajectories, and how to introduce the action--angle variables for the corresponding dynamical system.

\subsubsection{Complete set of integrals of motion}

We have learned in section~\ref{ssc:partcurvspc} that the motion of a free relativistic particle can be described as a dynamical system with the quadratic in momenta Hamiltonian \eqref{RPHam}.
Turning to the Kerr--NUT--(A)dS spacetime, the towers of Killing tensors \eqref{KTj} and Killing vectors \eqref{KVj} guarantee the existence of the following $D=2\dg+\eps$ integrals of geodesic motion, $\dg$ of which are  quadratic in momenta and $\dg+\eps$ of which are linear in momenta:
\begin{equation}\label{integemot}
 \begin{aligned}
    K_{j} &= \KTc{j}^{ab}\,p_a\,p_b\;,&\quad &j=0,\dots\dg-1\,,\\
    L_{j} &= \KVc{j}^{a}\,p_a\;,&\quad &j=0,\dots\dg-1+\eps\,.
\end{aligned}
\end{equation}
The observable $K_{0}$ is, up to a trivial multiplicative constant, equivalent to the Hamiltonian of the system
\begin{equation}\label{Hamiltonian}
     H= \frac12\, g^{ab} p_a p_b = \frac12 K_{0}\;.
\end{equation}
Thanks to the commutation relations \eqref{KTKVNS} all these observables are in {\em involution} \citep{Page:2006ka, Krtous:2006qy, Krtous:2007xf,Houri:2007uq}:
\begin{equation}\label{Killobsinv}
    \bigl\{K_{i},K_{j}\bigr\}=0\,,\quad\bigl\{K_{i},L_{j}\bigr\}=0\,,\quad\bigl\{L_{i},L_{j}\bigr\}=0\,,
\end{equation}
and in particular commute with the Hamiltonian. The motion of free particles in the curved Kerr--NUT--(A)dS spacetime is thus complete integrable in the Liouville sense, cf.\ section~\ref{ssc:complintgrb}.

For a particle with mass $m$ the value of the constant $K_0$ is $-m^2$. As we already explained, in the $\sigma$-parametrization, which we use, the above relations remain valid in the limit $m\to 0$, that is for massless particles. As mentioned in the remark after equation \eqref{KillObs} in section~\ref{ssc:KillingTower}, for a propagation of massless particles one can use a different set of observables $\{\tilde{K}_{j}, L_{j}\}$, where $\tilde{K}_{j}$ are generated from the conformal Killing tensors $\CKT{j}$,
\be\label{confintegemot}
\tilde{K}_{j} = \CKTc{j}^{ab}\,p_a\,p_b\;,\quad j=0,\dots\dg-1\, .
\ee
The new observables are conserved and in involution, provided that the momenta satisfy the zero-mass condition $\ts{p}^2=0$. Indeed, thanks to this constraint, the right-hand sides of commutation relations \eqref{KTKVNSnull} vanish, which implies the Poisson-bracket commutation of the observables in the new set.

\details{
It is interesting to note that the relations among the quadratic conserved quantities $K_{j}$ are highly symmetric. One could actually study a space with the (inverse) metric given by the Killing tensor $\KT{i}$, and all the tensors $\KT{j}$ would remain Killing tensors with respect to this new metric, e.g., \cite{Rietdijk:1995ye}. This fact is precisely expressed by the first condition \eqref{Killobsinv}, giving $[\KT{i}, \KT{j}]_\NS=0$ for the Nijenhuis--Schouten brackets among these tensors. Similarly, all the vectors $\KV{j}$ remain to be Killing vectors with respect to the new metric. The geodesic motion in any of the spaces with the metric given by $\KT{i}$ is thus also complete integrable.
However, in this context one should emphasize that only the space with $\ts{g}=\KT{0}$ is the Kerr--NUT--(A)dS spacetime. Spaces with the metric given by $\KT{i}$, $i>0$, neither possess the principal tensor and the associated towers of Killing--Yano tensors, nor are solutions of the vacuum Einstein equations. Moreover, although the geodesic motion is integrable in these spaces, this is no longer true for the corresponding fields; the symmetry among Killing tensors does not elevate to the symmetry of the corresponding symmetry operators for the test fields in these spaces, see section~\ref{ssc:SepWave}.
}

\subsubsection{Particle trajectories} 

Substituting the coordinate expressions \eqref{KTjcoor} and \eqref{KVjcoor} into \eqref{integemot} gives the following expressions for the integrals of motion in terms of momenta components $\px{\mu}=\ts{p}\cdot\cv{x_\mu}$ and $\ppsi{j}=\ts{p}\cdot\cv{\psi_j}$:
\begin{align}
  \label{Qocoor}
   K_{j} &= \sum_\mu\A{j}_\mu\Biggl(\frac{X_\mu}{U_\mu} \px{\mu}^2+\frac{U_\mu}{X_\mu}
     \Bigl(\sum_{k=0}^{\dg-1+\eps}\frac{(-x_\mu^2)^{\dg{-}1{-}k}}{U_\mu} \ppsi{k}\Bigr)^2\Biggr)
     +\eps\frac{\A{j}}{c\A{\dg}}\ppsi{\dg}^2\,,\\
  \label{Locoor}
   L_{j} &= \ppsi{j}\,.
\end{align}
These expressions can be `inverted' and solved for the particle momenta. Namely, summing equations \eqref{Qocoor} multiplied by $(-x_\mu^2)^{\dg-1-j}$ over values $j=0,\, \dots,\, \dg-1+\eps$, using the orthogonality relation \eqref{Aid1i}, and some additional manipulations, gives
\begin{align}
  \label{pmusol}
    \px{\mu} &= \pm\frac{\sqrt{X_\mu \Qofc_\mu-\Lofc_\mu^2}}{X_\mu} = \pm\frac{\sqrt{\Xfc_\mu}}{X_\mu}
     \,,\\
  \label{pjsol}
    \ppsi{j} &= L_{j}\, .
\end{align}
Here we have introduced auxiliary functions
\begin{align}
\Qofc_\mu &= \sum_{j=0}^{\dg-1+\eps} K_{j} (-x_\mu^2)^{\dg-1-j}\,,
\label{Qofcdef}\\
\Lofc_\mu &= \sum_{j=0}^{\dg-1+\eps} L_{j} (-x_\mu^2)^{\dg-1-j}\,,
\label{Lofcdef}
\end{align}
as well as their combination
\begin{equation}\label{Xfcdef}
    \Xfc_\mu = {X_\mu\Qofc_\mu-\Lofc_\mu^2}\,.
\end{equation}
In odd dimensions we set $K_{\dg} = {L_{\dg}^2}/{c}$.
Functions ${\Qofc_\mu}$ and ${\Lofc_\mu}$ for different ${\mu}$ are given by the same polynomial dependence and differ just by their argument, ${\Qofc_\mu=\Qofc(x_\mu)}$, ${\Lofc_\mu=\Lofc(x_\mu)}$. The coefficients ${K_{j}}$ and ${L_{j}}$
in these polynomials can be understood either as conserved observables on the phase space or as numeric values of these observables, i.e., constants characterizing the motion.

It is remarkable that  the expression \eqref{pmusol} for $\px{\mu}$ depends only on one variable $x_\mu$. This property stands behind the separability of the Hamilton--Jacobi equation discussed in the next section. Signs $\pm$ in equations \eqref{pmusol} are independent for different $\mu$ and indicate that for a given value of $x_\mu$ there exist two possible values of momentum $\px{\mu}$. We will return to this point below when discussing a global structure of the level set $\levset{(\tpd{K},\tpd{L})}$.

The trajectory of a particle can be found by solving the velocity equation,
\begin{equation}\label{HJvelRP}
    \dot{x}^a = \frac{\pa H}{\pa p_a} = g^{ab}p_b\;.
\end{equation}
Employing the inverse metric \eqref{KerrNUTAdSinvmetric}, we obtain the expressions for the  derivative of $x_\mu$ and $\psi_j$ with respect to the inner time $\sigma$
\begin{align}\label{xvel}
    &\;\dot{x}_\mu
      = \pm\frac{\sqrt{\Xfc_\mu}}{U_\mu}\;,\\
&\begin{aligned}\label{psivel}
    \dot{\psi}_j
      &= \sum_\mu \frac{(-x_\mu^2)^{\dg-1-j}}{U_\mu} \frac{\Lofc_\mu}{X_\mu}
        \;,&&\text{for ${j=0,\dots,\dg-1}$}\,,\\
    \dot{\psi}_\dg
      &= \frac{\Psi_\dg}{c \A{\dg}} - \sum_\mu\frac{1}{x_\mu^2 U_\mu}\frac{\Lofc_\mu}{X_\mu}
    \;,&&\text{for $D$ odd}\,.
\end{aligned}
\end{align}
Since the expressions for velocities ${\dot{x}_\mu}$ are independent of the Killing coordinates ${\psi_j}$, one can integrate the equations for $x_\mu$ and $\psi_j$ in two steps. Namely after solving equations \eqref{xvel}, finding ${x_\mu(\sigma)}$, one substitutes these into equations \eqref{psivel} and integrates the Killing coordinates ${\psi_j}$.

However, Eqs.~\eqref{xvel} for ${\dot{x}_\mu}$ are not decoupled since the factor ${U_\mu}$ mixes the equations. In four dimensions, ${\dg=2}$, these factors are, up to a sign, the same, ${U_1=-U_2=\Sigma}$, cf.~\eqref{DD}, and they can be eliminated by the time reparamatrization, cf.~\eqref{sigmarepar}. For general ${\dg}$ such a trick is not possible. However, the system can still be solved by an integration and algebraic operations. In four dimensions such a procedure was demonstrated by \cite{Carter:1968rr}, in the following we generalize it to an arbitrary dimension.

First, we rewrite \eqref{xvel} as\footnote{The factor $1/2$ in these expressions is chosen for convenience, in order the final formulae  can be directly translated to the action--angle expressions below. In the following we also assume the existence of turning points for the trajectories. This is automatically satisfied for $x_\mu$ describing the angle variables, and restricts the discussion to the bounded trajectories regarding the radial coordinate.}%
\begin{equation}\label{xvelj}
    \pm \frac{(-x_\mu^2)^{\dg-1-j}}{2\sqrt{\Xfc_\mu}}\, \dot{x}_\mu
      = \frac{(-x_\mu^2)^{\dg-1-j}}{2U_\mu}
      \;,
\end{equation}
where the factor in front of ${\dot{x}_\mu}$ on the l.h.s. is a function of ${x_\mu}$ only. Such a function can be, in principle, integrated
\begin{equation}\label{xintjdef}
    \int_{x_\mu^-}^{x_\mu} \frac{(-x_\mu^2)^{\dg-1-j}}{2\sqrt{\Xfc_\mu}}\, d x_\mu\;.
\end{equation}
The integral must be over an interval which  belongs to the allowed range of the coordinate $x_\mu$ and where $\Xfc_\mu>0$. This condition is satisfied between turning points $x_\mu^-$ and $x_\mu^+$, which are defined by $\Xfc_\mu=0$. It is natural to chose the lower integration limit to be the smaller turning point $x_\mu^-$. With this choice we have also chosen the plus sign in \eqref{xvelj}.

Next, we introduce a set of functions ${X^{j}(x_1,\dots,x_\dg)}$, ${j=0,\dots,\dg-1}$, given by the sum of integrals \eqref{xintjdef}:
\begin{equation}\label{xintdef}
    X^{j} = \sum_\mu \int_{x_\mu^-}^{x_\mu} \frac{(-x_\mu^2)^{\dg-1-j}}{2\sqrt{\Xfc_\mu}}\, d x_\mu\;.
\end{equation}
In terms of these functions, the sum of equations \eqref{xvelj} over ${\mu}$ gives
\begin{equation}\label{xinteq}
    \dot{X}^{j} = \frac12\sum_\mu \frac{(-x_\mu^2)^{\dg-1-j}}{U_\mu} = \frac12\delta^j_{0}\;,
\end{equation}
where the last equality follows from \eqref{Aid2}. We can now integrate over the time parameter ${\sigma}$, to get
\begin{equation}\label{xint=const}
\begin{aligned}
    &X^{0} = \frac12\sigma+X^0_{\mathrm{o}} = \frac{1}{2}\sigma + \text{const.}\;, & j&=0\;,\\
    &X^{j} = X^j_{\mathrm{o}} = \text{const.}\;, & j&=1,\dots,\dg-1\;.
\end{aligned}
\end{equation}
Inverting `known' relations \eqref{xintdef} between ${X^{0},\dots,X^{\dg-1}}$ and ${x_1,\dots,x_\dg}$, one obtains the time evolution of coordinates ${x_\mu(\sigma)}$ parametrized by constants ${K_{j}}$, ${L_{j}}$, and ${X^j_{\mathrm{o}}}$. Substituting into equations \eqref{psivel} and integrating, one gets the time evolution ${\psi_j(\sigma)}$ of the Killing coordinates parametrized by the same constants together with additional integration constants $\psi^j_{\mathrm{o}}$.

The procedure \eqref{xvelj}--\eqref{xint=const} may seem as an 
ad hoc manipulation. However, as we shall see below it is closely related to the Liouville construction for complete integrable systems.

We demonstrated that as a result of the complete integrability of geodesic equations in the higher-dimensional Kerr--NUT--(A)dS spacetimes, finding solutions of these equations reduces to the calculation of special integrals. This integral representation of the solution is useful for the study of general properties of particle and light motion in these metrics. However, it should be emphasized that only in some special cases these integrals can be expressed in terms of known elementary and special functions. Let us remind that in four dimensions a similar problem can be solved in terms of elliptic integrals, the properties of which are well known. The integrals describing particle and light motion in higher dimensions contain square roots of the polynomials of the order higher than four, and this power grows with the increasing number of spacetime dimensions. Another complication is that the higher-dimensional problem depends on a larger number of parameters. At present, the problem of classification of  higher-dimensional geodesics in Kerr--NUT--(A)dS metrics is far from its complete solution. Here we give some references on the publications connected with this subject. The particle motion in five-dimensional Kerr--(A)dS metrics was considered in \cite{FrolovStojkovic:2003b,Kagramanova:2012hw, Diemer:2014lba, Delsate:2015ina}. The papers \cite{Gooding:2008tf, Papnoi:2014aaa} discuss the shadow effect for five-dimensional rotating black holes. Different aspects of geodesic motion in the higher-dimensional black hole spacetimes were discussed in \cite{Hackmann:2009zz, Enolski:2010if}.

\subsubsection{Conjugate coordinates on the level set}

Having proved that the geodesic motion is completely integrable, let us now discuss the corresponding level sets (here) and the construction of the action--angle variables (below). For simplicity, in this exposition (till end of section~\ref{ssc:IntegrGeodMot}) we restrict ourselves to the case of even dimensions, ${D=2\dg}$.

Following the Liouville constructions, described in section~\ref{ssc:complintgrb}, let us obtain a generating function ${W(\tp{x},\tp{\psi};\tpd{K},\tpd{L})}$, which allows us to change the original phase space coordinates ${(x_\mu,\psi_j;\px{\mu},\ppsi{j})}$ to new canonically conjugate coordinates ${(X^j,\Psi^j;K_j,L_j)}$, where $K_j$ and $L_j$ are the integrals of motion \eqref{integemot}. It is given by the integral \eqref{genfcqP}, which now reads:
\begin{equation}\label{genfcW}
    W(\tp{x},\tp{\psi};\tpd{K},\tpd{L}) = \int_c \Bigl(
      \sum_\mu \px{\mu}(x_\mu;\tpd{K},\tpd{L})\, d x_\mu
      +\sum_j \ppsi{j}(L_j)\, d\psi_j\Bigr)\,.
\end{equation}
Here, the momenta $\px{\mu}$ and $\ppsi{j}$ are given by \eqref{pmusol} and \eqref{pjsol} as functions of old positions and new momenta and we have used the fact that $\px{\mu}$ depends only on ${x_\mu}$ and ${\ppsi{j}}$ is given by ${L_j}$. If we substitute \eqref{pmusol} and \eqref{pjsol} explicitly and use a curve ${c}$ that starts at $\psi_j=0$ and at turning values ${x_\mu^-}$ of variables ${x_\mu}$, we obtain
\begin{equation}\label{genfcWspec}
    W(\tp{x},\tp{\psi};\tpd{K},\tpd{L}) =
      \sum_\mu \int_{x_\mu^-}^{x_\mu} \frac{\sqrt{\Xfc_\mu}}{X_\mu}\,d x_\mu
      +\sum_j L_j \psi_j\,.
\end{equation}
As we shall see, this is precisely the separated Hamilton's function \eqref{Ssepansatz} and \eqref{Smuint}, for the Hamilton--Jacobi equation studied in the next section, recovering the general relation ${W=S}$, \eqref{ComplIntInt}.

The generating function $W$ defines new coordinates ${X^j}$ and ${\Psi^j}$, that are conjugate to observables ${K_{j}}$ and~${L_{j}}$, as follows:
\begin{equation}\label{conjKL}
\begin{aligned}
    X^j &= \frac{\pa W}{\pa K_j} =
    \sum_\mu \int_{x_\mu^-}^{x_\mu} \frac{(-x_\mu^2)^{\dg-1-j}}{2\sqrt{\Xfc_\mu}}\,d x_\mu\,,\\
    \Psi^j &= \frac{\pa W}{\pa L_j} = \psi^j +
      \sum_\mu \int_{x_\mu^-}^{x_\mu} \frac{\Lofc_\mu}{X_\mu} \frac{(-x_\mu^2)^{\dg-1-j}}{\sqrt{\Xfc_\mu}}\,d x_\mu\;.
\end{aligned}
\end{equation}
Since the integrand in \eqref{genfcWspec} vanishes at turning point ${x_\mu^-}$, we could ignore the derivative of the lower integral limit, despite the fact that ${x_\mu^-}$ depends on ${K_j}$.

Clearly, ${X^j}$ are exactly the integrals introduced in \eqref{xintdef}. However, ${\Psi^j}$ are not the same as the original ${\psi^j}$. Canonical Poisson brackets read
\begin{equation}\label{IMcanvar}
    \{X^{i},K_{j}\} = \delta^i_j\;,\quad
    \{\Psi^i,L_{j}\} = \delta^i_j\;,
\end{equation}
all other being zero.
Since the Hamiltonian is ${H=\frac12 K_0}$, the Hamilton equations for ${X^j}$ and ${\Psi^j}$ are just $\dot{X}^j=\frac12\delta^j_0$, cf.~\eqref{xinteq}, and ${\dot{\Psi}^j = 0}$. All $X^j$ and $\Psi^j$ are thus constants except $X^0=\frac{1}{2}\sigma+\text{const}$. Inverting the relations \eqref{conjKL} to $x_\mu(\tp{X},\tp{\Psi})$ and $\psi_j(\tp{X},\tp{\Psi})$ gives the trajectory of the particle.

\subsubsection{Action--angle variables}

Let us remind that for a completely integrable system with $D$ degrees of freedom there exists $D$ independent integrals of motion in involution $P_i$. We called a level set a $D$-dimensional submanifold of the phase space $\levset{P}$, where these integrals have fixed values. According to general theory \citep{Arnold:book,Goldstein:book} if this level set is compact, it has a structure of multi-dimensional torus with an affine structure, and one can introduce the so called action--angle variables.

\details{%
The affine structure of the level set $\levset{P}$ refers to the fact that coordinates $Q$, on the level set, which are conjugate to $P$, are given uniquely up to a linear transformation. In other words, if one uses a different combination of integrals of motions $\bar{P}$, the corresponding conjugate coordinates $\bar{Q}$ are related to $Q$ on the given level set by a linear transformation.
Action of the Hamilton flow associated with any of the conserved quantities is linear in the sense of affine structure---all conserved quantities generate Abelian group of translations on the level set.
The torus structure of the level set must be compatible with the affine structure.
However, its existence can be understood only after taking into account interpretation of the involved variables.
}\medskip

In the Kerr--NUT--(A)dS spacetime the conserved quantities are $P=(\tpd{K},\tpd{L})$ and the corresponding level set is ${\levset{(\tpd{K},\tpd{L})}}$. As in the general case, this set is a Lagrangian submanifold, where the momenta $p_a$ can be found as functions of the coordinates $(x_{\mu},\psi_j)$ and conserved quantities, cf.\ equations \eqref{pmusol} and \eqref{pjsol}. We see that relations \eqref{pmusol} are independent for each plane {${x_\mu}$--${\px{\mu}}$}. In each of these planes the condition \eqref{pmusol} defines a closed curve which spans the range ${(x_\mu^-,x_\mu^+)}$ between turning points ${x_\mu^\pm}$ for which ${\Xfc_\mu=0}$. The curve has two branches over this interval, one with ${\px{\mu}>0}$, another with ${\px{\mu}<0}$.

The turning points ${x_\mu^\pm}$ should exist for angular coordinates $x_\mu$ since the ranges of these coordinates are bound. Situation can be different for the radial coordinate ${r}$ (Wick rotated ${x_\dg}$). Depending on the values of conserved quantities, one can have two turning points (bounded orbits), one turning point (scattering trajectories), or no turning points (fall into a black hole). For simplicity, here we discuss only the case where there are two turning points for ${r}$. Thus, one has a full torus structure in the ${x}$-sector of the level set.
The torus structure in Killing coordinates ${\psi_j}$ is also present. Condition \eqref{pjsol} just fixes the momenta to be constant, but leaves the angles unrestricted. However, some linear combination of Killing angles $\psi_j$ defines angular coordinates ${\ph_\mu}$, which are periodic. In the maximally symmetric case or for the Myers--Perry solution these coordinates are simply ${\phi_\mu}$ discussed previously. In the periodic coordinates we get the explicit torus structure. The only exception is the time direction, for which one has an infinite range with a translation symmetry.

When the toroidal structure of the level set is identified, the angle variables are those linear coordinates adjusted to the torus which have period ${2\pi}$. The canonically conjugate coordinates can be calculated as integrals
\begin{equation}\label{actionint}
   \frac1{2\pi} \int_{c} p_a dx^a
\end{equation}
over a closed loop ${c}$ circling the torus exactly once in the direction of the angle variable. Similarly to discussion in section~\ref{ssc:complintgrb}, the integral does not depend on a continuous deformation of the curve. One can thus deform these curves either in such a way that they belong only to one of {${x_\mu}$--${\px{\mu}}$} planes, which defines the action variable ${I_\mu}$ conjugate to angle ${\alpha_\mu}$, or, one can use such a curve that only ${\ph_\mu}$ changes, which defines the action variable ${A_\mu}$ conjugate to the angle variable ${\ph_\mu}$. Thus we have
\begin{equation}\label{IAcoor}
\begin{aligned}
  I_\mu &= \frac1\pi \int_{x_\mu^-}^{x_\mu^+} \frac{\sqrt{\Xfc_\mu}}{X_\mu}\, d x_\mu\,,\\
  A_\mu &= \frac1{2\pi} \int_{0}^{2\pi} \sum_k L_k \frac{\pa\psi_k}{\pa\ph^\mu}\, d\ph_\mu
         = \sum_k L_k \frac{\pa\psi_k}{\pa\ph^\mu}\,.
\end{aligned}
\end{equation}
Here we have used that the integral over a loop in the {${x_\mu}$--${\px{\mu}}$} plane is twice the integral between the turning points, and that ${{\pa\psi_k}/{\pa\ph^\mu}}$ are constants.

These relations should be understood as relations between conserved quantities ${(\tp{I},\tp{A})}$ and ${(\tpd{K},\tpd{L})}$. Indeed, the action variables just give a different labeling of the level sets ${\levset{(\tpd{K},\tpd{L})}}$. These expressions can be, in principle, inverted, and substituted into the generating function \eqref{genfcWspec}, defining thus a generating function from original to the action-angle coordinates,
\begin{equation}\label{IAgenfc}
    W(\tp{x},\tp{\psi};\tp{I},\tp{A}) =
      W\bigl(\tp{x},\tp{\psi};\tpd{K}(\tp{I},\tp{A}),\tpd{L}(\tp{I},\tp{A})\bigr)\;.
\end{equation}
The angle variables can now be obtained by taking derivatives of $W$ with respect to ${\tp{I_\mu}}$ and ${A_\mu}$,
\begin{equation}\label{anglevar}
\begin{aligned}
    \alpha^\mu &= \frac{\pa W}{\pa I_\mu}
               = \sum_j X^j\frac{\pa K_j}{\pa I_\mu}\,,\\
    \Phi^\mu &= \frac{\pa W}{\pa A_\mu}
               = \sum_j X^j\frac{\pa K_j}{\pa A_\mu} + \sum_j \Psi^j \frac{\pa L_j}{\pa A_\mu}\,,
\end{aligned}
\end{equation}
where we used \eqref{conjKL}. As expected, the angle variables ${(\tp{\alpha},\tp{\Phi})}$ are just a linear combination of ${(\tp{X},\tp{\Psi})}$.
The constant coefficients ${{\pa K_j}/{\pa I_\mu}}$ and ${{\pa L_j}/{\pa A_\mu}}$ can be calculated as inverse matrices to ${{\pa I_\mu}/{\pa K_j}}$ and ${{\pa A_\mu}/{\pa L_j}}$, and
\begin{equation}\label{pKpA}
    \frac{\pa K_j}{\pa A_\mu} = - \sum_{\nu,k}
      \frac{\pa K_j}{\pa I_\nu}\frac{\pa I_\nu}{\pa L_k}\frac{\pa L_k}{\pa A_\mu}\;.
\end{equation}
The form of the inverse coefficients follows from \eqref{IAcoor},
\begin{equation}\label{IoverK}
\begin{aligned}
    \frac{\pa I_\mu}{\pa K_j} &=
      \frac1\pi \int_{x_\mu^-}^{x_\mu^+} \frac{(-x_\mu^2)^{\dg-1-j}}{2\sqrt{\Xfc_\mu}}\, d x_\mu
      \,,\\
    \frac{\pa I_\mu}{\pa L_j} &=
      \frac1\pi \int_{x_\mu^-}^{x_\mu^+} \frac{\Lofc_\mu}{X_\mu} \frac{(-x_\mu^2)^{\dg-1-j}}{\sqrt{\Xfc_\mu}}\, d x_\mu
      \,,\\
    \frac{\pa A_\mu}{\pa L_j} &= \frac{\pa\psi_j}{\pa\ph^\mu}\,,
\end{aligned}
\end{equation}
where, again, it is safe to ignore derivatives of integral limits ${x_\mu^\pm}$.

To summarize, the action variables ${(\tp{I},\tp{A})}$ are defined by \eqref{IAcoor}. The conjugate angle variables ${(\tp{\alpha},\tp{\Phi})}$ are related to  ${(\tp{X},\tp{\Psi})}$ by linear relations \eqref{anglevar}, and  ${(\tp{X},\tp{\Psi})}$ are defined in \eqref{conjKL}.
It should be mentioned that although in the definition of the action variable ${I_\mu}$ we used the loop circling the torus just in {${x_\mu}$--${\px{\mu}}$} plane, the conjugate angle variable ${\alpha^\mu}$ is not a function of just one coordinate ${x_\mu}$, it depends on all coordinates ${(x_1,\dots,x_\dg)}$.
Similarly, in the inverse relations, ${x_\mu}$ depends on all angles ${\alpha^\nu}$. However, the coordinates ${x_\mu}$ are multiply-periodic functions of angle variables. When any angle ${\alpha_\nu}$ changes by period ${2\pi}$, all ${x_\mu}$ return to their original values, cf. general discussion in Chapter~10 of \cite{Goldstein:book}.

\subsection{Separation of variables in the Hamilton--Jacobi equation}

As discussed in section~\ref{ssc:HamJacEq}, the particle motion can also be described in terms of the Hamilton--Jacobi equation. For an autonomous completely integrable system one can write down not only the Hamilton--Jacobi equation \eqref{HJeqH}
\begin{equation}\label{HJeqHschem}
    H(x,\grad S) = \text{const.}\;,
\end{equation}
but also the Hamilton--Jacobi equations \eqref{HJeqP} for all conserved quantities.

The relativistic particle is an autonomous system (physical observables do not depend explicitly on time parameter $\sigma$) and, as we have just seen, in the Kerr--NUT--(A)dS spacetime it is complete integrable. The Hamilton--Jacobi equations corresponding to the conserved quantities \eqref{integemot} read
\begin{gather}
    \grad S \cdot \KT{j} \cdot \grad S \equiv k_{(j)}^{ab} S_{,a}S_{,b}= K_j\,.\label{HJeqQj}\\
    \KV{j}\cdot\grad S \equiv l_{(j)}^{a} S_{,a} = L_j\,,\label{HJeqLj}
\end{gather}
The spacetime gradient $\grad S$ of the Hamilton--Jacobi function $S(x;\tpd{K},\tpd{L})$ contains information about partial derivatives with respect to spacetime coordinates $x_\mu,\,\psi_j$ of a spacetime point $x$
\begin{equation}\label{gradS}
    \grad S = \sum_\mu \frac{\pa S}{\pa x_\mu}(x;\tpd{K},\tpd{L})\, \grad x_\mu
      + \sum_j \frac{\pa S}{\pa \psi_j}(x;\tpd{K},\tpd{L})\, \grad \psi_j \,.
\end{equation}
Here $\tpd{K}=(K_0,\dots,K_{\dg-1})$ and $\tpd{L}=(L_0,\dots,L_{\dg-1+\eps})$ are constants labeling values of conserved quantities $K_{j}$ and $L_{j}$ for the induced particle motion. Explicit forms of the Hamilton--Jacobi equations are obtained by using coordinate expressions \eqref{KTjcoor} and \eqref{KVjcoor}, giving
\begin{align}
  \label{HJeqQjcoor}
   K_j &= \sum_\mu\A{j}_\mu\Biggl(
     \frac{X_\mu}{U_\mu} \Bigl(\frac{\pa S}{\pa x_\mu}\Bigr)^2
     \!{+}\frac{U_\mu}{X_\mu} \Bigl(\sum_{k=0}^{\dg{-}1{+}\eps}
     \!\frac{(-x_\mu^2)^{\dg{-}1{-}k}}{U_\mu} \frac{\pa S}{\pa \psi_j}\Bigr)^2\Biggr)
     {+}\eps\frac{\A{j}}{c\A{\dg}} \Bigl(\frac{\pa S}{\pa \psi_\dg}\Bigr)^2,\\
  \label{HJeqLjcoor}
  L_j &= \frac{\pa S}{\pa\psi_j}\,.
\end{align}

Since $H=\frac12K_{0}$, cf.~\eqref{Hamiltonian}, we do not have to consider the Hamilton--Jacobi equation \eqref{HJeqHschem} separately, it is part of the system \eqref{HJeqQj}--\eqref{HJeqLj}. It also implies that constant $K_0$ is given by the mass of the particle, $K_0=-m^2$.

The symmetry structure of the Kerr--NUT--(A)dS spacetime has a remarkable consequence. All the Hamilton--Jacobi equations  \eqref{HJeqQj} and \eqref{HJeqLj} can be solved using an additive separable ansatz
\begin{equation}\label{Ssepansatz}
    S = \sum_\mu S_\mu + \sum_{j=0}^{\dg-1+\eps} L_j\, \psi_j\;,
\end{equation}
where each $S_\mu\equiv S_\mu(x_\mu)$ is a function of just one variable $x_\mu$ (of course, $S_\mu$ depends also on constants ${K_j}$ and ${L_j}$).

Indeed, the linear dependence on the Killing coordinates $\psi_j$ directly solves equations \eqref{HJeqLjcoor}. Separability in the ${x_\mu}$ coordinate guarantees that $\frac{\pa S}{\pa x_\mu}=S_\mu'$. Upon multiplying the equations \eqref{HJeqQjcoor} by $(-x_\mu^2)^{\dg-1-j}$, and summing together, using relations \eqref{Aid1i}, gives the following equation for $S_\mu$:
\begin{equation}\label{HJsepSsol}
    (S_\mu')^2 = \frac{\Qofc_\mu}{X_\mu} - \frac{\Lofc_\mu^2}{X_\mu^2}
    = \frac{\Xfc_\mu}{X_\mu^2}\,,
\end{equation}
where the functions ${\Qofc_\mu}$, ${\Lofc_\mu}$, and ${\Xfc_\mu}$ are defined by \eqref{Qofcdef}--\eqref{Xfcdef}. Each of these functions, as well as the metric function~$X_\mu$, depend just on one variable $x_\mu$. The equation \eqref{HJsepSsol} is thus an ordinary differential equation in a single variable, which justifies the consistency of the ansatz \eqref{Ssepansatz}. Finding the Hamilton--Jacobi function ${S}$ is thus equivalent to integrating the ordinary differential equations \eqref{HJsepSsol}, giving
\begin{equation}\label{Smuint}
    S_\mu = \int_{x_\mu^-}^{x_\mu} \frac{\sqrt{\Xfc_\mu}}{X_\mu}\,d x_\mu\,,
\end{equation}
where, similar to \eqref{xintjdef}, we start the integration at the (smaller) turning point ${x_\mu^-}$, where ${\Xfc_\mu=0}$.

\details{In section~\ref{ssc:separabilitystructures} we have mentioned that the separability of the Hamilton--Jacobi equation can be characterized by the corresponding separability structure \citep{benenti1979remarks, BenentiFrancaviglia:1980, DemianskiFrancaviglia:1980, KalninsMiller:1981}.
The off-shell Kerr--NUT--(A)dS geometry possesses $(\dg+\eps)$-separability structure.
Indeed, we can identify the ingredients of the first theorem of section~\ref{ssc:separabilitystructures} as follows: in $D=2\dg+\eps$ dimensions, the Kerr--NUT--(A)dS geometry has $\dg+\eps$ Killing vectors $\KV{j}$, $j=0,\dots,\dg-1+\eps$, and $\dg$ Killing tensors $\KT{j}$, $j=0,\dots,\dg-1$. (i) All these objects commute in the sense of Nijenhuis--Schouten bracket, \eqref{KTKVNS}, and (ii) the Killing tensors have common eigenvectors $\cv{x_\mu}$ which obviously Lie-bracket commute with the Killing vectors $\KV{j}=\cv{\psi_j}$ and which are orthogonal to the Killing vectors. All the requirements of the theorem are thus satisfied and the result follows.}

\subsection{Separation of variables in the wave equation}
\label{ssc:SepWave}

The symmetry of the Kerr--NUT--(A)dS metric not only allows one to solve the particle motion,
but also provides separability of various test field equations. In this section we demonstrate that the massive scalar field equation
\be
(\Box-m^2)\phi=0\,
\ee
allows a complete separation of variables in the Kerr--NUT--(A)dS spacetime \citep{Frolov:2006pe}. Here, as earlier, we defined the scalar wave operator as
\begin{equation}\label{boxdef}
    \Box = g^{ab}\nabla_{\!a}\nabla_{\!b}\;.
\end{equation}

\details{
The box-operator in \eqref{boxdef} is, in a sense, a first quantized version of the Hamiltonian \eqref{Hamiltonian}. Similarly, one could define a second-order operator ${\mathcal{K}=-\nabla_{\!a} k^{ab}\nabla_{\!b}}$ for any symmetric second-rank tensor ${\ts{k}}$. This can be understood as a `heuristic first quantization' of a classical observable ${K=k^{ab} p_a p_b }$, using the rule ${\ts{p}\to-i\covd}$. Of course, when applying this rule one has to chose a particular operator ordering. In our example we have chosen the symmetric ordering. In principle, one  could also use a different (from the Levi-Civita) covariant derivative. However, all these alternative choices would lead to operators that differ in lower order of derivatives, which could be studied separately.
}

In the Kerr--NUT--(A)dS spacetime the tower of Killing tensors \eqref{KTj} and Killing vectors \eqref{KVj} defines a tower of the following associated second-order and first-order operators:
\begin{gather}
\label{Qopdef}
    \Qop{j} = - \nabla_{\!a} \KTc{j}^{ab} \nabla_{\!b}\,,\\
\label{Lopdef}
    \Lop{j} = - i\, \KVc{j}^{a} \nabla_{\!a}\,,
\end{gather}
with the wave operator \eqref{boxdef} equivalent to ${\Qop{0}}$. It is then natural to ask about the commutation properties of these operators, as an operator analogy to \eqref{Killobsinv}. In general, it can be shown \citep{Carter:1977pq, Kolar:2015cha} that two second-order operators constructed from the corresponding tensors ${\ts{k}_1}$ and ${\ts{k}_2}$ commute in the highest-order in derivatives provided the Nijenhuis--Schouten bracket of the two tensors vanishes, ${[\ts{k}_1,\ts{k}_2]_\NS=0}$. However, to guarantee the commutativity to all orders, some additional `{\em anomalous conditions}' must be satisfied, see \cite{Carter:1977pq, Kolar:2015cha}.

Since in the Kerr--NUT--(A)dS spacetimes all the operators \eqref{Qopdef} and \eqref{Lopdef} are generated by a single object, the principal tensor, it is not so surprising that the anomalous conditions hold and all these operators mutually commute \citep{Sergyeyev:2007gf,Kolar:2015cha}:
\begin{equation}\label{opcomut}
    \bigl[\Qop{k},\Qop{l}\bigr]=0\;,\quad
    \bigl[\Qop{k},\Lop{l}\bigr]=0\;,\quad
    \bigl[\Lop{k},\Lop{l}\bigr]=0\;.
\end{equation}
Commutativity can be also proved directly, by using the coordinate expressions for these operators \citep{Sergyeyev:2007gf}:
\begin{align}
 \Lop{j} &= -i\,\frac{\pa}{\pa\psi_j}\;,\label{Lopcoor}\\
 \Qop{j} &= \sum_\mu\frac{\A{j}_\mu}{U_\mu}\Qtop{\mu}\,, \label{Qopsplit}
\end{align}
where each ${\Qtop{\mu}}$ involves only one coordinate ${x_\mu}$ and Killing coordinates~${\psi_j}$:
\begin{equation}\label{Qopx}
\begin{split}
  &\Qtop{\mu}=
      -\frac{\pa}{\pa x_\mu}\biggl[X_\mu\frac{\pa}{\pa x_\mu}\biggr]
      -\eps \frac{ X_\mu}{x_\mu}\frac{\pa}{\pa x_\mu}\\
  &\mspace{100mu}-\frac{1}{X_\mu}\biggl[\sum_{k=0}^{\dg-1+\eps}(-x_\mu^2)^{\dg-1-k}\frac{\pa}{\pa\psi_k}\biggr]^2
      -\eps\frac{1}{c x_\mu^2}\,\biggl[\frac{\pa}{\pa\psi_\dg}\biggr]^2\;.
\end{split}
\end{equation}

The commutativity \eqref{opcomut} implies that the operators ${\Qop{j}}$ and ${\Lop{j}}$ have common eigenfunctions~${\phi}$,
\begin{equation}\label{eigenfc}
    \Qop{j}\phi = K_j\phi\;,\quad
    \Lop{j}\phi = L_j\phi\;,
\end{equation}
which can be labeled by the eigenvalues ${K_j}$ and ${L_j}$.
These eigenfunctions can be found by a separation of variables \citep{Frolov:2006pe, Sergyeyev:2007gf}. Namely, starting with the multiplicative separation ansatz
\begin{equation}\label{WOsepansatz}
    \phi = \prod_\mu R_\mu \prod_{k=0}^{\dg-1+\eps}\exp\bigl(i L_k\psi_k\bigr)\;,
\end{equation}
where each function ${R_\mu}$ depends only on one coordinate ${x_\mu}$, ${R_\mu=R_\mu(x_\mu)}$, one can show that equations \eqref{eigenfc} are equivalent to conditions
\begin{equation}\label{WOsepcond}
    (X_\mu R_\mu')' + \eps\frac{X_\mu}{x_\mu}R_\mu'
    + \frac{\Xfc_\mu}{X_\mu^2} R_\mu
    = 0\;.
\end{equation}
These are ordinary differential equations for functions ${R_\mu}$, which can be solved, at least in principle. Here,  functions ${\Xfc_\mu}$ are the same as before, defined by \eqref{Qofcdef}--\eqref{Xfcdef}.

\details{The separability of the wave equation is again in an agreement with the theory of separability structures mentioned in section~\ref{ssc:separabilitystructures}. In the previous section we have already shown that the Kerr--NUT--(A)dS geometry possesses $(\dg+\eps)$-separability structure. To fulfill the second theorem of section~\ref{ssc:separabilitystructures}, which guarantees the separability of the wave equation, one has to show that the eigenvectors $\cv{x_\mu}$ are eigenvectors of the Ricci tensor. However, the Ricci tensor is diagonal in the special Darboux frame, \eqref{Ricci}, and vectors $\cv{x_\mu}$ are just rescaled vectors $\env{\mu}$, cf.~\eqref{Darbouxvecfr}. This justifies the separability of the Klein--Gordon equation in off-shell Kerr--NUT--(A)dS spacetimes.}

Let us finally note that the Hamilton--Jacobi equations discussed in the previous section can be actually understood as a semiclassical approximation to the wave-like equations \eqref{eigenfc}. In such an approximation one looks for a solution in the form
\begin{equation}\label{PhiSrel}
    \phi = A \exp\Bigl(\,\frac{i}{\hbar}\,S\,\Bigr)\,,
\end{equation}
which when plugged into equations \eqref{eigenfc} with each derivative weighted by ${\hbar}$, and looking for the highest order in the limit ${\hbar\to 0}$, gives the Hamilton--Jacobi equations \eqref{HJeqQj}--\eqref{HJeqLj} for ${S}$, cf.~\cite{Frolov:2006pe, Sergyeyev:2007gf}.

\details{%
When discussing the complete integrability of geodesic motion in the previous section, we mentioned that the geodesic motion in spaces with the metric given by any of the Killing tensors ${\KT{i}}$ is also complete integrable. This property, however, does not elevate to the corresponding wave equations. Namely, the operators \eqref{Qopdef} given by the covariant derivative associated with the metric ${\KT{i}}$, ${i\neq0}$, no longer mutually commute; the anomalous conditions needed for the operator commutativity are satisfied only for the Levi-Civita derivative associated with the Kerr--NUT--(A)dS metric. In particular, the Ricci tensors associated with the metric given by higher ($i>0$) Killing tensors are not diagonal in the common frame of eigenvectors of all Killing tensors and the separability structure does not obey the extra condition needed for the separation of the Klein--Gordon field equation, see \cite{Kolar:masterthesis,Kolar:2015cha} for more details.
}

\subsection{Dirac equation}
\label{ssc:DiracEquation}

The solution of the massive Dirac equation in the Kerr--NUT--(A)dS spacetimes can be found in a special (pre-factor) separated form and the problem is transformed to a set of ordinary differential equations. Similar to the massive scalar equation, this solution is obtained as a common eigenfunction of a set of mutually commuting operators, one of which is the Dirac operator.

\subsubsection{Overview of results}

The study of Dirac fields in a curved spacetime has a long history. In 1973 Teukolsky rephrased the Dirac massless equation in Kerr spacetime in terms of a scalar `fundamental equation' which could be solved by a separation of variables. However, such an approach does not work for the massive Dirac equation and it is difficult to generalize it to higher dimensions. There is yet another method which goes along the lines we used for the scalar wave equation: one can postulate the multiplicative ansatz for the solution of the Dirac equation and obtain independent (but coupled) differential equations for each component in this ansatz.

This approach dates back to the seminal paper of Chandrasekhar who in 1976 separated and decoupled the Dirac equation in the Kerr background \citep{Chandrasekhar:1976}, see section~\ref{ssc:Separability4D}. A few years later, \cite{CarterMcLenaghan:1979}  demonstrated that behind such a separability stands a first-order operator commuting with the Dirac operator which is constructed from the Killing--Yano 2-form of \cite{Penrose:1973}. This discovery stimulated subsequent developments in the study of symmetry operators of the Dirac equation in curved spacetime.

In particular, the most general {\em first-order operator} commuting with the Dirac operator in four dimension was constructed by \cite{McLenaghanSpindel:1979}. This work was later extended by \cite{KamranMcLenaghan:1984} to {\em ${R}$-commuting} symmetry operators. Such operators map solutions of the massless Dirac equation to other solutions and correspond to  symmetries which are {conformal generalizations} of Killing vectors and Killing--Yano tensors.

With recent developments in higher-dimensional gravity, the symmetry operators of the Dirac operator started to be studied in spacetimes of an arbitrary dimension and signature. The first-order symmetry operators of the Dirac operator in a general curved spacetime has been identified by \cite{Benn:1996ia} and \cite{BennKress:2004}. The restriction to the operators commuting with the Dirac operator has been studied in \cite{Cariglia:2011yt}.

The higher-dimensional Dirac equation has been also studied in specific spacetimes. In the remarkable paper \cite{OotaYasui:2008} separated the Dirac equation in  the general off-shell Kerr--NUT--(A)dS spacetime, generalizing the results \citep{Chandrasekhar:1976,CarterMcLenaghan:1979} in four dimensions. The result of  \cite{OotaYasui:2008} has been reformulated in the language of a tensorial separability and related to the existence of the commuting set of operators in \cite{Wu:2008, Wu:2008b, Cariglia:2011qb}, and generalized to the presence of a weak electromagnetic field in \cite{CarigliaEtal:2012b}. Even more generally, separability of the torsion modified Dirac equation was demonstrated in the presence of $U(1)$ and torsion fluxes of the Kerr--Sen geometry and its higher-dimensional generalizations \citep{Houri:2010fr} as well as in the most general spherical black hole spacetime of minimal gauged supergravity \citep{Wu:2009cn, Wu:2009ug}, see also \cite{Kubiznak:2009qi, Houri:2010qc}. The Dirac symmetry operators in the presence of arbitrary fluxes were studied in \cite{Acik:2008wv, Kubiznak:2010ig}.

Before we review the results for Kerr--NUT--(A)dS spacetimes, let us make one more remark. Although the first-order symmetry operators are sufficient to justify separability of the massless Dirac equation in the whole Plebanski--Demianski class of metrics in four dimensions or separability of the massive Dirac equation in Kerr--NUT--(A)dS spacetimes in all dimensions, they are not enough to completely characterize all Dirac separable systems and one has to consider higher-order symmetry operators, e.g., \cite{McLenaghanEtal:2000}. In particular, there are known examples \citep{FelsKamran:1990} where the Dirac equation separates but the separability is related to an operator of the second-order. It means that the theory of separability of the Dirac equation must reach outside the realms of the so called {\em factorizable systems} \citep{Miller:1988}, as such systems are fully characterized by first-order symmetry operators.

In the following we review the separability results for the Dirac equation in the Kerr--NUT--(A)dS spacetime \citep{OotaYasui:2008,Cariglia:2011yt, Cariglia:2011qb}. A short overview of Dirac spinors in a curved spacetime of an arbitrary dimension can be found in section~\ref{ssc:DiracSpinors}. In the same appendix, in section~\ref{ssc:DiracSymOp}, one can also find a characterization of the first order operators that commute with the Dirac operator. Using these general results, we show below that the general off-shell Kerr--NUT--(A)dS spacetime admits a set of mutually commuting first-order operators including the Dirac operator, whose common eigenfunctions can be found in a tensorial ${R}$-separable form.

\subsubsection{Representation of Dirac spinors in Kerr--NUT--(A)dS spacetimes}

We want to study the Dirac operator and its symmetries in the Kerr--NUT--(A)dS spacetime. For that we have to specify the representation of the gamma matrices. As discussed in section~\ref{ssc:DiracSpinors}, the choice of representation is equivalent to the choice of a frame ${\ts{\tht}_E}$ in the Dirac bundle associated with the orthonormal Darboux frame \eqref{Darbouxvecfr} in the tangent space ${\mathbf{T}\,M}$ in such a way that components ${\gamma^a{}^A{}_B}$ of the gamma matrices are constant and satisfy
\begin{equation}\label{gammamtrxMT}
    \gamma^a\,\gamma^b + \gamma^b\,\gamma^a = 2\, g^{ab} I\;.
\end{equation}

Following \cite{OotaYasui:2008}, ${2^\dg}$-dimensional Dirac bundle can be chosen as the tensor product of ${\dg}$ two-dimensional bundles ${\mathbf{S}\,M}$, i.e., ${\mathbf{D}\,M = \mathbf{S}^\dg\,M}$. We use Greek letters ${\epsilon,\varsigma,\dots}$ for tensor indices in these 2-dimensional spaces and values ${\epsilon=\pm1}$ (or just ${\pm}$) to distinguish their components. In the Dirac bundle, we choose a frame ${\ts{\tht}_E}$  in a tensor product form:
\begin{equation}\label{Thetafr}
    \ts{\tht}_E = \ts{\tht}_{\epsilon_1\dots\epsilon_\dg}
      = \ts{\tht}_{\epsilon_1}\otimes\dots\otimes\ts{\tht}_{\epsilon_\dg}\;,
\end{equation}
where $\ts{\tht}_+$ and $\ts{\tht}_-$ form a frame in the 2-dimensional spinor space~${\mathbf{S}}$. With such a choice we have a natural identification of Dirac indices ${E}$ with the multi-index ${\{\epsilon_1,\dots,\epsilon_\dg\}}$.

A generic 2-dimensional spinor can thus be written as $\ts{\chi} = \chi^\epsilon \ts{\tht}_\epsilon = \chi^+ \ts{\tht}_+ + \chi^- \ts{\tht}_-$, with components being two complex numbers
${\left( \begin{smallmatrix} \chi^+ \\ \chi^- \end{smallmatrix} \right)}$.
Similarly, the Dirac spinors ${\ts{\psi}\in\mathbf{D} M}$ can be written as $\ts{\psi} = \psi^{\epsilon_1\dots\epsilon_\dg} \ts{\tht}_{\epsilon_1\dots\epsilon_\dg}$ with ${2^\dg}$ components ${\psi^{\epsilon_1\dots\epsilon_\dg}}$.

Before we specify the components of the gamma matrices in this frame, let us introduce some auxiliary notations. Let ${\ts{I}}$, ${\ts{\iota}}$, ${\ts{\sigma}}$, and ${\hat{\ts{\sigma}}}$ be the unit and respectively Pauli operators on ${\mathbf{S}\,M}$ with components
\begin{equation}\label{sigmamatrcomp}
    I^\epsilon{}_\varsigma =
    \left(\begin{array}{cc}
        1 & 0 \\
        0 & 1 \\
      \end{array}\right)
    \,,\quad
    \iota^\epsilon{}_\varsigma =
    \left(\begin{array}{cc}
        1 & 0 \\
        0 & -1 \\
      \end{array}\right)
    \,,\quad
    \sigma^\epsilon{}_\varsigma =
    \left(\begin{array}{cc}
        0 & 1 \\
        1 & 0 \\
      \end{array}\right)
    \,,\quad
    \hat\sigma{}^\epsilon{}_\varsigma =
    \left(\begin{array}{cc}
        0 & -i \\
        i & 0 \\
      \end{array}\right)
    \,.
\end{equation}
Next, for any linear operator ${\ts{\alpha}\in\mathbf{S}^1_1M}$ we denote by ${\ts{\alpha}_{\langle\mu\rangle}\in\mathbf{D}^1_1M}$ a linear operator on the Dirac bundle
\begin{equation}\label{alphaDB}
    \ts{\alpha}_{\langle\mu\rangle} = \ts{I}\otimes\dots\otimes \ts{I}\otimes
       \ts{\alpha}\otimes \ts{I}\otimes\dots\otimes \ts{I}\,,
\end{equation}
with ${\ts{\alpha}}$ on the ${\mu}$-th place in the tensor product. Similarly, for mutually different indices ${\mu_1,\dots,\mu_j}$ we define
\begin{equation}\label{multialphaDB}
    \ts{\alpha}_{\langle\mu_1\dots\mu_j\rangle} = \ts{\alpha}_{\langle\mu_1\rangle}\cdots\ts{\alpha}_{\langle\mu_j\rangle}\;,
\end{equation}
that means that ${\ts{\alpha}}$'s are on the positions ${\mu_1,\dots,\mu_j}$ in the product.

Equipped with this notation, we are now ready to write down the abstract gamma matrices with respect to the frame $(\enf{\mu},\,\ehf{\mu}\,,\ezf)$ given by \eqref{Darbouxvecfr} in the tangent space and ${\ts{\tht}_E}$ given by \eqref{Thetafr} in the Dirac bundle:
\begin{equation}\label{gammamatrmuhatmu}
\begin{gathered}
    \gamma^\mu = \iota_{\langle1\dots\mu{-}1\rangle}\,\sigma_{\langle\mu\rangle}\,,\quad
    \gamma^{\hat\mu} = \iota_{\langle1\dots\mu{-}1\rangle}\,\hat\sigma_{\langle\mu\rangle}\,,\quad
    \gamma^{0} = \iota_{\langle 1 \dots \dg\rangle} \, .
\end{gathered}
\end{equation}
The odd gamma matrix $\gamma^{0}$ is defined only in an odd dimension. It is straightforward to check that the matrices \eqref{gammamatrmuhatmu} satisfy the property \eqref{gammamtrxMT}.

In components, the action of these matrices on a spinor $\ts{\psi} = \psi^{\epsilon_1\dots\epsilon_\dg} \ts{\tht}_{\epsilon_1\dots\epsilon_N}$ is given as
\begin{equation}\label{gammaaction}
\begin{aligned}
    (\gamma^\mu\,\psi)^{\epsilon_1\dots\epsilon_\dg} &=
     \Bigl(\prod_{\nu =1}^{\mu-1} \epsilon_\nu\Bigr)\,\psi^{\epsilon_1\dots(-\epsilon_\mu)\dots\epsilon_\dg}\;,\\
    (\gamma^{\hat\mu}\,\psi)^{\epsilon_1\dots\epsilon_\dg} &=
     -i\epsilon_\mu\,\Bigl(\prod_{\nu=1}^{\mu-1} \epsilon_\nu\Bigr)\psi^{\epsilon_1\dots(-\epsilon_\mu)\dots\epsilon_\dg}\;, \\
    (\gamma^{0}\,\psi)^{\epsilon_1\dots\epsilon_\dg} &=
     \Bigl(\prod_{\nu =1}^{\dg} \epsilon_\nu\Bigr)\,\psi^{\epsilon_1\dots \epsilon_\dg}\;.
\end{aligned}
\end{equation}
Finally, we also use a shorthand
\begin{equation}\label{gammaprod}
   \gamma^{a_1\dots a_p} = \gamma^{[a_1}\dots\gamma^{a_p]}\,.
\end{equation}

\subsubsection{Dirac symmetry operators in Kerr--NUT--(A)dS spacetimes}

The Kerr--NUT--(A)dS spacetime is equipped with the full tower of Killing--Yano symmetry objects. As discussed in section~\ref{ssc:DiracSymOp}, such objects allow one to define first-order operators that commute with the Dirac operator. In fact, {as we now demonstrate}, it is possible to choose such a subset of Killing--Yano symmetries that yields a full set of ${D}$ first-order operators, one of which is the Dirac operator ${\sop{D}}$, that all mutually commute.

Namely, we can use ${\dg+\eps}$ explicit symmetries described by Killing vectors. Using \eqref{FODiraccomopK}, for each Killing vector ${\KV{j}}$ we thus have the corresponding operator ${\sop{L}_j}$,
{\begin{equation}\label{DsumopKj}
    \sop{L}_j = \sop{K}_{\KV{j}}
      = \KVc{j}^a\nabla_{\!a} \,+\, \frac{1}{4}\bigl(\nabla^{[a}\KVc{j}^{b]}\bigr)\gamma_{ab}
      \,, \qquad j= 0,1,\dots,\dg+\eps\,.
\end{equation}}%
These operators can be complemented with ${\dg}$ operators \eqref{FODiraccomopM} constructed from the even closed conformal Killing--Yano forms~${\CCKY{k}}$,
{\begin{equation}\label{DsumopMk}
\begin{split}
    \sop{M}_k &= \sop{M}_{\CCKY{k}}=\\
      &= \frac1{(2k)!}\,\gamma^{aa_1\dots a_{2k}}\;\CCKYc{k}_{a_1\dots a_{2k}}\nabla_{\!a}
       +\frac1{2(2k{-}1)!}\frac{D-2k}{D{-}2k{+}1}\,\bigl(\nabla^{c}\CCKY{k}_{ca_2\dots a_{2k}}\bigr)\,\gamma^{a_2\dots a_{2k}}\,,
\end{split}\raisetag{8ex}
\end{equation}}%
$k= 0,1,\dots,\dg$. In particular, for ${k=0}$ we get, as a special case, the Dirac operator itself, ${\sop{D}=\sop{M}_0}$. {(See Appendix~\ref{apx:spinor} for a more compact notation for these operators.)}

It turns out that the strong symmetry structure of the off-shell Kerr--NUT--(A)dS spacetime is sufficient to guarantee that these operators mutually commute \citep{Cariglia:2011qb}:
\begin{equation}\label{KMcommut}
    [\sop{L}_i,\,\sop{L}_j]=0\,,\qquad
    [\sop{M}_k,\,\sop{M}_l]=0\,,\qquad
    [\sop{L}_j,\sop{M}_k]=0\,.
\end{equation}
They thus have common spinorial eigenfunctions and one can hope that these can be found in a separable form.

To demonstrate that, we first write down the operators in an explicit coordinate form. The operators ${\sop{L}_j}$ are related to the explicit symmetry along the Killing vectors ${\KV{j}=\cv{\psi_j}}$. They thus have the following simple coordinate form:
\begin{equation}\label{opKexpl}
    \sop{L}_{j} = \frac{\pa}{\pa\psi_j}\;.
\end{equation}
The coordinate form of ${\sop{M}_k}$ is much more complicated, and in particular one needs to know the explicit form of the spin connection. This is listed in Appendix~\ref{ssc:spincon}. To illustrate the structure of ${\sop{M}_k}$, we just write it down in an even dimension, see \cite{Cariglia:2011qb} for the results in odd dimensions and their derivation. The even-dimensional  ${\sop{M}_k}$ reads
\begin{equation}\label{Mopexpl}
\begin{split}
&\sop{M}_j = i^j \sum_\mu \sqrt{\frac{X_\mu}{U_\mu}}\, B^{(j)}_\mu\Biggl(
  \frac{\pa}{\pa x_\mu}+\frac{X_\mu'}{4X_\mu}
  + \frac12 \sum_{\substack{\nu\\\nu\neq\mu}} \frac1{x_\mu{-}\iota_{\langle\mu\nu\rangle}x_\nu}\\
  &\mspace{270mu}
  - i\frac{\iota_{\langle\mu\rangle}}{X_\mu}\sum_k (-x_\mu^2)^{N{-}1{-}k}\frac{\pa}{\pa\psi_k}
  \Biggr)\,\gamma^\mu  \,,
\end{split}
\end{equation}
where the matrices ${B^{(k)}_\mu}$ are `spinorial analogues' of functions ${\A{j}_\mu}$, cf.~\eqref{AUdefs},
\begin{equation}\label{B_mu_def}
B^{(k)}_\mu =\hspace{-3mm}
    \sum_{\substack{\nu_1,\dots,\nu_k\\\nu_1<\dots<\nu_k,\;\nu_i\ne\mu}}\!\!\!\!\!
    \iota_{\langle\nu_1\rangle}x_{\nu_1}\cdots\iota_{\langle\nu_k\rangle}x_{\nu_k}\,.
\end{equation}

\subsubsection{Tensorial ${R}$-separability of common eigenfunctions}

Now we can formulate the desired result: the commuting symmetry operators ${\sop{L}_j}$ and ${\sop{M}_k}$ have common spinorial eigenfunctions ${\ts{\psi}}$
\begin{align}
    \sop{L}_j \ts{\psi} &= i\,L_j\ts{\psi}\;,\label{DireigenfcL}\\
    \sop{M}_k \ts{\psi} &= M_k\ts{\psi}\;,\label{DireigenfcM}
\end{align}
which can be found in the tensorial R-separated form
\begin{equation}\label{tensRsep}
    \ts{\psi} = \ts{R}\, \exp\bigl({\textstyle i\sum_j L_j \psi_j}\bigr)\,
           \bigotimes_\nu \ts{\chi}_\nu\;.
\end{equation}
Here, $\left\{\ts{\chi}_\nu \right\}$ is an $\dg$-tuple of 2-dimensional spinors and ${\ts{R}}$ is the Clifford-valued prefactor
\begin{equation}\label{Phidef}
  \ts{R} = \prod_{\substack{\kappa,\lambda\\\kappa<\lambda}}
    \Bigl(x_\kappa+\ts{\iota}_{\langle\kappa\lambda\rangle}x_\lambda\Bigr)^{-\frac12}\;.
\end{equation}
As a part of the separation ansatz we ask that $\ts{\chi}_\nu$ depends only on the variable ${x_\nu}$, $\ts{\chi}_\nu=\ts{\chi}_\nu(x_\nu)$.

In terms of components, this reduces to the ansatz made in \cite{OotaYasui:2008}:
\begin{equation}\label{compsep}
    \psi^{\epsilon_1\dots\epsilon_\dg} =
      \phi_{\epsilon_1\dots\epsilon_\dg}
      \exp\bigl({\textstyle i\sum_j L_j\psi_{j}}\bigr)
      \prod_\nu  \chi_\nu^{\epsilon_\nu} \;.
\end{equation}
Here, ${\phi_{\epsilon_1\dots\epsilon_\dg}}$ is a diagonal element of the prefactor ${\ts{R}}$,
\begin{equation}\label{phicompdef}
    \phi_{\epsilon_1\dots\epsilon_\dg} =
    \prod_{\substack{\kappa,\lambda\\\kappa<\lambda}}
    \Bigl(x_\kappa+\epsilon_\kappa\epsilon_\lambda x_\lambda\Bigr)^{-\frac12}\;.
\end{equation}

Plugging the multiplicative ansatz \eqref{tensRsep} into equations \eqref{DireigenfcL} and \eqref{DireigenfcM}, one finds that they are satisfied if each of the two-dimensional spinors ${\ts{\chi}_\mu}$ satisfies the ordinary differential equation in ${x_\mu}$ which, in an even dimension, reads
\begin{equation}\label{chieq}
\Biggl[\Bigl(
    \frac{d}{dx_\nu}+\frac{X_\nu'}{4X_\nu}
    +\frac{\Lofc_\nu}{X_\nu}\ts{\iota}_{\langle\nu\rangle}
    \Bigr)\,\ts{\sigma}_{\langle\nu\rangle}
    - \frac1{\sqrt{|X_\nu|}} {\bigl(- \ts{\iota}_{\langle\nu\rangle} \bigr)^{\dg{-}\nu}}
   \Mofc_{\nu}
   \Biggr]\ts{\chi}_\nu=0\,.
\end{equation}
Here, the function ${\Lofc_\nu}$ of a single variable ${x_\mu}$ is again given by \eqref{Lofcdef},
\begin{equation}\label{Lofcdef2}
    \Lofc_\mu = \sum_j L_{j} (-x_\mu^2)^{\dg{-}1{-}j}\,,
\end{equation}
and, similarly, we introduced the spinorial function ${\Mofc_\mu}$
\begin{equation}  \label{Mofcdef}
\Mofc_{\nu} =  \sum_{k} (-i)^k M_{k}\, (-\ts{\iota}_{\langle\nu\rangle} x_\nu )^{\dg{-}1{-}k} \, .
\end{equation}
Taking the component ${\varsigma=\pm}$ of the spinorial equation \eqref{chieq} we get
\begin{equation}\label{chieq_proj}
    \Bigr(\frac{d}{d x_\nu}+\frac{X_\nu'}{4X_\nu}
    -\varsigma\frac{\Lofc_\nu}{X_\nu}\Bigr)\,\chi_\nu^{-\varsigma}
    - \frac{\bigl(-\varsigma\bigr)^{\dg{-}\nu}}{\sqrt{|X_\nu|}}
    \tilde{M}_{\nu}^\varsigma \chi_\nu^{\varsigma}=0\,,
\end{equation}
with
\begin{equation}  \label{Mofcdefcomp}
\tilde{M}_{\nu}^\varsigma =  \sum_{k} (-i)^k M_{k}\, (-\varsigma x_\nu )^{\dg{-}1{-}k} \, .
\end{equation}
For each $\nu$, these are two coupled ordinary differential equations for components ${\chi_\nu^+}$ and ${\chi_\nu^-}$, which can be easily decoupled by substituting one into another. In other words, the problem of solving the massive Dirac equation in general Kerr--NUT--(a)dS spacetimes can be recast
as a problem of solving a number of decoupled ordinary differential equations for components of the corresponding multi-dimensional spinor.

\subsection{Tensor perturbations}%
\label{ssc:TensorPert}

The demonstrated separability of the Hamilton--Jacobi, Klein--Gordon, and Dirac equations in the general higher-dimensional Kerr--NUT--(A)dS spacetime created hopes that higher spin equations might also possess this property. In particular, there were hopes that the electromagnetic and gravitational perturbations can be solved by either a direct separation of the corresponding field equations, or by their reduction to a master equation, which, in its turn, is separable. In spite of many attempts, only partial results were obtained. In this section we briefly discuss the tensor perturbations and return to the electromagnetic fields in the next section.

The study of gravitational perturbations of black holes is key for understanding their stability, and is especially important in higher dimensions where many black holes are expected to be unstable and may (as indicated in recent numerical studies) branch to other black hole families, e.g., \cite{Choptuik:2003qd, Lehner:2010pn, Dias:2009iu, Dias:2010eu, Dias:2010maa, Dias:2014cia, Figueras:2015hkb}, or even result in a formation of naked singularities \citep{Figueras:2017zwa}.
The separability and decoupling of gravitational perturbations would also significantly simplify the study of quasi-normal modes of these black holes or  the study of Hawking radiation.

The gravitational perturbations have been analytically studied for higher-dimensional black holes with no rotation, e.g., \cite{Gibbons:2002pq, Kodama:2003jz, Ishibashi:2003ap}, or for black holes subject to restrictions on their rotation parameters, e.g., \cite{Kunduri:2006qa, Kodama:2007ph, Murata:2007gv, Murata:2008yx, Kodama:2009rq,Kodama:2009bf, Oota:2008uj, Murata:2011zz,Murata:2012ct}. Such black holes possess enhanced symmetries and are of a smaller co-homogeneity than the general Kerr--NUT--(A)dS spacetime. This allows one to decompose the corresponding perturbations into `tensor, vector and scalar' parts that can be treated separately, yielding the corresponding master equations, e.g., \cite{Kunduri:2006qa}.

By the time this review is written it is unknown whether there exists a method which would allow one to separate and decouple gravitational perturbations of the general Kerr---NUT--(A)dS spacetimes.
For example, as shown in papers by \cite{Durkee:2010qu, Durkee:2010ea} this goal cannot be achieved by following the `Teukolsky path', employing the higher-dimensional generalization of Newman--Penrose or Geroch's formalisms \citep{Pravdova:2008gp, Durkee:2010xq} building on \cite{Coley:2004jv, Pravda:2004ka, Ortaggio:2007eg}.

To conclude this section, let us briefly comment on a partial success by \cite{Oota:2008uj}, who demonstrated the separability of certain type of tensor perturbations in {\em generalized Kerr--NUT--(A)dS spacetimes}. In our discussion of the Kerr--NUT--(A)dS spacetimes we assumed that the principal tensor is non-degenerate, that is, it has $n$ functionally independent eigenvalues, that were used as canonical coordinates. One obtains a more general class of metrics once this assumption is violated. The corresponding metrics, called the \emph{generalized Kerr--NUT--(A)dS solutions}, were obtained in \cite{Houri:2008ng,Houri:2008th}, and we will discuss them in more detail in chapter~\ref{sc:aplgen}. Here we just describe some of their properties that are required for the formulation of the results of \cite{Oota:2008uj}. The generalized metric has $N$ essential coordinates which are non-constant eigenvalues of the principal tensor, $|m|$ parameters which are non-zero constant eigenvalues, and the degeneracy of a subspace responsible for the vanishing eigenvalue is $m_0$. The total number of spacetime dimensions is thus $D=2N+2|m|+m_0$. This space has a bundle structure. Its fiber is a $2n$-dimensional Kerr--NUT--(A)dS metric. All other dimensions form the base space. The tensor perturbations, analyzed in \cite{Oota:2008uj}, are those, that do not perturb the fiber metric and keep the bundle structure. These tensor perturbations admit the separation of variables and the corresponding field equations reduce to a set of ordinary second-order differential equations.

\subsection{Maxwell equations}
\label{ssc:AlignedEM}

The study of electromagnetic fields in general Kerr--NUT--(A)dS spacetimes is a complicated task. In particular, the  procedure leading to the Teukolsky equation in four dimension \citep{Teukolsky:1972,Teukolsky:1973} does not work in higher dimensions \citep{Durkee:2010qu, Durkee:2010ea}. However, recently there was an important breakthrough in the study of possible separability of higher-dimensional Maxwell equations in the Kerr--NUT--(A)dS spacetimes. Namely, \cite{Lunin:2017drx} succeeded to separate variables for some specially chosen polarization states of the electromagnetic field in the Myers--Perry metrics with a cosmological constant. In a general $D$-dimensional case, the number of polarizations of such a field is $D-2$. Lunin proposed a special ansatz for the field describing two special polarizations and demonstrated that it admits separation of variables. He also demonstrated how such a solution relates to the solution obtained in the Teukolsky formalism in four dimension. However, if one can obtain other components by a similar ansatz is still under investigation.

In the rest of this section we discuss yet other interesting test electromagnetic fields, namely fields aligned with the principal tensor. They include, for example, the field of weakly charged and magnetized black holes. It turns out, that they constitute the most general test electromagnetic field that preserves the integrability properties of the Kerr--NUT--(A)dS geometry.

\subsubsection{Wald's trick: electromagnetic fields from isometries}

The study of electromagnetic fields in the vicinity of (rotating) black holes in four dimensions has interesting astrophysical applications and
has been investigated by many authors, see e.g., \cite{Wald:1974np,King:1975tt, bicak1977stationary, bivcak1976stationary, bivcak1980stationary, bivcak1985magnetic, Aliev:1989wx, Penna:2014aza}. There is also a number of
exact solutions of the Einstein--Maxwell system, ranging from the Kerr--Newman solution for the charged black hole \citep{newman1965note, newman1965metric} to magnetized black holes of Ernst \citep{ernst1968new, ernst1976black}. However, it is often possible to restrict the description to a test field approximation assuming that the electromagnetic field obeys the Maxwell equations but does not backreact on the geometry.

A particularly elegant way for describing the behavior of certain test electromagnetic fields near a rotating
black hole is due to \cite{Wald:1974np}. The Wald approach is based on the well known fact
\citep{Papapetrou:1966zz} that any Killing vector field $\ts{\xi}$ obeys the following two equations:
\be\label{testMaxwell}
\nabla_{\!a}\xi^a=0\,,\quad \Box \xi^a+R^a{}_b\xi^b=0\,.
\ee
The first equation is an immediate consequence of the Killing equation \eqref{Kill}, whereas the latter follows from its integrability condition, cf. \eqref{KYBochnerLaplace} for $p=1$. These two equations are to be compared with the wave equation supplemented by the Lorenz gauge condition:
\be
\nabla_a A^a=0\,,\quad \Box A^a-R^a{}_b A^b=0\,,
\ee
for the electromagnetic vector potential $\ts{A}$.
This means that in a vacuum spacetime any Killing vector can serve as a vector potential for a test Maxwell field,
\begin{equation}\label{WaldF}
    \ts{A}=e\, \ts{\xi}\,,\quad\ts{F}=e\, \grad\ts{\xi}\,,
\end{equation}
where the constant $e$ governs the field strength. Therefore a special set of test electromagnetic fields in the background of vacuum spacetimes can be genarated simply by using the isometries of these spacetimes.
In such a way one can generate a weakly charged Kerr black hole, or immerse this black hole in a `uniform magnetic field' \citep{Wald:1974np}.

Of course, the same trick also works in higher dimensions. This fact was used in \cite{Aliev:2004ec} for a  study of the gyromagnetic ratio of a weakly charged five-dimensional rotating black hole in an external magnetic field. This was later generalized to the Myers--Perry spacetimes \citep{Aliev:2006yk}.

In the presence of a cosmological constant $\Lambda$ the Ricci tensor
\be
R_{ab}=\frac{2}{D-2}\Lambda g_{ab}
\ee
does not vanish and the Killing vector $\ts{\xi}$ can no longer be used as a vector potential for the test electromagnetic field. The situation improves when the spacetime possesses a closed conformal Killing--Yano 2-form $\ts{h}$ \citep{Frolov:2017bdq}. Namely, let $\ts{\xi}$ be a primary Killing vector,\footnote{%
A similar construction does not work for the secondary Killing vectors. One could try to use the Killing vector co-potentials \eqref{omegajjjjj} instead of the principal tensor as a correction term. However, they are not closed, $\ts{d\omega}^{(j)}\neq 0$ for $j>0$, and cannot thus play a role of the correction to the electromagnetic field based on $\ts{d}\KV{j}$.}
then the following {`improved' electromagnetic field}:
\be\label{FFhh}
\ts{F}=e\Bigl(\grad\ts{\xi}+\frac{4\Lambda}{(D-1)(D-2)}\ts{h}\Bigr)
\ee
satisfies the source-free Maxwell equations $\covd\cdot\ts{F}=0$.

As we mentioned earlier, in the Kerr--NUT--(A)dS spacetime in the canonical coordinates, the components of the principal tensor $h_{ab}$ do not depend on the metric parameters. Thus, the operation \eqref{FFhh} can be interpreted as a subtraction from $\grad\ts{\xi}$ a similar quantity, calculated for the corresponding (anti-)de~Sitter background metric. This prescription was used by \cite{Aliev:2006tt, Aliev:2007qi} for obtaining a weakly charged version of the Kerr--(A)dS black holes in all dimensions. The weakly charged and magnetized black rings were studied in \cite{Ortaggio:2006ng, Ortaggio:2004kr}.

\subsubsection{Aligned electromagnetic fields}

A wide class of test electromagnetic fields in the Kerr--NUT--(A)dS spacetimes has been constructed in \cite{Krtous:2007}. These fields are {\em aligned} with the geometry of the Kerr--NUT--(A)dS background: they are constant along the explicit symmetries of the spacetime and their Maxwell tensor commutes with the principal tensor. Concentrating again on even dimensions (see \cite{Krtous:2007} and \cite{CarigliaEtal:2012b} for the detailed discussion) such a field can thus be written as
\begin{equation}\label{alignedFgen}
    \ts{F} = \sum_\mu f_\mu\; \enf\mu\wedge\ehf\mu\;,
\end{equation}
where the components ${f_\mu=f_\mu(x_1,\dots,x_\dg)}$ are independent of Killing directions ${\psi_j}$. Since the Maxwell tensor must be closed, ${\grad\ts{F}=0}$, it is locally generated by the vector potential ${\ts{A}}$. The most general field with the structure \eqref{alignedFgen} then
corresponds to the vector potential given by\footnote{%
Various expressions below contain rescaled 1-forms
${ \sqrt{\frac{X_\mu}{U_\mu}}\enf\mu = \grad x_\mu }$
and
${ \sqrt{\frac{U_\mu}{X_\mu}}\ehf\mu = \sum_k \A{k}_\mu \grad\psi_k}$.
They have even simpler form than 1-forms ${\enf\mu}$ and ${\ehf\mu}$ themselves and they could be used as a natural frame. However, such a frame is not normalized and we do not introduce it here explicitly, although, in some expressions we keep these terms together.
}
\begin{equation}\label{alignedA}
    \ts{A} = \sum_\mu \frac{g_\mu x_\mu}{U_\mu}\; \sqrt{\frac{U_\mu}{X_\mu}}\ehf\mu\;,
\end{equation}
where each function ${g_\mu=g_\mu(x_\mu)}$ depends only on one variable ${x_\mu}$. In terms of these function, the components ${f_\mu}$ are
\begin{equation}\label{EMfE}
  f_\mu =  \frac{g_\mu}{U_\mu}+ \frac{x_\mu\,g_\mu'}{U_\mu}
  +2\,x_\mu\,\sum_{\substack{\nu\\\nu\neq\mu}}\frac{1}{U_\nu}\,\frac{x_\nu\,g_\nu - x_\mu\,g_\mu}{x_\nu^2-x_\mu^2}\;.
\end{equation}

This electromagnetic field represents the off-shell complement of the off-shell Kerr--NUT--(A)dS geometry. Its structure is sufficient to generalize most of the symmetry properties of the geometry to the situation with a background test electromagnetic field. However, the field \eqref{alignedFgen} with components ${f_\mu}$ given by \eqref{EMfE} does not necessary satisfy the source free Maxwell equations. The corresponding current $J^a = \nabla_{\!b}F^{ab}$ is%
\footnote{Notice that in \cite{Krtous:2007}, the relations (2.20) and (2.21) for the source in even dimensions have a wrong sign.}
\begin{equation}\label{EMJ}
  \tens{J} = -2\sum_{\mu} \frac\partial{\partial x_\mu^2}\Biggl[\sum_{\nu}\frac{x_\nu^2\,g_{\nu}'}{U_\nu}\Biggr]\;\sqrt{\frac{X_\mu}{U_\mu}}\,\ehv\mu\;.
\end{equation}

Imposing the vacuum Maxwell equations, ${\tens{J}=0}$, we find the on-shell field for which the functions ${g_\mu}$ integrate to
\begin{equation}\label{EMggenE}
  g_\mu=  e_\mu + \frac1{x_\mu}\, \sum_{k=0}^{\dg-1}\, \tilde{c}_k\, x_\mu^{2k}\;,
\end{equation}
with ${e_\mu}$ and ${\tilde{c}_k}$ being constants. Moreover, it turns out that the second term is a pure gauge and can be ignored. The on-shell aligned test electromagnetic field can thus be written as
\begin{equation}\label{alignedAonshell}
    \ts{A} = \sum_\mu \frac{e_\mu x_\mu}{U_\mu}\; \sum_k \A{k}_\mu\grad\psi_k\;.
\end{equation}
It is parameterized by $\dg$ constants ${e_\mu}$, ${\mu=1,\dots,\dg}$, which correspond to an electric charge and magnetic charges associated with rotations along different directions.
If we set all charges except one, say ${e_\nu}$, to zero, the Maxwell tensor ${\tens{F}}$ corresponds to the harmonic form ${\tens{G}^{(\nu)}_{(2)}}$ found in \cite{Chen:2007fs}, see also \cite{Chow:2008fe}.

Another special choice is obtained upon setting ${g_\mu=e\,X_\mu/x_\mu}$, with a constant $e$ characterizing the strength of the field. In this case the vector potential \eqref{alignedA} reduces to the primary Killing vector \eqref{primKV},
\begin{equation}\label{EMApropPKV}
  \ts{A} = e\, \ts{\xi}\;.
\end{equation}
The corresponding current reads ${\ts{J}=2(2\dg-1)e\lambda\,\ts{\xi}}$, where $\lambda$ is the cosmological constant parameter \eqref{Xsol2n}.
Thus, for the vanishing cosmological constant we recover the source-free electromagnetic field given by the Wald construction \eqref{WaldF}.

One can also recover the `improved' electromagnetic field \eqref{FFhh} which is source-free for the on-shell Kerr--NUT--(A)dS background with a non-vanishing cosmological constant, i.e., when $X_\mu$ is given by \eqref{Xsol2n}. The second term in \eqref{FFhh}  can be induced by adding the correction $-e\lambda(-x_\mu^2)^\dg$ to $x_\mu g_\mu=e X_\mu$. This cancels exactly the term with the highest power of $x_\mu$ in $X_\mu$. In fact, all other even-power terms in $X_\mu$ give only a gauge trivial contribution to the potential and do not contribute to the Maxwell tensor. The `improved' field \eqref{FFhh} is thus solely given by the linear terms in functions $X_\mu$, determining the charges of the source-free aligned field as
\begin{equation}\label{specg}
    e_\mu = -2 e b_\mu\;.
\end{equation}
The vector potential of the `improved' field reads
\begin{equation}\label{specvecpot}
    \ts{A} = -2\sum_\mu \frac{b_\mu x_\mu}{U_\mu}\; \sqrt{\frac{U_\mu}{X_\mu}}\ehf\mu\;.
\end{equation}

\subsubsection{Motion of charged particles}

Let us now investigate the motion of charged particles in the `weakly charged' Kerr--NUT--(A)dS spacetimes penetrated by the aligned electromagnetic field \eqref{alignedA}. A special case of the field \eqref{EMApropPKV} has been investigated in \cite{Frolov:2010cr} and \cite{CarigliaEtal:2012b}.

The following results have been shown in \cite{Kolar:2015hea}. The off-shell aligned electromagnetic field \eqref{alignedA} is the most general electromagnetic field in the Kerr--NUT--(A)dS background for which the motion of charged particles is integrable and the corresponding Hamilton--Jacobi equations for all conserved quantities are separable. The charged generalization of the conserved quantities \eqref{integemot} for the particle with a charge $q$ are
\begin{equation}\label{integemotEM}
 \begin{aligned}
    K_{j} &= \KTc{j}^{ab}\,(p_a-q A_a)\,(p_b-q A_b)\;,\\
    L_{j} &= \KVc{j}^{a}\,p_a\;.
\end{aligned}
\end{equation}
The solution $S$ of the Hamilton--Jacobi equations can be found again using the separability ansatz \eqref{Ssepansatz}, to obtain the following modified differential equations \eqref{HJsepSsol} for the functions $S_\mu$:
\begin{equation}\label{HJsepSsolEM}
    (S_\mu')^2 = \frac{\Qofc_\mu}{X_\mu} - \frac{(\Lofc_\mu-q g_\mu x_\mu)^2}{X_\mu^2}
    \,.
\end{equation}

\subsubsection{{Weakly charged operators}}%

Similarly, one can also study test scalar and Dirac fields in the weakly charged Kerr--NUT--(A)dS spacetimes.

Let us start by considering a charged scalar field, characterized by the charge $q$.
Then the requirement of commutativity of the following charged scalar operators:
\begin{equation}\label{QLopdefEM}
\begin{gathered}
    \Qop{j} = - [\nabla_{\!a}-iqA_a]\, \KTc{j}^{ab} \,[\nabla_{\!b}-iqA_b]\,,\\
    \Lop{j} = - i\, \KVc{j}^{a} \nabla_{\!a}\,,
\end{gathered}
\end{equation}%
constructed from the Killing tensors and Killing vectors of the Kerr--NUT--(A)dS spacetimes imposes severe conditions on the electromagnetic field \citep{Kolar:2015hea}. These conditions are satisfied for the off-shell aligned electromagnetic field \eqref{alignedA}. The corresponding charged operators thus have common eigenfunctions which can be written in a separated form \eqref{WOsepansatz}. The differential equation for the functions $R_\mu$ in the charged modify to
\begin{equation}\label{WOsepcondEM}
    (X_\mu R_\mu')'
    + \Bigl({\Qofc_\mu}-\frac1{X_\mu}({(\Lofc_\mu-q g_\mu x_\mu)^2})\Bigr) R_\mu'
    = 0\;.
\end{equation}%

Similarly, the symmetry operators of the Dirac operator can be generalized to the charged case \citep{CarigliaEtal:2012b} and the common eigenfunctions can be found in a tensorial separable form \eqref{tensRsep}. The equations \eqref{chieq} for the two-component spinor functions $\ts{\chi}_\mu$ again only modify by changing $\Lofc_\mu\to\Lofc_\mu-q g_\mu x_\mu$.

\subsubsection{On a backreaction of the aligned fields}

As demonstrated above, the aligned electromagnetic field \eqref{alignedA} extends naturally most of the properties of Kerr--NUT--(A)dS spacetimes based on their high symmetry to the charged case, albeit this electromagnetic field is only a test field and does not modify the geometry itself. A natural question arises: is it possible to backreact this electromagnetic field to obtain the full solution of the Einstein--Maxwell system?

To answer this question, it is interesting to note that the expressions \eqref{alignedFgen} for the Maxwell tensor, \eqref{alignedA} for the vector potential, and \eqref{EMJ} for the current do not contain a reference to the metric functions~${X_\mu}$. Indeed, the square roots of ${X_\mu}$ exactly compensate normalization factors included in the frame elements. It gives a hope that the metric functions could be chosen such that the geometry represents the gravitational back reaction of the aligned electromagnetic field. Even the stress-energy tensor of the electromagnetic field is diagonal in the Darboux frame \citep{Krtous:2007}, and corresponds thus to the structure of the Ricci tensor \eqref{Ricci}. Unfortunately, except for the case of four dimensions, the diagonal elements of the Einstein equations do not match and, therefore, the Einstein equations with the electromagnetic field as a source cannot be satisfied \citep{Krtous:2007} (see also \cite{Aliev:2004ec} for similar attempts).

Only in four dimensions the metric functions can be chosen so that the Einstein equations are fulfilled. The geometry then describes the charged Kerr--NUT--(A)dS spacetime \citep{Carter:1968pl,Plebanski:1975}. In higher dimensions, though, this is no longer possible within the realms of pure Einstein--Maxwell theory, see, however, \cite{Chong:2005hr}, and additional fields have to be introduced, e.g., \cite{Chow:2008fe}. In other words, the exact higher-dimensional analogue of the Kerr--Newman solution (without additional fields) remains elusive.

%% file: ch7-aplgen.tex

\section{Further developments}
\label{sc:aplgen}

In this chapter, we review several scattered results in the literature that are related to the existence of the principal tensor and its generalizations. Namely, we discuss the construction of parallel--transported frames along timelike and null geodesics, motion of classical spinning particles, and stationary configurations of strings and branes in the Kerr--NUT--(A)dS spacetimes. We then move beyond the Kerr--NUT--(A)dS spacetimes. Namely, we discuss what happens when some of the eigenvalues of the principal tensor become degenerate, which leads us to the generalized Kerr--NUT--AdS spacetimes. Some of these new spacetimes can be obtained by taking certain singular limits of the Kerr--NUT--(A)dS metric. The limiting procedure may preserve or even enhance the symmetries of the original metric.  Hidden symmetries of warped spaces and the corresponding `lifting theorems' are discussed next. We conclude this chapter by studying the  generalizations of Killing--Yano objects to spacetimes with torsion and their applications to various supergravity backgrounds where the torsion can be naturally identified with the 3-form flux present in the theory.

\subsection{Parallel transport}
\label{ssc:PralTransp}

In the previous chapters we have learned that the geodesic motion in general Kerr--NUT--(A)dS spacetimes is completely integrable.
In this section we show that the existence of the principle tensor $\ts{h}$ even allows one to construct a whole \emph{parallel-transported frame} along these geodesics.

Such a frame provides a useful tool for studying the behavior of extended objects in this geometry. For example, in the four-dimensional case it was employed for the study of tidal forces acting on a moving body, for example a star, in the background
of a massive black hole, e.g., \cite{luminet1985tidal, laguna1993tidal,
diener1997relativistic, ishii2005black}.
In quantum physics the parallel transport of frames is an important technical element of the point
splitting method which is used for calculating the renormalized values
of local observables (such as vacuum expectation values of currents,
stress-energy tensor etc.) in a curved spacetime. Solving the
parallel transport equations is also useful  when particles and fields with
spin are considered, e.g., \cite{christensen1978regularization}.

\subsubsection{Parallel-transported frame along timelike geodesics}

Consider a timelike geodesic $\gamma$ in the Kerr--NUT--(A)dS
spacetime and denote by $\ts{u}$ its normalized velocity.%
\footnote{In this section we assume a Lorentzian signature of the metric. Moreover, to simplify the formulas in this section we simply put the mass $m$ of a particle equal one. With this choice, $m=1$, the inner time $\sigma$ coincides with the proper time $\tau$, and the momentum of the particle $p_a$ is related to its velocity $u^a$ as follows $p_a=g_{ab}u^b$. We denote by dot the covariant derivative $\nabla_{\ts{u}}$.}
 Starting from the principal tensor $\ts{h}$, we may define the following 2-form:
\be\label{F2form}
F_{ab}=P^c_a P^d_b h_{cd}\,,
\ee
where $P^a_b=\delta^a_b+u^au_b$ is the projector along the geodesic. Referring to the discussion in section~\ref{ssc:CKYproperties}, we infer that the 2-form $\ts{F}$ is parallel-transported along $\gamma$, cf.\ \eqref{Fpform}. This property was originally used in \cite{Page:2006ka} to demonstrate the complete integrability of geodesic motion.
Since $\ts{F}$ is parallel-transported, so is any object constructed from $\ts{F}$ and the metric $\ts{g}$. In particular, this is true for the invariants constructed from $\ts{F}$, such as its eigenvalues. As we shall see below, for a generic geodesic it is possible to extract from $\ts{F}$ at least $n-1$ nontrivial independent eigenvalues, which together with the normalization of velocity, and other $n+\eps$ constants of motion due to Killing vectors, imply complete integrability.

\details{This idea was later formalized in \cite{Cariglia:2012qj}, where it was shown that the 2-form $F^a{}_b$ can be identified with the {\em covariant Lax tensor} \citep{Rosquist:1994yd, Rosquist:1997fn, Karlovini:1998kc, Baleanu:1999sq, Baleanu:2001ef,
Cariglia:2012qj}, whose covariant conservation, $\dot{F}^a{}_b=0$, can be rewritten as the standard Lax pair equation \citep{lax1968integrals}
\be
\dot{\mathsf{L}} = [\mathsf{L},\,\mathsf{M}]\;,
\ee
where $\mathsf{L}=[F^a{}_b],\, \mathsf{M} = \Bigl[\frac{\partial H}{\partial p_c}\Gamma_{\!cb}^a\Bigr]$, and
$H=\frac{1}{2}p^2$. Constants of motion are consequently generated from the traces of matrix powers of $\mathsf{L}$, $\Tr(\mathsf{L}^j)$.
If interested, see \cite{Cariglia:2012qj} for the construction of the corresponding {\em Clifford Lax tensor} and generalizations to a charged particle motion.
}\medskip

One can do even more. Namely, it is possible to use the 2-form $\ts{F}$
to explicitly construct a frame which is parallel-transported along the timelike geodesic.
To construct such a frame we use a method similar to the one developed by Marck for the four-dimensional Kerr metric \citep{marck1983solution}. For more details concerning the solution of the parallel transport equations in the higher-dimensional Kerr--NUT--(A)dS spacetime see \cite{Connell:2008vn}.

Let us denote $\ts{F}^2=\ts{F}\cdot\ts{F}$, or, in components, $({F}^2)_a^{\ b}=F_a^{\ c} F_c^{\ b}$, and consider the following eigenvalue problem:
{\be\label{FFF}
\ts{F}^2\cdot\ts{v}=-\lambda^2\,\ts{v}\, .
\ee}
It is easy to check that the following properties are valid:
\begin{itemize}
\item $\ts{F}^2\cdot\ts{u}=0$.
\item If $\ts{v}$ obeys \eqref{FFF} then the vector $\bar{\ts{v}}=\ts{F}\cdot \ts{v}$ obeys the same equation.
\item One also has $\ts{F}\cdot \bar{\ts{v}}=-\lambda^2\,\ts{v}$.
\item Eigenvectors {$\ts{v}_\mu$ and $\ts{v}_\nu$} of the operator $\ts{F}^2$ with different eigenvalues {$\lambda_\mu$ and $\lambda_\nu$} are orthogonal.
\item Denote  $\dot{\ts{v}}=\covd_{\!\ts{u}} \ts{v}$. Then since $\ts{F}$ is parallel-transported, one has
{\be\label{vdot}
\ts{F}^2\cdot\dot{\ts{v}} =\covd_{\!\ts{u}} (\ts{F}^2\cdot\ts{v})
  =\covd_{\!\ts{u}}(-\lambda^2\,\ts{v})=-\lambda^2\, \dot{\ts{v}}\,.
\ee}
\end{itemize}

Let us denote by $V_{\mu}$ a subspace spanned by the vectors with the eigenvalue $\lambda_{\mu}$.
Since the parallel-transported eigenvector remains to be an  eigenvector corresponding to the same eigenvalue, c.f. \eqref{vdot}, each subspace  $V_{\mu}$ is independently parallel-transported along the geodesic. These subspaces are enumerated by index $\mu$ which takes values $\mu=0,1,\ldots,p$; we assume that $V_0$ corresponds to zero eigenvalue: $\lambda_0=0$. We call $V_{\mu}$ \emph{a Darboux subspace} of $\ts{F}$ (or eigenspace of $\ts{F}^2$). The tangent vector space $T$ can thus be presented as a direct sum of independently parallel-transported Darboux subspaces $V_\mu$:
\be
T=V_0\oplus V_1\oplus\dots \oplus V_p\, .
\ee
It can be shown  that for a generic geodesic $\gamma$, the Darboux subspaces $V_\mu$ for $\mu\neq 0$ are two-dimensional \citep{Connell:2008vn}. This fact is directly linked to the non-degeneracy of the principal tensor $\ts{h}$. The 2-form $\ts{F}$ is simply a projection of $\ts{h}$ along a given geodesic. Since eigenspaces of $\ts{h}$ are non-degenerate and 2-dimensional, so will be the eigenspaces of $\ts{F}$, unless the direction determined by the geodesic is `special', see \cite{Connell:2008vn} for more details.
Moreover, one can show that in odd dimensions $V_0$ is one-dimensional, spanned by $\ts{u}$, whereas $V_0$ is two-dimensional in even dimensions, spanned by $\ts{u}$ and $\ts{z}$, where
\be
\ts{z}=\ts{u}\cdot (*\ts{h}^{\wedge (n-1)})=*(\ts{F}^{\wedge(n-1)}\wedge \ts{u})\, .
\ee
The vector $\ts{z}$ is orthogonal to $\ts{u}$ and,  after it is normalized, completes the orthonormal parallel-transported frame in $V_0$. It is easy to check that the number of Darboux subspaces $p$ in the odd-dimensional spacetime is $p=n$, while in even dimensions $p=n-1$.

\begin{figure}
\bigskip
\centerline{\includegraphics[width=6cm]{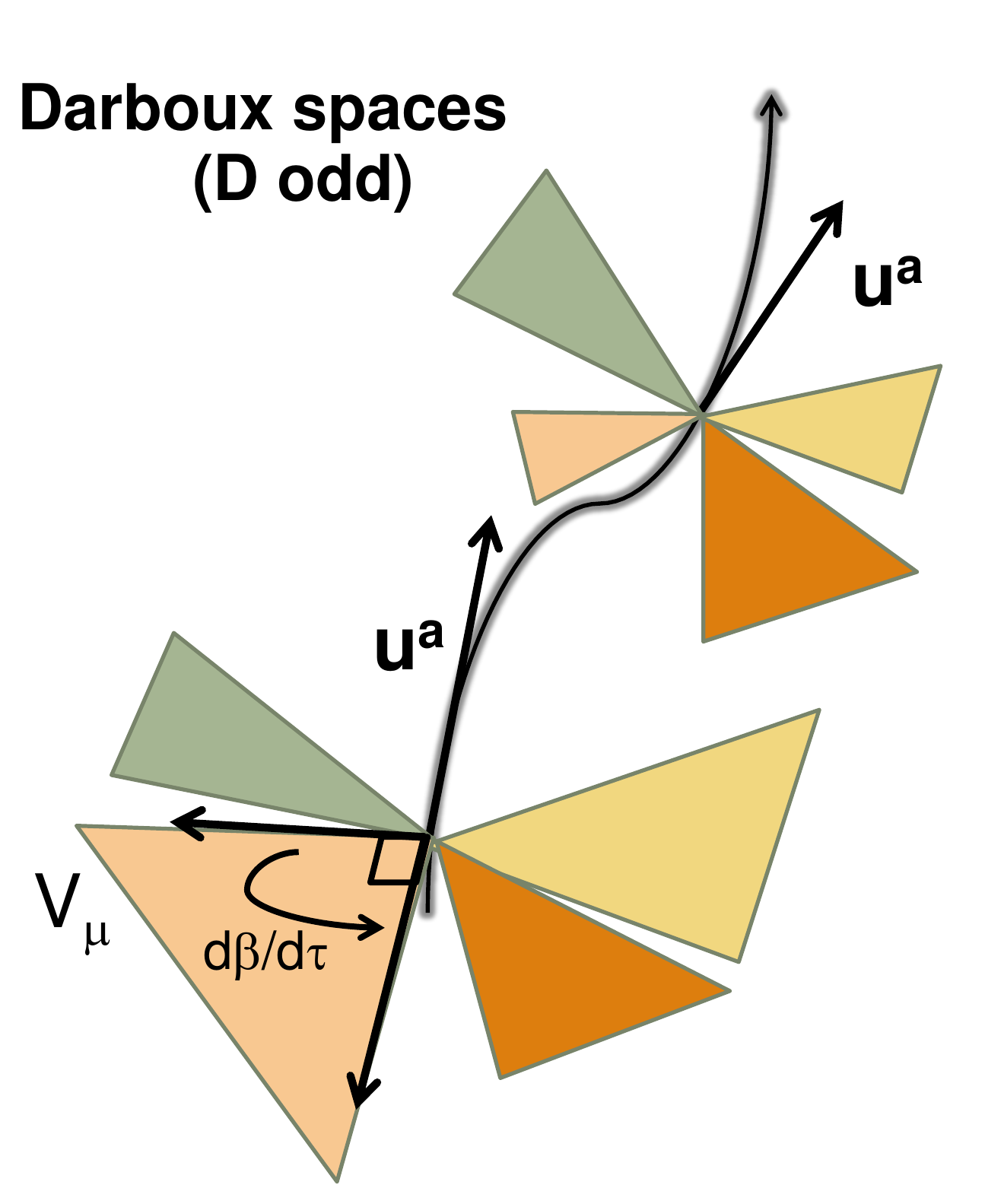}\quad\includegraphics[width=6cm]{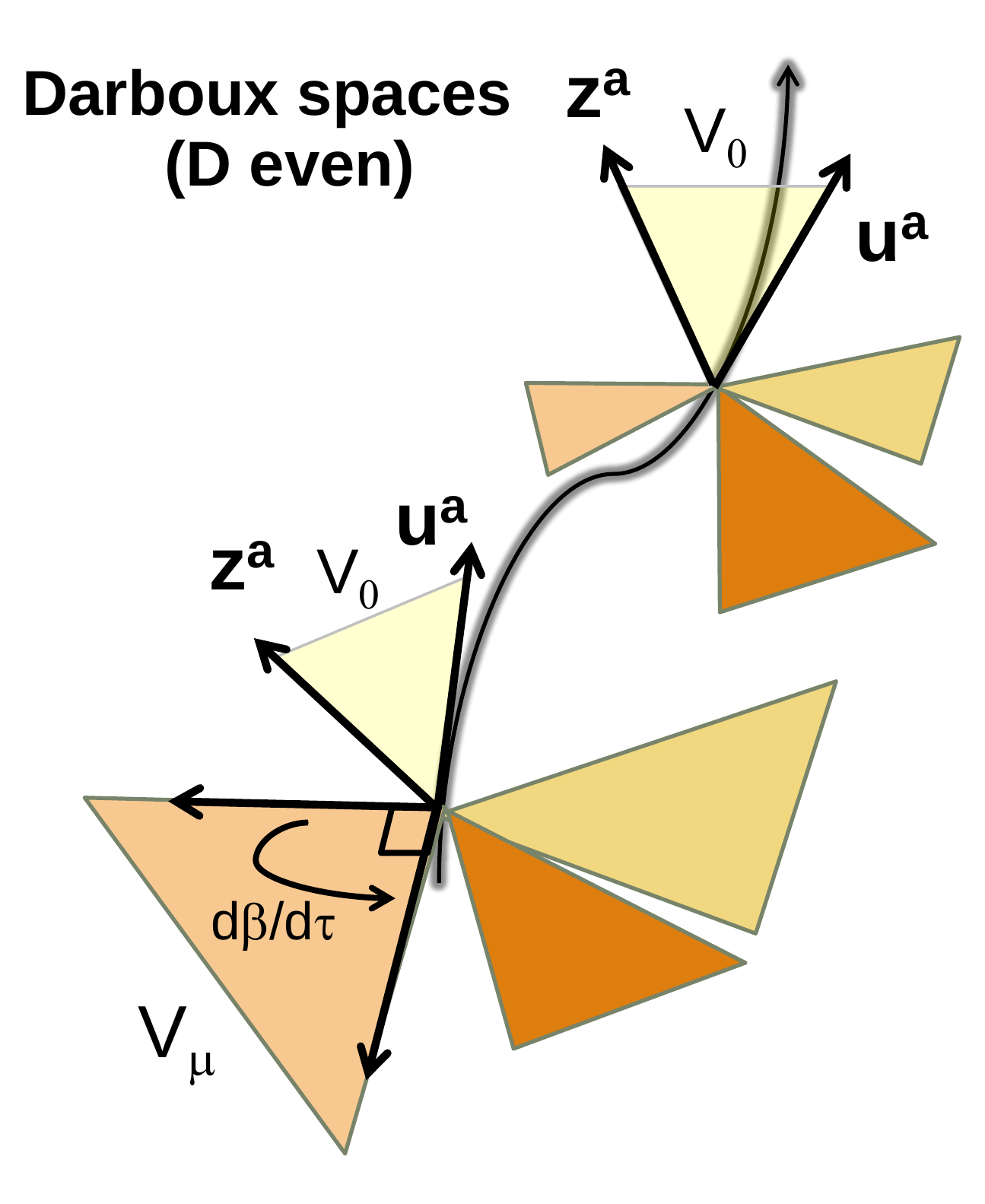}}
\caption{{\bf Parallel transport.} The figure schematically displays the construction of a parallel-transported frame along generic timelike geodesics in (left) odd dimensional and (right) even-dimensional Kerr--NUT--(A)dS spacetimes. The colored 2-planes correspond to orthogonal independently parallel-transported Darboux 2-planes $V_\mu$. }
\medskip
\label{Fig:PT}
\end{figure}

To construct a parallel-transported frame in a given two-dimensional Darboux subspace $V_{\mu>0}$ we proceed as follows. We choose a (not-necessarily par\-allel-transported) orthonormal basis spanning $V_\mu$: $\{\ts{n}_\mu, \bar{\ts{n}}_\mu\}$, and obtain a parallel-transported frame in $V_\mu$, $\{\ts{v}_\mu, \bar{\ts{v}}_\mu\}$, by a $\tau$-dependent rotation of this orthonormal basis,
\be
\ts{v}_\mu=\cos\beta_\mu \ts{n}_\mu-\sin\beta_\mu \bar{\ts{n}}_\mu\,,\quad
\ts{\bar v}_\mu=\sin\beta_\mu \ts{n}_\mu+\cos\beta_\mu \bar{\ts{n}}_\mu\,,
\ee
where the rotation angle $\beta_\mu$ obeys
\be\label{eqBeta}
\dot{\beta}_\mu=-\ts{n}_\mu\cdot  \dot{\bar{\ts{n}}}_\mu=\dot{\ts{n}}_\mu\cdot \bar{\ts{n}}_\mu\,.
\ee
The dot, as earlier, denote a derivative with respect to the proper time $\tau$.
If at the initial point $\tau=0$  bases $\{\ts{v},\bar{\ts{v}}\}$ and $\{\ts{n},\bar{\ts{n}}\}$ coincide, we have the following condition for the
above equations: $\beta_\mu|_{\tau=0}=0\,.$

The whole construction of the parallel-transported frame in Kerr--NUT--(A)dS spacetimes is schematically illustrated in figure~\ref{Fig:PT}. The procedure is algorithmic and the actual calculation can be technically simplified by using the so called velocity adapted basis. We refer the interested reader to \cite{Connell:2008vn} for more details.

\subsubsection{Parallel transport along null geodesics}
\label{ssc:NullGeod}

The described construction of parallel-transported frame does not straightforwardly apply to null geodesics.
In this section we show how to modify this construction and to obtain a parallel-transported frame along null geodesics in Kerr--NUT--(A)dS spacetimes, generalizing the results obtained by \cite{marck1983parallel} for the four-dimensional Kerr metric. The section is based on \cite{Kubiznak:2008zs} to where we refer the reader for more details.

The parallel-transported frame along null geodesics has applications in many physical situations. For example, it can be used for studying the polarized radiation of photons and gravitons in the geometric optics approximation, see, e.g., \cite{StarkConnors:1977, Connors:1977, ConnorsEtal:1980} and references therein.  It provides  a technical tool for the derivation of the equations for optical scalars \citep{Pirani:1965, Frolov:1977} and plays the role in the proof of the `peeling-off property' of the gravitational radiation \citep{Sachs:1961, Sachs:1962, NewmanPenrose:1962, Penrose:1965, KrtousPodolsky:2004b}.

We start with the following observation. Let us consider an affine parameterized null geodesic $\gamma$, with a tangent vector $\ts{l}$. We denote by dot the covariant derivative $\nabla_{\ts{l}}$. Then one has  $\dot{\ts{l}}=0$. Let $\ts{v}$ be a parallel-transported vector along $\gamma$, $\dot{\ts{v}}=0$, and $\ts{h}$ be the principal tensor. Then, defining
\be\label{waseed}
w_a=v^ch_{ca}+\beta l_a\,,
\ee
we find
\be
  \dot{\ts{w}}=\ts{v}\cdot\dot{\ts{h}}+\dot{\beta}\, \ts{l}=\ts{v}\cdot(\ts{l}\wedge\ts{\xi})+\dot{\beta}\, \ts{l}
  =\ts{\xi}\,(\ts{v}\cdot \ts{l})+\ts{l}\,(\dot \beta -\ts{v}\cdot \ts{\xi})\, .
\ee
Here we used the equation \eqref{defPKY} for the principal tensor, with $\ts{\xi}$ being the primary Killing vector associated with $\ts{h}$.
Hence, the vector $\ts{w}$ is parallel-transported provided the following conditions are satisfied:
\be\label{vcond}
\ts{v}\cdot \ts{l}=0\,,\quad \dot \beta=\ts{v}\cdot \ts{\xi}\,.
\ee

\begin{figure}
\bigskip
\centerline{\includegraphics[height=4cm]{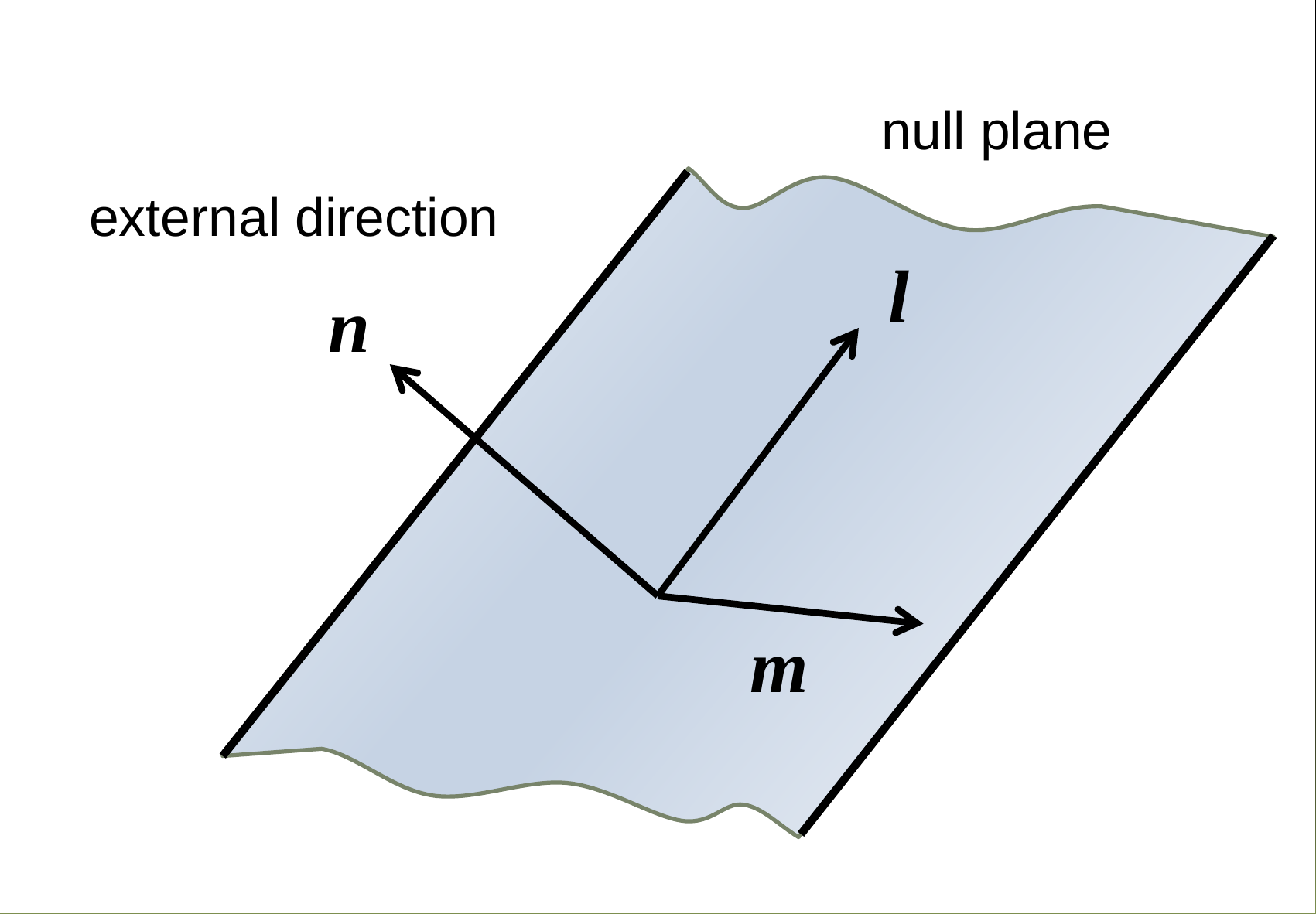}\quad\includegraphics[height=4cm]{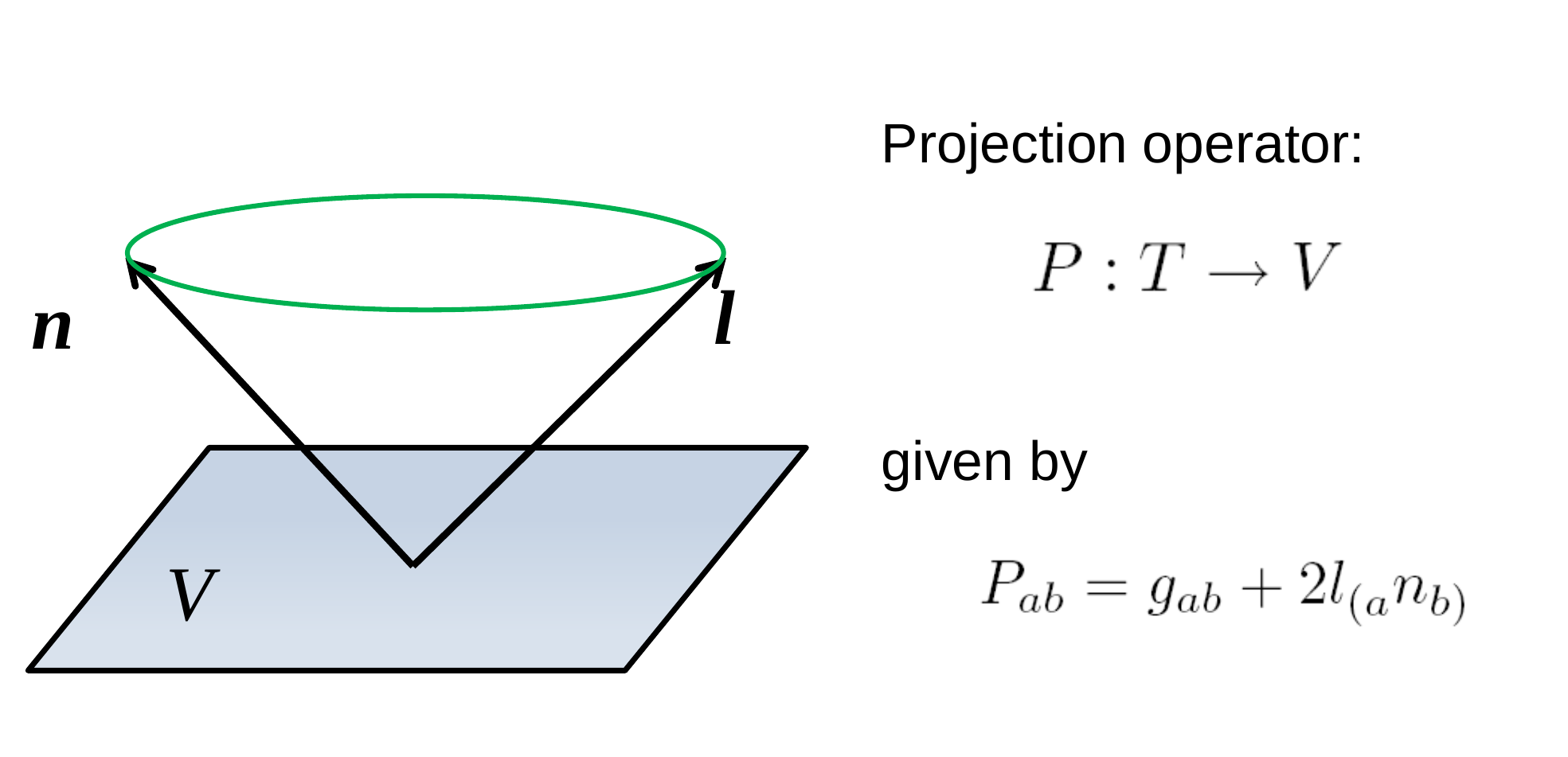}}
\caption{{\bf Geometry of null geodesics.} Left figure displays the geometry of the three parallel-transported vectors $\ts{l}, \ts{n}$ and $\ts{m}$.  Right figure demonstrates the action of the projection operator $P^a{}_b$ which now projects to a space $V$ that is  orthogonal to both $\ts{l}$ and $\ts{n}$.
}
\medskip
\label{Fig:PTP}
\end{figure}

This observation allows one to immediately construct two parallel-trans\-ported vectors, which we call $\ts{m}$ and $\ts{n}$. Namely, $\ts{m}$ is obtained by taking $\ts{v}=\ts{l}$ in \eqref{waseed};  the first condition in \eqref{vcond} is automatically satisfied and the second condition gives $\beta_{(l)}=\tau\,(\ts{l}\cdot \ts{\xi})$, since $\ts{\xi}$ is a Killing vector. Using next $\ts{v}=\ts{m}$ as a `seed' in \eqref{waseed}, we obtain the second vector $\ts{n}$.
The two vectors can be normalized so that
\be
\ts{n}\cdot \ts{l}=-1\,,\quad \ts{n}\cdot \ts{m}=0\,,\quad \ts{n}\cdot \ts{n}=0\, .
\ee
The vector $\ts{n}$ does not belong to a null plane of vectors orthogonal to $\ts{l}$, and, in this sense, it is `external' to it, see figure~\ref{Fig:PTP}.
For this reason one cannot use it as a new `seed' in \eqref{waseed}.

To generate additional parallel-transported vectors one  proceeds as follows.
We denote
\be\label{Ftilde}
\tilde F_{ab}=P^c_aP^d_b h_{cd}\,,\quad P_{ab}=g_{ab}+2l_{(a}n_{b)}\,.
\ee
Here, $P^a_b$ is a projector on a space  orthogonal to both $\ts{l}$ and $\ts{n}$ directions (see figure~\ref{Fig:PTP}). The 2-form $\ts{\tilde F}$ is
parallel-transported along $\ts{l}$. In particular, its eigenvalues are constant along the null rays and give the integrals of null geodesic motion.

Similar to the timelike case we may now consider the Darboux subspaces of $\ts{\tilde F}$. They are again independently parallel-transported. We denote by  $V_0$ the Darboux subspace corresponding to the zero eigenvalue. Its dimension depends on the dimension $D$ of the spacetime. Namely,
\begin{equation}
\begin{aligned}
&\text{for $D$ odd}:&   &V_0\ \text{is 3-dimensional and spanned by}\ \{\ts{l}, \ts{m},\ts{n}\}\,,\\
&\text{for $D$ even}:&  &V_0\ \text{is 4-dimensional and spanned by}\ \{\ts{l}, \ts{m},\ts{n}, \ts{z}\}\,,
\end{aligned}
\end{equation}
where as earlier $\ts{z}=\ts{l}\cdot (\ts{*h}^{\wedge(n-1)})$. These base vectors are parallel propagated by construction.

The other Darboux subspaces (with non-zero eigenvalues) are generically 2-dimensional. To construct the parallel-transported vectors that span them one can proceed as in the timelike case. Explicit expressions for the parallel transported frame along a null geodesic in the Kerr--NUT--(A)dS spacetimes can be found in \cite{Kubiznak:2008zs}.

Let us finally mention that the above construction does not work for the special null geodesics that are the eigenvectors of the principal tensor. It turns out that such directions describe the {\em principal null directions}, or WANDs (Weyl aligned null directions). Such directions play an important role in many physical situations, e.g., \cite{Coley:2004jv, Milson:2004jx,Coley:2007tp, Ortaggio:2012jd}. In Kerr--NUT--(A)dS spacetimes, these directions can be explicitly written down and the parallel-transported frame can be obtained by a set of local Lorentz transformations of the principal Darboux basis \citep{Kubiznak:2008zs}.

\details{It was shown in \cite{Mason:2010zzc} that the eigenvectors of a non-degenerate (not necessarily closed) conformal Killing--Yano 2-form are principal null directions and the corresponding spacetime is of the special algebraic type D. As we discussed in section~\ref{ssc:KerrSchild}, these special directions may also play a role in the Kerr--Schild construction of solutions of the Einstein equations.  }


\subsection{Classical spinning particle}
\label{ssc:ClSpinPart}

So far we have discussed the geodesic motion of point-like test particles as well as the propagation of test fields (possibly with spin) in the curved Kerr--NUT--(A)dS background.   An interesting problem is to consider the motion of {\em particles with spin}. There exist several proposals for describing spinning particles in general relativity, ranging from the traditional approach due to \cite{papapetrou1951spinning,corinaldesi1951spinning}, accompanied by a variety of supplementary conditions, e.g., \cite{semerak2015spinning, semerak2015spinning2}, to some more recent proposals e.g., \cite{Rempel:2016elp}.

In this section we concentrate on the spinning particle described by the worldline {\em supersymmetric} extension of the ordinary relativistic point-particle \citep{BerezinMarinov:1977,Casalbuoni:1976,BarducciEtal:1976,BrinkEtal:1976,%
BrinkEtal:1977,Rietdijk:1989qa,GibbonsEtal:1993,Tanimoto:1995,AhmedovAliev:2009,Ngome:2010gg}, where the spin degrees of freedom are described by Grassmann (anticommuting) variables. Such a model is physically very interesting as it provides a bridge between the semi-classical Dirac's theory of spin $\frac{1}{2}$ fermions and the classical Papapetrou's theory. Our aim is to show that the existence of the principal tensor provides enough symmetry to upgrade the integrals of geodesic motion to new bosonic integrals of spinning particle motion that are functionally independent and in involution. This opens a question of integrability of spinning particle motion in the Kerr--NUT--(A)dS spacetimes.

\subsubsection{Theory of classical spinning particles}

Let us start by briefly describing our model of a classical spinning particle.
To describe a motion of the particle in $D$ dimensions, we specify its worldline by giving the coordinates dependence on the proper time $\tau$: $x^a(\tau)$ ($a=1,\dots , D$). The particle's spin is given by the Lorentz vector of Grassmann-odd coordinates  $\theta^A(\tau)$ ($A = 1, \dots, D$). We denote by $A$ the vielbein index labeling an orthonormal vielbein $\{\ts{e}_{\!A}\}$ with components $e_{\!A}^a$. These components are used to change coordinate indices to vielbain ones and vice-versa, $v^a=v^A e_{\!A}^a$.

The motion of the spinning particle is governed by the following equations of motion:
\begin{align}
\frac{{\nabla}^2x^a}{d\tau^2}&=\ddot{x}^a + \Gamma^a_{bc}\, \dot{x}^b \dot{x}^c
  = \frac{i}{2} R^a{}_{bAB}\theta^{A}\theta^B\dot x^b\,,
\label{eq:Papapetrou}\\
\frac{{\nabla} \theta^A}{d\tau} &= \dot{\theta}^A + \omega_b{}^A{}_{\!B}\, \dot{x}^b \theta^B = 0 \, . \label{eq:covariantly_constant_spin}
\end{align}
Here $\Gamma^a_{bc}$ and $\omega_b{}^{A}{}_{\!B}$ are the Levi-Civita and spin connections, respectively, and
$R_{abcd}$ is the Riemann tensor.
The first equation is an analogue of the classical general-relativistic Papapetrou's equation.  It generalizes the geodesic equation for a point-like object to an extended object with spin. The latter
equation expresses the simple requirement that,
in the absence of interactions other than gravity, the spin vector is constant along the worldline of the particle.

The theory admits a Hamiltonian formulation, with the Hamiltonian $H$ given by
\begin{equation}
H=\frac{1}{2}\Pi_a\Pi_b\, g^{ab}\,,\ \
\Pi_a=p_a - \frac{i}{2}\,\omega_{a AB}\,\theta^A\theta^B\, ,
\end{equation}
where $p_a$ is the momentum canonically conjugate to $x^a$.
Velocity is related to the momentum as
\be\label{Qdefxdotspinp}
\dot{x}^a=\frac{\partial H}{\partial p_a} = g^{ab}\Pi_b = p^a - \frac{i}{2}\,\omega^a{}_{\!AB}\,\theta^A\theta^B\,.
\ee
The theory possesses a generic {\em supercharge} $Q$,
\be\label{Qdef}
Q=\theta^a \Pi_a\,,
\ee
obeying
\be\label{HQcond}
\{H,Q\}=0\,, \quad \{Q,Q\}=-2iH\, .
\ee
Here the super-Poisson brackets are defined as
\be\label{brackets}
\{F,G\}=
  \frac{\partial F}{\partial x^a}\frac{\partial G}{\partial p_a}
  -\frac{\partial F}{\partial p_a}\frac{\partial G}{\partial x^a}+
  i(-1)^{a_F}\frac{\partial F}{\partial \theta^A}\frac{\partial G}{\partial \theta_A}\,,
\ee
and $a_F$ is the Grassmann parity of $F$.
Equations of motion are accompanied by two physical (gauge fixing) constraints
\be\label{gaugecond}
2H=-1\,,\quad Q=0\,,
\ee
which state that $\tau$ is the proper time and the particle's spin is spacelike.

An important role for the spinning particle in curved spacetime is played by {\em non-generic superinvariants} which are quantities
that super-Poisson commute with the generic supercharge.
More specifically, a superinvariant $S$ is defined by the equation
\be\label{Q}
\{Q, S\}=0\,.
\ee
The existence of solutions of this equation imposes nontrivial conditions on the properties of the {geometry. The geometry} has to possess special symmetries such as
Killing vectors or Killing--Yano tensors,  for example. It follows from the Jacobi identity that any superinvariant is automatically a constant of motion,  $\{H, S\}=0$. At the same time quantity $\{S,S\}$ is a `new' superinvariant and a constant of motion (which may, or may not be equal to $H$).
Hence, superinvariants
correspond to an enhanced {\em worldline supersymmetry}.

{\em Linear in momentum} superinvariants were studied in \cite{GibbonsEtal:1993, Tanimoto:1995}, they are in one-to-one correspondence with Killing--Yano tensors and take the following form:
\be\label{ansatz}
{\mathcal{Q}}=\theta^{A_1}\!\dots \theta^{A_{p{-}1}} f^a{}_{A_1\dots A_{p{-}1}}\Pi_{a}
-\frac{i}{(p+1)^2}\theta^{A_1}\!\dots \theta^{A_{p{+}1}} (df)_{A_1\dots A_{p{+}1}}\,,
\ee
for a Killing--Yano $p$-form $\ts{f}$.
The Kerr--NUT--(A)dS spacetimes admit $n$ such superinvariants, associated with the
tower of Killing--Yano tensors \eqref{KYj}. However, such superinvariants are (i) not `invertible' for velocities \citep{Kubiznak:2011ay} and (ii) not in involution. In fact one can show that in even dimensions, where such superinvariants are fermionic, their Poisson brackets are not closed and generate  an extended superalgebra \citep{AhmedovAliev:2009}.

\subsubsection{Bosonic integrals of motion}

It turns out that for the Kerr--NUT--(A)dS spacetimes one can construct $D$ functionally independent and mutually commuting bosonic integrals of motion \citep{Kubiznak:2011ay}. Namely, in addition to
$(n+\varepsilon)$ bosonic linear in momenta superinvariants \eqref{ansatz} corresponding to the isometries $\KV{j}$, \eqref{KVj}:
\be\label{Qk}
{\mathcal{Q}_{(j)}}=\KVc{j}^a\Pi_a-\frac{i}{4}\theta^A\theta^B \bigl(d\KVc{j}\bigr)_{AB}\,,
\ee
one can also construct the following $n$ {\em quadratic in momenta} bosonic superinvariants $\mathcal{K}_{(j)}$, whose leading term contains no $\theta$'s and is completely determined by the Killing tensors $\KT{j}$, \eqref{KTj}:
\be\label{quadr}
\mathcal{K}_{(j)}=\KTc{j}^{ab} \Pi_a\Pi_b+\mathcal{L}_{(j)}^a\Pi_a+\mathcal{M}_{(j)}\,,
\ee
where
\begin{equation}\label{solution}
\begin{gathered}
\mathcal{L}_{(j)}^a=\theta^A\theta^B L_{(j)}{}_{AB}{}^a,\quad
\mathcal{M}_{(j)}=\theta^A\theta^B\theta^C\theta^D M_{(j)}{}_{ABCD}\,,\\
\KTc{j}^{ab}=\frac1{(p-1)!}\KYc{j}{}^{a k_2\dots k_p}\KYc{j}{}^b{}_{k_2\dots k_p}\,,\quad
\\
L_{(j)}{}_{ab}{}^c=-\frac{2ip}{(p+1)!}\Bigl(
  \KYc{j}{}_{[a}{}^{k_2\dots k_p}\bigl(d\KYc{j}\bigr){}_{b]}{}^c{}_{k_2\dots k_p}
  +\bigl(d\KYc{j}\bigr){}_{ab k_2\dots k_p}\KYc{j}{}^{ck_2\dots k_p}\Bigr)\,, 
\\
M_{(j)}{}_{abcd}=-\frac{i}{4} \nabla{}_{\!\![a} L_{(j)}{}_{bcd]}\,.
\end{gathered}
\end{equation}
Here, $\KY{j}$ is the Killing--Yano $p$-form \eqref{KYj} with $p=D-2j$.
In the absence of spin, such quantities reduce to the quadratic integrals of geodesic motion, responsible for its complete integrability.

In other words, the following quantities:
\be\label{integrals}
H, \mathcal{K}_{(1)},\dots ,\mathcal{K}_{(n-1)},
\mathcal{Q}_{(0)},\dots ,\mathcal{Q}_{(n-1+\eps)}\,,
\ee
form a complete set of bosonic integrals of motion for the spinning particle in Kerr--NUT--(A)dS spacetimes \citep{Kubiznak:2011ay}, which are functionally independent and in involution,
\be\label{BR}
\{\mathcal{Q}_{(i)}, \mathcal{Q}_{(j)}\}=0\,,\
\{\mathcal{Q}_{(i)}, \mathcal{K}_{(j)}\}=0\,,\
\{\mathcal{K}_{(i)}, \mathcal{K}_{(j)}\}=0\,,
\ee
making the `bosonic part' of the spinning particle motion integrable.

\subsubsection{Concluding remarks}

Let us stress that the above results regard the bosonic sector and have not dealt with the fermionic part of the motion, whose integrability would require a separate analysis. For this reason, the question of complete integrability of the whole (bosonic and fermionic) system of equations of motion of the spinning particle remains open. However, there are reasons to expect that this system might be fully integrable. Perhaps the most suggestive one is the observation that the Dirac equation, that corresponds to the quantized system and can be formally recovered by replacing $\theta$'s with $\gamma$ matrices and $\Pi$'s with the spinorial derivative, admits a separation of variables in these spacetimes, see section~\ref{ssc:DiracEquation}.  To achieve such separation it is enough to use a set of $D$ mutually-commuting operators, as many as the Poisson commuting functions that have been found for the motion of spinning particle.

Let us finally discuss some important differences between the  supersymmetric description presented in this section and the Papapetrou's theory. Formally, the Papapetrou's equations can be obtained by replacing $- i \theta^A\theta^B$ with the spin tensor $S^{ab}$. (It can be shown that the object $-i \theta^A\theta^B$ satisfies the correct Lie algebra of the Lorentz group under Poisson brackets.)
After this identification Eqs. \eqref{eq:Papapetrou} and \eqref{eq:covariantly_constant_spin} become Papapetrou's equations with the particular choice of supplementary condition:
\be\label{Pap}
\frac{{\nabla}^2x^a}{d\tau^2}= - \frac{1}{2} R^a{}_{bcd}S^{cd}\dot x^b\,,\quad
\frac{{\nabla}S^{ab}}{d\tau}=0\,.
\ee
Under such a transition, linear superinvariants \eqref{Qk} translate into the full integrals of motion for Papapetrou's equations \eqref{Pap}. However,
the quadratic superinvariants \eqref{quadr} become only approximate integrals---valid to a linear order in the spin tensor $S^{ab}$.
An interesting open question is whether such broken integrals of motion originate the
chaotic behavior of the spinning particle motion described by Papapetrou's theory in black hole spacetimes, e.g., \cite{suzuki1997chaos, semerak1999spinning, semerak2010free, semerak2012free, sukova2013free, witzany2015free}.

\subsection{Stationary strings and branes}
\label{ssc:StringsBranes}

\subsubsection{Dirac--Nambu--Goto action for extended objects}

There are interesting cases when the principal tensor allows one to integrate equations for some extended test objects, such as strings and branes. Such objects play a fundamental role in string theory. At the same time cosmic strings and domain walls are topological defects, which can be naturally created during phase transitions in the early Universe, e.g., \cite{vilenkin2000cosmic, Polchinski:2004ia, Davis:2005dd}, and their interaction with astrophysical black holes may
result in interesting observational effects, e.g., \cite{Gregory:2013xca}. Another motivation for studying these objects is connected with the brane-world models. For example, the interaction of a bulk black hole with a brane representing our world \citep{Emparan:1999wa, Frolov:2003mc, Frolov:2004wy, Frolov:2004bq, Majumdar:2005ba} can be used as a toy model for the study of (Euclidean) topology change transitions \citep{Frolov:2006tc}, see also \cite{Kobayashi:2006sb, Albash:2006ew, Hoyos:2006gb} for the holographic interpretation of this phenomenon. This model demonstrates interesting scaling and self-similarity properties during the phase transition that are similar to what happens in the Choptuik critical collapse  \citep{Choptuik:1992jv}.

In this section we study strings and branes  in the higher-dimensional  Kerr--NUT--(A)dS spacetimes. A worldsheet of a $p$-brane is a $(p+1)$-dimensional submanifold of the $D$-dimensional spacetime with metric $g_{ab}$. We assume that $\zeta^A$, $(A=0,1,\ldots ,p)$ are coordinates on the brane submanifold, and equations $x^a=x^a(\zeta^A)$ define the  embedding of the brane in the bulk spacetime. This embedding induces the metric $\gamma_{AB}$ on the brane
\be\label{gammaAB}
\gamma_{AB}=\frac{\partial x^a}{\partial{\zeta^A}}\frac{\partial x^a}{\partial{\zeta^B}}\,g_{ab}\, .
\ee
In the `test field approximation', that is when one neglects the effects connected with the thickness and tension, the evolution of the $p$-brane is described by the Dirac--Nambu--Goto action
\be\label{DNG}
I=-\mu\int d^{p+1}\zeta\sqrt{\det(\gamma_{AB})}\,,
\ee
where $\mu$ is the brane tension. The equations obtained by variation of this action with respect to $x^a(\zeta)$, which describe the brane motion, are non-linear and in general it is very difficult to solve them \citep{Stepanchuk:2012xi}.
However, there exists a remarkable exception, that of the {\em stationary strings} and {\em $\xi$-branes}. In what follows we concentrate on the equations for these objects in the the Kerr--NUT--(A)dS spacetimes. See \cite{Kozaki:2009jj, Kozaki:2014aaa} for a detailed discussion of the motion of these objects restricted to the  Minkowski space.

\subsubsection{Killing reduction of action for a stationary string}

Following \cite{Kubiznak:2007ca}, we will first discuss stationary strings.
Consider a stationary spacetime and denote by $\ts{\xi}$ its Killing vector.
A {\em stationary string} is a string whose worldsheet $\Sigma_\xi$ is aligned with this vector. In other words, the surface
$\Sigma_{\xi}$ is generated by a 1-parameter family of the Killing
trajectories (the integral lines of $\ts{\xi}$).

A general formalism for studying a stationary spacetime, based on its foliation by Killing trajectories, was developed by \cite{Geroch:1970nt}. In this approach, one considers a congruence $S$ of all Killing orbits as a quotient space and introduces the structure of the differential Riemannian manifold on it.
A tensor
\be\label{2}
q_{ab}=g_{ab}-\xi_{a}\xi_b/\xi^2\, .
\ee
plays the role of the metric on $S$.

Let us introduce coordinates $x^a=(t,y^i)$, so that the Killing vector $\ts{\xi}=\ts{\partial}_t$, and $y^i$ are coordinates that are constant along the Killing trajectories (coordinates in $S$). In these coordinates
{$q_{ib}\xi^b=0$}, and so
\be
\ts{q}= q_{ij} \ts{d}y^i \ts{d}y^j\, .
\ee
Thus one has
\be
\ts{g} =-F(\ts{d}t+A_i \ts{d}y^i)^2+\ts{q}\,,
\ee
where $F=g_{tt}=-\xi_a\xi^a$ and $A_i=g_{ti}/g_{tt}$ .

In this formalism, a stationary string is uniquely determined by a curve in $S$. Choosing coordinates on the string worldsheet $\zeta^0=t$ and $\zeta^1=\sigma$, the string configuration is determined by $y^i=y^i(\sigma)$, and the induced metric \eqref{gammaAB} reads
\begin{equation}
\ts{\gamma}=\gamma_{AB}\ts{d}\zeta^A\ts{d}\zeta^B=-F(\ts{d}t+A \ts{d}\sigma)^2+\ts{d}l^2\,,
\end{equation}
where
\be
\ts{d}l^2=q \ts{d}\sigma^2\,,\quad {A}=A_i \frac{dy^i}{d\sigma}\,,\quad
q=q_{ij}\frac{dy^i}{d\sigma}\frac{dy^j}{d\sigma}\, .
\ee
The metric $\ts{\gamma}$ has the following determinant: $\det(\gamma_{AB})=-F q$. The Dirac--Nambu--Goto action \eqref{DNG} then reads
\be\label{E}
I=-\Delta t E\,,\quad
E=\mu \int\sqrt{F}dl= \mu\int d\sigma
\sqrt{F{q_{ij}}\frac{dy^i}{d\sigma}\frac{dy^j}{d\sigma}}\,\,.
\ee
Note that, in a stationary spacetime,
the energy density of a string is proportional to its proper length $dl$ multiplied
by the red-shift factor $\sqrt{F}$.
The problem of finding a stationary string configuration therefore reduces to solving a geodesic equation in the  $(D-1)$-dimensional  background with the metric
\begin{equation}\label{H}
\tilde{\ts{q}}=\tilde q_{ij} \ts{d}y^i \ts{d}y^j=F\, q_{ij} \ts{d}y^i \ts{d}y^j\, .
\end{equation}

\subsubsection{Solving stationary string equations in Kerr--NUT--(A)dS spacetimes}

Stationary strings in the four-dimensional Kerr spacetime were studied in \cite{Frolov:1988zn,Carter:1989bs}. It was demonstrated that the effective metric ${\tilde{\ts{q}}}$ inherits symmetry properties of the Kerr metric and the stationary string equations are completely integrable.
The same was found true in \cite{Frolov:2004qw} for the five-dimensional Myers--Perry spacetime.  It was shown in \cite{Kubiznak:2007ca} that these results can  be extended to all higher dimensions. Namely, the equations for a stationary string in the Kerr--NUT--(A)dS spacetime, that is a string aligned along the primary Killing vector $\ts{\xi}=\ts{l}_{(0)}$, are completely integrable in all dimensions.

The integrability follows from the existence of a sufficient number of explicit and hidden symmetries of the $(D{-}1)$-dimensional effective metric ${\tilde{\ts{q}}}$, \eqref{H}.
By construction this metric possesses $(n-1+\eps)$ Killing vectors, $\ts{l}_{(j)}\  (j=1,\dots, n-1+\eps)$.
Let us denote by $\ts{c}_{(k)}$  natural projections of the Killing tensors $\ts{k}_{(k)}$ of the Kerr--NUT--(A)dS spacetime along the primary Killing vector trajectories:
\begin{equation}\label{KTstring}
  \ts{c}_{(k)} =\sum_{\mu=1}^\dg\; \A{k}_\mu\biggl[\; \frac{X_\mu}{U_\mu}\,{\cv{x_{\mu}}^2}
  + \frac{U_\mu}{X_\mu}\,\Bigl(\,\sum_{j=1}^{\dg-1+\eps}
    {\frac{(-x_{\mu}^2)^{\dg-1-j}}{U_{\mu}}}\,\cv{\psi_j}\Bigr)^{\!2}\;\biggr]
  +\eps\,\frac{\A{k}}{\A{\dg}}\cv{\psi_n}^2\,.
\end{equation}
Note that when compared to \eqref{KTjcoor}, the $j=0$ direction $\cv{\psi_0}$  is omitted.
One can check that these objects are Killing tensors for the induced ($D-1)$-dimensional metric $\ts{q}$.
Let us denote
\be
\ts{\tilde k}_{(k)}=\ts{c}_{(k)}-F_{(k)}\ts{\tilde q}^{-1}\, ,
\ee
where
\be
F_{(k)}=\sum_{\mu=1}^n \frac{X_\mu A_\mu^{(k)}}{U_\mu}+\eps\frac{c A^{(k)}}{A^{(n)}}\,.
\ee
Then it is possible to check that these $(n-1)$ objects $\ts{\tilde k}_{(k)}$ $(k=1, \dots, n-1)$ are irreducible Killing tensors for the the metric $\tilde{\ts{q}}$.

%

The Killing tensors $\ts{\tilde k}_{(k)}$, together with the metric $\ts{\tilde q}$ and the Killing vectors $\ts{l}_{(j)}$ all mutually Nijenhuis--Schouten commute. Their existence therefore implies a complete set of mutually commuting constants of geodesic trajectories {in the geometry $\ts{\tilde q}$}. Hence, the stationary string configurations in the Kerr--NUT--(A)dS spacetimes are completely integrable.

Let us conclude with the following remarks: (i) Although stationary string configurations in Kerr--NUT--(A)dS spacetimes are completely integrable, this is not true for strings aligned along other (rotational) Killing directions; the primary Killing vector is very special in this respect; (ii) A stationary string near a five-dimensional charged Kerr-(A)dS black hole was discussed in \cite{Ahmedov:2008pq}; (iii) The presented formalism of Killing reduction of the Dirac--Nambu--Goto action has been generalized to the case of spinning strings in \cite{Ahmedov:2009ni};  (iv) More recently, the notion of a stationary string has been generalized to the so called self-similar strings in \cite{Igata:2016uvp}.

\subsubsection{$\xi$-branes}

The notion of a stationary string readily generalizes to that of \emph{$\xi$-branes} \citep{Kubiznak:2007ca} which are $p$-branes
formed by a 1-parametric family of Killing surfaces.
Suppose a $D$-dimensional spacetime  admits $p$ mutually commuting Killing vectors $\ts{\xi}_{(M)}$ ($M=1,\ldots,p$).  According to the Frobenius theorem the set of $p$ commuting vectors defines a $p$-dimensional submanifold, which has the property that vectors $\ts{\xi}_{(M)}$  are tangent to it. We call such a submanifold \emph{a Killing surface}.

Similarly to Geroch formalism for one Killing vector, one can define a quotient space $S$, determined by the action of the isometry group generated by the Killing vectors $\ts{\xi}_{(N)}$ on $M$. In other words, $S$ is the space of Killing surfaces. The spacetime metric $\ts{g}$ then splits into a part $\ts{\Xi}$ tangent to Killing surfaces and a part $\ts{q}$ orthogonal to them
\be\label{gsplitxibr}
g_{ab}=q_{ab}+\Xi_{ab}\,.
\ee
The tangent part $\ts{\Xi}$ can be written as \citep{Mansouri:1983jj}
\be
\Xi_{ab}=\sum_{M,N=1}^p \Xi^{-1MN}\,\xi_{(M)a}\,\xi_{(N)b}\, ,
\ee
where $\Xi_{MN}=\xi_{(M)}^a\xi_{(N)}^b g_{ab}$ is a $(p\times p)$ matrix and $\Xi^{-1MN}$ is its inverse, $\Xi^{-1MN}\Xi_{NK}=\delta^K_M$.

Let us introduce adjusted coordinates $x^a=(y^i, \psi^M)$ such that $y^i$ ($i=1,\dots,D-p$) are constant along the Killing surfaces, and Killing coordinates $\psi^M$ ($M=1,\dots,p$) are defined as $\ts{\xi}_{(M)} = \cv{\psi^M}$. {Since $q_{a\psi^N}=q_{ab}\xi^b_{(N)}=0$, one has $\ts{q}=q_{ij} \ts{d}y^i \ts{d}y^i$.}
On other hand, vectors $\cv{y^i}$ are not, in general,  orthogonal to the Killing surfaces. It means that $\xi_{(M)a}\equiv g_{ab}\xi_{(M)}^b$ and $\Xi_{ab}$ have both tangent and orthogonal components.
In other words we have
\be
\ts{g}=q_{ij}\ts{d}y^i \ts{d}y^j+\Xi_{ab} \ts{d}x^a \ts{d}x^b\, .
\ee

The configuration of a $\xi$-brane is defined by giving  functions $y^i=y^i(\sigma)$. Denoting by
\be
q=q_{ij}\frac{dy^i}{d\sigma}\frac{dy^j}{d\sigma}\,,
\ee
the Dirac--Nambu--Goto action \eqref{DNG} then reduces to the following expression:
\be
I=- V  E\,,\quad  E=\mu\int \sqrt{q   F} d\sigma\,,
\ee
where $V=\int d^p \psi$ and $F=\det(\Xi_{MN})$.

Thus after the dimensional reduction the problem of finding a
configuration of a $\xi$-brane  reduces to a problem of solving a
geodesic equation in the  reduced $(D-p)$-dimensional space
with the effective metric
\begin{equation}\label{HH}
\tilde{\ts{q}}=F\, q_{ij} \grad y^i\grad y^j\, .
\end{equation}

In general, the integrability of $\xi$-branes is not obvious; see \cite{Kubiznak:2007ca} for a discussion of  special integrable cases.

\subsection{Generalized Kerr--NUT--(A)dS spacetimes}
\label{ssc:GenKNA}

So far our discussion was mostly concentrated on Kerr--NUT--(A)dS spacetimes. Such spacetimes represent a unique geometry admitting the principal tensor which is a closed conformal Killing--Yano 2-form whose  characteristic feature is that it is {\em non-degenerate}. However, it is very constructive to relax the last requirement and consider more general geometries that admit a possibly degenerate closed conformal Killing--Yano 2-form. Such geometries are now well understood and are referred to as the {\em generalized Kerr--NUT--(A)dS spacetimes} \citep{Houri:2008th, Houri:2008ng, Oota:2008uj, Yasui:2011pr}.
These metrics describe a wide family of geometries, ranging from the K\"ahler metrics, Sasaki--Einstein geometries, generalized Taub-NUT metrics, or rotating black holes with some equal and/or some vanishing rotation parameters.

\subsubsection{General form of the metric}
The generalized Kerr--NUT--(A)dS spacetime possesses a bundle structure. The fiber is the $2N$-dimensional Kerr--NUT--(A)dS metric. The base $B$ takes a form of the product space $B=M^1\times M^2\times\dots M^I\times M^0$, where the manifolds $M^i$ are $2m_i$-dimensional K\"ahler manifolds with metrics $\tens{g}^i$ and K\"ahler 2-forms $\tens{\omega}^i=\tens{dB}^i$, and $M^0$ is an `arbitrary' manifold of dimension $m_0$ and a metric $\tens{g}^0$. This means that the total number of dimensions $D$ decomposes as
\be
D=2N+2|m|+m_0\,,\quad |m|=\sum_{i=1}^I m_i\, .
\ee
The generalized Kerr--NUT--(A)dS metric takes the following form:
\be\label{GeneralizedKerrNUTAdS}
\tens{g}=\sum_{\mu=1}^N \frac{\tens{d}x_\mu^2}{P_\mu(x)}+\sum_{\mu=1}^N P_\mu(x)\Bigl(\sum_{k=0}^{N-1}A_\mu^{(k)}\tens{\theta}_k\Bigr)^2
+\sum_{i=1}^I\prod_{\mu=1}^N(x_\mu^2-\xi_i^2)\tens{g}^i+A^{(N)}\tens{g}^0\,,
\ee
where
\begin{equation}
\begin{aligned}
\tens{\theta}_k&=\tens{d}\psi_k-2\sum_{i=1}^I(-1)^{n-k}\xi_i^{2(N-k)-1}\tens{B}^i\,,\\
P_\mu&=X_\mu(x_\mu)\Bigl[x_\mu^{{m}_0}\prod_{i=1}^I(x_\mu^2-\xi_i^2)^{m_i}(-1)^N
U_\mu\Bigr]^{-1}\,.
\end{aligned}
\end{equation}
The corresponding closed conformal Killing--Yano 2-form is degenerate and reads
\be
\tens{h}=\sum_{\mu=1}^N x_\mu \tens{d} x_\mu\wedge\Bigl(\sum_{k=1}^{N-1}A_\mu^{(k)}\tens{\theta}_k\Bigr)
+\sum_{i=1}^I \xi_i \prod_{\mu=1}^N (x_\mu^2-\xi_i^2)\tens{\omega}^i\,.
\ee
Here, the quantities $A_\mu^{(k)}\,, A^{(k)}$ and $U_\mu$ are defined in terms of coordinates $x_\mu$ exactly in the same way as in the Kerr--NUT--(A)dS case, with $n$ replaced by $N$ in the sums/products. Note also that besides the familiar coordinates $x_\mu$ and $\psi_k$, the generalized Kerr--NUT--(A)dS spacetimes also possess a number of coordinates that implicitly characterize the base manifolds.

Coordinates $x_\mu$ are the non-constant functionally independent eigenvalues of $\tens{h}$, whereas parameters $\xi_i$ stand for the non-zero
constant eigenvalues of $\tens{h}$, each having multiplicity $m_i$ that determines the dimension of K\"ahler manifolds $M^i$. The dimension
$m_0$ of the manifold $M^0$ equals the multiplicity of the zero value eigenvalue of $\tens{h}$.
For $m_0=1$, the metric $\tens{g}^0$ can take a special form
\be\label{special}
A^{(N)}\tens{g}^0=\frac{c}{A^{(N)}}\Bigl(\sum_{k=0}^N A^{(k)}\tens{\theta}_k\Bigr)^2\,.
\ee

Let us stress that the generalized Kerr--NUT--(A)dS metrics do not necessarily admit the Killing tower of symmetries. The presence of a degenerate closed conformal Killing--Yano tensor is not enough to generate this full tower and much smaller subset of symmetries exists in these spacetimes. In particular, metrics $\tens{g}^i$ are in general `arbitrary' K\"ahler metrics without any additional symmetries.

With a proper choice of the metric functions $X_\mu(x_\mu)$ and the base metrics, the generalized Kerr--NUT--(A)dS spacetimes become solutions of the Einstein equations.
Namely, assuming that the base metrics $\tens{g}^0$ and $\tens{g}^i$ are Einstein spaces with cosmological constants $\lambda^0$ and $\lambda^i$, respectively, the generalized Kerr--NUT--(A)dS metric solves the vacuum Einstein equations  with the cosmological constant, $\Ric_{ab}=\lambda \tens{g}_{ab}$,
provided the metric functions $X_\mu$ take the following form:
\be
X_\mu=x_\mu\Bigl(b_\mu+\int\chi(x_\mu)x_\mu^{m_0-2}\prod_{i=1}^I(x_\mu^2-\xi_i^2)^{m_i}dx_\mu\Bigr)\,,
\ee
where
\be
\chi(x)=\sum_{i=-\eta}^N\alpha_i x^{2i}\,,\quad \alpha_n=-\lambda\,.
\ee
Here $b_\mu$ and $\alpha_i$ are constant parameters. For convenience, we also introduced a parameter $\eta$ which takes a value $\eta=0$ for a general $\tens{g}^0$ and $\eta=1$ for the special choice of $\tens{g}^0$ given by \eqref{special}.
The constants $\alpha_i$ are constrained by the requirement that $\lambda^i$ are given by $\lambda^i=(-1)^{N-1}\chi(\xi_i)$.
Moreover, for $\eta=0$ we have $\alpha_0=(-1)^{N-1}\lambda^0$, while for $\eta=1$ one has
\be
\alpha_0=(-1)^{n-1}2c\sum_{i=1}^I\frac{m_i}{\xi_i^2}\,,\quad \alpha_{-1}=(-1)^{N-1}2c\,.
\ee


\subsubsection{{Concrete examples}}

The on-shell generalized Kerr--NUT--(A)dS metrics \eqref{GeneralizedKerrNUTAdS} describe a {\em large family} of vacuum (with cosmological constant) geometries of mathematical and physical interest. To obtain concrete examples one may simply specify the base metrics and the parameters of the solution.

To illustrate, a subfamily of solutions with vanishing NUT charges, describing the {\em Kerr-(A)dS black holes} \citep{Gibbons:2004js,Gibbons:2004uw} with partially equal and some vanishing angular momenta, has been identified in \cite{Oota:2008uj}. Namely, in odd dimensions the general-rotating Kerr-(A)dS spacetime \citep{Gibbons:2004js,Gibbons:2004uw} has an isometry $\mathbb{R}\times U(1)^{n}$ and corresponds to identifying the base space with a product of the 2-dimensional Fubini--Study metrics, $B=\mathbb{CP}^{1}\times \dots \times \mathbb{CP}^{1}$. When some of the rotation parameters become equal, the symmetry is enhanced and the dimension of the corresponding Fubini--Study metric enlarges. In particular, equal spinning Kerr-(A)dS black hole has $B=\mathbb{CP}^{n-1}$ and its symmetry is $\mathbb{R}\times U(n)$, see \cite{Oota:2008uj,Yasui:2011pr} for more details.

Another example is that of `{\em NUTty spacetimes}' describing twisted and/or deformed black holes has been studied more recently in \cite{Krtous:2015zco}. Such black holes correspond to the even-dimensional `warped structure' where all the K\"ahler metrics $\tens{g}^i$ identically vanish and the metric $\tens{g}^0$ becomes again the Kerr--NUT--(A)dS spacetime. As discussed in \cite{Krtous:2015zco}, these solutions have a full Killing tower of symmetries.

\subsubsection{Special Riemannian manifolds}

There is yet another, very effective, method for obtaining concrete examples of generalized Kerr--NUT--(A)dS metrics: the method of taking special limits of the original (possibly off-shell) Kerr--NUT--(A)dS spacetimes \eqref{KerrNUTAdSmetric}. Especially interesting are the `singular limits' where some of the originally functionally independent eigenvalues of the principal tensor become equal, constant, or vanish, or some of the original parameters of the Kerr--NUT--(A)dS metrics take special values/coincide. In what follows we shall give several examples of such limits that lead to interesting geometries.

As shown by \cite{Geroch:1969ca}, limiting procedures of this kind are generally non-unique.  This is related to a well known ambiguity in constructing the limiting spaces when some of the parameters limit to zero: there is always a possibility to make a coordinate transformation depending on the chosen parameters, before taking the limit. As we shall see on concrete examples below, to escape the pathology and to achieve a well defined limit, one should properly rescale both the metric parameters and the coordinates.

An important class of metrics that can be obtained by a certain scaling (supersymmetric) limit \citep{Martelli:2005wy, Chen:2006xh, Hamamoto:2006zf, Kubiznak:2009ad} of the Kerr--NUT--(A)dS metrics \eqref{KerrNUTAdSmetric} and belongs to the generalized spacetimes discussed in this section is that of {\em special Riemannian manifolds}.
In even dimensions, the corresponding limit is achieved by setting
\be
x_\mu\to 1+\epsilon x_\mu\,,
\ee
followed by taking $\epsilon\to 0$, which effectively amounts to
setting all the functionally independent eigenvalues of the principal tensor equal to one. When accompanied by an appropriate singular rescaling of Killing coordinates, see  \cite{Kubiznak:2009ad}, the principal tensor becomes completely degenerate and yields the K\"ahler 2-form.

In this way one can obtain the (most general explicitly known) K\"ahler metric $\ts{g}_{\rmlab{K}}$, together with the associated K\"ahler potential $\ts{B}$, and K\"ahler 2-form $\ts{\omega}=\ts{dB}$,
\begin{equation}\label{Kahler}
\begin{aligned}
\ts{g}_{\rmlab{K}}&=
  \sum_{\mu=1}^n\Bigl[\frac{\Delta_\mu}{X_\mu(x_\mu)}\tens{d}x_\mu^2+\frac{X_\mu(x_\mu)}{\Delta_\mu}
  \Bigl(\sum_{j=0}^{n-1}\sigma_{\mu}^{(j)}\tens{d}\psi_j\Bigr)^2\Bigr]\,,\\
{ \tens{B}} &=\sum_{k=0}^{n-1} \sigma^{(k+1)}\tens{d}\psi_k\,,
\end{aligned}
\end{equation}
where
\be
\Delta_{\mu}=\prod_{\nu\ne\mu}(x_{\nu}-x_{\mu})\,,\quad
\sigma^{(k)}_\mu=\sum_{\substack{\nu_1<\dots<\nu_k\\\nu_i\ne\mu}}\!\!\!\!\!x_{\nu_1}\dots x_{\nu_k},\quad
\sigma^{(k)}=\sum_{\nu_1<\dots<\nu_k}\!\!\!\!\!x_{\nu_1}\dots x_{\nu_k}\;\label{vztahy2}.
\ee
With the following choice of metric functions $X_\mu$:
\be\label{X_EK}
X_\mu=-4\prod_{i=1}^{n+1}(\alpha_i-x_\mu)-2b_\mu\,,
\ee
where $\alpha_i$ and $b_\mu$ are free parameters, we recover the Einstein--K\"ahler manifold, obeying
\be\label{Ric_EK}
\Ric_{\rmlab{K}}=(2n+2)\tens{g}_{\rmlab{K}}\,.
\ee

The metric is identical to the Einstein--K\"ahler metric admitting the non-degenerate Hamiltonian 2-form constructed in \cite{apostolov2006hamiltonian}, or to the metric constructed by the requirements of separability in \cite{Kolar:2015hea}.
The K\"ahler manifold \eqref{Kahler} is Ricci flat provided instead of \eqref{X_EK} we set
$X_\mu=-4\prod_{i=1}^{n}(\alpha_i-x_\mu)-2b_\mu$; see \cite{Chen:2006xh} where such a metric is derived by taking the BPS limit of the even-dimensional Kerr--NUT--(A)dS spacetime.

Having obtained the Einstein--K\"ahler manifold \eqref{Kahler}, \eqref{X_EK}, one can apply the procedure \citep{Gibbons:2002th} to construct
the most general known Einstein--Sasaki space \citep{Chen:2006xh}, constructed as a $U(1)$ bundle over the Einstein--K\"ahler base:
\be\label{bundle}
\tens{g}_{\rmlab{ES}}=\tens{g}_{\rmlab{K}}+\ts{\eta}\,\ts{\eta}\,,
\ee
where $\ts{\eta}=2\tens{B}+\tens{d}\psi_n$ is the Sasakian 1-form, and the new $(2n+1)$-dimensional Einstein--Sasaki space obeys
\be\label{Ric_ES}
\Ric_{\rmlab{ES}}=2n \tens{g}_{\rmlab{ES}}\,.
\ee
By restricting the parameters in \eqref{X_EK}, one can obtain a complete and non-singular manifold, see e.g., \cite{Yasui:2011pr} for an example.

\subsubsection{Partially rotating deformed black holes}

Another class of generalized Kerr--NUT--(A)dS metrics is obtained when one tries to `switch off' some of the rotation parameters of the canonical metric \eqref{KerrNUTAdSmetric}. In the Lorentzian signature this yields partially rotating black holes that are deformed by the presence of NUT charges.
Similar to the special Riemannian manifolds above, these metrics possess enough explicit and hidden symmetries, inherited from the original Kerr--NUT--(A)dS spacetime, to guarantee the complete integrability of geodesic equations.
In the following we sketch the idea of the corresponding limit, generalizing the procedure performed in \cite{Oota:2008uj} for the case of vanishing NUT parameters. The details of the construction can be found in \cite{Krtous:2015zco}.

For simplicity let us concentrate on the even-dimensional case, $D=2n$.
{We start with the Kerr--NUT--(A)dS spacetime, \eqref{KerrNUTAdSmetric},  where  $n$ coordinates} $x_{\mu}$ are eigenvalues of the principal tensor $\ts{h}$, while other $n$ coordinates, $\phi_k$, are Killing parameters. For the black hole case one of the coordinates, $x_n$ is identified with the radial coordinate $r$, while the other $n-1$ coordinates $x_1,\ldots,x_{n-1}$ are `angle coordinates'. Besides the cosmological constant, the metric contains $2n-1$ {arbitrary} parameters, describing the mass, $n-1$ rotation parameters $a_{\mu}$, and $n-1$ NUT parameters. As we described in section~\ref{ssc:SubcasesLorentzian}, we assume the following ordering of coordinates $x_{\mu}$ and rotational parameters $a_{\mu}$:
\be\label{xxaa}
{}^- x_1 < x_1<a_1<x_2<a_2< \ldots <x_{n-1}<a_{n-1}\, ,
\ee
cf.\ also figure~\ref{fig:XrangesBH}. Lower bound ${}^- x_1$ has property that when $a_1\to 0$ its value also tends to 0.

It is obvious from this ordering that in the limit when the first $p$ rotation parameters $\{ a_1,\ldots, a_p\}$ tend to zero, the first $p$
angle coordinates, grasped between them, must tend to zero as well. In other words,
to preserve the regularity of the metric one needs to, besides rescaling the rotation parameters, also
properly rescale the first $p$ angle coordinates. As shown in \cite{Krtous:2015zco} this can be consistently done.

As a result, the principal tensor $\ts{h}$ becomes degenerate and its {matrix} rank becomes $2(n-p)$.
The number of the rank-2 Killing tensors, generated from $\ts{h}$ is reduced to $n-p$. At the same time, the limiting procedure generates new additional hidden symmetries, which provide one with additional $p$ quadratic in momenta integrals of geodesic motion. The number of the first order in momenta integrals of motion, associated with Killing vectors, remains the same: $n$. Thus the total number of the integrals of motion, $2n$, is sufficient to guarantee complete integrability of geodesics in the limiting spacetime.

The resulting metric has one less parameter and is a special case of the generalized Kerr--NUT--(A)dS metric with $N=n-p$, $|m|=0$, and $m_0=2 p$. It has a warped structure: both components in the warped product are lower-dimensional Kerr--NUT--(A)dS metrics. We refer to \cite{Krtous:2015zco} for more details and explicit formulas.
A symmetry structure of warp product metrics has been studied in \cite{Krtous:2015ona} and we will return to it in the next section.

\subsubsection{NUTty spacetimes and near horizon geometries}

Final interesting limiting cases of the Kerr--NUT--(A)dS metric that we are going to discuss in this section are those of NUTty spacetimes and near horizon geometries. They can be obtained as follows.
Consider a coordinate $x_{\mu}$. It belongs to an interval given by the roots of the metric function, see sections~\ref{ssc:SubcasesEuclidean} and \ref{ssc:SubcasesLorentzian}. Now, we want to study a `double-root' limit of this metric function.  In such a limit, the end points of the interval tend one to the other and the value of the coordinate $x_{\mu}$,  which is grasped between them,  becomes in general a non-vanishing constant. This implies the degeneracy of the principal tensor. Such double-root limits generalize two interesting cases known from four dimensions: the Taub--NUT limit and the near-horizon limit of the extremal Kerr black hole. As earlier, the corresponding limiting procedure has to be accompanied by a proper rescaling of coordinates.

It was shown in \cite{Kolar:2017vjl}, that when the double-root limit is taken for all angular coordinates $x_{\mu}$, it leads to the `multiply-NUTty spacetime', obtained by \cite{Mann:2005ra,Mann:2003zh}.

If the double-root limit is taken for the metric function governing the position of the horizons, it leads to the near-horizon limit of the extremal black hole metrics, which is similar to the extreme Kerr throat geometry in four dimensions \citep{Bardeen:1999px}. This higher-dimensional limiting spacetime geometry has enhanced symmetry, while some of the hidden symmetries of the original spacetime encoded by Killing tensors become reducible.

\medskip
\details{In the near horizon limit of an extremal Myers--Perry black hole in an arbitrary dimension the isometry group of the metric is enhanced to include the conformal factor $SO(2,1)$. In particular, when all $n$ parameters of the rotation are equal this group is $SO(2,1)\times U(n)$ \citep{Galajinsky:2012vh}. For the near horizon extremal Myers--Perry metric one of the rank 2 Killing tensors  decomposes into a quadratic combination of the Killing vectors corresponding to the conformal group, while the remaining ones are functionally independent \citep{Chernyavsky:2013hia}. Similar result is valid for the Kerr--NUT--(A)dS metric. 
Namely, for the near horizon extremal Kerr--NUT--(A)dS geometry only one rank-2 Killing tensor decomposes into a quadratic combination of the Killing vectors, which are generators of conformal group, while the others are functionally independent \citep{Xu:2015dxa}.}
\medskip

Additional details and the discussion of various  limiting geometries corresponding to double root limits can be found in \cite{Kolar:2017vjl}.

\subsection{Lifting theorems: hidden symmetries on a warped space}
\label{ssc:Warping}

As we have seen in the previous chapters, the existence of hidden symmetries imposes strong restrictions on the background geometry.
Consequently, not every {geometry} admits such symmetries. Even if the symmetries are present, finding their explicit form, by solving the corresponding differential equations, is a formidable task.  For this reason, it is of extreme value to seek alternative ways for finding such symmetries.
In this section we proceed in this direction. Namely, we study hidden symmetries on a {\em warped space}, formulating various criteria under which the Killing--Yano and Killing tensors on the base space can be {\em lifted} to symmetries of the full warped geometry.
This decomposes a task of finding such symmetries to a {simpler} problem (that of finding hidden symmetries for a smaller seed metric) and opens a way towards extending the applicability of hidden symmetries to more complicated spacetimes.

To illustrate this on a simple example, let us consider the {\em rotating black string} in five dimensions whose metric can be written in the form
$\ts{g}=\ts{\bar g}+\ts{d}z^2$, where $\ts{\bar g}$ is the Kerr metric:
\begin{equation}\label{Kerr}
\begin{aligned}
\ts{\bar g}&=
-\frac{\Delta}{\rho^2}\left[\grad t - {a} \sin^2\theta \grad\phi \right]^2 + \frac{\rho^2}{\Delta}\grad r^2 + {\rho^2} \grad\theta^2
+ \frac{\sin^2\theta }{\rho^2} \left[a \grad t - {(r^2+a^2)} \grad \phi \right]^2\,,\\
\Delta &= r^2+a^2 - 2mr\,,\quad \rho^2 = r^2+a^2\cos^2\theta\,.
\end{aligned}
\end{equation}
As shown in chapter~\ref{sc:Kerr}, the Kerr metric \eqref{Kerr} admits a non-trivial Killing--Yano 2-form \citep{Penrose:1973}
\be
\tens{\bar f}=a\cos\theta \grad r \wedge \bigl(\grad t -a\sin^2\!\theta \grad \phi\bigr)-r\sin\theta \grad \theta \wedge \bigl(a\grad t-(r^2+a^2)\grad \phi\bigr)\,.
\ee
One can show that $\ts{\bar f}$  immediately lifts to the Killing--Yano 2-form $\tens{f}=\tens{\bar f}$ of the black string in five dimensions.

More generally, following \cite{Krtous:2015ona}, let us consider a warped space $M$, realized as a direct product $M=\tilde{M}\times\bar{M}$ of two manifolds of arbitrary dimensions $\tilde D$ and $\bar D$, with the metric
\begin{equation}\label{warpedmtrc}
    \ts{g} = \ts{\tilde g}+\tilde w^2\ts{\bar g}\,,
\end{equation}
where $\ts{\tilde g}$ is called the base metric, $\ts{\bar g}$ is the seed metric, and
$\tilde w$ is the warp factor. The corresponding Levi-Civita tensor splits as $\ts{\eps} = \tilde w^{\bar D}\ts{\tilde \eps}\wedge \ts{\bar \eps}\,.$ Here we assume that tilded objects ${\tilde{\tens{A}}}$ are non-trivial only in `tilded directions' and depend only on a position in  $\tilde{M}$, and similarly, barred objects ${\bar{\tens{A}}}$ are non-trivial only in `barred directions' and depend on positions in  $\bar{M}$.
Then one can {prove} the following {\em lifting theorems} for various hidden symmetries \citep{benn2006geodesics, Kubiznak:2009sm, Krtous:2015ona}.\\[0.5ex]
\textbf{Theorem:}
{\it Let the seed  metric $\ts{\bar g}$ of the warped geometry \eqref{warpedmtrc} admits a Killing--Yano $p$-form $\ts{\bar f}$ and/or a closed conformal Killing--Yano $q$-form $\ts{\bar h}$.
Then the following forms:
\be
\tens{f} = \tilde w^{p{+}1}\ts{\bar f}\,, \quad \tens{h} = \tilde w^{q{+}1}\ts{\tilde \eps}\wedge \ts{\bar h}\,, \label{CCKYliftb}
\ee
are the Killing--Yano $p$-form and/or the closed conformal Killing--Yano ${(\tilde D{+}q)}$-form
of the full warped geometry \eqref{warpedmtrc}.}\\[1ex]
\textbf{Theorem:}
{\it If ${\bar{\tens{k}}}$ is a rank $r$ Killing tensor of the metric $\ts{\bar g}$, then
\begin{equation}\label{KTliftb}
    {k}^{a_1\dots a_r} = \bar{{k}}^{a_1\dots a_r}
\end{equation}
is a Killing tensor of the full warped geometry $\ts{g}$.}\\[1ex]
\textbf{Theorem:}
{\it Let $\ts{\tilde f}$ be a Killing--Yano ${p}$-form of the seed metric $\ts{\tilde g}$ and let the warped factor $\tilde w$ satisfies
$\tilde{\grad} \bigl(\tilde w^{-(p{+}1)}\ts{\tilde f}\bigr)=0\;.$ Then
\begin{equation}\label{KYliftt}
    \tens{f}=\tilde w^{\bar D}\ts{\tilde f}\wedge \ts{\bar \eps}
\end{equation}
is a Killing--Yano ${(\bar D+p)}$-form of the full metric \eqref{warpedmtrc}.
Similarly, let $\ts{\tilde h}$ be a closed conformal Killing--Yano \mbox{${q}$-form} of $\ts{\tilde g}$ and the the warp factor satisfies
$\ts{\tilde \nabla}\cdot \bigl(\tilde w^{-(\tilde D{+}q{+}1)}\ts{\tilde h}\bigr)=0\;.$
Then
\begin{equation}\label{CCKYliftt}
    \tens{h}=\ts{\tilde h}
\end{equation}
is a closed conformal Killing--Yano ${q}$-form of the metric \eqref{warpedmtrc}.}\\[1ex]
\textbf{Theorem:}
{\it Let ${\tilde{\tens{q}}}$ be a rank 2 conformal Killing tensor of the metric $\ts{\tilde g}$ with its symmetric derivative given by vector $\tilde{\tens{\sigma}}$,
$\tilde \nabla^{(a} \tilde q^{bc)}=\tilde g^{(ab}\tilde\sigma^{c)}$, and the logarithmic gradient $\ts{\tilde \lambda}= \tilde w^{-1}\ts{\tilde d}\tilde w$ of the warp factor satisfies $\tilde{\tens{\sigma}}=2\,\tilde{\tens{q}}\cdot\ts{\tilde \lambda}.$
Then
\begin{equation}\label{CKTliftb}
    {q}^{ab} = \tilde{{q}}^{ab}
\end{equation}
is a conformal Killing tensor of the warped metric $\ts{g}$ and its symmetric derivative is given by vector ${\sigma}^a=\tilde{{\sigma}}^a$.}

{There exist a number of  examples,} e.g., \cite{Krtous:2015ona},  where these theorems  can be applied and exploited for finding hidden symmetries of complicated metrics. For example, a very non-trivial application happens for the NUTty spacetimes \citep{Krtous:2015zco, Krtous:2015ona} which inherit the full tower of hidden symmetries lifted from their two off-shell Kerr--NUT--(A)dS bases $\ts{\tilde g}$ and~$\ts{\bar g}$.

Let us finally mention that the lifting theorems presented in this section are not the only possibility for lifting hidden symmetries to higher-dimensional geometries. For example, a completely {different approach, the so called {\em Eisenhart lift},} \citep{eisenhart1928dynamical} was recently used to construct spacetimes with higher-rank Killing tensors \citep{Gibbons:2011hg} and subsequently applied to more complicated situations, e.g., \cite{Cariglia:2012fi, Galajinsky:2012vn, Cariglia:2012fi, Cariglia:2013efa, Cariglia:2014dfa, Cariglia:2014dwa, Cariglia:2014ysa, Cariglia:2015fva, Galajinsky:2016zer}.

\subsection{Generalized Killing--Yano tensors}
\label{ssc:GenCCKY}

\subsubsection{{Motivation}}


{Till now we have discussed mainly vacuum solutions of the higher-dimensional Einstein equations with or without the cosmological constant. However, we already mentioned that, for example, in the four-dimensional case there exist the charged versions of the Kerr--NUT--(A)dS metric which are solutions of the Einstein--Maxwell equations and which also admit the Killing--Yano tensor (see e.g., \cite{Keeler:2012mq}.) A natural question is how far can one  generalize the presented in this review theory of hidden symmetries to non-vacuum solutions of the Einstein equations. For example, there are known solutions, describing  black holes with non-trivial gauge fields, such as those of various supergravity theories which arise in low energy limits of string theory compactifications. It is also well known that some of these solutions, that can be thought of as generalizations of Kerr--NUT--(A)dS metrics, possess Killing tensors (see, e.g., \cite{Emparan:2008eg} and references therein) and allow separability of the Hamilton--Jacobi and Klein--Gordon equations \citep{Chow:2008fe, Chow:2016hdg}. In fact this is how some of these solutions were `constructed'.

In this section we demonstrate that the properties of such non-vacuum black holes can be explained by the existence of a deeper structure associated with the {\em generalized Killing--Yano tensors}.
}

\subsubsection{Systematic derivation}

The generalized Killing--Yano tensors can be systematically derived by studying symmetry operators of the Dirac operator with fluxes \citep{Houri:2010qc,Kubiznak:2010ig}. The idea of the construction is as follows. In the backgrounds of superstring or supergravity theories, the metric is often supplemented by other fields or fluxes which couple to the spinor field and modify the Dirac equation, which now reads
\be
{{\cal D}}\psi=0\,,\quad {\cal D}=\gamma^a\nabla_a+\sum_p\frac{1}{p!}B_{a_1\dots a_p}\gamma^{a_1}\dots \gamma^{a_p}\,.
\ee
This includes the case of a massive Dirac operator, the Dirac operator minimally coupled to a Maxwell field, the Dirac operator in the presence of torsion, as well as more general operators.

The generalized Killing--Yano tensors are then in one-to-one correspondence with the first-order symmetry operators of this modified Dirac operator ${\cal D}$. In the notations reviewed in appendix~\ref{apx:spinor}, in analogy with section~\ref{ssc:DiracSymOp} such operators can be written as \citep{Benn:1996ia, BennKress:2004, Acik:2008wv, Houri:2010qc, Kubiznak:2010ig}
\be
{\cal L}=\ts{\omega}\cdot \ts{\nabla}+\ts{\Omega}\,,
\ee
where $\ts{\omega}$ and $\ts{\Omega}$ are inhomogeneous forms to be determined. The requirement that this operator is a symmetry operator of  ${\cal D}$ results in a $\ts{B}$-dependent system of differential equations for $\ts{\omega}$, called the {\em generalized Killing--Yano system}. Once $\ts{\omega}$ is known, $\ts{\Omega}$ can also be determined, cf. \eqref{FODiraccomopK}.

In general, the generalized Killing--Yano system couples various homogeneous parts of inhomogeneous form $\ts{\omega}$, and these only decouple for a special form of the flux $\ts{B}$. In particular, this happens for $\ts{B}=i\ts{A}-\frac{1}{4}\ts{T}$, with a 1-form $\ts{A}$ and a 3-form $\ts{T}$, in which case the Killing--Yano system reduces to the torsion generalization of the conformal Killing--Yano equation \eqref{generalizedCKY} below. We refer to \cite{Kubiznak:2010ig} for more details.

\subsubsection{{Killing--Yano tensors in a spacetime with torsion}}

In what follows,  let us focus on a specific {\em `torsion generalization'} of Killing--Yano tensors which finds its applications for a variety of supergravity black {hole solutions}. We assume that the torsion is completely antisymmetric and described by a 3-form $\tens{T}$. It is related to the standard torsion tensor as $T^{d}_{ab}=T_{abc}g^{cd}$.
Let us define a torsion connection $\tens{\nabla}^T$ acting on a vector field $\ts{X}$ as
\be\label{j1-m}
\nabla^T_{\!a} X^b = \nabla_{\!a} X^b + \frac{1}{2}\,T_{ac}^b X^c\,,
\ee
where $\tens{\nabla}$ is the Levi-Civita (torsion-free) connection. Connection $\covd^T$ satisfies the metricity condition, $\covd^T \tens{g}=0$, and has the same geodesics as $\covd$.

The connection \eqref{j1-m} induces a connection {acting on forms. Namely, let $\tens{\Psi}$ be a $p$-form, then}
\be\label{j5-m}
\covd^T_{\!\ts{X}} \tens{\Psi}=\covd_{\!\ts{X}} \tens{\Psi}
  -\frac{1}{2} \bigl(\ts{X}\cdot\tens{T}\bigr)\underset{1}{\wedge} \tens{\Psi}\,,
\ee
{using the contracted wedge product introduced in \eqref{ContrProd}.}
One can then define the following two operations:
\begin{align}
\tens{d}^T \tens{\Psi}&\equiv \covd^T \wedge \tens{\Psi}=\tens{d\Psi}-\tens{T}\underset{1}{\wedge}\tens{\Psi}\,,
\label{j6-m}\\
\coder^T \tens{\Psi}&\equiv-\covd^T\cdot\tens{\Psi}=\coder\tens{\Psi}-\frac{1}{2}\,\tens{T}\underset{2}{\wedge} \tens{\Psi}\,.\label{j7-m}
\end{align}

A {\em generalized conformal Killing--Yano} (GCKY) tensor $\tens{k}$  is a $p$-form satisfying for any vector field $\tens{X}$ \citep{Kubiznak:2009qi}
\be\label{generalizedCKY}
\nabla^T_X \tens{k}-\frac{1}{p+1}\tens{X}\cdot \tens{d}^T \tens{k}+\frac{1}{D-p+1}
\tens{X} \wedge \tens{\delta}^T \tens{k}=0\,.
\ee
In analogy with the Killing--Yano tensors defined with respect to the Levi-Civita connection,
a GCKY tensor $\tens{f}$ obeying
$\tens{\delta}^T \tens{f}=0$ is called a {\em generalized Killing--Yano} (GKY) tensor, and a GCKY $\tens{h}$ obeying
$\tens{d}^T \tens{h}=0$ is a {\em generalized closed conformal Killing--Yano} (GCCKY) tensor.

\details{Interestingly, the GKY tensors were first discussed from a mathematical point of view in \cite{yano1953curvature} many years ago, and rediscovered more recently in \cite{Rietdijk:1995ye, Kubiznak:2009qi} in the framework of black hole physics. The GCKY generalization \eqref{generalizedCKY} has been first discussed in \cite{Kubiznak:2009qi}}.

The following properties, generalizing the properties of conformal Killing--Yano tensors, have been shown in
\cite{Kubiznak:2009qi,Houri:2010fr} for the  GCKY tensors:

\begin{enumerate}
\item A GCKY 1-form is {identical} to a conformal Killing 1-form.
\item The Hodge star $\tens{*}$ maps GCKY $p$-forms to GCKY $(D-p)$-forms. In particular, the Hodge star of a GCCKY $p$-form is a GKY $(D-p)$-form and vice versa.
\item GCCKY tensors form a (graded) algebra with respect to a wedge product, i.e.,  when $\tens{h}_1$ and $\tens{h}_2$ is a GCCKY $p$-form and $q$-form, respectively, then $\tens{h}_3=\tens{h}_1 \wedge \tens{h}_2$ is a GCCKY $(p+q)$-form.
\item Let $\tens{k}$  be a GCKY $p$-form for a metric $\tens{g}$ and a
torsion 3-form $\tens{T}$. Then, $\tens{\tilde k}=\Omega^{p+1} \tens{k}$ is a GCKY $p$-form for the metric $\tens{\tilde g}=\Omega^2 \tens{g}$ and the torsion $\tens{\tilde T}=\Omega^2 \tens{T}$.
\item Let $\tens{\xi}$ be a conformal Killing vector, $\lied_{\ts{\xi}} \tens{g}=2f\tens{g}$, for some function $f$, and $\tens{k}$ a GCKY $p$-form with torsion $\tens{T}$, obeying
$\lied_{\ts{\xi}} \tens{T}=2f\tens{T}$. Then
 $\tens{\tilde k}=\lied_{\ts{\xi}} \tens{k} -(p+1)f\tens{k}$ is a GCKY $p$-form with $\tens{T}$.
 \item
 Let $\tens{h}$ and $\tens{k}$ be two generalized (conformal) Killing--Yano tensors of rank $p$. Then
\be
K_{ab}=h_{(a |c_1\ldots c_{p-1}|}k_{b)}{}^{c_1\ldots c_{p-1}}
\ee
is a (conformal) Killing tensor of rank 2.
\end{enumerate}

The generalized Killing--Yano tensors naturally appear in black hole spacetimes in supergravity theories, where the torsion may be identified with a 3-form field strength. For example, a non-degenerate GCCKY 2-form exists \citep{Kubiznak:2009qi} in the black hole spacetime of \cite{Chong:2005hr}, which is a doubly spinning black hole solution of 5-dimensional minimal supergarvity, described by the Lagrangian density
\be\label{L}
\mathcal{L}=\tens{*}(R+\Lambda)-\frac{1}{2}\tens{F}\wedge \tens{*F}\!+
\frac{1}{3\sqrt{3}}\,\tens{F} \wedge \tens{F}\wedge \tens{A}\,.
\ee
In this case the torsion can be identified with the Maxwell field strength
\be
\ts{T}=\frac{1}{\sqrt{3}}\ts{*F}\,,
\ee
and is, due to the Maxwell equations `harmonic', $\ts{\delta}^T\ts{T}=0, \ts{d}^T\ts{T}=0$. The GCCKY tensor guarantees separability of the Hamilton--Jacobi and Klein--Gordon equations \citep{Davis:2005ys}, as well as the `torsion modified' Dirac equation \citep{Wu:2009cn, Wu:2009ug} in this spacetime.

Another example \citep{Houri:2010fr} is provided by the Kerr--Sen black hole \citep{Sen:1992ua} and its higher-dimensional generalizations \citep{Cvetic:1996dt, Chow:2008fe}, which are solutions to the following action:
\be\label{S}
S=\int_{M^D} e^{\phi\sqrt{D/2-1}}\Bigl(\tens{*}R+\frac{D-2}{2}\,\tens{*d}\phi\wedge \tens{d}\phi
   -\tens{*F}\wedge \tens{F}-\frac{1}{2}\tens{*H}\wedge \tens{H}\Bigr) \,,
\ee
where $\tens{F}=\tens{dA}$ and $\tens{H}=\tens{dB}-\tens{A}\wedge \tens{dA}$. The general multiply-spinning black hole solution admits a non-degenerate GCCKY 2-form which, upon identifying the torsion with the 3-form field strength
\be
\ts{T}=\ts{H}\,,
\ee
is responsible for complete integrability of geodesic
motion and separability of the scalar and Dirac equations. 

The metrics admitting a non-degenerate GCCKY 2-form have been locally classified in \cite{Houri:2012eq}.
In general such metrics admit a tower of Killing tensors but no additional explicit symmetries.
A subfamily of these metrics provided a new class of Calabi--Yau with torsion metrics \citep{Houri:2012eq}, see also \cite{Houri:2012zv} for the generalized Sasaki--Einstein metrics, and \cite{Hinoue:2014zta} for a generalization of the Wahlquist metric.
Further developments on the GKY tensors can be found in \cite{Chow:2015sfk, Chow:2016hdg}. We also refer to the wonderful review on applications of Killing--Yano tensors to string theory by \cite{Chervonyi:2015ima}.

\subsection{Final remarks}
\label{ssc:summary}

This Living Review was mainly devoted to two subjects: hidden symmetries and higher-dimensional black holes. Black holes in higher dimensions find applications in many physical situations. They naturally appear in low energy approximations of string theory, play an important role in brane-world scenarios, as well as provide a window to the nature of gravitational theory in four and higher-dimensions. As we explained in this review, all higher-dimensional Kerr--NUT--(A)dS black holes possess a set of explicit and hidden symmetries, which is sufficient to guarantee complete integrability of geodesic equations and separation of variables in physical field equations. The origin and seed of all these symmetries is a single very special object, called the principal tensor. This is a non-degenerate closed conformal Killing--Yano 2-form. The existence of this object makes properties of higher-dimensional black holes very similar to the properties of the four-dimensional Kerr metric.

During ten years that have passed since the discovery of the principal tensor,  there have been  published many papers devoted to hidden symmetries of higher-dimensional black holes. In the present review, we collected the obtained results and provided the references to the main publications on this subject. It should be mentioned that during the work on the review we also obtained a number of new, yet unpublished, results that fill some loopholes in the literature. For example, we discussed in detail the solution of geodesic equations in terms of the action--angle variables, provided a direct proof of the commutation relations of the objects in the Killing tower without using the explicit form of the metric, studied a possibility of understanding the principal tensor as a symplectic form on the spacetime, or systematically discussed the meaning of coordinates and special cases of the Kerr--NUT--(A)dS metrics.

Let us mention several open problems that are immediately connected to the results presented in this review. For example, we showed that the geodesic equations in rotating black hole spacetimes  are completely integrable in all dimensions.  This provides a highly non-trivial infinite set of completely integrable dynamical systems. This might be of interest to researchers who study (finite-dimensional) dynamical systems. In particular, we demonstrated how the action-angle variables approach can be developed for studying the particle and light motion. This opens an interesting possibility of applying the fundamental theorem of Kolmogorov--Arnold--Moses \citep{Arnold:book} to develop a perturbation theory for slightly distorted geodesics in such spacetimes.
Another interesting mathematical problem,  waiting for its solution, is the study of properties of the solutions of the ordinary differential equations which arise in the separation of variables of the Klein--Gordon and other field equations in the background of higher-dimensional black holes. In particular, it is important to describe properties of higher-dimensional {\em spin-weighted spheroidal harmonics} \citep{Berti:2005gp, Kanti:2010mk, Cho:2011xi, Brito:2012gw, Kanti:2012ac, Kanti:2014vsa}. These functions are defined as solutions of the Sturm--Liouville eigenvalue problem for the second-order ordinary differential equation with polynomial coefficients, see section~\ref{ssc:SepWave}.

There is a number of interesting possible extensions of the presented in this review subjects, which are still waiting for their study. These problems include, for example, the classification and  complete study of  metrics obtained from the Kerr--NUT--(A)dS metrics by different limiting procedures  and, more generally, a thorough study of the generalized Kerr--NUT--(A)dS solutions. It would also be interesting to extend the applicability of hidden symmetries to non-empty and supersymmetric generalizations of the higher-dimensional Einstein equations. More generally, the subject of hidden symmetries has many interesting applications that go well beyond the realms of black hole physics.
It casts a new light on (integrable) dynamical systems, advances mathematical techniques, provides new tools for constructing solutions of Einstein's equations, is related to special Riemannian manifolds, symmetry operators, and the Dirac theory.
We refer to a beautiful review on hidden symmetries in classical and quantum physics \citep{Cariglia:2014ysa}.

We would like to conclude this review by the following remark. The principal tensor, which exists in higher-dimensional black holes, provides us with powerful tools that allow us to study these spacetimes. Why at all the Nature `decided' to give us such a gift?

%% file: apx-notation.tex

\section{Notation and conventions}
\label{apx:notation}

\subsection{Tensor notation}
\label{ssc:TensNot}

We denote tensors, vectors and forms in bold, e.g., metric $\ts{g}$, vector $\ts{X}$, or antisymmetric $p$-form $\ts{\omega}$. Their components are then $g_{ab}$, $X^a$, or $\omega_{a_1\dots a_p}$, respectively. By dot ${\cdot}$ we denote a contraction of two tensors in adjacent indices. For example, $\ts{h}\cdot\ts{X}$ denotes 1-form with components $h_{ac} X^c$ while $\ts{X}\cdot\ts{h}$ has components $X^c h_{ca}$.

We use the spacetime metric for implicit raising and lowering indices, $X^a$ are thus components of a vector and $X_a = g_{ac}X^c$ are components of a corresponding 1-form. In the `index-free' notation the difference is not so clear. In the mathematical literature, it is custom to use  special symbols $\sharp$ and $\flat$ to distinguish related objects: $\ts{X}^{\flat}$ is the 1-form associated with a vector $\ts{X}$ and $\ts{\omega}^{\sharp}$ is the vector associated with a 1-form $\ts{\omega}$. To simplify our notations we do not use these symbols since the meaning is usually obvious from the context.

We use the signature $(-++\dots+)$ for Lorentzian metrics and the sign conventions of \cite{MTW} for the curvature tensors. We use the Einstein summation convention for generic coordinate and tensor indices on any space. However, we do not employ this convention for indices connected with special coordinate charts or vector frames (typically such indices do not run over the whole dimension of space; see, for example, the Greek indices in Kerr--NUT--(A)dS spacetime).

\subsection{Exterior calculus}
\label{ssc:ExtCalc}

In the following we overview various operations with differential forms mainly to fix sign and normalization conventions.

A $p$-form $\ts{\alpha}$ is a completely antisymmetric tensor of rank $(0,p)$. The \defterm{exterior product} of a ${p}$-form ${\ts{\alpha}}$ with a ${q}$-form ${\ts{\beta}}$ is denoted by $\wedge$. Up to a normalization, it is given by the antisymetrization of the tensor product
\begin{equation}\label{wedgedef}
    (\alpha\wedge\beta)_{a_1\ldots a_pb_1\ldots b_q} =
    \frac{(p+q)!}{p!\,q!}\; \alpha_{[a_1\ldots a_p}\,\beta_{b_1\ldots b_q]}\;.
\end{equation}
An \defterm{insertion of a vector} ${\ts{X}}$ into the first slot of a form $\ts{\alpha}$ (the operation which is in the literature often written as ${i_{\!\ts{X}}\ts{\omega}}$) is denoted by ${\ts{X}\cdot\ts{\omega}}$, and given by
\begin{equation}\label{insertiondef}
    (X\cdot\alpha)_{a_2\ldots a_p} = X^a \alpha_{aa_2\ldots a_p}\;.
\end{equation}
The two operations obey the following properties:
\begin{gather}
  \ts{\alpha}\wedge \ts{\beta} = (-1)^{pq}\ts{\beta} \wedge \ts{\alpha}\, ,\label{wedgecom}\\
  \ts{X}\cdot (\ts{\alpha}\wedge \ts{\beta}) =
    ( \ts{X}\cdot\ts{\alpha})\wedge \ts{\beta}
    +(-1)^p \ts{\alpha} \wedge (\ts{X}\cdot \ts{\beta})\, .\label{insertionLeibniz}
\end{gather}%
Using the metric, we can also introduce the scalar product ${\ts{\omega}\bullet\ts{\sigma}}$ of two ${p}$-forms:
\begin{equation}\label{bulletdef}
    \ts{\omega}\bullet\ts{\sigma} = \frac1{p!}\,\omega_{c_1\dots c_p}\,\sigma^{c_1\dots c_p}\;.
\end{equation}

A \defterm{Hodge dual} of a $p$-form $\ts{\alpha}$ is a $(D-p)$-form $*\ts{\alpha}$, defined as
\begin{equation}\label{Hodgedef}
  (* \alpha)_{a_{p+1}\ldots a_{D}}=
  \frac{1}{p!}\, \alpha^{a_1\ldots a_p}\,\eps_{a_1\ldots a_p a_{p+1}\ldots a_{D}}\, ,
\end{equation}
where $\ts{\eps}$ is the totally antisymmetric tensor.
For an arbitrary $p$-form $\ts{\alpha}$ and a vector $\ts{X}$, we have
\begin{gather}
  * * \ts{\alpha} = \epsilon_p\,\ts{\alpha}\, ,\qquad
    \epsilon_p = (-1)^{p(D-p)}\frac{\det g}{\left|\det g\right|}\, ,\label{sss}\\
  *(\ts{\alpha}\wedge \ts{X}) = \ts{X}\cdot (*\ts{\alpha})\,,\qquad
  *(\ts{\alpha}\cdot \ts{X}) = \ts{X}\wedge (*\ts{\alpha})\,,\label{Xhook}
\end{gather}
where ${\ts{\alpha}\cdot \ts{X}=(-1)^{p-1}\ts{X}\cdot\ts{\alpha}}$.

The \defterm{exterior derivative} $\grad $ maps $p$-forms to $(p+1)$-forms. It is defined by the following properties:
\begin{equation}\label{extderdef}
\begin{aligned}
    &\text{(i)}   &&\grad(\ts{\alpha}+\ts{\beta})=\grad\ts{\alpha}+\grad\ts{\beta}\,,\\
    &\text{(ii)}  &&\grad(\ts{\alpha}\wedge \ts{\beta})
      =(\grad\ts{\alpha})\wedge\ts{\beta}+(-1)^p \ts{\alpha}\wedge(\grad\ts{\beta})\,,\\
    &\text{(iii)} &&\grad\grad\alpha = 0\,,\\
    &\text{(iv)}  &&\text{it maps a function $f$ to its differential ${\grad f}$}\;.
\end{aligned}
\end{equation}
The \defterm{co-derivative} $\coder\ts{\alpha}$ of a ${p}$-form ${\ts{\alpha}}$ is the dual operation to the exterior derivative,
\be\label{ddd}
\coder\ts{\alpha} = (-1)^p \epsilon_{p-1}*\grad*\ts{\alpha}\,,
\ee
with ${\epsilon_p}$ given by \eqref{sss}. The exterior derivative can be also expressed in terms of  the antisymmetric part of the metric covariant derivative
\begin{equation}\label{extdcovd}
    (d\alpha)_{a_0\ldots a_p} = (p+1) \nabla_{\![a_0}\alpha_{a_1\ldots a_p]}\;.
\end{equation}
Similarly, the co-derivative can be expressed using the covariant divergence,
\begin{equation}\label{codercovd}
    (\delta\alpha)_{a_2\ldots a_p} = -\nabla^a\alpha_{aa_2\ldots a_p}\;.
\end{equation}
These relations can also be written using the wedge and the dot operations
\begin{equation}\label{extdercodercovd}
    \grad\ts{\alpha}=\covd\wedge\ts{\alpha} \,,\qquad
    \coder\ts{\alpha} = - \covd\cdot\ts{\alpha}\;.
\end{equation}
The duality relations \eqref{Xhook} then read
\begin{equation}\label{covdwegedot}
    \covd\wedge(*\ts{\alpha}) = (-1)^{p-1} * (\covd\cdot\ts{\alpha})\;,\qquad
    \covd\cdot(*\ts{\alpha}) = (-1)^{p} * (\covd\wedge\ts{\alpha})\;.
\end{equation}

We also use the \defterm{inhomogeneous forms}. In terms of its homogenous $p$-form parts ${}^{p\!}\ts{\alpha}$, an inhomogeneous form $\ts{\alpha}$ is given by
\be
\ts{\alpha}=\sum_{p=0}^D \,{}^{p\!}\ts{\alpha}\,.
\ee
The operations $\ts{d}$ and $\ts{\delta}$ act naturally on such forms. In addition, we can introduce \defterm{rank} $(\pi)$ and \defterm{parity} $(\eta)$ operators by:
\be\label{pieta}
\pi\, \ts{\alpha}=\sum_{p=0}^D p\;{}^{p\!}\ts{\alpha}\,,\quad
\eta\, \ts{\alpha}=\sum_{p=0}^D(-1)^p\;{}^{p\!}\ts{\alpha}\,.
\ee

%% file: apx-pscint.tex

\section{Phase space formalism and complete integrability}
\label{apx:pscint}

In this appendix, we briefly describe some general properties of Hamiltonian systems, introduce integrals of motion, and discuss the notion of complete integrability and its relation to separability of the Hamilton--Jacobi equation. We refer the reader to standard books on Hamiltonian dynamics and symplectic geometry, for example \cite{Arnold:book,Goldstein:book}, for further exposure.

\subsection{Symplectic geometry}
\label{ssc:symplgeom}

In Hamiltonian mechanics, a dynamical system is described in terms of the phase space whose geometric representation can be given in terms of  the symplectic geometry, which we now briefly review.

\subsubsection{Symplectic structure}
Let $\Gamma$ be a ${2N}$-dimensional manifold. A \defterm{symplectic structure} on $\Gamma$ is a 2-form $\ts{\Omega}$ which is:
\begin{equation}\label{symplstr}
\begin{aligned}
&i)&&\mbox{\em closed:}          &&\grad\ts{\Omega}=0\,,\\
&ii)&&\mbox{\em non-degenerate:} &&
  \text{for any ${\ts{X}}$, there exists ${\ts{Y}}$ such that
  ${\ts{\Omega}(\ts{X},\ts{Y})\neq 0}$\,.}
\end{aligned}
\end{equation}
The pair $(\Gamma, \ts{\Omega})$ is called a \defterm{symplectic manifold}. It describes a dynamical system with ${N}$ \defterm{degrees of freedom}.

Let ${z}^A$, $A=1,\ldots, 2N$, be coordinates on $\Gamma$. Then, the components $\Omega_{AB}$ of the symplectic structure form an antisymmetric non-degenerate matrix. We can define an \defterm{inverse symplectic form} $\ts{\Omega}^{-1}$, with components  $\Omega^{AB}$, by relations
\be\label{OOcoor}
\Omega_{AC} \Omega^{BC}=\delta_A^B\, .
\ee
The last relation can be briefly written as $\ts{\Omega}\cdot \ts{\Omega}^{-1}=-\ts{I}$, where $\ts{I}$ is a unit tensor with components $\delta_A^B$ and the dot ${\cdot}$ indicates the contraction in adjacent indices.

Note also that the very existence of a non-degenerate 2-form implies that $\Gamma$ has to be even-dimensional. At the same time the closeness of $\ts{\Omega}$  means that (at least locally) there exists a 1-form \defterm{symplectic potential} $\ts{\theta}$, such that~$\ts{\Omega}=\grad\ts{\theta}$.

\subsubsection{Hamiltonian vector flow}

A scalar function on $\Gamma$ is called an \defterm{observable}. We focus on {\em autonomous systems} whose observables do not explicitly depend on time. Given an observable $F$, the symplectic structure defines the corresponding \defterm{Hamiltonian vector field} ${\hamv{F}}$, given by
\begin{equation}\label{hamvect}
    \hamv{F} = \ts{\Omega}^{-1}\cdot\grad F\, \quad \Leftrightarrow\quad  \hamv{F}\cdot \ts{\Omega}=\ts{d}F\,,
\end{equation}
or, in components, ${X_F^A=\Omega^{AB}F_{,B}}$.

The Hamiltonian vector fields preserve the symplectic structure, ${\lied_{\!\hamv{F}}\ts{\Omega}=0}$, where $\lied_{\hamv{F}}$ stands for the {\em Lie derivative} with respect a vector field $\hamv{F}$. We also have $\lied_{\hamv{F}}\ts{\Omega}^{-1}=0$. Employing the Leibnitz property for the Lie derivative, one can easily derive the {\em Liouville's theorem}: The Hamiltonian vector field ${\hamv{F}}$ preserves the phase space volume element ${\ts{\Omega}^{\wedge N}}$ induced by the symplectic structure, ${\lied_{\hamv{F}}\ts{\Omega}^{\wedge N}=0}$.

The integral curves of $\hamv{F}$ determine a map of the phase space into itself, called a \defterm{Hamiltonian flow}. Parameterizing by parameter $\tau$, the integral curves $\gamma(\tau)$ with coordinates $z^A(\tau)$ are given by
\be\label{dz1}
\hamv{F}= \frac{\ts{d}\gamma}{\ts{d}\tau}=\frac{d{z}^A}{d\tau}\,\cv{z^A}\,.
\ee
Since the Hamiltonian flow preserves the symplectic structure, any observable $F$ induces a \emph{symplectomorphism} of the phase space.

\subsubsection{Poisson brackets}

Given two observables $F$ and $G$, the symplectic structure defines another observable, called the \defterm{Poisson bracket} $\{F,G\}$, given by
\be\label{PB}
\{F,G\}=\grad F\cdot \ts{\Omega}^{-1} \cdot \grad G = F_{,A}\,\Omega^{AB}\,G_{,B}\,,
\ee
or equivalently, using \eqref{hamvect},
\be\label{FG}
\{F,G\}
   =\hamv{G} \cdot \grad F
   = - \hamv{F} \cdot \grad G
   = \hamv{F} \cdot \ts{\Omega} \cdot \hamv{G}\, .
\ee
The closeness of $\ts{\Omega}$ implies that  for any three functions $F$, $G$, and $H$ on the
symplectic manifold, we get the {\em Jacobi identity}:
\be\label{JID}
\{\{F,G\},H\}+\{\{G,H\},F\}+\{\{H,F\},G\}=0\, .
\ee
Observables thus form a {\em Lie algebra} with respect to the Poisson bracket. This algebra is related to the Lie algebra of Hamiltonian vector fields by the relation
\be\label{LB-PB}
[\hamv{F},\hamv{G}]= - \hamv{\{F,G\}}\, .
\ee
Here $[\ts{X}, \ts{Y}]$ stands for the {\em Lie bracket} (commutator) of two vector fields $\ts{X}$ and $\ts{Y}$.
We note that the relation \eqref{LB-PB} can be used to prove the Jacobi identity for Poisson brackets \eqref{JID}  by rewriting it into the Jacobi identity for the corresponding Lie brackets.

\subsubsection{Canonical coordinates}

Until now we have not made any particular choice of coordinates. However, it turns out that the symplectic structure allows one to identify a special class of coordinates, in which most of the equations simplify significantly and take a form familiar from the basic courses on theoretical mechanics. The existence of such coordinates follows from the following:\\[0.5ex]
\noindent{\bf Darboux theorem}:
{\it Let $\ts{\Omega}$ be a symplectic structure. Then in a vicinity of a phase space point it is possible to choose coordinates $(q^1,\ldots,q^N,p_1,\ldots,p_N)$, called the \defterm{canonical coordinates}, in which $\ts{\Omega}$ and ${\ts{\Omega}^{-1}}$ take the following canonical forms:
\be\label{Omqp}
\ts{\Omega}=\sum_{i=1}^N \grad q^i\wedge \grad p_i\,, \quad
\ts{\Omega}^{-1}=\sum_{i=1}^N \Bigl(\cv{q^i}\, \cv{ p_i}-\cv{p_i}\, \cv{q^i}\Bigr)\, ,
\ee
and the corresponding symplectic potential reads
\be\label{thpq}
\ts{\theta}=-\sum_{i=1}^N p_i \grad q^i\, .
\ee
}\\[0.5ex]
The components of the symplectic structure and of its inverse thus are
\be
z^A=\left(\begin{array}{c}
                   q^i \\
                   p_i
                 \end{array}\right)\, ,\quad
\Omega_{AB}=\left(\begin{array}{cc}
                                      0 & \delta^j_i \\
                                      -\delta^i_j & 0
                 \end{array}\right)
                \,,\qquad
\Omega^{AB}=\left(\begin{array}{cc}
                                      0 & \delta^i_j \\
                                      -\delta^j_i & 0
                 \end{array}\right)\, .
\ee

Using the canonical coordinates, the Hamiltonian vector field and the Poisson bracket take the familiar forms
\begin{align}
  \hamv{F} &= \sum_{i=1}^{N}
    \Bigl(\frac{\pa F}{\pa p_i}\,\cv{q^i}-\frac{\pa F}{\pa q^i}\,\cv{p_i} \Bigr)\,,\label{HVFC}\\
  \{F,G\}&=\sum_{i=1}^{N}
    \Bigl(\frac{\pa F}{\pa q^i}\frac{\pa G}{\pa p_i}
   -\frac{\pa F}{\pa p_i}\frac{\pa G}{\pa q^i}\Bigr)\, .\label{P.6}
\end{align}
In particular, one has
\begin{equation}\label{cancoordPB}
  \{q^i,p_j\}=\delta^i_j\,.
\end{equation}

A choice of the canonical coordinates is not unique. There exist transformations to different canonical coordinates that preserve the canonical form of the symplectic structure $\ts{\Omega}$. Such transformations are called \defterm{canonical transformations}. In  general, more than one canonical coordinate chart is required to cover a complete symplectic manifold. A transition between such two charts covering a vicinity of some point is given by a canonical transformation. The complete set of canonical charts covering the symplectic manifold is called a {\em symplectic atlas}. The Darboux theorem guarantees the existence of this atlas.

\subsubsection{Time evolution}

The {\em dynamics} is specified by the \defterm{Hamiltonian} $H$, a given scalar function on the phase space. Since we concentrate on \emph{autonomous systems}, we assume that $H$ is time independent. The time evolution in the phase space is then determined by the Hamiltonian flow corresponding to the Hamiltonian ${H}$. In other words, the \defterm{dynamical (phase-space) trajectories} are integral curves of the Hamiltonian vector field ${\hamv{H}}$,
\be
\ts{X}_H=\sum_{i=1}^N\left( \dot q^i\cv{q^i} + \dot p_i\cv{p_i}\right)\,.
\ee
Comparing with \eqref{HVFC}, we arrive at the \defterm{Hamilton canonical equations}:
\be\label{canEqs}
\dot q^i=\frac{\partial H}{\partial p_i}\,,\quad \dot p_i=-\frac{\partial H}{\partial q^i}\,.
\ee
Here, for any observable $F$, $\dot F$ represents its time derivative, i.e., a derivative along the dynamical trajectories:
\begin{equation}\label{timeder}
    \dot F = \frac{dF}{d\tau} = \hamv{H}\cdot\grad F=\{F,H\}\,.
\end{equation}

\subsubsection{Integrals of motion}

An observable $F$, which remains constant along the dynamical trajectories, is called a \defterm{conserved quantity} or an \defterm{integral/constant of motion}. Clearly, it commutes with the Hamiltonian~$H$,
\be
\{F,H\}=0\,.
\ee

It follows from the Jacobi identity \eqref{JID} that given two integrals of motion $F$ and $G$, their Poisson bracket, $K=\{F,G\}$, is also a (not necessarily non-trivial) constant of motion. Two observables $F$ and $G$ are said to \defterm{Poisson commute} provided their Poisson bracket vanishes, $\{F,G\}=0$. Observables that mutually Poisson commute are called \defterm{in involution}.

We have already seen that an observable induces a transformation of the phase space, see \eqref{dz1}. If observable $F$ is a conserved quantity, this transformation commutes with the time evolution, $ [\hamv{F},\hamv{H}]=0$, as follows from the identity \eqref{LB-PB}. Any trajectory in the phase space satisfying the equation of motion can thus be shifted using the transformation generated by $F$ into another trajectory satisfying the equation of motion. This means that the conserved quantities generate symmetries of the time evolution of a dynamical system. This relation is one-to-one: any symmetry of the time evolution is generated by an integral of motion. We can formulate the following:\\[0.5ex]
\noindent{\bf Theorem}:
{\it Let $\ts{Y}$ preserves both the symplectic 2-form, ${\lied_\ts{Y}\ts{\Omega}=0}$, and the Hamiltonian, ${\lied_\ts{Y}H=0}$. Then there exists an integral of motion $I$, such that $\ts{Y}=\hamv{I}$.
}\\[0.5ex]
\details{
This theorem can be viewed as a phase space version of the famous \emph{Noether's theorem} about the correspondence of continuous symmetries and conserved quantities. The Noether's theorem is usually formulated on a configuration space and refers to the symmetries of the action. In the present version, we have rather stated the correspondence  between conserved quantities of a dynamical system and symmetries of the time evolution on the phase space.
}

\subsection{Complete integrability}
\label{ssc:complintgrb}

Dynamical systems may admit more than one symmetry. An important situation occurs when these symmetries commute among each other. A system with the maximal possible number of independent mutually commuting symmetries is called completely integrable. The evolution of such systems is highly `ordered' in the phase space: the trajectories remain in well-defined submanifolds and can be found by a well-defined procedure. Global integrability and chaotic motion are thus in some sense two opposite properties of dynamical systems. In this comparison, the global complete integrability is rare and exceptional, while the chaotic nature is generic. Although exceptionally rare, integrable systems are solvable by analytic methods and play thus a very important role in the study of dynamical systems.

\subsubsection{Liouville's integrability}

In our application we are interested mainly in a regular ordered evolution contingent on the complete integrability of the system. We focus on the local aspects of integrability and will not discuss its global issues. Namely, we concentrate on the local notion of complete Liouville integrability.

In its original sense, integrability means that a system of differential equations can be solved by `{\em quadratures}', that is, its solution can be found in a finite number of well-defined steps involving algebraic operations and integrations of given functions. Thanks to this `prescription', integrable systems are often solvable by analytic methods and thus play a very important role in the study of dynamical systems.

Nowadays, integrability of finite-dimensional dynamical systems is usually characterized by the existence of conserved quantities:\\[0.5ex]
\noindent{\bf Complete Liouville integrability}:
{\it The dynamical system with $N$ degrees of freedom is \defterm{completely (Liouville) integrable} if it admits ${N}$ functionally independent integrals of motion $P_i$ that are in involution:
\begin{equation}\label{ComplInt}
  \{P_i,H\}=0\,,\quad  \{P_i,P_j\}=0\,,\quad i,j=1,\dots, N\,.
\end{equation}
}\\[0.5ex]
Since the total number of independent integrals of motion in involution cannot be larger than $N$, the Poisson-commutation of ${H}$ with all ${P_i}$ implies that the Hamiltonian is function of ${\tpd{P}=(P_1,\dots,P_N)}$,
\be\label{HofP}
H=H(\tpd{P})\,.
\ee%
Note also that for autonomous systems, the Hamiltonian ${H}$ as well as the conserved quantities ${P_i}$ should not explicitly depend on the time parameter, they are just functions on the phase space.

The relation between the existence of conserved quantities and the original notion of integrability was established by \cite{liouville1855note} who proved the following theorem:\\[0.5ex]
\noindent{\bf Liouville's theorem}:
{\it A solution of equations of motion of a completely integrable system can be obtained by {\em quadratures}, that is, by a finite number of algebraic operations and integrations.
}

\subsubsection{Level sets}

%
Before we hint on how the solution by quadratures proceeds, we make a couple of related geometrical comments. First, we introduce the \defterm{level set} ${\levset{\tpd{\Phi}}}$ as a subspace of the phase space ${\Gamma}$ given by fixing the conserved quantities ${\tpd{P}=(P_1,\dots,P_N)}$ to values ${\tpd{\Phi}=(\Phi_1,\dots,\Phi_N)}$,
\be\label{LevSet}
P_{i} = \Phi_i\,, \quad i=1,\ldots ,N\, .
\ee
The functional independence of observables  ${P_i}$ means that the gradients $\grad P_{i}$ are linearly independent at each point of $\levset{\tpd{\Phi}}$, which implies that each level set $\levset{\tpd{\Phi}}$ is an $N$-dimensional submanifold of $\Gamma$.

The involution conditions ${\{P_i,P_j\}=0}$ imply
\be\label{P.12a}
\hamv{P_j} \cdot \grad P_i =  0 \, ,
\ee
which means that the Hamiltonian vectors ${\hamv{P_j}}$ are tangent to the level set ${\levset{\tpd{\Phi}}}$ and thus the level set is invariant under the Hamiltonian flows generated by ${P_i}$. Since the Hamiltonian depends just on ${\tpd{P}}$, the dynamical trajectories (orbits of ${\hamv{H}}$) remain also in the level set ${\levset{\tpd{\Phi}}}$.

Finally, the components of the symplectic structure ${\ts{\Omega}}$ restricted to the \mbox{${N}$-dimensional} level set ${\levset{\tpd{\Phi}}}$ can be evaluated as ${\Omega_{ij} = \hamv{P_i}\cdot\ts{\Omega}\cdot\hamv{P_j}}$. However, using \eqref{FG} we find that they identically vanish on ${\levset{\tpd{\Phi}}}$, i.e., ${\ts{\Omega}|_{\levset{\tpd{\Phi}}}=0}$. The ${N}$-dimensional submanifold of the ${2N}$-dimensional symplectic space ${\Gamma}$ on which the restriction of the symplectic form vanishes is called a \defterm{Lagrangian submanifold}. We have thus found that conserved quantities $P_i$ define the foliation of the phase space into Lagrangian submanifolds ${\levset{\tpd{\Phi}}}$.

\subsubsection{Liouville's procedure}
Let us now return back to the Liouville's integrability theorem. The main idea behind it is that one can use the ${N}$ independent integrals of motion ${\tpd{P}=(P_1,\dots,P_N)}$ as new momentum coordinates and supplement them with ${N}$ new canonically conjugate position coordinates ${\tpu{Q}=(Q^1,\dots,Q^N)}$. Since $H=H(\tpd{P})$, the dynamical equations \eqref{canEqs} become trivial
\be\label{QPeqmot}
\dot{Q}^i =\frac{\partial H}{\partial P_i}(\tpd{P})\, ,\quad
\dot{P}_i =-\frac{\partial H}{\partial Q^i}(\tpd{P})=0\, .
\ee
A solution to these equations is
\be\label{solPQ}
P_i=\text{const}\,,\quad Q^i=\omega^i \tau+\text{const} ,
\ee
where `frequencies'
\be
\omega^i=\frac{\partial H}{\partial P_i}(\tpd{P})
\ee
are constant along dynamical trajectories.
To obtain the solution in original coordinates ${(\tpu{q},\tpd{p})}$, one has to substitute these trivial solutions back into relations between ${(\tpu{q},\tpd{p})}$ and new coordinates ${(\tpu{Q},\tpd{P})}$.

The key problem of the Liouville procedure is thus finding new coordinates ${\tpu{Q}}$ which are canonically conjugate to the integrals of motion ${\tpd{P}}$. To define these coordinates, we start by inverting the original expressions for integrals of motion in terms of original canonical coordinates,
\begin{equation}\label{Pofqp}
    P_i = P_i(\tpu{q},\tpd{p})\,,
\end{equation}
with respect to the momenta,
\begin{equation}\label{pofqP}
    p_i = p_i(\tpu{q},\tpd{P})\,.
\end{equation}
A canonical transformation between the original coordinates ${(\tpu{q},\tpd{p})}$ and new coordinates ${(\tpu{Q},\tpd{P})}$ can be defined by using a generating function ${W(\tpu{q},\tpd{P})}$ obeying the following conditions:
\begin{equation}\label{cantransfgenfc}
    p_i(\tpu{q},\tpd{P}) = \frac{\pa W}{\pa q^i}(\tpu{q},\tpd{P})\,,\quad
    Q^i(\tpu{q},\tpd{P}) = \frac{\pa W}{\pa P_i}(\tpu{q},\tpd{P})\,.
\end{equation}
The first condition is automatically satisfied by the following generating function: 
\begin{equation}\label{genfcqP}
    W(\tpu{q},\tpd{P}) = \int_{\tpu{q}_0}^{\tpu{q}}
       \sum_i  p_i(\tpu{\bar{q}},\tpd{P})\, d\bar{q}^i\,,
\end{equation}
where the integration is performed for fixed values of ${\tpd{P}}$ and starts at an arbitrary chosen origin ${\tpu{q}_0}$ of coordinates ${\tpu{q}}$. The key observation is that this integral does not depend on a path of integration, as follows from the fact that ${\tpd{P}}$'s are in involution.

The second equation in \eqref{cantransfgenfc}, with ${\tpd{P}}$ given by \eqref{Pofqp}, provides a desired definition of coordinates ${\tpu{Q}}$. Indeed, relations \eqref{cantransfgenfc} imply
\be
\grad W=\sum_i \bigl( p_i\, \grad q^i + Q^i\, \grad P_i\bigr)\, .
\ee
Employing $\grad^2 W=0$ and expression \eqref{Omqp} for the symplectic structure in canonical coordinates, one gets
\be
\ts{\Omega}=\sum_i \grad q^i\wedge \grad p_i=\sum_i \grad Q^i\wedge \grad P_i\, .
\ee
This proves that ${(\tpu{Q},\tpd{P})}$ defined in this way are canonical coordinates, and
concludes the construction.

Summarizing, the Liouville's construction consists of: (i) inverting the relations \eqref{Pofqp} for the conserved quantities to obtain \eqref{pofqP} (algebraic operations) (ii) integrating  the generating function \eqref{genfcqP} (a quadrature) (iii) defining the new canonical variables ${(\tpu{Q},\tpd{P})}$ by \eqref{cantransfgenfc} and \eqref{Pofqp} (derivatives and algebraic operations). A solution of the equations of motion is then concluded by (iv) writing down the (in new coordinates trivial) solution \eqref{solPQ} and by (v) inverting the relations \eqref{cantransfgenfc} and \eqref{Pofqp}, which gives the solution in terms of the original coordinates (possibly highly nontrivial algebraic operations).

\subsubsection{Action-angle variables}

It can be demonstrated that for a given foliation of the phase space into its level sets, the coordinates ${\tpu{Q}}$ on the level set are given uniquely up to an affine transformation. The level sets thus possess an affine structure.

When the level set is compact and connected, the affine structure implies that it is isomorfic to an ${N}$-dimensional torus \citep{Arnold:book}. In such a case, coordinates $\tpu{Q}$ can be linearly mixed to form cyclic coordinates ${\tpu{\alpha}=(\alpha_1,\ldots,\alpha_N)}$ on the torus, i.e., \defterm{angle variables} (with a period ${2\pi}$) along main circles of the torus. In order to complement these angles by canonically conjugate coordinates, one needs to perform a transformation of momenta ${\tpd{P}}$ into new conserved quantities $\tpd{I}=\tpd{I}(\tpd{P})$, which label the toroidal level sets in a slightly different way. The so called \defterm{action variables} can be defined as integrals analogous to \eqref{genfcqP}, integrated along the main circles ${\ell_i}$ of the torus,
\begin{equation}\label{Idef}
    I_i(\tpd{P}) = \frac1{2\pi} \int_{\ell_i} p_i(\tpu{\bar{q}},\tpd{P})d\bar{q}^i\;.
\end{equation}
Similarly to \eqref{genfcqP}, the integrals are independent of continuous deformations of the path of integration. They thus measure `invariant' sizes of the toroidal level set ${\levset{\tpd{P}}}$. Together, $(\tpu{\alpha},\tpd{I})$ form the canonical coordinates called the \defterm{action--angle variables} \citep{Arnold:book}.

\subsection{Hamilton--Jacobi equation}
\label{ssc:HamJacEq}

\subsubsection{Time-dependent case}

An alternative method for solving the dynamical system is via the \defterm{Hamilton--Jacobi equation}. This is a partial differential equation for the \defterm{Hamilton's principal function} ${\bar{S}(\tpu{q};\tau)}$, given by
\begin{equation}\label{HJeqt}
    \frac{\pa{\bar{S}}}{\pa \tau}(\tpu{q};\tau) + H\bigl(\tpu{q},\tpd{p};\tau\bigr)\Big|_{\tpd{p}=\frac{\pa{\bar{S}}}{\pa\tpu{q}}(\tpu{q};\tau)} = 0\,,
\end{equation}
where, for a moment, we allowed the Hamiltonian ${H(\tpu{q},\tpd{p};\tau)}$ to depend explicitly on time $\tau$.
It turns out that solving this equation allows one to find a family of trajectories ${(\tpu{q}(\tau),\tpd{p}(\tau))}$ in the phase space ${\Gamma}$ which satisfy the Hamilton canonical equations \eqref{canEqs}. Such trajectories are given by identifying the momenta with the derivatives of Hamilton's principal function
\begin{equation}\label{HJmom}
    p_j(\tau) = \frac{\pa{\bar{S}}}{\pa{q^j}}\bigl(\tpu{q}(\tau);\tau\bigr)\;.
\end{equation}
Plugging these to equations \eqref{canEqs} for the velocities:
\begin{equation}\label{HJvel}
    \dot{q}^i(\tau) = \frac{\pa H}{\pa p_i}\bigl( \tpu{q}(\tau),\tpd{p};\tau\bigr)
    \Big|_{\tpd{p}=\frac{\pa{\bar{S}}}{\pa{\tpu{q}}}\bigl(\tpu{q}(\tau);\tau\bigr)}\;,
\end{equation}
one gets ${N}$ coupled first-order differential equations for positions ${\tpu{q}(\tau)}$. Solutions ${\tpu{q}(\tau)}$ of these equations together with the momenta ${\tpd{p}(\tau)}$ defined by \eqref{HJmom} give the trajectories which satisfy all the Hamilton canonical equations \eqref{canEqs}.

\subsubsection{Autonomous systems}

As in the previous discussion, let us now restrict our attention to the autonomous systems for which the Hamiltonian does not depend explicitly on time. In such a case, the time dependence of the Hamilton's principal function can be solved by the following ansatz:
\begin{equation}\label{HJfc}
    \bar{S}(\tpu{q};\tau) = S(\tpu{q}) - E\tau\,,
\end{equation}
where the function ${S}$ is called the \defterm{Hamilton's characteristic function} \citep{Goldstein:book} and the constant ${E}$ is an \defterm{energy}. Substituting this ansatz into the Hamilton--Jacobi equation \eqref{HJeqt}, it turns out that ${E}$ coincides with the value of the Hamiltonian, and  clearly remains conserved along the trajectory. We thus obtained the following \defterm{time-independent Hamilton--Jacobi equation}:
\begin{equation}\label{HJeq}
    H(\tpu{q},p)\big|_{p=\frac{\pa{S}}{\pa\tpu{q}}(\tpu{q})} = E\,.
\end{equation}
Similarly to the time-dependent case, the function $S(\tpu{q})$ generates a family of trajectories by
\begin{equation}\label{HJvelstat}
    \dot{q}^i(\tau) =
    \frac{\pa H}{\pa p_i}\bigl( \tpu{q}(\tau),\tpd{p}\bigr)
    \Big|_{\tpd{p}=\frac{\pa{S}}{\pa{\tpu{q}}}\bigl(\tpu{q}(\tau)\bigr)}\;.
\end{equation}
As discussed below, different solutions ${S}$ of \eqref{HJeq} generate different families of trajectories.

\subsubsection{Connections to Liouville's integrability}

A solution ${S(\tpu{q},\tpd{\Phi})}$ of the time-independent Hamilton--Jacobi equation \eqref{HJeq} that depends on ${N}$ independent constants ${\tpd{\Phi}}$ is called a \defterm{complete integral}. In this definition, the notion of independence of the constants means that when these constants are varied, they generate dynamical trajectories which fill up the whole phase space.

For completely integrable systems one can show that the generating function ${W(\tpu{q},\tpd{P})}$ defined in \eqref{genfcqP} solves the Hamilton--Jacobi equation in variables ${\tpu{q}}$, provided that the new momenta ${\tpd{P}}$ are kept constant, equal to ${\Phi}$. That is, we have the following complete integral:
\begin{equation}\label{ComplIntInt}
    S(\tpu{q},\tpd{\Phi})=W(\tpu{q},\tpd{\Phi})\;
\end{equation}
which satisfies the time-independent Hamilton--Jacobi equation
\begin{equation}\label{HJeqH}
    H\Bigl(\tpu{q},\frac{\pa{S}}{\pa\tpu{q}}(\tpu{q},\tpd{\Phi})\Bigr) = E\,,
\end{equation}
where ${E}$ is the value of the Hamiltonian given by the values of conserved quantities~${\tpd{P}=\tpd{\Phi}}$,
\begin{equation}\label{Econst}
  E=H(\tpd{\Phi})\,.
\end{equation}
The same complete integral ${S(\tpu{q},\tpd{\Phi})}$ also satisfies the analogous Hamilton--Jacobi equations corresponding to other conserved quantities ${\tpd{P}}$:
\begin{equation}\label{HJeqP}
    P_i\Bigl(\tpu{q},\frac{\pa{S}}{\pa\tpu{q}}(\tpu{q},\tpd{\Phi})\Bigr) = \Phi_i\,.
\end{equation}
In other words, the complete integral $S(\tpu{q},\tpd{\Phi})$ is a common solution of all the Hamilton--Jacobi equations \eqref{HJeqH} and \eqref{HJeqP}.


We have thus demonstrated that {\em complete integrable systems always have the complete integral} of the time-independent Hamilton--Jacobi equation.
In fact, the opposite statement is also true: {\em the existence of the complete integral of the Hamilton--Jacobi equation is a sufficient condition for the system to be completely integrable.}

To prove the latter statement let us assume that the complete integral ${S(\tpu{q},\tpd{\Phi})}$ exists. If so, it can be used as  a generating function ${ W(\tpu{q},\tpd{P}) = S(\tpu{q},\tpd{P})}$ for a canonical transformation $(\tpu{q},\tpd{p})\to(\tpu{Q},\tpd{P})$, i.e., for the transformation given implicitly by \eqref{cantransfgenfc}. One has to solve the left set of equations with respect to new momenta ${\tpd{P}}$. Clearly, these are conserved quantities since they corresponds to the constants that appear in the complete integral. Therefore, observables ${\tpd{P}}$ commute with the Hamiltonian and, since they have been generated by the canonical transformation as new momenta, they are in involution. The system is hence completely integrable.

Finally, let us note that for every choice of constants ${\tpd{\Phi}}$ the Hamilton's characteristic function ${S(\tpu{q},\tpd{\Phi})}$ generates a different family of trajectories in the space of coordinates ${\tpu{q}}$ solving \eqref{HJvelstat}. When lifted to the phase space through \eqref{HJmom}, the trajectories in such a family belong to the level set ${\levset{\tpd{\Phi}}}$. Varying values ${\tpd{\Phi}}$, the trajectories fill up the whole phase space.

\subsubsection{Additive separability of the Hamilton--Jacobi equation}

There exists a powerful method  that turns out to be very useful for finding integrable systems. It is the {\em method of separation of variables} in the Hamilton--Jacobi equation, see e.g., \cite{Carter:1968rr, Carter:1968cmp} for an application of this method in the context of black hole physics.

Specifically, let us consider the time-independent Hamilton--Jacobi equation \eqref{HJeq} and seek, in a given coordinate system, its solution in the form of the following {\em additive separation ansatz}:
\begin{equation}\label{HJsepansatz}
    S(\tpu{q}) = \sum_{i=1}^N S_i(q^i)\,,
\end{equation}
where each function $S_i$ depends only on the corresponding coordinate $q^i$. The ansatz is consistent if its substitution into the Hamilton--Jacobi equation leads to ${N}$ independent ordinary differential equations for functions~${S_i}$. If these functions are labeled by ${N}$ independent constants (which also determine $E$), we have found a complete integral and the system is completely integrable.

Let us stress, however, that whether or not the separation ansatz works depends on the choice of coordinates and for a given dynamical system there is no general prescription for how to seek the convenient coordinates. For this reason, the route from separability of the Hamilton--Jacobi equation to complete integrability, although often fruitful, is more or less a route of trial and error.

Let us assume that the complete integral ${S(\tpu{q},\tpd{\Phi})}$ can be written in the separated form \eqref{HJsepansatz}.
Taking into account that components of the momentum are given by derivatives of Hamilton's characteristic function with respect to ${q}$'s \eqref{HJmom}, the additive separation ansatz requires that
\begin{equation}\label{HJsepps}\notag
    p_i=S_i'(q^i,\tpd{\Phi})\;.
\end{equation}%
In other words, the ${i}$-th component of the momentum has to depend only on one coordinate ${q^i}$ and not on the remaining coordinates ${q^j}$, ${j\neq i}$.  Since the constants ${\tpd{\Phi}}$ can be understood as values of the integrals of motions ${\tpd{P}}$, we actually require that
\begin{equation}\label{pofqPsep}
    p_i = p_i(q^i,\tpd{P})\;,
\end{equation}
that is, each relation \eqref{pofqP} for momentum ${p_i}$ depends only on \emph{one variable} ${q^i}$.
{This is a sufficient and necessary condition for the additive separation of variables in the Hamilton--Jacobi equation \eqref{HJeqH} using the ansatz \eqref{HJsepansatz}.}

Solving \eqref{pofqPsep} with respect to $\tpd{P}$, the Hamilton's characteristic function $S$ in the form \eqref{HJsepansatz} also satisfies all the Hamilton--Jacobi equations \eqref{HJeqP}.

As we demonstrate in section~\ref{ssc:separabilitystructures}, very explicit conditions for separability of the Hamilton--Jacobi equation can be formulated for geodesic motion in a curved spacetime. This is described by a theory of {\em separability structures}, and will be exploited in chapter~\ref{sc:intsep} for the study of geodesics in higher-dimensional black hole spacetimes, generalizing the classic works of \cite{Carter:1968rr, Carter:1968cmp} on geodesic motion around four-dimensional black holes.

\subsection{Covariant formalism on a cotangent bundle}
\label{ssc:PhaseSpcCotBundle}

Dynamical systems are most commonly formulated in terms of the motion on a configuration space $M$. As described in section~\ref{ssc:partcurvspc}, the phase space $\Gamma$ is the cotangent bundle over the configuration space. It has natural symplectic structure \eqref{natsymplstr} and each coordinates $x^a$ on the configuration space generate canonical coordinates $(x^a,p_a)$ on the phase space. We can thus apply all the formalism discussed in the previous sections, simply replacing $(q^i,p_i)\to (x^a,p_a)$.


The Poisson brackets and other quantities on the cotangent space can be written in a more covariant way by using the covariant derivative induced from the configuration space. Given a torsion-free covariant derivative $\covd$ on $M$, one can define the following tensorial quantities: (i) \defterm{covariant partial derivative} $\frac{\covd F}{\pa x}$ with respect to position $x$ and (ii) \defterm{momentum partial derivative} $\frac{\pa F}{\pa \ts{p}}$ with respect to momentum of the observable $F(x,\ts{p})$. {It corresponds to splitting a phase-space tangent vector $\ts{X}$ into its position and momentum parts,
\begin{equation}\label{phspcvectsplt}
    \ts{X} = \ts{u}\cdot\frac{\covd}{\pa x} + \ts{f}\cdot\frac{\pa}{\pa \ts{p}}\;,
\end{equation}}%
see appendix of~\cite{CarigliaEtal:2012b} for more details. The covariant partial derivative essentially `covariantly ignores' the momentum variable. In particular, the partial derivatives of a monomial observable \eqref{monobs}~read
\begin{equation}\label{CovDerTensPow}
    \frac{\nabla_{\!a} F}{\pa x} = (\nabla_{\!a} f^{c_1\dots c_r} )\, p_{c_1}\dots p_{c_r}\;,\quad
    \frac{\pa F}{\pa p_a} = r f^{a c_2\dots c_r} p_{c_2}\dots p_{c_r}\,.
\end{equation}

With such a machinery, the Poisson bracket can be written as
\begin{equation}\label{Pbrcov}
    \{F,G\} = \frac{\nabla_{\!a} F}{\pa x} \frac{\pa G}{\pa p_a}
      - \frac{\pa F}{\pa p_a} \frac{\nabla_{\!a} G}{\pa x}\;.
\end{equation}
Employing this, one can easily derive the expression \eqref{NSbrcov} for the Nijenhuis--Schouten bracket. In particular, on readily gets the expression \eqref{KKKPPP}:
\begin{equation}\label{KTcovder}
    \{K,H\} =
      \frac{\nabla_{\!a} K}{\pa x} \frac{\pa H}{\pa p_a}
      - \frac{\pa K}{\pa p_a} \frac{\nabla_{\!a} H}{\pa x}
      = (\nabla^{a_0}k^{{a}_1\ldots{a}_s})\,
        p_{{a}_0}p_{{a}_1}\ldots p_{{a}_s}\, ,
\end{equation}
where we employed rules \eqref{CovDerTensPow} and that for the Hamiltonian \eqref{RPHam} one has ${\frac{\covd H}{\pa x}=0}$ and ${\frac{\pa H}{\pa p_a}=g^{ab}p_b}$.

%% file: apx-cky.tex

\section{Integrability conditions for conformal Killing--Yano forms}
\label{apx:cky}

\subsection{Laplace operator and conformal Killing--Yano forms}
\label{asc:Laplace}

A particular consequence of the integrability conditions for the conformal Killing--Yano forms is that the action of the Laplace operator on such objects takes a special form. Since this can be shown in a rather elegant way, we start with the discussion of the Laplace operators.

\subsubsection{Laplace operators and Weitzenb\"ock identity}

Let us first list some well-known definitions and identities. For antisymmetric forms one can introduce two Laplace-like operators. The \defterm{de~Rham--Laplace operator} $\Delta\ts{\omega}$ is defined using the exterior derivative and the divergence:
\begin{equation}\label{DeRhamLaplace}
    \Delta\ts{\omega} = - \covd\wedge(\covd\cdot\ts{\omega}) - \covd\cdot(\covd\wedge\ts{\omega})\;,
\end{equation}
while the \defterm{Bochner--Laplace operator} is just a contraction with the second covariant derivative:
\begin{equation}\label{BochnerLaplace}
    \nabla^2\ts{\omega} = \covd\cdot\covd\ts{\omega}\;.
\end{equation}
The two operators differ by terms that are linear in curvature. The so called \defterm{Weitzenb\"ock identity} reads
\begin{equation}\label{WeitzenbockId}
    \Delta\ts{\omega}=-\nabla^2\ts{\omega}+ W\ts{\omega}\;,
\end{equation}
where the \defterm{Weitzenb\"ock operator} $W$ acts on a $p$-form as
\begin{equation}\label{WeitzenbockOp}
    (W \omega)_{a_1a_2\dots a_p} = p\, R_{c[a_1}\,\omega^{c}{}_{a_2\dots a_p]}
      - \frac{p(p-1)}{2}R_{cd[a_1a_2}\,\omega^{cd}{}_{\dots a_p]}\;.
\end{equation}

\subsubsection{Action of Laplace operators on conformal Killing--Yano forms}

Let us start with a general conformal Killing--Yano form $\ts{\omega}$ obeying the conformal Killing--Yano condition \eqref{CKYeq}. Tearing off the vector ${\ts{X}}$, it can be written in a form, which respects the duality between the dot and wedge operations,
\begin{equation}\label{CKYdefforint}
    \covd\ts{\omega} = \frac1{p+1}\,\ts{g}\cdot (\covd\wedge\ts{\omega}) + \frac1{D-p+1}\,\ts{g}\wedge(\covd\cdot\ts{\omega})\;.
\end{equation}
Here we have slightly abused the notation in the second term by understanding that only the second index of the metric participates in the wedge operation (since $\ts{g}$ is symmetric, it could not be otherwise, anyway). By applying $\covd\cdot$ to \eqref{CKYdefforint}, we obtain the Bochner--Laplace operator,
\begin{equation}\label{CKYBochnerLaplace}
    \nabla^2\ts{\omega} = \frac1{p+1}\,\covd\cdot (\covd\wedge\ts{\omega}) + \frac1{D-p+1}\,\covd\wedge(\covd\cdot\ts{\omega})\;.
\end{equation}
Comparing \eqref{CKYBochnerLaplace} with the definition of the de~Rham--Laplace operator \eqref{DeRhamLaplace},
\begin{equation}\label{CKYDeRhamLaplace}
  -\Delta\ts{\omega} = \covd\cdot (\covd\wedge\ts{\omega}) + \covd\wedge(\covd\cdot\ts{\omega})
\end{equation}
and employing the Weitzenb\"ock identity \eqref{WeitzenbockId} we obtain the relation for the Weitzenb\"ock operator
\begin{equation}\label{CKYWeitzenbock}
    -W\ts{\omega} =
    \frac{p}{p+1}\,\covd\cdot (\covd\wedge\ts{\omega})
    + \frac{D-p}{D-p+1}\,\covd\wedge(\covd\cdot\ts{\omega})\;.
\end{equation}

The above relations simplify for Killing--Yano and closed conformal Killing--Yano forms since in these cases only one of the terms on the right-hand side survives. Indeed, for a Killing--Yano $p$-form $\ts{f}$ we have $\covd\cdot\ts{f}=0$. When we use this in \eqref{CKYBochnerLaplace}, \eqref{CKYDeRhamLaplace}, and \eqref{CKYWeitzenbock}, we obtain
\begin{equation}\label{KYLaplace}
    \Delta\ts{f}=-(p+1)\nabla^2\ts{f}=-\covd\cdot(\covd\wedge\ts{f})=\frac{p+1}{p}\,W\ts{f}\;.
\end{equation}
Similarly, for a closed conformal Killing--Yano $p$-form $\ts{h}$ we have $\covd\wedge\ts{h}=0$, and
\begin{equation}\label{CCKYLaplace}
    \Delta\ts{h}=-(D-p+1)\nabla^2\ts{h}=-\covd\wedge(\covd\cdot\ts{h})=\frac{D-p+1}{D-p}\,W\ts{h}\;.
\end{equation}
In both cases we see that the action of Laplace operators is given by the algebraic Weitzenb\"ock operator which involves only the curvature. The same restrictions can be obtained directly from the integrability conditions as we will see below, cf.\ \eqref{KYBochnerLaplace} and \eqref{CCKYBochnerLaplace}.

\subsection{Integrability conditions}
\label{asc:IntegrCond}

As we already mentioned, the conformal Killing--Yano equation \eqref{CKYeq} is over-determined \citep{dunajski2008overdetermined}. The existence of the conformal Killing--Yano forms imposes severe restrictions on the geometry. These are called in general the \defterm{integrability conditions}. They have been studied from the very beginnings of the study of Killing--Yano objects \citep{Yano:1952, yano1953curvature, tachibana1969integrability} till recent works \citep{Houri:2014hma,Batista:2014fpa}. (See also \cite{Houri:2017tlk} for a recent progress on integrability conditions for the Killing tensors.) We give here the review of the integrability conditions for Killing--Yano equation and closed conformal Killing--Yano equation.

The basic common restriction is that the second covariant derivative of such a form can be expressed in terms of the curvature and the form itself. It allows one to derive an algebraic condition for solutions of the given equation expressed in terms of the curvature. From these conditions one can derive particular consequences which play a role of necessary integrability conditions. Examples of these are the expressions for the Laplace operator and the Weitzenb\"ock operator encountered above.

\subsubsection{Preliminaries}

The \defterm{Ricci identity} for an antisymmetric ${p}$-form ${\ts{\omega}}$ reads
\begin{equation}\label{RicciIdASF}
   \nabla_{\!a}\nabla_{\!b}\,\omega_{c_1c_2\dots c_p}
   - \nabla_{\!b}\nabla_{\!a}\,\omega_{c_1c_2\dots c_p} =
   - p\,R_{ab}{}^{e}{}_{[c_1} \omega_{|e|c_2\dots c_p]}\;.
\end{equation}
For the purpose of the proofs below, let us write explicitly the consequences of the Leibniz rule \eqref{insertionLeibniz} for a ${p}$-form ${\ts{\omega}}$ and a form ${\ts{\sigma}}$ of rank ${1}$ and ${2}$,
\begin{gather}
\label{wedge1p}
  (p+1)\,\sigma_{[a}\omega_{c_1c_2\dots c_p]} =
    \sigma_a \omega_{c_1c_2\dots c_p} - p\, \sigma_{[c_1}\omega_{|a|c_2\dots c_p]}\;,\\
\label{wedge2p}
  (p+2)\,\sigma_{[ac_0}\omega_{c_1c_2\dots c_p]} =
    2\,\sigma_{a[c_0} \omega_{c_1c_2\dots c_p]} + p\, \sigma_{[c_0c_1}\omega_{|a|c_2\dots c_p]}\;.
\end{gather}

\subsubsection{Killing--Yano forms}

The Killing--Yano condition
\begin{equation}\label{KYdefforint}
    \nabla_{\!a}f_{a_1\dots a_p} =  \nabla_{\![a}f_{a_1\dots a_p]}
\end{equation}
is rather restrictive. It implies that the second derivatives of Killing--Yano forms are algebraically related to the form itself. Namely, any Killing--Yano ${p}$-form ${\ts{f}}$ satisfies
\begin{equation}\label{KY2ndder}
    \nabla_{\!a}\nabla_{\!b} f_{c_1c_2\dots c_p}
      = -(p+1) R_{a[b}{}^{e}{}_{c_1}\,f_{|e|c_2\dots c_p]}
      = \frac{p+1}{2} R_{ea[bc_1}\,f^{e}{}_{c_2\dots c_p]}
      \;.
\end{equation}
Contracting in indices ${a}$ and ${b}$, one obtains the Bochner--Laplace operator acting on the Killing--Yano form ${\ts{f}}$,
\begin{equation}\label{KYBochnerLaplace}
-\nabla^2f_{c_1c_2\dots c_p} =
    R_{e[c_1}\, f^{e}{}_{c_2\dots c_p]}
      - \frac{p{-}1}{2} R_{de[c_1c_2}\,f^{de}{}_{\dots c_p]}\;.
\end{equation}
On the right-hand side we can recognize the action of the Weitzenb\"ock operator. We thus showed ${-\nabla^2\ts{f} = \frac1{p}W\ts{f}}$, cf.~\eqref{KYLaplace}. We will return to this fact at the end of this section.

Let us prove now the relations \eqref{KY2ndder}. The second equality follows from the cyclic property of the Riemann tensor (the first Bianchi identity). The proof of the first equality starts with a  trivial property of the exterior derivative ${\grad\grad\ts{f}}=0$. Rewriting it using the covariant derivative and the identity \eqref{wedge1p}, we obtain
\begin{equation}
    (p+2)\nabla_{\![a}\nabla_{\!b} f_{c_1c_2\dots c_p]}
    = \nabla_{\!a}\nabla_{\!b}f_{c_1\dots c_p}
       - (p+1) \nabla_{\![b}\nabla_{\!|a|}f_{c_1\dots c_p]}=0\;,
\end{equation}
where we used that ${\nabla_{\!a}f_{c_1\dots c_p}}$ is antisymmetric in all indices, cf.~\eqref{KYdefforint}. Employing the Ricci identities \eqref{RicciIdASF} to the second term we find
\begin{equation}
    - p \nabla_{\!a}\nabla_{\!b}f_{c_1c_2\dots c_p}
       - (p+1) p\, R_{a[b}{}^{e}{}_{c_1}f_{|e|c_2\dots c_p]}=0\;,
\end{equation}
which proves \eqref{KY2ndder}.

The existence of a Killing--Yano form is a non-trivial property of the geometry. It requires a consistency of the curvature with the Killing--Yano form, which can be written as
\begin{equation}\label{KYcurvaturecond1}
    R^{e[a}{}_{[c_1}{}^{b]}\,f_{|e|c_2c_3\dots c_p]}
    + R^{e}{}_{[c_1}{}^{[a}{}_{c_2}\, f_{|e|}{}^{b]}{}_{c_3\dots c_p]} = 0\;.
\end{equation}
Using the cyclic property of the Riemann tensor in both terms it can also be written as
\begin{equation}\label{KYcurvaturecond2}
    R^{abe}{}_{[c_1}\,f_{|e|c_2c_3\dots c_p]}
    + R^{e[a}{}_{[c_1c_2}\, f_{|e|}{}^{b]}{}_{c_3\dots c_p]} = 0\;.
\end{equation}
The integrability condition \eqref{KYcurvaturecond1} can be obtained by applying the expansion \eqref{wedge2p} to the right-hand side of \eqref{KY2ndder}, taking antisymmetrization of both sides in indices ${a}$ and ${b}$, and using the Ricci identity \eqref{RicciIdASF} to the left-hand side. The cyclic property of the Riemann tensor is needed to reshuffle appropriately the indices.

Taking the contraction of these identities in indices ${b}$ and ${c_1}$ one can get another necessary condition between the curvature and the Killing--Yano form,
\begin{equation}\label{KYcurvaturetrcond1}
    R_{e}{}^{a}\, f^{e}{}_{c_2c_3\dots c_p} - R_{e[c_2}\, f^{ae}{}_{c_3\dots c_p]} =
    \frac{p-2}{2} \Bigl( R_{de}{}^{a}{}_{[c_2}\,f^{de}{}_{c_3\dots c_p]}
    -R_{de[c_2c_3}\,f^{ade}{}_{\dots c_p]} \Bigr)\;.
\end{equation}
With the help of \eqref{wedge2p} this can be also re-arranged to the form
\begin{equation}\label{KYcurvaturetrcond2}
\begin{split}
    &p\, R_{ec_1}\, f^{e}{}_{c_2\dots c_p}
      - \frac{p(p{-}1)}{2} R_{dec_1[c_2}\,f^{de}{}_{\dots c_p]} =\\
    &\mspace{180mu}=p\, R_{e[c_1}\, f^{e}{}_{c_2\dots c_p]}
      - \frac{p(p{-}1)}{2} R_{de[c_1c_2}\,f^{de}{}_{\dots c_p]} \;.
\end{split}
\end{equation}
The right-hand side is actually the action ${W\ts{f}}$ of the Weitzenb\"ock operator \eqref{WeitzenbockOp} on the Killing--Yano form.

\subsubsection{Closed conformal Killing--Yano forms}

Let us now turn to the closed conformal Killing--Yano forms. Similar to the Killing--Yano equation, the associated condition
\begin{equation}\label{CCKYdefforint}
    \nabla_{\!a}h_{a_1\dots a_p} = p g_{a[a_1}\,\xi_{a_2\dots a_p]}\;,\qquad
    \xi_{a_2\dots a_p} = \frac1{D-p+1}\nabla_{\!c} h^{c}{}_{a_2\dots a_p}\;,
\end{equation}
is also restrictive. The second covariant derivative of the closed conformal Killing--Yano form can be expressed using the curvature and the form itself:
\begin{equation}\label{CCKY2ndder}
    \nabla^{a}\nabla^{b} h_{c_1c_2\dots c_p}
      = -\frac{p}{D-p}\Bigl(
      R^a{}_e\,\delta^b_{[c_1}\,h^e{}_{c_2\dots c_p]}
      +\frac{p-1}2 R_{de}{}^a{}_{[c_1}\,\delta^b_{c_2}\,h^{de}{}_{\dots c_p]}\Bigr)
      \;.
\end{equation}
Taking the contraction in indices $a$ and $b$, one gets the expression for the Bochner--Laplace operator of the closed conformal Killing--Yano form
\begin{equation}\label{CCKYBochnerLaplace}
-\nabla^2h_{c_1c_2\dots c_p} = \frac1{D-p}\Bigl(
    p\,R_{e[c_1}\, h^{e}{}_{c_2\dots c_p]}
      - \frac{p(p{-}1)}{2} R_{de[c_1c_2}\,h^{de}{}_{\dots c_p]}\Bigr)\;.
\end{equation}
Here, we can again identify the action of the Weitzenb\"ock operator, thus ${-\nabla^2\ts{h}= \frac{1}{D{-}p}W\ts{h}}$.

The proof of \eqref{CCKY2ndder} starts by applying the Ricci identity \eqref{RicciIdASF} to ${\ts{h}}$ upon which the use of the closed conformal Killing--Yano condition gives
\begin{equation}\label{CCKYint1}
    R_{abe[c_1}\,h^{e}{}_{c_2\dots c_p]}
    + g_{b[c_1}\nabla_{\!|a|}\xi_{c_2\dots c_p]}
    - g_{a[c_1}\nabla_{\!|b|}\xi_{c_2\dots c_p]}
    = 0\;.
\end{equation}
Contracting in indices ${b}$ and ${c_1}$, and using repeatedly \eqref{wedge1p} and the cyclic property of the Riemann tensor, one can express the covariant derivative of ${\ts{\xi}}$:
\begin{equation}\label{covderxi}
    \nabla_a\xi_{c_2\dots c_p} = \frac1{D-p}\Bigl(
    -R_{ea}\,h^{e}{}_{c_2\dots c_p} + \frac{p-1}{2} R_{dea[c_2}\, h^{de}{}_{\dots c_p]}
    \Bigr)\;.
\end{equation}
Substituting \eqref{covderxi} into the covariant derivative of the closed conformal Killing--Yano condition \eqref{CCKYdefforint}
\begin{equation}\label{covdCCKYcond}
    \nabla^a\nabla^b h_{c_1\dots c_p} = p\, \delta^b_{[c_1}\nabla^a\xi_{c_2\dots c_p]}\;,
\end{equation}
we obtain the desired expression \eqref{CCKY2ndder} for the second derivative of the closed conformal Killing--Yano form.

Substituting \eqref{covderxi} into \eqref{CCKYint1}, a bit of work leads to another necessary consistency condition between the curvature and the
closed conformal Killing--Yano form
\begin{equation}\label{CCKYcurvaturecond}
    2 R^{[a}{}_{e}\,\delta^{b]}_{[c_1}\, h^{e}{}_{c_2\dots c_p]}
    - (D-p) R^{ab}{}_{e[c_1}\,h^{e}{}_{c_2\dots c_p]}
    +(p-1)R_{de}{}^{[a}{}_{[c_1}\,\delta^{b]}_{c_2}\,h^{de}{}_{\dots c_p}
    =0\;.
\end{equation}

%% file: apx-mtrfcs.tex

\section{Kerr--NUT--(A)dS metric related quantities}
\label{apx:mtrfcs}

\subsection{Properties of metric functions}
\label{ssc:mtrfcs}

In chapter~\ref{sc:hdbh} we have introduced the auxiliary functions $\A{k}$  of variables ${x_\mu^2}$, and the analogous polynomials $\Aa{k}$ of parameters ${a_\mu^2}$. They can be defined using the generating functions ${J(a^2)}$ and ${\Ja(x^2)}$ as follows%
\footnote{Let us remind that, if not indicated otherwise, the sums (and products) run over `standard' ranges of indices:
\[
\sum_\mu \equiv \sum_{\mu=1}^\dg\;,\qquad
\sum_k \equiv \sum_{k=0}^{\dg-1}\;.
\]
}
\begin{align}
\label{metricpolysx}
    J(a^2)&=\prod_{\nu}(x_\nu^2-a^2) =\sum_{k=0}^\dg \A{k} (-a^2)^{\dg{-}k}\;,
    \\
\label{metricpolysa}
    \Ja(x^2)&=\prod_{\nu}(a_\nu^2-x^2) =\sum_{k=0}^\dg \Aa{k} (-x^2)^{\dg{-}k}\;.
\end{align}
These definitions imply
\begin{align}
    \A{k} &=\sum_{\substack{\mu_1,\dots,\mu_k\\\mu_1<\dots<\mu_k}} x_{\mu_1}^2\dots x_{\mu_k}^2\;,\label{Ax}\\
    \Aa{k} &=\sum_{\substack{\mu_1,\dots,\mu_k\\\mu_1<\dots<\mu_k}} a_{\mu_1}^2\dots a_{\mu_k}^2\;.\label{Aa}
\end{align}
Similarly, we define the functions $J_\mu(a^2)$, $\A{j}_\mu$, $\Ja_\mu(x^2)$, and $\Aa{j}_\mu$, which skip the $\mu$-th variables $x_\mu$ and $a_\mu$ as follows
\begin{align}
    J_\mu(a^2)&=\prod_{\substack{\nu\\\nu\neq\mu}}(x_\nu^2-a^2)
       =\sum_{k} \A{k}_\mu (-a^2)^{\dg{-}1{-}k}\;,\label{metricpolysmux}\\
    \Ja_\mu(x^2)&=\prod_{\substack{\nu\\\nu\neq\mu}}(a_\nu^2-x^2)
       =\sum_{k} \Aa{k}_\mu (-x^2)^{\dg{-}1{-}k}\;,\label{metricpolysmua}
\end{align}
with
\begin{align}
    \A{k}_\mu &=\sum_{\substack{\nu_1,\dots,\nu_k\\\nu_1<\dots<\nu_k\\\nu_i\neq\mu}}
        x_{\nu_1}^2\dots x_{\nu_k}^2\;,\label{Amux}\\
    \Aa{k}_\mu &=\sum_{\substack{\nu_1,\dots,\nu_k\\\nu_1<\dots<\nu_k\\\nu_i\neq\mu}}
        a_{\nu_1}^2\dots a_{\nu_k}^2\;.\label{Amua}
\end{align}
These functions satisfy
\begin{equation}\label{J0}
\begin{gathered}
    J(x_\mu^2)=0\;,\qquad  \Ja(a_\mu^2)=0\;,\\
    J_\mu(x_\nu^2)=0\;,\qquad  \Ja_\mu(a_\nu^2)=0\;,\qquad\text{for $\nu\neq\mu$}\;.
\end{gathered}
\end{equation}
Finally, we define
\begin{align}
    U_\mu = J_\mu(x_\mu^2)= \prod_{\substack{\nu\\\nu\neq\mu}}(x_\nu^2-x_\mu^2)\;,\label{Uxdef}\\
    \Ua_\mu = \Ja_\mu(a_\mu^2)= \prod_{\substack{\nu\\\nu\neq\mu}}(a_\nu^2-a_\mu^2)\;.\label{Uadef}
\end{align}
The polynomials $\A{k}$ and $\A{k}_\mu$ satisfy the following identities:
\begin{equation}\label{Aid0}
    \A{k}=\A{k}_\mu+x_\mu^2\A{k{-}1}_\mu\;,
\end{equation}
\begin{gather}
\label{Aid1}
    \sum_k\A{k}_\mu \frac{(-x_\nu^2)^{\dg{-}1{-}k}}{U_\nu}=\delta^\nu_\mu\;,\\
\label{Aid1i}
    \sum_\mu\A{k}_\mu \frac{(-x_\mu^2)^{\dg{-}1{-}l}}{U_\mu}=\delta^k_l\;,\\
\label{Aid2}
    \sum_\mu\A{k}_\mu \frac{(-x_\mu^2)^{\dg}}{U_\mu}=-\A{k{+}1}\;,\\
\label{Aid7}
    \sum_{\mu} \frac{\A{k}_\mu}{x_\mu^2 U_\mu} = \frac{\A{k}}{\A{\dg}}\;,\\
\label{Aid3}
    \sum_\mu\A{k}_\mu = (\dg-k)\A{k}\;,\\
\label{Aid4}
    \sum_k(\dg-k)\A{k} \frac{(-x_\nu^2)^{\dg{-}1{-}k}}{U_\nu} = 1\;,\\
\label{Aid5}
    \sum_{k=0,\dots,\dg} \A{k} (-x_\nu^2)^{\dg{-}k} = 0\;,\\
\label{Aid6}
    \sum_{l=0,\dots,k} \A{l} (-x_\nu^2)^{k{-}l} = \A{k}_\mu\;.
\end{gather}
Analogous identities hold also for the complementary polynomials $\Aa{k}$ and $\Aa{k}_\mu$.

For the functions $J(a^2)$ and $\Ja(x^2)$ we can write
\begin{equation}\label{Jid1}
\begin{gathered}
    \prod_\mu J(a_\nu^2) = (-1)^\dg \prod_\nu \Ja(x_\mu^2)\;,\\
    \prod_{\substack{\mu\\\mu\neq\kappa}} J_\kappa(a_\nu^2) =
       (-1)^{\dg{-}1} \prod_{\substack{\nu\\\nu\neq\kappa}} \Ja_\kappa(x_\mu^2)\;.
\end{gathered}
\end{equation}
These functions satisfy important orthogonality relations
\begin{equation}\label{Jortrel1}
    \sum_\alpha \frac{J_\nu(a_\alpha^2)}{\Ua_\alpha}\frac{\Ja_\alpha(x_\mu^2)}{U_\mu} = \delta^\mu_\nu\;,
\end{equation}
\begin{equation}\label{Jortrel2}
    \sum_\alpha \frac{J_\mu(a_\alpha^2)J_\nu(a_\alpha^2)}{J(a_\alpha^2)\Ua_\alpha} =
      - \frac{U_\mu}{\Ja(x_\mu^2)} \delta_{\mu\nu}\;,
\end{equation}
\begin{equation}\label{Jortrel3}
    \sum_\mu J_\mu(a_\alpha^2)J_\mu(a_\beta^2)\frac{\Ja(x_\mu^2)}{U_\mu} =
      - J(a_\alpha^2) \Ua_\alpha \delta^{\alpha\beta}\;.
\end{equation}

\subsection{Spin connection}
\label{ssc:spincon}

In this section we present the spin connection for the Kerr--NUT--(A)dS spacetimes written in the frame \eqref{Darbouxformfr}.
In even dimension the only non-zero connection coefficients with respect to the frame $({\enf\mu,\,\ehf\mu})$ are:
\begin{equation}\label{conncoef}
\begin{split}
  &\omega_{\mu\mu\nu} = -\omega_{\mu\nu\mu}
    =\sqrt{\frac{X_\nu}{U_\nu}}\,\frac{x_\nu}{x_\nu^2-x_\mu^2}\;,\quad
  \omega_{\mu\hat\mu\hat\nu} = -\omega_{\mu\hat\nu\hat\mu}
    =\sqrt{\frac{X_\nu}{U_\nu}}\,\frac{x_\mu}{x_\nu^2-x_\mu^2}\;,\nonumber\\
  &\omega_{\hat\mu\hat\mu\nu} = -\omega_{\hat\mu\nu\hat\mu}
    =\sqrt{\frac{X_\nu}{U_\nu}}\,\frac{x_\nu}{x_\nu^2-x_\mu^2}\;,\quad
  \omega_{\hat\mu\hat\nu\mu} = -\omega_{\hat\mu\mu\hat\nu}
    =\sqrt{\frac{X_\nu}{U_\nu}}\,\frac{x_\mu}{x_\nu^2-x_\mu^2}\;,\nonumber\\
  &\omega_{\hat\mu\nu\hat\nu} = -\omega_{\hat\mu\hat\nu\nu}
    =\sqrt{\frac{X_\mu}{U_\mu}}\,\frac{x_\nu}{x_\nu^2-x_\mu^2}\;,\\
  &\omega_{\hat\mu\hat\mu\mu} = -\omega_{\hat\mu\mu\hat\mu}
    =\frac12\sqrt{\frac{X_\mu}{U_\mu}}\frac{X_\mu'}{X_\mu}+
           \sqrt{\frac{X_\mu}{U_\mu}}\sum_{\substack{\nu\\\nu\neq\mu}}\frac{x_\mu}{x_\nu^2-x_\mu^2}\;.
           \nonumber
\end{split}\raisetag{25ex}
\end{equation}
Here, indices ${\mu}$ and ${\nu}$ are different.
In odd dimension the same spin coefficients apply, plus the following extra terms:
\begin{equation}\label{conncoef_odd}
\begin{split}
  &\omega_{\mu \hat\mu \hat 0} = -\omega_{\mu \hat 0 \hat\mu} = - \sqrt{\frac{c}{A^{(n)}}}\frac{1}{x_\mu}  \; , \quad
  \omega_{\hat\mu \mu \hat 0} = -\omega_{\hat\mu \hat 0 \mu} = \sqrt{\frac{c}{A^{(n)}}}\frac{1}{x_\mu} \; , \\
  &\omega_{\hat 0 \mu \hat 0} = -\omega_{\hat 0 \hat 0 \mu} = - \sqrt{\frac{X_\mu}{U_\mu}} \frac{1}{x_\mu} \; , \quad
  \omega_{\hat 0\hat\mu\mu} = -\omega_{\hat 0\mu\hat\mu} = - \sqrt{\frac{c}{A^{(n)}}}\frac{1}{x_\mu} \; .
\end{split}\raisetag{25ex}
\end{equation}

%% file: apx-MP.tex

\section{Myers--Perry metric}
\label{apx:MyersPerry}

In section~\ref{ssc:SubcasesLorentzian} we have recovered the higher-dimensional rotating black hole metric of \citep{MyersPerry:1986} as a subcase of the Kerr--NUT--(A)dS metric of \citep{Chen:2006xh}. In this appendix we give a short overview of the Myers--Perry metric in its original coordinates and its related Kerr--Schild form.

\subsection{Tangherlini solution}
\label{ssc:Tangherlini}

The simplest higher-dimensional solution of the Einstein equations describing a static spherically symmetric black hole in a $D$-dimensional asymptotically flat spacetime is the Tangherlini metric \citep{tangherlini1963schwarzschild}
\be\label{TTT}
\ts{g}=-F \grad t^2 +\frac1F \grad r^2 + r^2 \ts{d\omega}_{D-2}^2\, ,
\ee
where $\ts{d\omega}_{D-2}^2$ is a metric on a unit $(D-2)$-dimensional sphere, and
\be
F=1-\left(\frac{2M}{r}\right)^{D-3}\,.
\ee
The constant $M$ is related to the physical mass of the spacetime. The corresponding relation can be found by either calculating the asymptotic integrals or by comparing the metric at far distance to the gravitational field of a static source in the Newtonian theory. Either procedure yields the following physical mass $\mathcal{M}$:
\be
\mathcal{M}=\frac{(D-2)\omega_{D-2}}{8\pi} M\,,
\ee
where $\omega_{d}$ is the area of a unit $d$-dimensional sphere
\be\label{volunitsphr}
\omega_{d}=\frac{2\pi^{d+\frac12}}{\Gamma({d+\frac12})}\, .
\ee

\subsection{Myers--Perry solution}
\label{ssc:MyersPerry}

\subsubsection{Angular momentum in higher dimensions}

\begin{figure}[ht]
\bigskip
\centerline{\includegraphics[width=6cm]{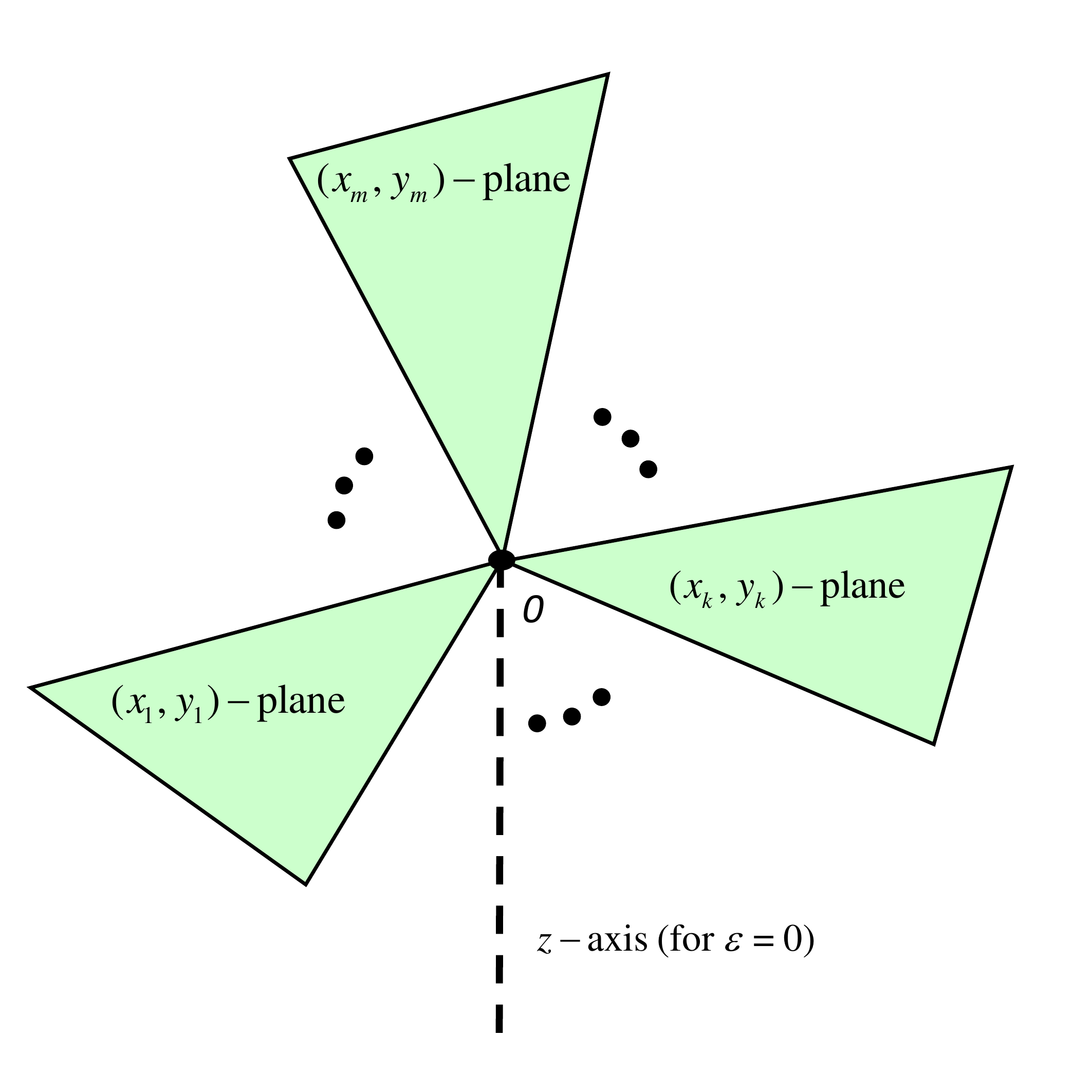}}
\caption{\label{fig:angmomhd}%
{{\bf 2-planes of rotation.}
A schematic illustration of $m$ mutually orthogonal 2-planes in a spatial section of the flat $D$-dimensional spacetime, {$D=2m+2-\eps$}.}}
\end{figure}

If a stationary higher-dimensional black hole rotates, its metric becomes more complicated. A higher-dimensional generalization of the Kerr metric was obtained by \cite{MyersPerry:1986}. To get a feeling for the properties of the Myers--Perry solution, let us first consider a {\em flat spacetime} in
\be\label{Dm}
  {D=2m+2-\eps}
\ee
number of spacetime dimensions, {with $\eps=0$ for even and $\eps=1$ for odd dimensions}. Obviously, the number of spatial dimensions is {$2m+1-\eps$} and the space contains $m$  mutually orthogonal {\em spatial 2-planes}, see figure~\ref{fig:angmomhd}

The global angular momentum of matter is characterized by an antisymmetric matrix $J_{ij}$, where $i$ and $j$ are spatial indices. It is well known that such a tensor can be transformed to a special canonical form, by performing rigid spatial rotations. The corresponding matrix contains $m$ 2-dimensional block matrices at its diagonal, while other components vanish. These 2-dimensional block matrices have the form
\be\label{JJJ}
\left( \begin{array}{cc}
~0&J_i\\
-J_i&~0
\end{array}
\right)\, ,
\ee
where $J_i$ are the `components' of the angular momentum.

\subsubsection{Myers--Perry form of the metric}

The Myers--Perry spacetime describes a stationary vacuum isolated rotating black hole in a $D$-dimensional asymptotically flat spacetime \citep{MyersPerry:1986}. Since the  metric is stationary and asymptotically flat, one can expect it to be described by $m+1$ parameters: the mass $M$ and $m$ independent components of the angular momentum $J_i$, related to the rotation parameters $a_i$ in each of the rotation 2-planes.

The Myers--Perry metric reads
\be\label{MPM}
\begin{split}
\tens{g}=
  &-\grad t^2+\frac{U\grad r^2}{V{-}2M}+
  \frac{2M}{U}\Bigl(\grad t+\sum_{i=1}^{m}a_i\mu_i^2 \grad \phi_i\Bigr)^2\\
  &+\sum_{i=1}^{m}(r^2+a_i^2)(\grad \mu_i^2+\mu_i^2\grad \phi_i^2)
  +(1-\eps) r^2 \grad \mu_0^2\, ,
\end{split}
\ee
where
\be\label{MPmfc}
V = \frac{1}{r^{1+\eps}}\prod_{i=1}^{m}(r^2+a_i^2)\,,\quad
U = V \Bigl(1-\sum_{i=1}^m\frac{a_i^2 \mu_i^2}{r^2+a_i^2}\Bigr)\, .
\ee
{We call coordinates $(t,r,\mu_i,\phi_j)$, $i=\eps,\dots,m$, $j=1,\dots,m$ the {\em Myers--Perry coordinates}. They are not all independent, namely coordinates $\mu_i$ obey a constraint}
\begin{equation}\label{constr}
\sum_{i=\eps}^m\mu_{i}^2 = 1\,.
\end{equation}

The metric admits $m+1$ Killing vectors. The vector {$\ts{\xi}_{(t)}=\ts{\partial}_t$} is a generator of time translations, while {$\ts{\xi}_{(j)}=\ts{\partial}_{\phi_j}$} generate rotations in $m$ independent 2-planes. The coordinates {$\mu_i$ (including $\mu_0$} in even dimensions) are direction cosines with respect to these planes and have the range $0\leq \mu _{i}\leq 1$.
The angle coordinates $\phi_j$ take values $-\pi\leq \phi_j\leq \pi$ \citep{Myers:2011yc}.
In $D=4$ dimensions, the metric reduces to the Kerr spacetime, \eqref{kerr1}.

\subsubsection{Basic properties}

As expected, the metric contains $m+1$ parameters: {$M$ and $a_i$}. These parameters
are related to the {physical} mass $\mathcal{M}$ and the angular momentum components {$J_i$, $i=1,\dots,m$} as follows
\be\label{MMMM}
\mathcal{M}=\frac{(D-2)\omega_{D-2}}{8\pi} M\,,\qquad
J_i=\frac{2}{D-2} \mathcal{M} a_i\, ,
\ee
where $\omega_{d}$ is again the area of a unit $d$-dimensional sphere \eqref{volunitsphr}.

A surface where $\ts{\xi}_{(t)}^{2}=0$ is called a surface of infinite redshift or an ergosurface. The equation of this surface is
\be
U-2M=0\, . \label{13.4.23}
\ee
We denote
\be
\ts{\eta}=\ts{\xi}_{(t)}+\sum_{i=1}^{m}\Omega _{i}\, \ts{\xi}_{(i)}\,,\qquad \Omega_{i}=\frac{a_{i}}{r^{2}_{+}+a^{2}_{i}}\, .
\label{KHH}
\ee
Then the surface
\be
V-2M=0\, ,
\ee
located at $r=r_+$ where the Killing vector $\ts{\eta}$ becomes null, is a Killing horizon, and coincides with the position of the event horizon. The domain between the event horizon and the ergosurface is an ergosphere. The parameters $\Omega_j$ are components of the angular velocity of the black hole. Similar to the angular momentum, the angular velocity has $m$ independent components, corresponding to the same number of 2-planes of rotation.

We also get the following expressions for the surface gravity $\kappa$ and the horizon area ${\mathcal A}$
\begin{align}
\kappa&=\sum_{i=1}^{m} \frac{r_+}{a_i^2+r_+^2}-\frac{2^{\eps-1}}{r_+}\,,\\
{\mathcal A}&=\frac{\omega _{D-2}}{r_+^{\eps}}\prod_{i=1}^m (a_i^2+r_+^2)\,.
\end{align}
This yields the following higher-dimensional {\em first law} of black hole mechanics:
\be
\delta \mathcal{M}=\frac{\kappa}{2\pi}\frac{\delta {\mathcal A}}{4}+\sum_{i=1}^m\Omega_i \delta J_i
\ee
for the  Myers--Perry black hole \citep{MyersPerry:1986}. See \cite{Altamirano:2014tva} for interesting thermodynamic phase transitions these black holes can demonstrate.

It turns out that 5D rotating black holes are quite similar to the Kerr metric, see, e.g., \cite{deFreitas:2015pda}. However, in dimensions $D\geq 6$ there are important differences. For $D\ge 6$ and fixed black hole mass there exist solutions with arbitrary large angular momentum. Such black holes are called {\em ultra-spinning}. However such ultra-spinning black holes are dynamically unstable \citep{Emparan:2003sy, Dias:2010eu, Dias:2010maa, Figueras:2017zwa}. Further discussion of the properties of the Myers--Perry metric can be found in reviews by \cite{Emparan:2008eg} and \cite{Myers:2011yc}. Metrics obtained by an analytical continuation of the Myers--Perry and their properties are discussed in \cite{Dowker:1995gb}.

It was shown in \cite{Frolov:2007nt} that the Myers--Perry metric admits a principal tensor. It can be generated from a 1-form potential $\ts{b}$,
\be
\ts{b}=\frac{1}{2}\Bigl[\Bigl(r^2+\sum_{i=1}^{m} a_i^2\mu_i^2\Bigr) \, \grad t+
\sum_{i=1}^{m}a_i\mu_i^2(r^2+a_i^2)\, \grad \phi_i\Bigr]\,,
\ee
and reads
\be\label{pcky_MP}
\tens{h}=
  \sum_{i=1}^{m} a_i\mu_i \grad \mu_i\wedge
    \Bigl[a_i\grad t+(r^2+a_i^2) \grad\phi_i\Bigr]
  + r\grad r\wedge \Bigl(\grad t+\sum_{i=1}^{m}a_i\mu_i^2\grad \phi_i \bigr)\, .
\ee
This tensor generates the Killing tower of symmetries, discussed in a more general case in chapter~\ref{sc:hshdbh}.

\subsubsection{The flat space limit: $M=0$}

When the mass parameter in the Myers--Perry metric vanishes, $M=0$, the metric simplifies to
\be\label{MPMF}
\begin{gathered}
\ts{g}=-\,\grad t^{2} + \grad r^2
  +\sum _{i=1}^{m}
  \Bigl[ (r^{2}{+}a^{2}_{i})(\grad \mu^{2}_{i}+\mu ^{2}_{i} \,\grad \phi ^{2}_{i})
  -\frac{a_i^2 \mu_i^2}{r^2{+}a_i^2} \grad r^2\Bigr]
  +(1{-}\eps) r^2 \grad \mu_0^2\, .
\end{gathered}
\ee
Although not obvious from this expression, the metric is flat. An explicit transformation to the Cartesian coordinates $(x_i,y_i)$ (and $z$ in even dimensions) reads\footnote{\label{ftn:coorx}%
{Coordinates $x_i$ introduced here are not directly related to coordinates $x_\nu$ used in the main text. For relation of $\mu$'s to $x_\nu$'s see \eqref{mudef} in chapter~\ref{sc:hdbh}.}}
\begin{equation}\label{flatMPCart}
    x_i=\mu_i\sqrt{r^2+a_i^2}\cos\phi_i\,,\quad
    y_i=\mu_i\sqrt{r^2+a_i^2}\sin\phi_i\,,\quad
    z=\mu_0 r\,,
\end{equation}
upon which one recovers the standard Minkowski metric in $D$ dimensions, c.f. also \eqref{KS_MP}--\eqref{k_MP} in the next section. The constraint \eqref{constr} now relates $r$ to the Cartesian coordinates.

The spacetime \eqref{MPMF} is invariant under the action of a cyclic group of {$m$-dimensional torus}. The Killing vector {$\ts{\xi}_{(i)}$} vanishes when $\mu_i=0$, that is at the center $x_i=y_i=0$ of the $(x_i,y_i)$ plane. Relations \eqref{flatMPCart} show that this is a regular $(D-2)$-dimensional geodesic submanifold, called {${i}$-th} axis of rotation. This conclusion remains  valid for the Myers--Perry metric \eqref{MPM}. Taking {$\phi_i$'s} to be periodic coordinates with  period $2\pi$ makes this metric axisymmetric. The integral lines for each of the Killing vectors {$\ts{\xi}_{(i)}$} are closed cycles.

\subsection{Kerr--Schild form}
\label{ssc:MPKerrSchield}

Similar to the Kerr metric, the  Myers--Perry metric can be written in the Kerr--Schild form \citep{MyersPerry:1986} that is intrinsically related to the special algebraic type of the Weyl tensor \citep{Ortaggio:2008iq} and the existence of hidden symmetries. It will also allow us to easily understand the flat space limit of the principal tensor $\ts{h}$.

To obtain the Kerr--Schild form of the metric, let us start with the transformation
\begin{equation}
\grad t=\grad \tau-\frac{2M}{V-2M}\,\grad r\,,\quad
\grad \phi_j=\grad \varphi_j+\frac{V}{V-2M} \frac{a_j}{r^2+a_j^2}\,\grad r\,,
\end{equation}
which transforms the metric element \eqref{MPM} into the `Eddington-like' form.
We further introduce the Kerr--Schild coordinates in analogy with \eqref{flatMPCart} above\textsuperscript{\ref{ftn:coorx}}
\be\label{xyk}
\begin{gathered}
x_i = \mu_i\sqrt{r^2{+}a_i^2}
   \cos\bigl(\varphi_i{-}\arctan\frac{a_i}{r}\bigr)\,,\\
y_i = \mu_i\sqrt{r^2{+}a_i^2}
   \sin\bigl(\varphi_i{-}\arctan\frac{a_i}{r}\bigr)\,,\quad
z = \mu_0 r\,,
\end{gathered}
\ee
where $i$ runs from 1 to $m$ and the last coordinate $z$ is introduced only in an even number of spacetime dimensions. The inverse transformation reads
\begin{equation}
\mu_i^2=\frac{x_i^2+y_i^2}{r^2+a_i^2}\,,\quad
\varphi_i=\arctan\frac{a_i}{r}+\arctan\frac{y_i}{x_i}\,,\quad
\mu_0=\frac{z}{r}\,.
\end{equation}
These relations imply
\begin{equation}
\begin{gathered}
\mu_i \grad \mu_i=
  \frac{x_i\grad x_i+y_i\grad y_i}{r^2+a_i^2}
  -\frac{(x_i^2+y_i^2)r\grad r}{(r^2+a_i^2)^2}\,,\\
\grad \varphi_i=\frac{x_i\grad y_i-y_i\grad x_i}{x_i^2+y^2_i}
  -\frac{a_i\grad r}{r^2+a_i^2}\,.
\end{gathered}
\end{equation}
The constraint \eqref{constr} defines coordinate $r$ in terms of $(x_i,y_i,z)$%
\footnote{In the original paper \cite{MyersPerry:1986} the authors derive the {Myers--Perry form
\eqref{MPM} of the metric} from the Kerr--Schild ansatz \eqref{KS_MP}. We are now going backwards.
}
\begin{equation}\label{defr_MP}
\sum_{i=1}^m\frac{x_i^2+y_i^2}{r^2+a_i^2}+(1{-}\eps)\frac{z^2}{r^2}=1\,.
\end{equation}
Differentiating this expression we find
\begin{gather}
\partial_{x_i}r = \frac{rx_i}{F(r^2+a_i^2)}\,,\quad
\partial_{y_i}r=\frac{ry_i}{F(r^2+a_i^2)}\,,\quad
\partial_zr=(1{-}\eps)\frac{z}{Fr}\,,\label{ru_MP}\\
F = \frac{U}{V}=1-\sum_{i=1}^m\frac{a_i^2(x_i^2+y_i^2)}{(r^2+a_i^2)^2}=
r^2\sum_{i=1}^m\frac{x_i^2+y_i^2}{(r^2+a_i^2)^2}+(1{-}\eps)\frac{z^2}{r^2}\,,
\label{Fff_MP}
\end{gather}
and therefore
\begin{equation}\label{dr_MP}
\grad r=\frac{r}{F}\sum_{i=1}^m\frac{x_i\grad x_i+y_i\grad y_i}{r^2+a_i^2}
  +(1{-}\eps)\frac{z\grad z}{Fr}\,.
\end{equation}
Using these relations we find that
the metric \eqref{MPM} takes the Kerr--Schild form
\be\label{KS_MP}
\tens{g}=\tens{\eta}+H\,\ts{l}\,\ts{l}\,,
\ee
where $\ts{\eta}$ is the flat metric, $H$ is a scalar function linear in $M$, and $\ts{l}$ is a null vector (with respect to both $\ts{g}$ and $\ts{\eta}$), given by:
\begin{gather}
\tens{\eta}=-\grad \tau^2+ \sum_{i=1}^m(\grad x_i^2+\grad y_i^2)+(1{-}\eps) \grad z^2\,,\quad
H=\frac{2M}{U}\,,\\
\tens{l}=\grad \tau+\sum_{i=1}^m\frac{r(x_i\grad x_i+y_i\grad y_i)+a_i(x_i\grad y_i-y_i\grad x_i)}{r^2+a_i^2}+
(1{-}\eps)\frac{z\grad z}{r}\,.\label{k_MP}
\end{gather}
The principal tensor \eqref{pcky_MP} reads
\begin{equation}\label{PCKY_KS}
\tens{h}=\sum_{i=1}^m\Bigl[(x_i\grad x_i+y_i\grad y_i)\wedge\grad \tau+a_i\grad x_i\wedge\grad y_i\Bigr]+(1{-}\eps) z\grad z\wedge\grad \tau\,.
\end{equation}

Written in the form \eqref{KS_MP}, it is now straightforward to take the flat space limit, $M\to 0$. We recover the standard Minkowski metric $\ts{\eta}$, while the principal tensor is still given by \eqref{PCKY_KS}. Note that it has exactly the same structure as for the Kerr metric, \eqref{tshf}, only now it `spreads through' $m$ rotation 2-planes.


\subsection{Special spinning black holes}
\label{ssc:SpecSpinBH}

The group of isometries of a general multiply-spinning Myers--Perry solution is $\realn\times
U(1)^m$. If $p$ rotation parameters are equal and non-vanishing, the subgroup $U(1)^p$ for the corresponding rotation planes is enhanced to a non-Abelian group  $U(p)$. If $\,p$ rotation parameters vanish the corresponding subgroup $U(1)^p$ is enhanced and becomes $SO(2p+1-\eps)$ \citep{Emparan:2008eg}.


An interesting special case happens for the odd-dimensional  Myers--Perry metric with all equal angular momenta, $a_i=a$ for all $i$. The enhanced symmetry group of such a spacetime is   ${\realn\times U(n)}$, where $\realn$ denotes time translations. In this case the metric depends only on one essential coordinates, $r$. The constant $r$ surface is a homogeneous space. One says, that such a spacetime is of cohomogeneity-1%
\footnote{The cohomogeneity of a $D$-dimensional spacetime is $p$ if there exist a group of symmetry acting on the spacetime with orbits having dimensionality $D-p$.}.

Another interesting case is a black hole with a single rotation parameter. These black holes are sometimes called \emph{simply rotating}.  The symmetry of the solution is ${\realn\times U(1)\times SO(D-3)}$, where $\realn$ denotes time translations and $U(1)$ corresponds to the 2-plane with a  single rotation.  The spacetime is of cohomogeneity-2. The $(D-2)$-dimensional sections where $r$ and $\theta$ are constant are homogeneous.

%% file: apx-spinor.tex

\section{{Spinors in curved space}}
\label{apx:spinor}

In this appendix we give a short overview of the Dirac theory in a curved space. After discussing the general properties of Dirac spinors, we describe the most general linear symmetry operators that commute with the Dirac operator. We conclude by introducing the special Killing--Yano forms and discuss their relationship to twistor and Killing spinors. For a more thorough exposure to the subject, we refer the interested reader to the books by \cite{BennTucker:book,Cartan:book,Cnops:book,LawsonMichelsohn:book} or the discussion in papers \cite{Benn:1996ia, BennKress:2004, Cariglia:2011yt,Cariglia:2011qb,Trautman:2008, semmelmann2003conformal,Houri:2012eq}.

\subsection{{Dirac spinors}}
\label{ssc:DiracSpinors}

\subsubsection{{Clifford objects}}

The Dirac spinors can be understood as a vector bundle $\mathbf{D}M$ over the spacetime manifold $M$. The fibers of it serve as representation spaces of the Clifford algebra. The irreducible representation of the Clifford algebra in $D=2\dg+\eps$ dimensions is realized on a $2^\dg$-dimensional space, the fibers $\mathbf{D}_x M$ thus have the dimension $2^\dg$. If necessary, we shall use capital Latin indices $A,B,\dots$ to indicate components of the Dirac spinors, but usually we omit the spinorial indices.\footnote{\label{fnt:spinorindices}%
We follow a common practice of omitting spinor indices even when writing components. A spinor $\ts{\psi}$ has components $\psi^A$ which we collect to the `column' $\psi$. Similarly, a Clifford object $\slash\!\!\!\ts{\omega}$ has components $\slash\!\!\!\omega^A{}_B$, which we collect to the matrix $\slash\!\!\!\omega$, \eqref{Clliford}. In other words, $\slash\!\!\!\ts{\omega}\in\mathbf{D}^1_1\,M$ is a tensorial object, and $\slash\!\!\!\omega$ is a shorthand for its components. The gamma matrices $\gamma^a$ are shorthands for $\gamma^a{}^A{}_B$, components of the abstract generators of the Clifford algebra $\ts{\gamma}\in\mathbf{T}^1_0\otimes\mathbf{D}^1_1\,M$. We also assume the implicit matrix multiplication denoted by juxtaposition of spinor matrices.}


The Clifford algebra is realized as operators on the Dirac bundle. It is generated by the abstract gamma matrices, i.e., by tensors $\ts{\gamma}$ with components $\gamma^a{}^A{}_B$ which satisfy
\begin{equation}\label{gammamtrx}
    \gamma^a\,\gamma^b + \gamma^b\,\gamma^a = 2\, g^{ab} I\;.
\end{equation}
Here, $I$ is the identity matrix and the implicit matrix multiplication is assumed. A general element of the Clifford algebra is a linear combination of products of gamma matrices with all spacetime indices contracted. Using the property \eqref{gammamtrx},
one can always eliminate symmetric products of gamma matrices and a general element of the Clifford algebra can be represented as
\begin{equation}\label{Clliford}
    \slash\!\!\!{\omega} = \sum_p \frac1{p!}\; {}^p\omega_{a_1\dots a_p}\gamma^{a_1\dots a_p}\;,
\end{equation}
where
\begin{equation}\label{multigamma}
   \gamma^{a_1\dots a_p} = \gamma^{[a_1}\dots\gamma^{a_p]}
\end{equation}
and ${}^p\omega_{a_1\dots a_p}$ are components of antisymmetric $p$-forms ${}^p\ts{\omega}$.
In other words, the Clifford objects are in one-to-one correspondence with inhomogeneous antisymmetric forms
\begin{equation}\label{nonhomAF}
    \ts{\omega}=\sum_p {}^p\ts{\omega}\;,
\end{equation}
through an isomorphisms $\gamma_*$,
\begin{equation}\label{CliffordExterior}
    \gamma_* \,:\quad
    \ts{\omega}\in\mathbf{\Lambda}\,M \quad \leftrightarrow \quad
    \slash\!\!\!\ts{\omega}\in\mathbf{D}^1_1\,M\;.
\end{equation}

This isomorphism induces also a new multiplication ``$\circ$'' on the exterior algebra $\mathbf{\Lambda}\,M$, the so called Clifford multiplication, which corresponds to the matrix multiplication in $\mathbf{D}^1_1\,M$,
\begin{equation}\label{CliffordProd}
    \gamma_* \,:\quad \ts{\alpha}\circ\ts{\beta} \quad\leftrightarrow\quad
    \slash\!\!\!\!\ts{\alpha} \; \slash\!\!\!\!\ts{\beta}\;.
\end{equation}
It can be shown \citep{BennTucker:book} that for homogeneous forms $\ts{\alpha}\in\mathbf{\Lambda}^p M$ and $\ts{\beta}\in\mathbf{\Lambda}^q M$, $p\le q$, the Clifford product reads
\begin{equation}\label{CliffordProdHom}
    \ts{\alpha}\circ\ts{\beta} =
    \sum_{m=0}^{p}\frac{(-1)^m(p{-}m{+}1)+[m/2]}{m!}\;\ts{\alpha}\underset{m}{\wedge}\ts{\beta}\;,
\end{equation}
where $\ts{\alpha}\underset{m}{\wedge}\ts{\beta}$ is $m$-times contracted wedge product introduced in \cite{Houri:2010qc},
\begin{equation}\label{ContrProd}
(\alpha\underset{m}{\wedge}\beta)_{a_1\dots a_{p{-}m}b_1\dots b_{q{-}m}}=
\frac{(p+q-2m)!}{(p-m)!(q-m)!}\,
\alpha_{c_1\dots c_m[a_1\dots a_{p{-}m}}\beta^{c_1\dots c_m}{}_{b_1\dots b_{q{-}m}]}\;.
\end{equation}
In particular, for a 1-form ${\ts{\alpha}}$ and a general ${p}$-form ${\ts{\omega}}$ one obtains
\begin{equation}\label{CliffordProd1form}
\begin{aligned}
    \ts{\alpha}\circ\ts{\omega} &= \ts{\alpha}\wedge\ts{\omega} + \ts{\alpha}\cdot\ts{\omega}\,,\\
    \ts{\omega}\circ\ts{\alpha} &= (-1)^p\, ( \ts{\alpha}\wedge\ts{\omega} - \ts{\alpha}\cdot\ts{\omega})\;.
\end{aligned}
\end{equation}
These relations in terms of gamma matrices (cf.\ \eqref{multigamma}) read
\begin{equation}\label{multigammaProd}
\begin{aligned}
    \gamma^a\, \gamma^{a_1\dots a_p} &= \gamma^{aa_1\dots a_p} + p\, g^{a[a_1}\gamma^{a_2\dots a_p]}\,,\\
    \gamma^{a_1\dots a_p}\, \gamma^a &=
      (-1)^p\,\bigl(\gamma^{aa_1\dots a_p} - p\, g^{a[a_1}\gamma^{a_2\dots a_p]}\bigr)\;.
\end{aligned}
\end{equation}

We shall also work with inhomogeneous forms \eqref{nonhomAF} and employ the operators $\pi$ and $\eta$ introduced in section~\ref{ssc:ExtCalc} by \eqref{pieta}. We say that an inhomogeneous form ${\ts{\omega}}$ is {\em even} if ${\eta\,\ts{\omega} = \ts{\omega}}$ and it is {\em odd} if ${\eta\,\ts{\omega} = -\ts{\omega}}$.

\subsubsection{{Invariant products on the spinor space}}

The universality of the Clifford algebra and the irreducibility of its spinor representation generated by abstract gamma matrices \eqref{gammamtrx} imply that the spinor space possesses two natural products \citep{Trautman:2008,LawsonMichelsohn:book,BennTucker:book}: the {\em Dirac scalar product} $\langle\ts\psi,\ts\ph\rangle$ (antilinear in the first and linear in the second argument) and the {\em real product} $(\ts\psi,\ts\ph)$ (linear in both arguments). The first is related to the (antilinear) Dirac conjugation $\dirconj{\ }:\mathbf{D}M\to \mathbf{D}^*M$ and the second to the (antilinear) charge conjugation ${\ }^\chrgconj:\mathbf{D}M\to \mathbf{D}M$ as
\begin{gather}
   \langle\ts\psi,\ts\ph\rangle = \dirconj\psi_A\, \ph^A\,,\label{dirconj}\\
   (\ts\psi^\chrgconj,\ts\ph) = \langle\ts\psi,\ts\ph\rangle\,.\label{chrgconj}
\end{gather}
The symmetry properties, positivity and relations to the abstract gamma matrices of these products depend on dimensionality and signature. In general one has
\begin{align}
   \langle\ts\gamma\ts\psi,\ts\ph\rangle &=
     \eps_{\rmlab{A}}\langle\ts\psi,\ts\gamma\ts\ph\rangle \,,&
   \dirconj{\ts\gamma} &= \eps_{\rmlab{A}}\,\ts\gamma \,,    \label{dirconjgamma}\\
   (\ts\gamma\ts\psi,\ts\ph) &= \eps_{\rmlab{B}}(\ts\psi,\ts\gamma\ts\ph) \,,&
   \ts\gamma^\chrgconj &= \eps_{\rmlab{C}}\,\ts\gamma \,,    \label{chrgconjgamma}
\end{align}
and
\begin{equation}
   \langle\ts\psi,\ts\ph\rangle = \langle\ts\ph,\ts\psi\rangle^*\,,\qquad
   (\ts\psi,\ts\ph) = \sigma_{\rmlab{B}}\, (\ts\ph,\ts\psi)\;,\qquad
   {\ts\psi}^{\chrgconj\chrgconj} = \sigma_{\rmlab{C}}\,\ts\psi\;,\label{symsigns}
\end{equation}
where all $\eps$'s and $\sigma$'s are just signs.

\details{
These products are related to the intertwiners between different representations of the Clifford algebra on the spinor spaces. Following the notation of \cite{Trautman:2008} (where one can understand Trautman's representation $\rho$ as our map $\ts\gamma_*$ generated by abstract gamma matrices $\ts\gamma$), one can introduce three intertwiners $A_{\bar{K}L}$, $B_{KL}$, and $C^{\bar{K}}{\!}_{L}$. $B_{KL}$ relates the representation on $\mathbf{D}M$ and $\mathbf{D}^*M$, while $C^{\bar{K}}{\!}_{L}$ relates the representation on $\mathbf{D}M$ and $\bar{\mathbf{D}}M$, and \begin{equation}\label{A=BC}
   A_{\bar{K}L}=\bar{B}_{\bar{K}\bar{N}}C^{\bar{N}}{\!}_{L}\,.
\end{equation}
Here $\bar{\ }: \mathbf{D}M\leftrightarrow\bar{\mathbf{D}}M$ is the conjugation between the spinor space and its conjugate.

The products and conjugations \eqref{dirconj} and \eqref{chrgconj} are then defined as follows:
\begin{align}
    \langle\ts\psi,\ts\ph\rangle &= \bar\psi^{\bar{K}}\,A_{\bar{K}L}\,\ph^{L}\,,&
    {\dirconj\psi}_K &= {\bar\psi}^{\bar{N}}A_{\bar{N}K}\,,\\
    (\ts\psi,\ts\ph) &= \psi^{K}\,B_{LK}\,\ph^{L}\,,&
    \psi^{\chrgconj K} &= \bar{C}^K{\!}_{\bar{N}}\,\bar{\psi}^{\bar{N}}\,.
\end{align}
These imply relations
\begin{equation}
    \dirconj{\ts\gamma} = \ts{A} \bar{\ts\gamma} \ts{A}^{-1}\,,\qquad
    {\ts\gamma}^\chrgconj = \bar{\ts{C}}\bar{\ts\gamma}\bar{\ts{C}}^{-1}\,,\qquad
    {\ts\psi}^{\chrgconj\chrgconj} = \bar{\ts{C}}\ts{C}\ts\psi\,.
\end{equation}
Using these definitions and relations one can read out $\eps$'s and $\sigma$'s signs from the properties (6)--(11) of \cite{Trautman:2008}.

The products and the conjugations are fixed by the spinor representation up to a freedom of one complex number at each spacetime point, see discussion in \cite{Trautman:2008}. We assume that this freedom is fixed and the products are chosen as a part of the definition of the spinor bundle.
}

\subsubsection{{Covariant derivative}}
The metric covariant derivative on the tangent space can be extended to spinors. It satisfies
\begin{equation}\label{DerGamma}
    \covd\ts{\gamma} = 0\;,
\end{equation}
and the invariant products are covariantly constant, i.e.,
\begin{equation}\label{DerProd}
\begin{gathered}
  \covd\langle\ts\psi,\ts\ph\rangle = \langle\covd\ts\psi,\ts\ph\rangle+\langle\ts\psi,\covd\ts\ph\rangle\,,\\
  \covd(\ts\psi,\ts\ph) = (\covd\ts\psi,\ts\ph)+(\ts\psi,\covd\ts\ph)\,.
\end{gathered}
\end{equation}
These conditions fix the extension of the covariant derivative on spinors unique\-ly, see \cite{Trautman:2008}.

Choosing a frame $\ts{e}_a$ in the tangent space and a frame $\ts{\vartheta}_A$ in the spinor space, such that the components $\gamma^a{}^A{}_B$ are constant,\footnote{%
This choice is usually done by choosing an orthonormal frame $\ts{e}_a$ in tangent space (but in the context of Lorentzian geometry a choice of a null frame is also common) and by specifying a particular form of the components of the gamma matrices $\gamma^a{}^A{}_B$. Different realizations of the gamma matrices which can be found in the literature can thus be understood as a different choice of the frame $\ts{\vartheta}_A$ associated with the spacetime frame $\ts{e}_a$. See \eqref{Thetafr} for a particular choice of the frame $\ts{\vartheta}_A$ in the Kerr--NUT--(A)dS spacetime.}
the covariant derivative on spinors and Clifford objects writes as
\begin{align}\label{CovDerSpinors}
    \nabla_{\!a}\psi &= \partial_a \psi + \frac14\omega_{abc}\;\gamma^b\,\gamma^c\,\psi\;,\\
    \nabla_{\!a}\,\slash\!\!\!\omega &= \partial_a \slash\!\!\!\omega
      + \Bigl[\,\frac14\omega_{abc}\;\gamma^b\,\gamma^c,\,\slash\!\!\!\omega\,\Bigr]\;,
    \label{CovDerClifford}
\end{align}
with the standard spin coefficients $\omega_a{}^b{}_c$ given by
\begin{equation}\label{spincoef}
    \covd_{\!a} \ts{e}_b = \omega_a{}^c{}_b\; \ts{e}_c\;,
\end{equation}
and $[\,\slash\!\!\!\sigma,\,\slash\!\!\!\omega\,]=\slash\!\!\!\sigma\,\slash\!\!\!\omega
  -\slash\!\!\!\omega\,\slash\!\!\!\sigma$ being the commutator.
If one introduces 1-forms of the curvature as $\ts{\omega}^b{}_c = \ts{e}^a \omega_{a}{}^b{}_c$, these satisfy the Cartan equations
\be
\grad\ts{e}^a + \ts{\omega}^a{}_b \wedge \ts{e}^b=0\,.
\ee
Here $\ts{e}^a$ is the frame of 1-forms dual to $\ts{e}_a$.

\subsubsection{{Dirac operator}}

The Dirac operator operator on spinors is defined by ${\sop{D}\ts{\psi} = \slash\!\!\!\!\covd\ts{\psi}}$, that is,
\begin{equation}\label{DirOpSpin}
    \sop{D}\psi = \gamma^a\nabla_{\!a}\psi\;.
\end{equation}
When applied on a Clifford object ${\slash\!\!\!\ts{\omega}}$, its action ${\sop{D}\,\slash\!\!\!\ts{\omega}=\slash\!\!\!\!\covd\slash\!\!\!\ts{\omega}}$ can be translated to the action on the corresponding antisymmetric form ${\ts{\omega}}$. We denote it by the same symbol ${\sop{D}}$. Taking into account the definition of the Clifford multiplication \eqref{CliffordProd} and its relation to the wedge and the dot operations \eqref{CliffordProd1form}, we can write the Dirac operator on the exterior algebra $\mathbf{\Lambda}M$ as
\begin{equation}\label{DirOpAF}
    \sop{D}\ts{\omega} = \covd\circ\ts{\omega} = \covd\wedge\ts{\omega}+\covd\cdot\ts{\omega}\;.
\end{equation}

\subsection{{Symmetry operators of the Dirac operator}}
\label{ssc:DiracSymOp}

For solving the Dirac equation and for a discussion of its properties it is useful to know the symmetry operators of the Dirac operator,  and in particular those that commute with ${\sop{D}}$. A general first-order symmetry operator ${\sop{S}}$ satisfying ${\sop{D} \sop{S} = \sop{R} \sop{D}}$ for some ${\sop{R}}$ has been constructed in \cite{Benn:1996ia, BennKress:2004}. It is in one-to-one correspondence with conformal Killing--Yano tensors. Based on this result it is possible to characterize the operators that commute with ${\sop{D}}$---they are related to the Killing--Yano and closed conformal Killing--Yano forms \citep{Cariglia:2011yt}.

\subsubsection{{First-order operators that commute with the Dirac operator}}

Using the exterior algebra representation, the most general first-order operator ${\sop{S}}$ that commutes with the Dirac operator, ${[\sop{D},\sop{S}]=0}$, takes the following form:
\begin{equation}\label{FODiraccomop}
    \sop{S} = \sop{K}_{\ts{f}} + \sop{M}_{\ts{h}}\;,
\end{equation}
\begin{align}
   \sop{K}_{\ts{f}} &= \ts{f}\cdot\covd \,+\, \Bigl(\frac{\pi-1}{2\pi}\covd\wedge\ts{f}\Bigr)\,,
     \label{FODiraccomopK}\\
   \sop{M}_{\ts{h}} &= \ts{h}\wedge\covd \;+\; \Bigr(\frac{D-\pi-1}{2(D-\pi)}\covd\cdot\ts{h}\Bigr)\,,
     \label{FODiraccomopM}
\end{align}
where ${\ts{f}}$ is an (in general inhomogeneous) odd Killing--Yano form and ${\ts{h}}$ is an (inhomogeneous) even closed conformal Killing--Yano form. 

\details{
To illuminate the condense notation for the operators, we write explicitly the action of the operators ${\sop{K}_{\ts{f}}}$ and ${\sop{M}_{\ts{h}}}$ on a form ${\ts{\omega}}$ in the case when ${\ts{f}}$ and ${\ts{h}}$ is a homogeneous odd ${p}$-form and a homogeneous even ${q}$-form, respectively,
\begin{align}
   \sop{K}_{\ts{f}}\,\ts{\omega} &= [\ts{f}\cdot\covd]\circ\ts{\omega} \,+\, \frac{p}{2(p+1)}(\covd\wedge\ts{f})\circ\ts{\omega}\,,
     \label{FODiraccomopKom}\\
   \sop{M}_{\ts{h}}\,\ts{\omega} &= [\ts{h}\wedge\covd]\circ\ts{\omega} \,+\, \frac{D-q}{2(D-q+1)}(\covd\cdot\ts{h})\circ\ts{\omega}\,.
     \label{FODiraccomopMom}
\end{align}
The derivatives ${[\ts{f}\cdot\covd]}$ and ${[\ts{h}\wedge\covd]}$ enter into the Clifford multiplication as forms with indices
${[-f\cdot\nabla]_{a_2\dots a_p}} = {f_{a_2\dots a_p}{}^{a}\nabla_{\!a} = f^{a}{}_{a_2\dots a_p}\nabla_{\!a}}$ (${p}$ odd) and
${[h\wedge\nabla]_{a_0a_1\dots a_q}}={(q+1)h_{[a_1\dots a_q}\nabla_{\!a_0]}}$ (${q}$ even).
}\medskip

Writing these operators directly on the Dirac bundle, one gets
\begin{align}
  \sop{K}_{\ts{f}}&=\sum_{\text{${p}$ odd}}
    \Bigl[\frac1{(p{-}1)!}\,\gamma^{a_1\dots a_{p{-}1}}\;{}^p\!f^a_{\ a_1\dots a_{p{-}1}}\nabla_a
    +\frac{p}{2(p{+}1)!}\,\gamma^{a_1\dots a_{p{+}1}}\;{}^p\!\kappa_{a_1\dots a_{p{+}1}}\Bigr]\,,\\
  \sop{M}_{\ts{h}}&=\sum_{\text{${p}$ even}}
    \Bigl[\frac1{p\,!}\,\gamma^{aa_1\dots a_{p}}\;{}^ph_{a_1\dots a_{p}}\nabla_a
    +\frac{D-p}{2(p{-}1)!}\,\gamma^{a_1\dots a_{p{-}1}}\;{}^p\xi_{a_1\dots a_{p{-}1}}\Bigr]\,.
\end{align}
Here, the inhomogeneous forms are written as
\begin{equation}\label{fhnonhom}
    \ts{f} = \sum_{\text{${p}$ odd}} {}^p\!\ts{f}\;,\qquad
    \ts{h} = \sum_{\text{${p}$ even}} {}^p\ts{h}\;,
\end{equation}
where $p$-forms ${}^p\!\ts{f}$ are odd Killing--Yano tensors, and ${}^p\ts{h}$ are even closed conformal Killing--Yano tensors, and satisfy
\begin{align}
  \nabla_{\!a}\, {}^p\!f_{a_1\dots a_p} &= {}^p\!\kappa_{aa_1\dots a_p}\,,&
    &{}^p\!\kappa_{a_0a_1\dots a_p} = \nabla_{\![a_0}\, {}^p\!f_{a_1\dots a_p]}\,,
    \label{pfKYkappa}\\
  \nabla_{\!a}\; {}^ph_{a_1\dots a_p} &= p\,g_{a[a_1}\,{}^p\xi_{a_2\dots a_p]}\,,&
    &{}^p\xi_{a_2\dots a_p} = \frac1{D-p+1}\nabla_{\!c}\;{}^ph^c{}_{a_2\dots a_p}\;.
    \label{phCCKYxi}
\end{align}

To summarize, the operators commuting with the Dirac operator are in one-to-one correspondence with odd Killing--Yano forms and even closed conformal Killing--Yano forms.

\subsubsection{Killing--Yano bracket}

The conditions when the above operators also commute among each other have been studied in \cite{Cariglia:2011yt}. Having two operators ${\sop{S}_1}$, ${\sop{S}_2}$ of the type \eqref{FODiraccomopK} or \eqref{FODiraccomopM}, their commutator is of the first-order only if the corresponding Killing--Yano and closed conformal Killing--Yano forms satisfy certain algebraic conditions. Provided these conditions are satisfied, the commutator ${[\sop{S}_1,\sop{S}_2]}$ is again an operator that commutes with the Dirac operator and thus it has the form \eqref{FODiraccomop}. This fact can be exploited to define a new operation on the Killing--Yano and closed conformal Killing--Yano tensors, called the {\em Killing--Yano bracket} \citep{Cariglia:2011yt}.

To illustrate the action of Killing--Yano brackets, let us consider two odd Killing--Yano forms ${\ts{\kappa}}$ and ${\ts{\lambda}}$. The Killing--Yano bracket is defined by the requirement that
\begin{equation}\label{KYbracket}
    [\sop{K}_{\ts{\kappa}},\,\sop{K}_{\ts{\lambda}}] = \sop{K}_{[\ts{\kappa},\ts{\lambda}]_\KYbr}\;,
\end{equation}
which is true provided the following necessary algebraic conditions are satisfied:
\begin{equation}\label{algcondforKY}
    \sum_{k=1,\dots} \frac{(-1)^k}{(2k-1)!} \ts{\kappa}\underset{2k}{\wedge}\ts{\lambda}=0\,,
\end{equation}
in which case the bracket can be explicitly written as \citep{Cariglia:2011yt}
\begin{equation}\label{KYbracketpot}
    [\ts{\kappa},\ts{\lambda}]_\KYbr = \frac1{\pi}\covd\cdot\!\!\!\sum_{k=0,\dots}
      \frac{(-1)^k}{(2k+1)!} (\pi\ts{\kappa})\underset{2k}{\wedge}(\pi\ts{\lambda})\,.
\end{equation}
In particular, for homogeneous forms of rank ${p}$ and ${q}$, respectively, the consistency conditions splits into a set of conditions
\begin{equation}\label{algcondforKYhom}
    \ts{\kappa}\underset{2k}{\wedge}\ts{\lambda}=0\qquad\text{for ${k=1,2,\dots}$}\;,
\end{equation}
and the Killing--Yano bracket simplifies to
\begin{equation}\label{KYbracketpothom}
    [\ts{\kappa},\ts{\lambda}]_\KYbr = \frac{p\,q}{p+q-1}
      \covd\cdot(\ts{\kappa}\wedge\ts{\lambda})\,.
\end{equation}
As discussed in the main text, the algebraic conditions \eqref{algcondforKYhom} (and the analogous conditions for the closed conformal--Killing--Yano tensors) are automatically satisfied for the Killing--Yano tower of hidden symmetries generated from the principal tensor of the Kerr--NUT--(A)dS geometry.

\subsection{Killing--Yano tensors and Killing spinors}
\label{ssc:KillingSpinors}

In this section we review the connections between Killing--Yano tensors and various Killing spinors, based on works of \cite{semmelmann2003conformal}, \cite{Cariglia:2003kf}, and \cite{Houri:2012eq}.

\subsubsection{Conformal Killing spinors}

To motivate the definition of conformal Killing spinors, let us recall that in section~\ref{ssc:KYfamily} we have introduced conformal Killing--Yano forms as those annihilated by the twistor operator \eqref{twistop} whose definition is based on splitting the space ${\mathbf{T}^*\!\otimes\mathbf{\Lambda}\,M}$ into the subspaces given by the projector $\mathcal{A}+\mathcal{C}$ and the projector $\mathcal{T}$, cf. \eqref{DECOMP}. The covariant derivative $\covd\ts{\omega}$ of an antisymmetric form $\ts{\omega}$ can be thus split into the antisymmetric plus divergence parts and the part given by the twistor operator. If the twistor operator $\twist\ts{\omega}$ yields zero, $\ts{\omega}$ is the conformal Killing--Yano form.

Similarly, a 1-form spinor ${\ts\alpha\in\mathbf{T}^*\!\otimes\mathbf{D}\,M}$ can be split into the parts given by the projector $\mathcal{B}$ and the projector $\mathcal{T}$,
\begin{align}\label{spinDECOMP}
    (\mathcal{B}\alpha)_{a} &= \frac1D \gamma_a \gamma^n \alpha_n\,, \\
    (\mathcal{T}\alpha)_{a} &= \alpha_a - \frac1D \gamma_a \gamma^n \alpha_n\,.
\end{align}
Obviously, $\mathcal{B}$ and $\mathcal{T}$ are complementary projectors as can be seen by using $\gamma^a\mathcal{B}\alpha_a = \gamma^a\alpha_a$ and $\gamma^a\mathcal{T}\alpha_a=0$. Applying these projectors on the covariant derivative $\covd\ts\psi$, one gets
\begin{equation}\label{covdpsisplit}
    \covd\ts\psi = \mathcal{B}\covd\ts\psi + \mathcal{T}\covd\ts\psi
      = \frac1D\, \ts\gamma\, \sop{D}\ts\psi + \twist\ts\psi\,,
\end{equation}
where the (spinorial) twistor operator $\twist\ts\psi$ reads
\begin{equation}\label{spintwistop}
    \twistc_{\!a} \psi = \mathcal{T}\nabla_{\!a}\psi
     = \nabla_{\!a}\psi - \frac1D\, \gamma_a \sop{D}\psi\;.
\end{equation}
Spinors for which the twistor operator vanishes are called \emph{twistor spinors} or \emph{conformal Killing spinors}. They satisfy
\begin{equation}\label{CKspinors}
    \covd\ts\psi = \frac1D\, \ts\gamma\,\sop{D}\ts\psi
    \,.
\end{equation}

\subsubsection{Relation between conformal Killing spinors and conformal Killing--Yano tensors}

There is a natural connection between the twistor spinors and the conformal Killing--Yano tensors:

\medskip
\noindent{\bf Theorem}:
{\it
Let ${\ts\psi_1}$ and ${\ts\psi_2}$ be two twistor spinors. Then the following $p$-forms, defined using the invariant product \eqref{dirconj} for any $p=0,\dots,D$:
\be\label{CKYfromCKspin}
\omega_{a_1\dots a_p} = \langle\psi_1, \gamma_{a_1\dots a_p}\psi_2\rangle\,,
\ee
are the conformal Killing--Yano tensors.
}\medskip

To prove the theorem, we compute the covariant derivative $\covd\ts\omega$. Employing \eqref{DerProd}, \eqref{DerGamma} with \eqref{multigamma} and the assumption that ${\ts\psi_1}$ and ${\ts\psi_2}$ satisfy \eqref{CKspinors}, we obtain
\begin{equation}\label{proofCKYCKspin1}
\begin{split}
  \nabla_{\!a}\omega_{a_1\dots a_p}
    &= \langle\nabla_{\!a}\psi_1, \gamma_{a_1\dots a_p}\psi_2\rangle
      + \langle\psi_1, \gamma_{a_1\dots a_p}\nabla_{\!a}\psi_2\rangle\\
    &= \frac1D\langle\gamma_{a}\sop{D}\psi_1, \gamma_{a_1\dots a_p}\psi_2\rangle
      + \frac1D\langle\psi_1, \gamma_{a_1\dots a_p}\gamma_{a}\sop{D}\psi_2\rangle
\end{split}
\end{equation}
Applying \eqref{dirconjgamma} in the first term and using \eqref{multigammaProd}, we get
\begin{equation}\label{proofCKYCKspin2}
\begin{split}
  \nabla_{\!a}\omega_{a_1\dots a_p}
    &= \frac1D\Bigl( \eps_{\rmlab{A}}\langle \sop{D}\psi_1, \gamma_{aa_1\dots a_p}\psi_2\rangle
      + (-1)^p\langle\psi_1, \gamma_{aa_1\dots a_p}\sop{D}\psi_2\rangle\Bigr)\\
    &+\frac{p}{D}\,g_{a[a_1}
      \Bigl(\eps_{\rmlab{A}}\langle \sop{D}\psi_1, \gamma_{a_2\dots a_p]}\psi_2\rangle
      - (-1)^p\langle\psi_1, \gamma_{a_2\dots a_p]}\sop{D}\psi_2\rangle\Bigr)\,.
\end{split}
\end{equation}
Clearly, the derivative $\covd\ts\omega$ has the form \eqref{CKYeqkapxiidx} and $\ts\omega$ is thus a conformal Killing--Yano form.

Recalling \eqref{kappaxidefidx}, we can also read out
\begin{equation}
\begin{gathered}
   (\nabla\wedge \omega)_{aa_1\dots a_p}
   = \frac{p+1}D\Bigl( \eps_{\rmlab{A}}\langle \sop{D}\psi_1, \gamma_{aa_1\dots a_p}\psi_2\rangle
      + (-1)^p\langle\psi_1, \gamma_{aa_1\dots a_p}\sop{D}\psi_2\rangle\Bigr)\,,\\
   (\nabla\cdot \omega)_{a_2\dots a_p}
   =\frac{D{-}p{+}1}{D}\,
      \Bigl(\eps_{\rmlab{A}}\langle \sop{D}\psi_1, \gamma_{a_2\dots a_p}\psi_2\rangle
      - (-1)^p\langle\psi_1, \gamma_{a_2\dots a_p}\sop{D}\psi_2\rangle\Bigr)\,.
\end{gathered}
\end{equation}

\subsubsection{Killing spinors}

We call a spinor $\ts\psi$ obeying
\be\label{Kspinors}
\covd \ts\psi = \frac{\mu}{D}\, \ts\gamma\, \ts\psi
\ee
for some $\mu\in \mathbb{C}$ a \emph{Killing spinor}.

If a spinor $\ts\psi$ is both the twistor spinor and the Killing spinor, it follows that it satisfies the massive Dirac equation
\begin{equation}\label{muDiracEq}
    \sop{D}\ts\psi = \mu\,\ts\psi\,.
\end{equation}
The relation is symmetric, any two of the conditions \eqref{CKspinors}, \eqref{Kspinors}, and \eqref{muDiracEq} imply the third.

\subsubsection{Special conformal Killing--Yano tensors}

Let as define a {\em special Killing--Yano tensor $\ts\omega$} to be a $p$-form which obeys
\be\label{SpecKY}
\covd_{\!\ts{X}} \ts\omega=\frac{1}{p+1}\ts{X}\cdot (\covd\wedge \omega)\,,\qquad
\covd_{\!\ts{X}}(\covd\wedge\ts\omega)=c\, \ts{X} \wedge \ts\omega\,,
\ee
for any vector field $\ts{X}$ and some constant $c$. Obviously, the special Killing--Yano tensors are subclass of Killing--Yano tensors. They have been introduced by \cite{TachibanaYu:1970} and exploited by \cite{semmelmann2003conformal}.

Substituting Killing--Yano equation into the second condition of \eqref{SpecKY}, we obtain
\begin{equation}\label{SpecKYalt}
    \nabla_{\!a}\nabla_{\!a_0}\omega_{a_1\dots a_p} = c\, g_{a[a_0}\omega_{a_1\dots a_p]}\;.
\end{equation}
Using \eqref{SpecKY} we also find that $\ts\omega$ is an eigenform of the de~Rham--Laplace operator defined in \eqref{DeRhamLaplace}, ${\Delta\ts\omega = - \covd\wedge(\covd\cdot\ts\omega)-\covd\cdot(\covd\wedge\ts\omega)}$, \be
  \Delta\ts\omega= - c(D-p)\,\ts\omega\,.
\ee
Moreover, when $\ts\omega$ is an odd-rank special Killing--Yano tensor, so is
\be
\ts\omega\wedge (\covd\wedge\ts\omega)^{\wedge k}\,,
\ee
for any $k=0,1,\dots$.

Similarly, one can define a {\em special closed conformal Killing--Yano tensor} $\ts\omega$ to be a $p$-form obeying
\be\label{SpecCCKY}
\covd_{\!\ts{X}} \ts\omega = \frac{1}{D{-}p{+}1}\ts{X}\wedge (\covd\cdot\ts\omega)\,,\qquad
\covd_{\!\ts{X}}(\covd\cdot\ts\omega)= c\, \ts{X} \cdot \ts\omega\,,
\ee
for any vector field $\ts{X}$ and some constant $c$.

Again, such $\ts\omega$ is an eigenform of the de~Rham--Laplace operator,
\begin{equation}
 \Delta\omega=- c p\, \ts\omega\,.
\end{equation}

\subsubsection{Relation between Killing spinors and special Killing--Yano tensors}

Similar to twistor spinors, the Killing spinors also generate conformal Killing--Yano forms:

\medskip
\noindent{\bf Theorem}:
{\it
Let ${\ts\psi_1}$ and ${\ts\psi_2}$ be two Killing spinors,
\be\label{SpecKspin12}
\covd \ts\psi_1 = \frac{\mu_1}{D}\, \ts\gamma\, \ts\psi_1\,,\qquad
\covd \ts\psi_2 = \frac{\mu_2}{D}\, \ts\gamma\, \ts\psi_2\,.
\ee
Then the following $p$-forms:
\be\label{CKYfromKspin}
\omega^{(p)}_{a_1\dots a_p} = \langle\psi_1, \gamma_{a_1\dots a_p}\psi_2\rangle\,,
\ee
$p=0,\dots,D$, are conformal Killing--Yano forms.
}\medskip

The same reasoning as in \eqref{proofCKYCKspin1} and \eqref{proofCKYCKspin2}, just using \eqref{SpecKspin12} instead \eqref{CKspinors}, yields
\begin{equation}\label{proofCKYKspin1}
\begin{split}
  \nabla_{\!a}\omega^{(p)}_{a_1\dots a_p}
    &= \frac1D\bigl( \eps_{\rmlab{A}}\mu_1+(-1)^p\mu_2\bigr)\,
       \langle \psi_1, \gamma_{aa_1\dots a_p}\psi_2\rangle\\
    &\quad+\frac{p}{D}\bigl(\eps_{\rmlab{A}}\mu_1-(-1)^p\mu_2\bigr)\,
       g_{a[a_1}\, \langle \psi_1, \gamma_{a_2\dots a_p]}\psi_2\rangle\\
    &= \frac{\mu^{(p)}_+}D\,\omega^{(p{+}1)}_{aa_1\dots a_p}
    +\frac{\mu^{(p)}_-}{D}\,p\,g_{a[a_1}\, \omega^{(p{-}1)}_{a_2\dots a_p]}\,,
\end{split}\raisetag{4ex}
\end{equation}
i.e.,
\begin{equation}\label{proofCKYKspin2}
   \covd_{\!\ts{X}}\ts\omega^{(p)}
     = \frac{\mu^{(p)}_+}D\,\ts{X}\cdot\ts\omega^{(p{+}1)}
     + \frac{\mu^{(p)}_-}D\,\ts{X}\wedge\ts\omega^{(p{-}1)}\,,
\end{equation}
where we have set
\begin{equation}\label{mup+-}
    \mu^{(p)}_\pm = \eps_{\rmlab{A}}\bar\mu_1 \pm (-1)^p\mu_2\,.
\end{equation}
We see that the derivative $\covd\ts\omega^{(p)}$ has the form \eqref{CKYeq2} and $\ts\omega$ is thus a conformal Killing--Yano form, which concludes the proof of the theorem.

From \eqref{kappaxidef} we obtain
\begin{equation}\label{covdomp}
\begin{gathered}
  \covd\wedge\ts\omega^{(p)} = (p+1)\frac{\mu^{(p)}_+}D\,\ts\omega^{(p{+}1)}\,,\\
  \covd\cdot\ts\omega^{(p)} = (D-p+1)\frac{\mu^{(p)}_-}D\,\ts\omega^{(p{-}1)}\,.
\end{gathered}
\end{equation}
Taking the covariant derivative of these expressions while applying \eqref{proofCKYKspin2} and ${\mu^{(p)}_\pm=\mu^{(p{+}1)}_\mp=\mu^{(p{-}1)}_\mp}$, we obtain
\begin{equation}\label{covdcovdomp}
\begin{gathered}
  \covd_{\!\ts{X}}(\covd\wedge\ts\omega^{(p)})
    =\frac{p+1}{D^2} \mu^{(p)}_+\mu^{(p)}_- \ts{X}\cdot\ts\omega^{(p{+}2)}
    +\frac{p+1}{D^2} (\mu^{(p)}_+)^2 \ts{X}\wedge \ts\omega^{(p)}\,,\\
  \covd_{\!\ts{X}}(\covd\cdot\ts\omega^{(p)})
    =\frac{D{-}p{+}1}{D^2} \mu^{(p)}_-\mu^{(p)}_+ \ts{X}\wedge\ts\omega^{(p{-}2)}
    +\frac{D{-}p{+}1}{D^2} (\mu^{(p)}_-)^2 \ts{X}\cdot \ts\omega^{(p)}\,.
\end{gathered}
\end{equation}
Inspecting \eqref{covdomp} and \eqref{covdcovdomp}, we can formulate:

\medskip
\noindent{\bf Theorem}:
{\it
Under assumptions of the previous theorem we recognize two special cases:
\begin{align}
  &\mu^{(p)}_-=0\;\Rightarrow
    &&\text{$\ts\omega^{(p)}$ is a special Killing--Yano tensor with}\nonumber\\
    &&&\qquad\quad c=\frac{p+1}{D}(\mu^{(p)}_+)^2\,,\\
  &\mu^{(p)}_+=0\;\Rightarrow
    &&\text{$\ts\omega^{(p)}$ is a special closed conformal Killing--Yano tensor with}\nonumber\\
    &&&\qquad\quad c=\frac{D{-}p{+}1}{D}(\mu^{(p)}_-)^2\,.
\end{align}
}\medskip

\noindent
If we assume $\mu_1=\mu_2=\mu$, the conditions $\mu^{(p)}_\pm=0$ require that $\mu$ must be either real or imaginary, depending on the dimension, signature, and order $p$, cf.~\eqref{mup+-}.

Let us finally note that all the results of this section can be straightforwardly generalized to the case of a covariant derivative with torsion. The torsion generalized twistor/Killing spinors, which find applications in various supergravity theories, are then related to the torsion generalized conformal Killing--Yano tensors discussed in section~\ref{ssc:GenCCKY}. We refer the interested reader to appendix~A in \cite{Houri:2012eq} for more details.